\documentclass[usenatbib,useAMS,usegraphicx,a4paper]{mn2e}

\pdfoutput=1

\usepackage{graphicx}
\usepackage{footmisc}
\DefineFNsymbols*{mn2e}{{\mbox{${^{\star}}$}}\dagger\ddagger\S\P\|{\mbox{$^{\star\star}$}}{\dagger\dagger}
   {\ddagger\ddagger}{\S\S}{\P\P}{\|\|}}
\setfnsymbol{mn2e}
\usepackage{placeins}
\usepackage{lscape}
\usepackage{txfonts}
\usepackage{color}
\usepackage{mathrsfs}
\usepackage{pstricks}

\title[The XMM-LSS Class 1 sample over the initial 5~\dd]{
   \vspace{-0.6cm}The XMM-LSS survey: the Class 1 cluster sample over the initial 5~deg$^2$ 
   and its cosmological modelling\thanks{Based on data collected with XMM, VLT, Magellan, 
   NTT, and CFH telescopes; ESO programme numbers are: 070.A-0283, 070.A-907 (VVDS), 
   072.A-0104, 072.A-0312, 074.A-0360, 074.A-0476}\vspace{-0.2cm}}
\author[F.~Pacaud et al.]{F.~Pacaud$^{1,2,3}$\thanks{E-mail:pacaud@astro.uni-bonn.de},
M.~Pierre$^{1,2}$, C.~Adami$^4$, B.~Altieri$^5$,
S.~Andreon$^6$, L.~Chiappetti$^7$,
\newauthor A.~Detal$^8$, P.-A.~Duc$^2$, G.~Galaz$^9$, A.~Gueguen$^2$,
J.-P. Le F\`evre$^{10}$, G. Hertling$^9$,
\newauthor C.~Libbrecht$^8$, J.-B. Melin$^{11}$, T.~J.~Ponman$^{12}$, H.~Quintana$^9$,
A.~Refregier$^{1,2}$,
\newauthor P.-G.~Sprimont$^8$, J.~Surdej$^8$, I.~Valtchanov$^5$, J.~P.~Willis$^{13}$, D.~Alloin$^{2}$,
M.~Birkinshaw$^{14}$,
\newauthor M.~N.~Bremer$^{14}$, O.~Garcet$^8$, C.~Jean$^8$, L.~R.~Jones$^{12}$, O.~Le~F\`evre$^4$,
D.~Maccagni$^7$,
\newauthor A.~Mazure$^4$, D. Proust$^{15}$, H.~J.~A.~R\"ottgering$^{16}$, G.~Trinchieri$^6$\vspace{0.15cm}\\
$^1$DSM/DAPNIA/SAp, CEA Saclay, F-91191 Gif-sur-Yvette, France.\\
$^2$AIM - Unit\'e Mixte de Recherche CEA - CNRS - Universit\'e Paris VII - UMR  715\\
$^3$Argelander-Institut f\"ur Astronomie, University of Bonn, Auf dem H\"ugel 71, 53121 Bonn, Germany\\
$^4$Laboratoire d'Astrophysique de Marseille, BP8, F-13376 Marseille Cedex 12, France.\\
$^5$ESA, Villafranca del Castillo, Spain.\\
$^6$INAF-Osservatorio Astronomico di Brera, via Brera 28, I-20121 Milano, Italy.\\
$^7$INAF-IASF Milano, via Bassini 15, I-20133 Milano, Italy.\\
$^8$Institut d'Astrophysique et de G\'eophysique, Universit\'e de Li\`ege, All\'ee du 6 Ao\^ut, 17, B5C, 4000 Sart Tilman, Belgium.\\
$^9$Departamento de Astronom\'ia y Astrof\'isica, Pontificia Universidad Cat\'olica de Chile, Casilla 306, Santiago 22, Chile.\\
$^{10}$DSM/DAPNIA/SEDI, CEA Saclay, F-91191 Gif-sur-Yvette, France.\\
$^{11}$DSM/DAPNIA/SPP, CEA Saclay, F-91191 Gif-sur-Yvette, France.\\
$^{12}$School of Physics and Astronomy, University of Birmingham, Edgbaston, Birmingham, B15 2TT, UK.\\
$^{13}$Department of Physics and Astronomy, University of Victoria, Elliot Building, 3800 Finnerty Road, Victoria, V8V 1A1, BC, Canada.\\
$^{14}$Department of Physics, University of Bristol, Tyndall Avenue, Bristol BS8 1TL,UK.\\
$^{15}$GEPI, Observatoire de Paris-Meudon, F-92195 Meudon CEDEX, France.\\
$^{16}$Leiden Observatory, P.O. Box 9513, 2300 RA Leiden, The Netherlands.\vspace{-0.35cm}}

\begin{document}

\date{Accepted 2007 September 12. Received 2007 September 12; in original form 2007 March 7\vspace{-0.35cm}}

\pagerange{\pageref{firstpage}--\pageref{lastpage}} \pubyear{2007}

\newcommand{\dd}{deg$^{2}$}
\newcommand{\flux}{~$\rm erg\,s^{-1}cm^{-2}$}
\newcommand{\lum}{~$\rm erg\,s^{-1}$}
\newcommand{\countr}{$\rm cts\,s^{-1}$}

\maketitle

\label{firstpage}

\begin{abstract}
We present a sample of 29 galaxy clusters from the XMM-LSS survey
over an area of some 5 \dd\ out to a redshift of $z=1.05$. The
sample clusters, which represent about half of the X-ray clusters 
identified in the region, follow well defined X-ray selection criteria and
are all spectroscopically confirmed. For all clusters, we provide
X-ray luminosities and temperatures as well as masses, obtained
from dedicated spatial and spectral fitting.  The cluster
distribution peaks around $z=0.3$ and $T$=1.5~keV, half of the
objects being groups with a temperature below 2~keV. Our
$L_X-T(z)$ relation points toward self-similar evolution, but does
not exclude other physically plausible models. Assuming 
that cluster scaling laws follow self-similar evolution, our
number density estimates up to $z$=1 are compatible with the 
predictions of the concordance cosmology and with the findings 
of previous ROSAT surveys. Our well monitored selection function 
allowed us to demonstrate that the inclusion of selection effects is 
essential for the correct determination of the evolution of the 
$L_X-T$ relation, which may explain the contradictory results from 
previous studies.
Extensive simulations show that extending the survey area to
10 \dd\ has the potential to exclude the non-evolution hypothesis,
but that constraints on more refined ICM models will probably
be limited by the large intrinsic dispersion of the $L_X-T$ 
relation, whatever the sample size. We further demonstrate that
increasing the dispersion in the scaling laws increases the number
of detectable clusters, hence generating further degeneracy [in
addition to $\sigma_8, \Omega_m, L_X-T(z)$] in the cosmological
interpretation of the cluster number counts. We provide useful
empirical formulae for the cluster mass$\,-\,$flux and 
mass$\,-\,$count-rate relations as well as a comparison between 
the XMM-LSS mass sensitivity  and that of forthcoming SZ surveys.
\end{abstract}

\begin{keywords}
surveys - X-rays: galaxies: clusters - large-scale structure of Universe
- cosmological parameters\vspace{-1.35cm}
\end{keywords}

\section{Introduction}

Along with Cosmic Microwave Background (CMB) measurements and
supernova (SN) observations, clusters of galaxies provide key
cosmological information. It is especially instructive to
cross-check the constraints from these three classes of data, since 
they originate from different physical processes. Moreover, since
theory and numerical simulations allow us to follow cluster
formation from the initial power spectrum, which is directly
measured from the CMB, it is critical to test that the ``CMB
WMAP concordance cosmology'' is consistent with the observed
properties of clusters in the low-$z$ Universe.

In the framework of hierarchical cosmic structure formation
involving Cold Dark Matter type scenarios (CDM), where the
smallest perturbations collapse first, clusters correspond to the
mass scale that entered the non-linear regime between
redshift three and the present epoch. In this sense, the most
massive galaxy clusters in the local Universe represent the
largest virialised structures. This property of being
both ``relaxed'' and rare $3\sigma$ events regime has been
extensively exploited through formalisms like that of Press \&
Schechter (1974) in connection with the spherical collapse model,
for the general case of Gaussian random field fluctuations. This
connects, in an analytically tractable manner and for any
redshift, both the cluster abundance as a function of mass and 
the cluster spatial distribution, to the properties of the initial
fluctuation spectrum -- in particular its normalisation, $\sigma_{8}$, its shape,
$\Gamma$, as a function of the density of the Universe,
$\Omega_{m}$, and the equation of state of Dark Energy (e.g.,
\citealt{refregier02}; \citealt{majumdar03}). This first-order
approach is well supported by numerical CDM simulations: clusters
lie at nodes of the cosmic network, have virialized cores, and are
still growing by accretion along filaments at a rate that depends
on the cosmology.  However, at the same time as clusters started
being used as cosmological tools, it was realized that the
interpretation of their observed abundance as a function of time
is actually very much dependent on the evolution of the observable
cluster properties themselves. In order to break 
this latent degeneracy in such a way that clusters can effectively 
be used as cosmological candles, it is essential to understand 
how cluster properties impact on their delectability at any epoch.

While galaxies constitute only a few per cent of the total cluster
mass, about 80\% of the baryonic mass resides in the tenuous X-ray
emitting intra-cluster medium (ICM), settled in the cluster
gravitational potential.
Because cluster X-ray emission is extended, clusters are readily 
identified among the high-galactic latitude X-ray population, which 
is dominated by point-like AGN. 
However, whatever the detection method -- optical
or X-ray -- the fundamental question of {\em how to relate
observable quantities to cluster masses} remains. This is
crucial because it is the cluster masses that enter the theory of
structure formation as generated by theoretical cosmological 
calculations. This issue
becomes particularly important outside the local Universe, since
the evolution of cluster properties is not well established.

The Einstein Medium Sensitivity Survey (EMSS), followed by REFLEX
based on the ROSAT All-Sky Survey, and a number of serendipitous
clusters surveys (RDCS, SHARC, MACS, $160deg^2$ etc) from ROSAT deep
pointings, have provided the first `cosmological cluster
samples' (see the synoptic plot in \citet{xmmlss} and references therein).
The mass-observable relations used in the analysis of data from these
surveys relied on the assumption of hydrostatic equilibrium and on 
the (mostly local) observed $L_X-T$ relation. 
The cluster selection function -- a key ingredient -- was modelled 
using a variable flux limit across the survey area. Under these 
hypotheses, the derived cosmological constraints appeared to be 
in agreement with the concordance model [see review in \citet{rosati02}].

Within the same period, deep GINGA, ROSAT and ASCA observations 
of nearby clusters revealed that the $L_X-T$ relation is 
significantly steeper than expected from purely gravitational 
heating \citep{arnaudevrard}, hence suggesting the presence of 
other heating/cooling sources such as feedback from star 
formation or AGN, in addition to the effects of cooling flows. 
This particularly affects the low-mass end of the cluster population 
-- groups with temperatures [$0.5-2$]~keV where the 
gravitational binding energies are low.

Nearby cluster observations at high spatial resolution by Chandra
have also shown that the ICM is not the well-relaxed medium previously
assumed: shocks, cold fronts, and bulk velocities are seen even in
apparently relaxed clusters (e.g. \citealt{mazzotta2002}, 
\citealt{dupke2006}). 
High spectral resolution XMM pointings have led to a totally new 
version of the putative ``cooling flow'' scenario, where episodic
heating/accretion by the central AGN could play a key role in the
ICM, preventing any central ``cooling catastrophe''. Measurements of the
$L_X-T$ relation for distant ($z > 0.5$) massive clusters are in
progress, whether the evolution of scaling laws follows simple
self-similar expectations is still hotly debated
[see review in \citet{marnaud05}]. All these results currently
pertain to the upper end of the cluster mass function. They
present a new challenge for high-resolution numerical simulations,
which, in turn, should quantify deviations from hydrostatic
equilibrium.

In parallel, the building of large serendipitous XMM and 
Chandra cluster samples - from public archive data - has been 
initiated by a number of groups: the XMM cluster survey (XCS, 
\citealt{Romer01}), SEXCLAS \citep{kolokotronis06} and ChaMP 
\citep{Barkhouse06}.  
Preliminary results over areas $<\,10~{\rm deg}^2$, give 
cluster densities of the order $5 {\rm deg}^{-2}$ for objects detected 
independently in both X-ray and optical wavebands. 
No cosmological analysis has been performed on these samples so 
far, but these searches have enabled the detection of the most distant 
X-ray clusters to date at $z=1.4$ \citep{mullis05} and $z=1.45$ 
\citep{stanford06}. 

In addition to these large surveys, a number of contiguous surveys, such
as COSMOS \citep{finoguenov07} and XBootes \citep{XBootes}, are also
being conducted.
Following on from the REFLEX cluster survey \citep{reflex}, and
exploiting the unrivalled sensitivity of the XMM-Newton X-ray
observatory, the XMM wide area survey [XMM-LSS, \citet{xmmlss}] 
is the largest contiguous X-ray cluster survey being undertaken
at the present time. It has been designed
to investigate the large scale structure of the
Universe as traced by galaxy clusters to redshifts $z = 1$ and
beyond. The XMM-LSS sensitivity limit is $\sim 1000$ times deeper than
REFLEX i.e. $\sim 4 \times 10^{-15}$ \flux\ in the [$0.5-2$]~keV
band for point sources. Moreover, the XMM-LSS is able to make a
systematic exploration for massive clusters out to at least 
$z \sim 1.5$ [\citet{refregier02}, \citet{bremer06}, \citet{pipeline}]. 
Another key improvement is the spatial resolution of XMM: 
$\sim 6^{\prime\prime}$ FWHM on axis compared to $\sim 20^{\prime\prime}$ 
for ROSAT. 

The two major requirements of the X-ray processing were to reach 
the sensitivity
limit of the data in a statistically tractable manner in terms of
cluster detection efficiency, and so to provide the selection
function of the detected objects. To achieve these goals, it was
necessary to design a new two-step X-ray pipeline,  combining
wavelet multi-resolution analysis with maximum likelihood fitting;
Poisson statistics being used in both steps \citep{pipeline}. We stress here
that at our sensitivity and spatial resolution, the survey is primarily 
limited by surface brightness, rather than by flux as assumed by
past generations of surveys. This led us to define cluster
selection criteria in a two-dimensional parameter space,
corresponding to specific levels of contamination and
completeness, as discussed below.

Building on  the preliminary results from the first square degree
of the survey \citep{D1paper}, we have defined a well-controlled
cluster sample  covering all currently available XMM-LSS
observations, i.e. about 5 \dd. The sample comprises some 30
clusters with fluxes in the [0.5-2]~keV band ranging from 1 to 50
$\times 10^{-14}$\flux. They are referred to as `First Class', or C1
clusters, because the criteria used to construct the sample from a
two-dimensional X-ray parameter space guarantee no contamination
by point-like sources. The observations were performed in a rather
homogeneous way (10-20 ks exposures) and enable us, in addition,
to estimate temperatures for all the sample clusters. This is a
key datum which, along with careful modelling of the
survey selection effects, allow us to tackle for the first time on real
data the calibration of the mass-observable
relations in a cluster survey -- something which has been mostly
addressed in a formal way thus far [e.g. \cite{majumdar03} and
references therein]. This leads  us to quantitatively investigate
some of the degeneracies affecting the cosmological interpretation
of the redshift distribution, $dn/dz$, of cluster number counts.

The paper is organized as follows. In section 2, we summarise the
principles of our X-ray analysis,  provide a detailed
calculation of the selection function, and subsequently present the
cluster sample. The next section describes the spectroscopic
confirmation of the clusters, and the determination of their X-ray
temperature and luminosity. Section 4 provides a thorough discussion
of the $L_X-T$ relation: special care is given to the modelling of
the selection effects in the  analysis of possible evolutionary
trends. This is an especially informative exercise as it is the
first time that the behaviour of the relation has been explored for a
well controlled sample. Section 5 is devoted to the modelling of the
sample;  starting from the linear spectrum of the initial density
fluctuations, we compute the dark matter halo mass spectrum and
infer the halo X-ray luminosity and temperature using empirical
scaling laws compatible with our data. The cluster population thus 
obtained is then folded through the XMM-LSS selection function. The
derived $dn/dz$ is compared with that observed, and the model
provides us with mass-observable relations. The model also allows us to
further explore (Section 6) the impact of several cluster parameters
on classical cosmological tests; for instance, we examine the
sensitivity of our observed $dn/dz$ to the shape of the M-T
relation in the group regime, to the amount of scatter in the
scaling laws, and to changes in cosmological evolution resulting 
from different equations of state for dark energy.

\vspace{.5cm}

Throughout this article, data are analyzed using the cosmological
parameters estimated by \cite{wmap}, namely:
$H_0=73$~km~s$^{-1}$~Mpc$^{-1}$, $\Omega_m=0.24$, $\Omega_\Lambda=0.76$,
$\Omega_b=0.041$, $n_s=0.95$ and $\sigma_8=0.74$.

\section{Construction of the sample}
\FloatBarrier
  \subsection{The dataset}
  \label{layout}
  To date, the X-ray data received consist of a single region
  of roughly 6~$deg^2$ covered by 51 XMM-Newton
  pointings\footnote{These are currently being complemented during
  XMM-Newton AO5 so as to cover the full 10~deg$^2$ of the Spitzer/SWIRE
  area surveyed in the XMM-LSS field.}. Most of the observations
  (32 pointings, hereafter B pointings) were obtained through
  XMM AO1/AO2, and have a nominal exposure time of $10^4s$,
  whilst 19 (hereafter G pointings) are deeper
  $2\times10^4s$ guaranteed time observations. The latter constitute
  the XMM Medium Deep Survey (XMDS, \citealt{xmds}).

  \begin{figure}
    \begin{center}
    \includegraphics[width=8.3cm]{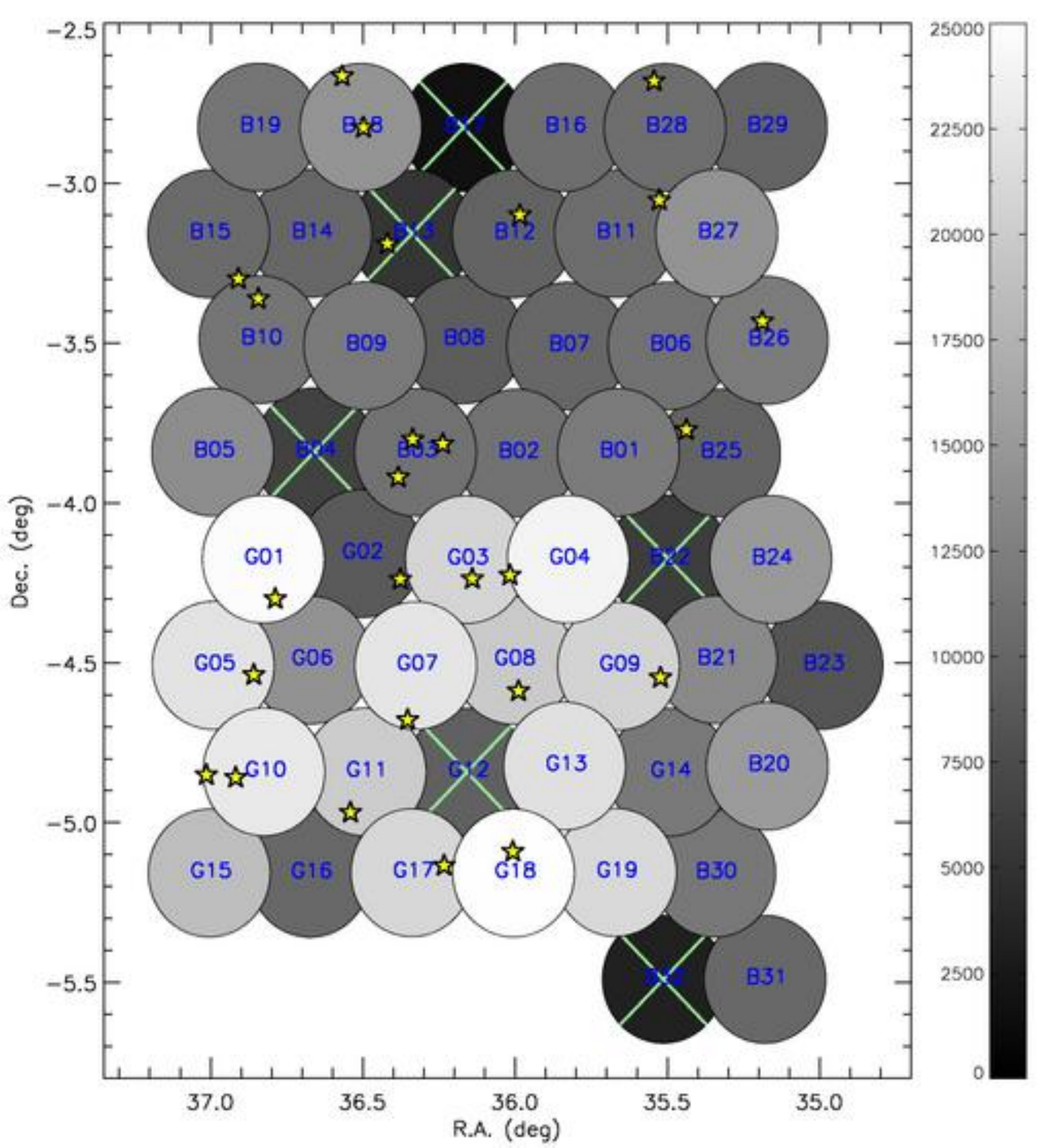}
    \caption{Sky distribution of the C1 clusters. The colour scale
    indicates the on-axis exposure time of each pointing in seconds
    (after particle flare filtering). Pointings marked by a cross
    are strongly affected by flares and will be re-observed during
    AO5.\label{skydist}}
    \end{center}
  \end{figure}

  Calibrated event lists were created using the
  \texttt{XMM-SAS}\footnote{XMM Science Analysis System,
  http://xmm.vilspa.esa.es/sas/} tasks \texttt{emchain} and
  \texttt{epchain}, and periods of intense proton flares
  were filtered out using the method suggested by \cite{pratt2002}.
  Out of the 51 observations, 6 (including one G pointing) were
  too strongly contaminated to properly monitor and remove the
  high flare periods. These pointings are scheduled for re-observation
  during XMM AO5 and, consequently, are not included in the analysis
  of the present paper, which covers only 5~deg$^2$.

  Details of the observations can be found in the publication of the
  full X-ray catalogue pertaining to this first 5~deg$^2$ \citep{FullCatalog}.

  \subsection{Data pre-analysis}

  For each pointing, images were generated in the [0.5-2]~keV band
  from the filtered event lists using the \texttt{XMM-SAS} task
  \texttt{evselect}.

  These were subsequently analyzed using the XMM-LSS source
  detection pipeline (described in detail by \citealt{pipeline}):
  First, the images are filtered using a wavelet
  multi-resolution algorithm \citep{starckpierre} that was specifically
  designed to properly account for the Poisson noise in order to smooth
  the background and lower the noise level, while keeping unchanged the
  relevant information.
  Then \texttt{SExtractor} is used on the filtered images to
  extract a very deep primary source catalogue.
  Finally, detailed properties of each detected source are
  assessed by {\sc Xamin}, a maximum likelihood profile fitting
  algorithm that we have developed to characterize extended
  sources in XMM-Newton images.

  \subsection{Source selection process}

  \subsubsection{Definition of the sample}
  As shown by \cite{pipeline}, the probability of detecting an
  extended source at the XMM-LSS sensitivity and resolution depends
  on both its flux and angular extent. This led us to abandon
  the simple concept of a flux limited cluster survey and to define a
  system of extended source classes: the final source selection is
  performed in the X{\sc{amin}} output parameter space, and several
  samples are defined, allowing for various amounts of contamination
  from point sources and spurious detections. The sub-samples, or
  classes, have been defined by means of extensive `in situ' simulations,
  involving a large range of cluster fluxes and apparent sizes, for the
  nominal exposure of the XMM-LSS, i.e. $10^4s$. In this paper, we
  focus on the Class 1 sample (hereafter C1) which was defined from
  our simulations to be the largest uncontaminated extended 
  source sample available. It is obtained by selecting candidates with 
  {\tt extension} $>$ 5\arcsec, {\tt extension likelihood} $>$ 33, 
  and {\tt detection likelihood} $>$ 32.
  Given that no spurious extended detection was found in our whole 
  simulated dataset, we are confident that the false detection rate is
  close to zero, and visual inspection of the X-ray/optical 
  overlays for the C1 sources showed the existence of a cluster
  of galaxies or of a nearby galaxy as an optical counterpart in every case. 
  Furthermore, we find that the C1 sample coincides with clusters for which 
  a reliable temperature can be obtained using X-ray spectroscopy from 
  the survey data (section \ref{spec}).
  
  Detailed information on the C1 selection process can be found in 
  \cite{pipeline}.

  \subsubsection{Selection function}
\begin{figure}
    \begin{center}
    \includegraphics[width=9cm]{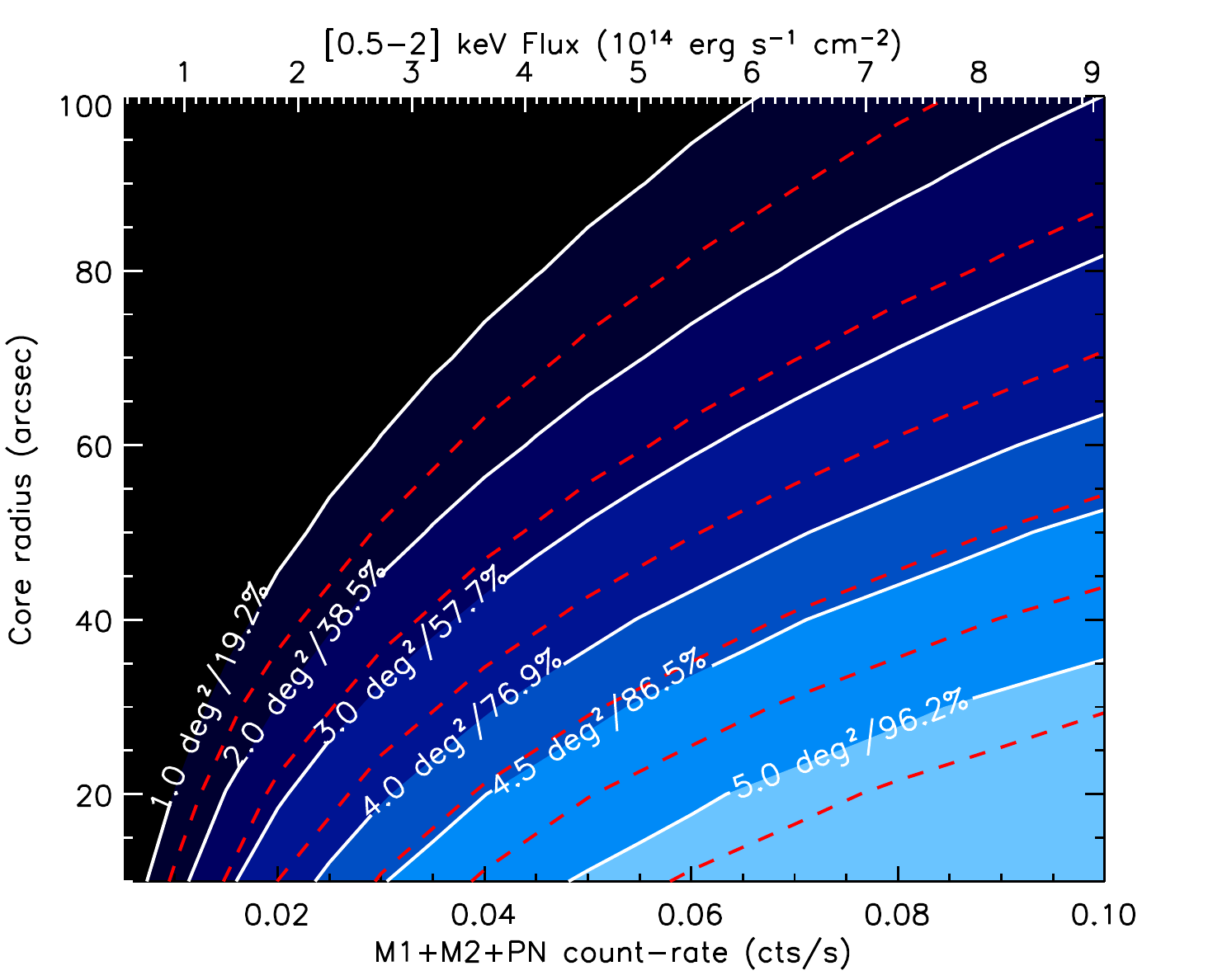}
        \caption{The survey coverage is displayed in a 2D parameter 
	space: the sky coverage is a function of both cluster flux and extent. 
	The dashed lines are the result of extensive 10ks simulations.
    The slightly shifted white lines are the analytical corrections accounting 
    for exposure variations across the surveyed area, hence indicating the 
    effective selection function of the current data set. The sample 
    completeness corresponding to each line is indicated in terms of both 
    the sky coverage and the percentage of detected sources.
    Extent values correspond to the core radius of a $\beta$-model with 
    $\beta=2/3$. The count-rate to flux conversion assumes a 2~keV spectrum 
    at $z=0$.
      \label{skycov} }
    \end{center}
   \end{figure}

  Given the varying effective exposure time per pointing, the survey coverage 
  is not uniform ($7<t_{exp}/10^3s<20,$ Fig.~\ref{skydist}). 
  Consequently, similar likelihood values do not exactly pertain to the same 
  objects across the survey.
  Rather than lowering all exposures to $7\times10^3$s, we apply the C1 criteria to
  every pointing in order to maximize the size of the cluster sample.
  The probability of detecting a cluster of a given flux and extent for any
  pointing is subsequently derived by applying an analytic correction to the
  $10^4s$ simulations, scaling as a function of exposure time the S/N produced by
  such an object\footnote{The mean background values measured by \cite{read2003} are
  assumed}. This is justified by the fact that the range spanned by the exposure
  times is modest ($1/2<t_{exp}/10^4s<2$), implying also that source confusion
  does not change significantly. The survey selection function,
  i.e. the probability to detect a source of a given flux and apparent
  size as a C1 cluster, is obtained by integrating the contributions from all
  pointings. It is this function that is used in the following for cosmological
  applications. It is displayed in the form of a sky coverage
  in Fig.~\ref{skycov}. The maximum area covered reaches 5.2 deg$^2$. 
  This happens e.g. for sources with 20\arcsec core radius, and fluxes above 
  $5\times10^{-14}$\flux.
  Our sensitivity drops by roughly a factor of two for sources with
  flux around $1.5\times10^{-14}$ \flux\ and $20\arcsec$
  core radius, or $5\times10^{-14}$\flux\ flux and
  $75\arcsec$ core radius.

 \subsection{The sample}
  Applying the C1 selection criteria to the inner 11.5\arcmin
  of the 45 valid observations yields 37 detections, among which 4 are
  duplicated sources detected on adjacent pointings. Optical/X-ray overlays
  reveal that 5 of the remaining 33 sources are actually nearby X-ray
  emitting galaxies whose properties are presented in Appendix \ref{C1g}.
  Hence we have a total of 28 galaxy clusters. An additional unambiguous
  C1 source is detected in pointing B13 (highly contaminated by
  flares); we provide the basic properties of this source in the present
  article, but it is not used in the scientific analysis as we cannot
  accurately compute the selection function for this pointing. The
  cluster sample is listed on Table \ref{cluslist}. We note
  that a complementary cluster sample of about the same size
  has also been identified at lower significance in the same area; 
  these clusters belong to our C2 and C3
  classes, for which the selection criteria are much less well
  defined. They will be published in a separate paper (Adami et al.,
  in prep.).

\section{Assessing individual source properties}

  \subsection{Spectroscopic validation}

Each C1 X-ray source was confirmed as a bona-fide galaxy cluster
at specific redshift via optical photometry and spectroscopy.  

\begin{landscape}

\begin{table}
\begin{center}
\vspace{.4cm}
\hspace{-3.6cm}
\begin{minipage}{20cm}
\begin{center}
\caption{List of the C1 galaxy clusters and their basic optical/X-ray properties. More information on peculiar
individual clusters is provided in Appendix \ref{C1app}.
\label{cluslist}}
\begin{tabular}{cccccccccccccccc}
\hline
Source name        &  XLSSC  &  Point.\footnote{XMM pointing identifiers from the XMM-LSS internal labelling as defined in Table 1 of \protect\cite{FullCatalog}, and displayed in Fig. \ref{skydist}.}         &
            R.A.  &  Dec.  &   Redshift\hspace{-0.1cm} & $N_{gal}$\footnote{number of spectroscopically confirmed cluster members}    &
            $R_{spec}$   &     Counts       &     T    &   $R_{fit}$       &    Counts    &      F$_X$\footnote{flux in $10^{-14}$ \flux\ in the [0.5-2]~keV band.} in      &
            r$_{500}$ & L$_{500}$\footnote{L$_{500}$ is the bolometric luminosity.} & $M_{500}$\footnote{these are just rough estimates based on the isothermal assumption.}\\
                   &  number &  &
           (J2000)& (J2000)& & &
             ($\arcsec$)   &  in $R_{spec}$   &  ($keV$) &    (\arcsec / Mpc) & in $R_{fit}$ &
           0.5~Mpc     &  (Mpc)  & \hspace{-0.1cm}($10^{43}$\lum)\hspace{-0.1cm} & \hspace{-0.1cm}($10^{13}\,h^{-1}M_{\odot}$)\hspace{-0.1cm}\\
\hline
 XLSS J022023.5-025027 &   039   &     B29    & 35.0983 & -2.8409 &  0.23 &3   &          66.5    &  144  & $1.3^{+0.3}_{-0.1}$    &      54 / 0.198       &    112       & $ 2.5^{+0.9}_{-0.9}$ & 0.417 & $ 0.9^{+0.2}_{-0.3}$ & $2.4$\\
 XLSS J022045.4-032558 &   023   &     B26    & 35.1894 & -3.4328 &  0.33  &7   &        55.0    &  338  & $1.7^{+0.3}_{-0.2}$    &      84 / 0.399       &    384       & $ 4.4^{+0.4}_{-0.4}$ & 0.457 & $ 3.8^{+0.3}_{-0.3}$ & $3.1$\\
 XLSS J022145.2-034617 &   006   &     B25    & 35.4385 & -3.7715 &  0.43\footnote{already published in \protect\cite{Willis}.\label{fn:willis}} &26
                                         &         80.0    & 1304  & $4.8^{+0.6}_{-0.5}$    &     606 / 3.399       &   2099       & $25.0^{+0.8}_{-0.7}$ & 0.838 & $60.3^{+1.8}_{-1.8}$ & $19$\\
 XLSS J022205.5-043247 &   040   &     G09    & 35.5232 & -4.5464 &  0.32 &3    &          41.5    &  116  & $1.6^{+1.1}_{-0.3}$    &     117 / 0.544       &    273       & $ 2.1^{+0.3}_{-0.2}$ & 0.442 & $ 1.6^{+0.2}_{-0.2}$ & $3.4$\\
 XLSS J022206.7-030314 &   036   &     B11    & 35.5280 & -3.0539 &  0.49 &3    &          46.5    &  507  & $3.6^{+0.6}_{-0.4}$    &     213 / 1.286       &    659       & $10.7^{+0.6}_{-0.5}$ & 0.676 & $28.9^{+1.5}_{-1.5}$ & $11$\\
 XLSS J022210.7-024048 &   047   &     B28    & 35.5447 & -2.6801 &  0.79 &12    &         60.0    &  114  & $3.9^{+2.8}_{-1.4}$    &      75 / 0.561       &     89       & $ 1.6^{+0.3}_{-0.2}$ & 0.592 & $13.2^{+2.2}_{-2.3}$ & $12$\\
 XLSS J022253.6-032828 &   048   &     B07    & 35.7234 & -3.4745 &  1.00 &3    &           55.0    &  137  & $1.8^{+0.7}_{-0.4}$    &      66 / 0.529       &     94       & $ 1.5^{+0.2}_{-0.2}$ & 0.327 & $17.4^{+3.1}_{-3.1}$ & $2.9$\\
 XLSS J022348.1-025131 &   035   &     B16    & 35.9507 & -2.8588 &  0.17 &5    &       60.0    &  145  & $1.2^{+0.1}_{-0.1}$    &     162 / 0.469       &    334       & $ 4.8^{+0.5}_{-0.5}$ & 0.394 & $ 0.77^{+0.07}_{-0.07}$ & $1.5$\\
 XLSS J022356.5-030558 &   028   &     B12    & 35.9857 & -3.0997 &
 0.30\footnote{two foreground  galaxies in the centre of the X-ray
 emission have a redshift of 0.08 - see Appendix \ref{C1app}} &8

                                             &    60.0    &  144  & $1.3^{+0.2}_{-0.2}$    &     60 / 0.267       &    118   & $ 2.7^{+0.4}_{-0.4}$ & 0.399 & $ 1.5^{+0.2}_{-0.2}$ & $1.5$\\
 XLSS J022357.4-043517 &   049   &     G08    & 35.9892 & -4.5883 &  0.49 &4    &         25.0    &   76  & $2.2^{+0.9}_{-0.5}$    &    456 / 2.753       &    722   & $ 1.9^{+0.2}_{-0.2}$ & 0.493 & $ 4.3^{+0.4}_{-0.4}$ & $3.4$\\
 XLSS J022402.0-050525 &   018   &     G18    & 36.0087 & -5.0904 &  0.32\footref{fn:willis} &14
                                            &    35.0    &  203  & $2.0^{+0.7}_{-0.4}$    &     75 / 0.349       &    245   & $ 1.5^{+0.2}_{-0.2}$ & 0.521 & $ 1.3^{+0.2}_{-0.2}$ & $4.7$\\
 XLSS J022404.1-041330 &   029   &     G04    & 36.0172 & -4.2251 &  1.05\footnote{already published in \protect\cite{D1paper}.\label{fn:D1}} &5
                                         &       33.5    &  310  & $4.1^{+0.9}_{-0.7}$    &     60 / 0.486       &    355   & $ 3.1^{+0.2}_{-0.2}$ & 0.524 & $48.3^{+3.7}_{-3.4}$ & $8.9$\\
 XLSS J022433.8-041405 &   044   &     G03    & 36.1411 & -4.2347 &  0.26\footref{fn:D1} &10
                                         &       54.5    &  319  & $1.3^{+0.2}_{-0.1}$    &    294 / 1.183       &    544   & $ 2.8^{+0.2}_{-0.2}$ & 0.399 & $ 1.2^{+0.1}_{-0.1}$ & $2.0$\\
 XLSS J022456.2-050802 &   021   &     G17    & 36.2345 & -5.1339 &  0.08 &7    &      28.5    &  265  & $0.68^{+0.04}_{-0.02}$ &      75 / 0.113       &    372   & $ 3.3^{+0.6}_{-0.5}$ & 0.297 & $ 0.11^{+0.01}_{-0.01}$ & $0.9$\\
 XLSS J022457.1-034856 &   001   &     B03    & 36.2381 & -3.8157 &  0.61\footnote{already published in \protect\cite{ivanhz}.\label{fn:hz}} &23
                                         &     60.0    &  730  & $3.2^{+0.4}_{-0.3}$    &    129 / 0.869       &    770   & $ 7.6^{+0.3}_{-0.4}$ & 0.584 & $33.2^{+1.5}_{-1.5}$ & $8.3$\\
 XLSS J022520.8-034805 &   008   &     B03    & 36.3370 & -3.8015 &  0.30\footref{fn:willis} &11
                                         &      45.0    &   99  & $1.3^{+0.7}_{-0.2}$    &      57 / 0.254      &    104   & $ 2.1^{+0.3}_{-0.3}$ & 0.396 & $ 1.2^{+0.2}_{-0.2}$ & $1.4$\\
 XLSS J022524.7-044039 &   025   &     G07    & 36.3531 & -4.6776 &  0.26\footref{fn:D1} &15
                                         &      35.0    &  661  & $2.0^{+0.2}_{-0.2}$    &     153 / 0.615      &    925   & $ 8.7^{+0.4}_{-0.4}$ & 0.533 & $ 4.6^{+0.2}_{-0.2}$ & $4.1$\\
 XLSS J022530.6-041420 &   041   &     G02    & 36.3777 & -4.2391 &  0.14\footref{fn:D1} &15
                                         &       45.0    &  523  & $1.34^{+0.14}_{-0.06}$ &    159 / 0.392       &    785   & $20.2^{+1.0}_{-1.1}$ & 0.440 & $ 2.4^{+0.1}_{-0.1}$ & $2.2$\\
 XLSS J022532.2-035511 &   002   &     B03    & 36.3844 & -3.9200 &  0.77\footref{fn:hz} &11
                                         &    37.5    &  225  & $2.8^{+0.8}_{-0.5}$    &     66 / 0.489       &    200   & $ 2.8^{+0.3}_{-0.3}$ & 0.493 & $19.6^{+1.8}_{-1.8}$ & $5.8$\\
 XLSS J022540.6-031121 &   050   &     B13    & 36.4195 & -3.1894 &  0.14 &13    &      86.0    & 1386  & $3.5^{+0.6}_{-0.5}$    &    177 / 0.436       &   1509   & $48.2^{+2.2}_{-2.4}$ & 0.804 & $ 9.3^{+0.5}_{-0.6}$ & $16$\\
 XLSS J022559.5-024935 &   051   &     B18    & 36.4982 & -2.8265 &  0.28 &11    &        60.0    &  224  & $1.2^{+0.1}_{-0.1}$    &     57 / 0.242       &    169   & $ 1.5^{+0.6}_{-0.3}$ & 0.384 & $ 0.9^{+0.2}_{-0.1}$ & $4.4$\\
 XLSS J022609.9-045805 &   011   &     G11    & 36.5413 & -4.9682 &  0.05\footref{fn:D1} &9
                                         &     67.5    &  424  & $0.64^{+0.06}_{-0.04}$ &     465 / 0.454       &   1045   & $11.6^{+1.1}_{-1.1}$ & 0.290 & $ 0.11^{+0.01}_{-0.01}$ & $0.6$\\
 XLSS J022616.3-023957 &   052   &     B18    & 36.5681 & -2.6660 &  0.06 &2    &   46.5    &  529  & $0.63^{+0.03}_{-0.03}$ &    102 / 0.118       &    802   & $14.0^{+1.7}_{-1.4}$ & 0.285 & $ 0.25^{+0.02}_{-0.02}$ & $0.9$\\
 XLSS J022709.2-041800 &   005   &     G01    & 36.7885 & -4.3000 &  1.05\footref{fn:hz}$^,$\footref{fn:D1} &9
                                         &    35.0    &  164  & $3.7^{+1.5}_{-1.0}$    &     96 / 0.777       &    179   & $ 1.1^{+0.1}_{-0.1}$ & 0.489 & $17.1^{+2.0}_{-2.0}$ & $8.9$\\
 XLSS J022722.4-032144 &   010   &     B10    & 36.8435 & -3.3623 &  0.33\footref{fn:willis} &9
                                         &    60.0    &  452  & $2.4^{+0.5}_{-0.4}$    &     96 / 0.456       &    440   & $ 6.3^{+0.4}_{-0.4}$ & 0.574 & $ 6.1^{+0.5}_{-0.5}$ & $6.1$\\
 XLSS J022726.0-043216 &   013   &     G05    & 36.8586 & -4.5380 &  0.31\footref{fn:willis}$^,$\footref{fn:D1} &18
                                         &      30.0    &  160  & $1.0^{+0.1}_{-0.1}$    &    417 / 1.899       &    536   & $ 2.1^{+0.2}_{-0.2}$ & 0.340 & $ 1.3^{+0.1}_{-0.1}$ & $1.3$\\
 XLSS J022738.3-031758 &   003   &     B15    & 36.9098 & -3.2996 &  0.84\footref{fn:hz} &13
                                         &     48.0    &  231  & $3.3^{+1.1}_{-0.7}$    &    393 / 2.998       &    523   & $ 4.2^{+0.4}_{-0.4}$ & 0.518 & $37.8^{+3.5}_{-3.3}$ & $6.3$\\
 XLSS J022739.9-045127 &   022   &     G10    & 36.9165 & -4.8576 &  0.29\footref{fn:D1} &9
                                         &      39.5    & 1305  & $1.7^{+0.1}_{-0.1}$    &    171 / 0.744       &   1791   & $ 9.8^{+0.3}_{-0.3}$ & 0.471 & $ 6.2^{+0.2}_{-0.2}$ & $3.2$\\
 XLSS J022803.4-045103 &   027   &     G10    & 37.0143 & -4.8510 &  0.29 &7    &      60.0    &  438  & $2.8^{+0.6}_{-0.5}$    &    123 / 0.535       &    577   & $ 6.1^{+0.4}_{-0.4}$ & 0.653 & $ 4.8^{+0.4}_{-0.3}$ & $9.1$\\
\hline
\end{tabular}
\end{center}
\vspace{-.2cm}
\end{minipage}
\end{center}
\end{table}

\end{landscape}

\begin{table*}
\begin{center}
\begin{minipage}{140mm}
\begin{center}
\caption{Instrumental characteristics for each spectrograph
configuration employed during the observations. All spectral
observations were performed with a slit width of between
1\farcs0 and 1\farcs4.}
\label{SpectrObs}
\begin{tabular}{lcccccc}
\hline
Telescope & Instrument & $\rm Grism + Filter$ & Wavelength &
Pixel sampling  & Spectral & Identifier\\
&&& interval (\AA) & ($\rm \AA \, {pix}^{-1}$) &
resolution\footnote{Estimated via the mean full-width at
half-maximum of the bright, isolated arc emission lines.} (\AA) \\
\hline
VLT     & FORS2 & $\rm 300V + GG435$ & 4000--9000 & 3.2 & 14 & 1\\
VLT     & FORS2 & $\rm 600RI+ GG435$ & 5000--8500 & 1.6 & 7  & 2\\
VLT     & FORS2 & $\rm 600z + OG590$ & 7500--10000& 1.6 & 7  & 3\\
Magellan    & LDSS2 & medium--red    & 4000--9000 & 5.1 & 14 & 4\\
NTT     & EMMI  & Grism \#3      & 4000--9000 & 3.0 & 8  & 5\\
\hline
\end{tabular}
\end{center}
\vspace{-0.7cm}
\end{minipage}
\end{center}
\end{table*}

The overall procedure used to confirm individual clusters is very
similar to the approach taken in previously published XMM-LSS
papers (\citealt{ivanhz}; \citealt{Willis}; \citealt{D1paper}) 
and is summarised below.

A combination of either CTIO/MOSAIC II $Rz^\prime$ \citep{andreon2004} or 
CFHT/MEGACAM\footnote{Data are taken from the CFHT Wide
Synoptic Legacy Survey. See the URL {\tt
www.cfht.hawaii.edu/Science/CFHTLS/} for further details.} $ugriz$
imaging was used to associate the location of each X-ray source
with the spatial barycentre of a significant overdensity of
galaxies displaying characteristically red colours. Based on the CTIO
data, galaxies lying within a given colour tolerance of this ``red
sequence'' were then flagged as candidate cluster members and
given a high priority in subsequent multi-object spectroscopic
observations.
A small number of  low X-ray temperature ($\lesssim 1$~keV) groups
at moderate redshift ($z>0.2$) and local ($z<0.2$) compact groups
did not display a statistically significant red galaxy
overdensity. These systems were inspected visually and
spectroscopic targets were assigned manually.

Due to the moderately large size and the extended redshift range
covered by the sample, cluster targets were observed using a
number of facilities over several observing semesters. Details of
the observing configurations can be found in Table
\ref{SpectrObs}.

Spectroscopic data were reduced using standard procedures
described in detail in previously published XMM-LSS publications
(\citealt{ivanhz}; \citealt{Willis}; \citealt{D1paper}).
Where possible, redshift values and associated uncertainties of
individual galaxy targets were computed via cross-correlation with
galaxy reference templates. In the remaining cases redshifts were
assigned manually, from emission features.

The nominal condition adopted by XMM-LSS to confirm a cluster
redshift is to observe three concordant redshifts (typically
$\Delta z \lesssim 0.01$) within  a projected scale of about 500
kpc of the cluster X-ray centre. Only cluster XLSSC-052 does not
fulfill this condition: the X-ray emission is associated with a
pair of $z=0.06$ galaxies very close to a bright star. 
Except for this specific system, the number of galaxy members 
confirmed per cluster ranges between 3 and 15, with a typical 
cross-correlation velocity error of order 50-150 km s$^{-1}$ per galaxy. 
Despite this velocity accuracy, we are limited for 
a large fraction of our sources by small-number statistics and 
consequently quote redshifts only to $\pm 0.01$. In each case, the 
final cluster redshift is computed from the unweighted mean of all 
galaxies lying within $\pm 3,000 ~\rm km.s^{-1}$ of the visually 
assigned redshift peak.

  \subsection{Spectral analysis}
  \label{spec}

  In order to measure the temperature of the intra-cluster gas, X-ray
  spectra were extracted in a circular aperture around each source.
  The corresponding background emission was estimated within a surrounding
  annulus having inner radius large enough for the cluster contribution to be
  considered  negligible. Preliminary modelling of the cluster surface
  brightness profile allowed the determination of the optimal extraction
  radii in terms of S/N.

  The resulting spectra were fitted using X{\sc{spec}}\footnote{http://heasarc.gsfc.nasa.gov/docs/xanadu/xspec/}
  to a thermal plasma model (APEC) assuming a fixed hydrogen column
  density set to the Galactic value as derived from \mbox{H\,\textsc{I}}
  observations by \cite{nh}. The metal abundance of the gas was held fixed
  during the fitting process at 0.3 times the solar abundance, as estimated
  by \cite{GRSA}.
  As explained in \cite{Willis}, the cluster spectra were constructed
  imposing  a minimum requirement of 5 background photons per bin
  in order to avoid  the apparent bias we identified in X{\sc{spec}} temperature
  estimates when using the Cash statistic on very sparse spectra.
  Our simulations \citep{Willis} showed that  this procedure
  provides quite reliable temperature measurements ($\pm 10-20\%$)
  for $\sim$1-3~keV clusters having only  a few hundred counts.
  We further investigated the impact of fixing the metal abundance
  at $0.3~Z_\odot$, by computing best fit temperatures obtained using
  extreme mean abundances of $0.1$ and $0.6~Z_\odot$.
  In  most  cases, the  temperatures fell within the
  $1\sigma$ error bars from our initial fit.
  For five systems (namely XLSSC-008, XLSSC-028, XLSSC-041, XLSSC-044
  and XLSSC-051) one of these two extreme measurements just fell a few
  percent outside of our error range.
  The measured temperatures are presented in Table \ref{cluslist}.

  \subsection{Spatial analysis}

\subsubsection{Surface brightness modelling}

  To accurately determine the cluster fluxes and luminosities we modelled
  the observed photon spatial distribution in the [0.5-2]~keV band with a
  radial $\beta$-profile:
  \begin{equation}
    S(r) = \frac{S_0}{[1+(r/R_{c})^{2}]^{3\beta-1/2}},
  \end{equation}
  using a refined version of the method described in \cite{D1paper}.

For each source, we started by fitting the mean background levels
  (vignetted and particle components) over the inner 13\arcmin\ of
  the pointing, excluding all sources detected by the pipeline.
  Then, for each EPIC camera, the photons were binned within 3\arcsec\
  annuli centred on the cluster peak. 
The resulting profiles were
  subsequently rebinned imposing a minimum $S/N$ of 3 with respect
  to the estimated background level, weighted by the annular exposure
  times, and finally co-added to build a MOS1+MOS2+PN count-rate
  profile.
The fitted model is constructed by convolving the circular $\beta$-profile
  with an analytical parametrization of the PSF \citep{ghizzardi01}, as
  implemented by \cite{arnaudPSF}. The $\chi^2$ statistic is computed over
  a discrete grid of $\beta$ and $R_c$ values, with
  the value of the normalization coefficient, $S_0$, optimised analytically. 

As already discussed in \cite{D1paper}, the majority of the
clusters are faint, and the detected photon distribution
in many cases represents only a fraction of the extended X-ray
surface brightness distribution. Under such conditions the
parameters $\beta$ and $R_c$ are degenerate when fitted
simultaneously, limiting the extent to which `best-fitting'
parameters can be viewed as a physically realistic measure of the
cluster properties, although they provide a useful ad hoc
parametrisation (Appendix \ref{lumacc}). For this reason, we do
not quote here best fitting values of $\beta$ and $R_c$ derived
for each confirmed cluster. A dedicated analysis of the mean
cluster  profiles, obtained by stacking the data as a function of
redshift and temperature, is underway (Alshino et al in prep.).

The photon count-rate within a specified radius is obtained by
integrating the best-fitting spatial profile.
The conversion into [0.5-2]~keV flux and unabsorbed rest-frame
bolometric ([0.001-50]~keV) luminosity is performed via X{\sc{spec}} 
using the cluster temperatures previously derived.

\subsubsection{Luminosity and flux determination}

  Luminosities are integrated within $r_{500}$ i.e. the radius at which
  the cluster mass density reaches 500 times the critical density of the
  Universe at the cluster redshift. As in \cite{Willis} and \cite{D1paper},
  this radius is estimated from the cluster mean temperature using the M-T
  relation of \citet{FRB}, converted to ${\Lambda}CDM$ cosmology, which 
  gives:
  \begin{equation}
    r_{500} = 0.375~T^{0.63}~h_{73}(z)^{-1}~Mpc,\label{eq:r500}
  \end{equation}
  where T is expressed in keV and $h_{73}$ is the hubble constant in units 
  73~km s$^{-1}$Mpc$^{-1}$.
Although ``total'' fluxes are often quoted in the cluster
literature, our simulations show that the present data do not
allow us to reliably perform such measurements (see Appendix
\ref{lumacc}).
To limit extrapolation uncertainties, aperture flux values are computed 
by integrating within a fixed radius of 0.5~Mpc. 
As conspicuous in Table \ref{cluslist}, 0.5~Mpc is generally smaller 
than $2\times R_{fit}$ except for the nearby low temperature groups 
XLSSC-039, XLSSC-021 and XLSSC-052. 
We chose to compute the flux within 0.5 Mpc (rather than $R_{500}$, which 
is generally smaller), since this is a similar approach to aperture 
photometry. For all clusters it corresponds to about 
2/3 of the total flux, i.e. integrated to infinity, assuming a
profile defined by $\beta = 2/3,~ R_c = 180$ kpc. We emphasise
that these fluxes are only used in the comparative
analysis of the log(N)-log(S) relation, while $L_{500}$ and $M_{500}$ 
(along with $T$) are the actual physical quantities used in our 
cosmological modelling and in the subsequent discussion.

Finally, the $1\sigma$ errors on the extrapolated fluxes and luminosities
are computed by identifying the region of the ($S_0$, $R_c$,$\beta$) 
parameter space where $\Delta\chi^2\leq1$, and computing the extreme values of
the extrapolated count-rates allowed by these models (see Appendix
\ref{lumacc} for further discussion on the flux measurement accuracy).

\subsubsection{Mass determination}

Based on our best spatial fit profile, we estimate cluster
masses, assuming that the gas is isothermal (the limited number of
photons  does not allow  us to derive temperature profiles). Under
these assumptions, for $M_{500}$ in $M_\odot$, the hydrostatic
equilibrium assumption yields (see e.g. \citealt{Ettori2000}):
  \begin{equation}
  M_{500}=1.11\,10^{14}\times \beta R_{c} T\frac{x_{500}^3}{1+x_{500}^2},
  \end{equation}
where $x_{500}=r_{500}/R_c$ (with $r_{500}$ given by Eq. \ref{eq:r500}),
$R_{c}$ is expressed in Mpc, and $T$ in keV.

Values of $r_{500}$, flux and luminosity for the C1 clusters are listed in
Table \ref{cluslist}, together with   $M_{500}$.

  \begin{figure}
    \begin{center}
    \includegraphics[width=8.6cm]{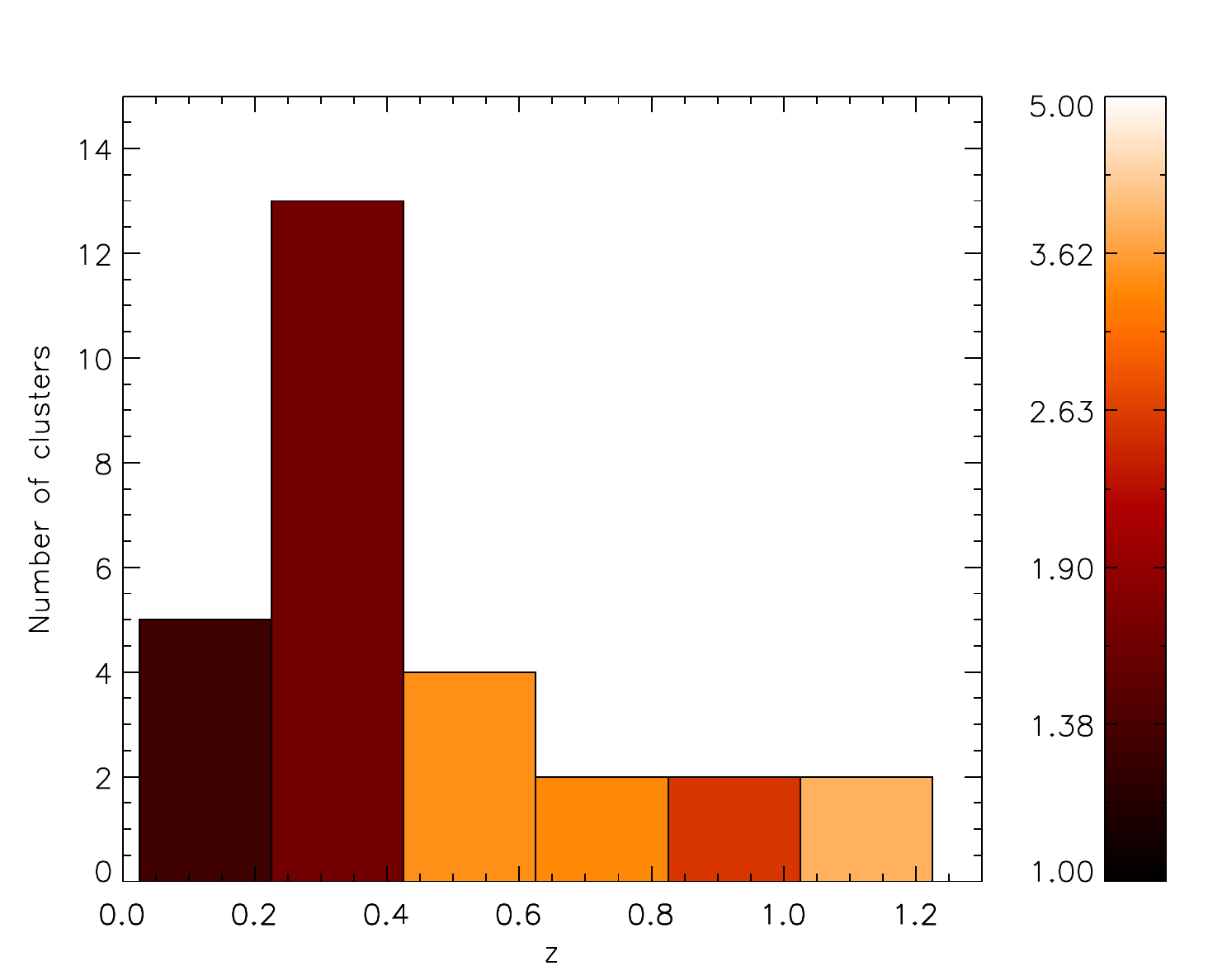}
    \caption{Redshift distribution of the C1 sample. The colour scale
    indicates the cluster mean temperature  for each bin (unweighed mean of the
    individual cluster temperatures in keV)  .\label{zt_trend}}
    \end{center}
  \end{figure}

\section{Results}
  \subsection{Global properties of the sample}
The detected clusters span the [0.05-1.05] redshift range with a
pronounced peak  around $z\sim0.3$
(Fig.~\ref{zt_trend}). Roughly half of the objects
have a temperature lower than 2~keV, pointing to a large
fraction of groups in our sample. As a natural consequence of
its  sensitivity, good PSF and dedicated source detection procedure, the XMM-LSS systematically
unveils for the first time the $z\sim0.3$, $T\lesssim2$~keV
cluster population on large scales.
We note also that none of the objects detected so far exhibits
strong lensing features.
In the following, we will generally use  the word ``cluster''
regardless of the temperature, while ``group'' specifically refers
to objects below 2~keV. A noticeable feature of the sample, illustrated by
Fig.~\ref{zt_trend},
is  the increasing mean cluster temperature with redshift.
Given the strong correlation between X-ray luminosity
and temperature (Sec. \ref{ltsec}) this can readily be
understood as the unavoidable  Malmquist bias. Moreover,
the current sky coverage of the survey makes
the sample likely to be affected by cosmic variance. Both
issues will be given special attention in the following
cosmological analysis.

Additional information on individual sources, including possible contamination
by AGN emission, are discussed in Appendix~\ref{C1app}. 
We also display the optical images of each cluster with confirmed member 
galaxies and X-ray contours overlayed.

  \subsection{The $M_{500}$-$L_X$ relation}

\begin{figure}
    \begin{center}
    \includegraphics[width=8.6cm]{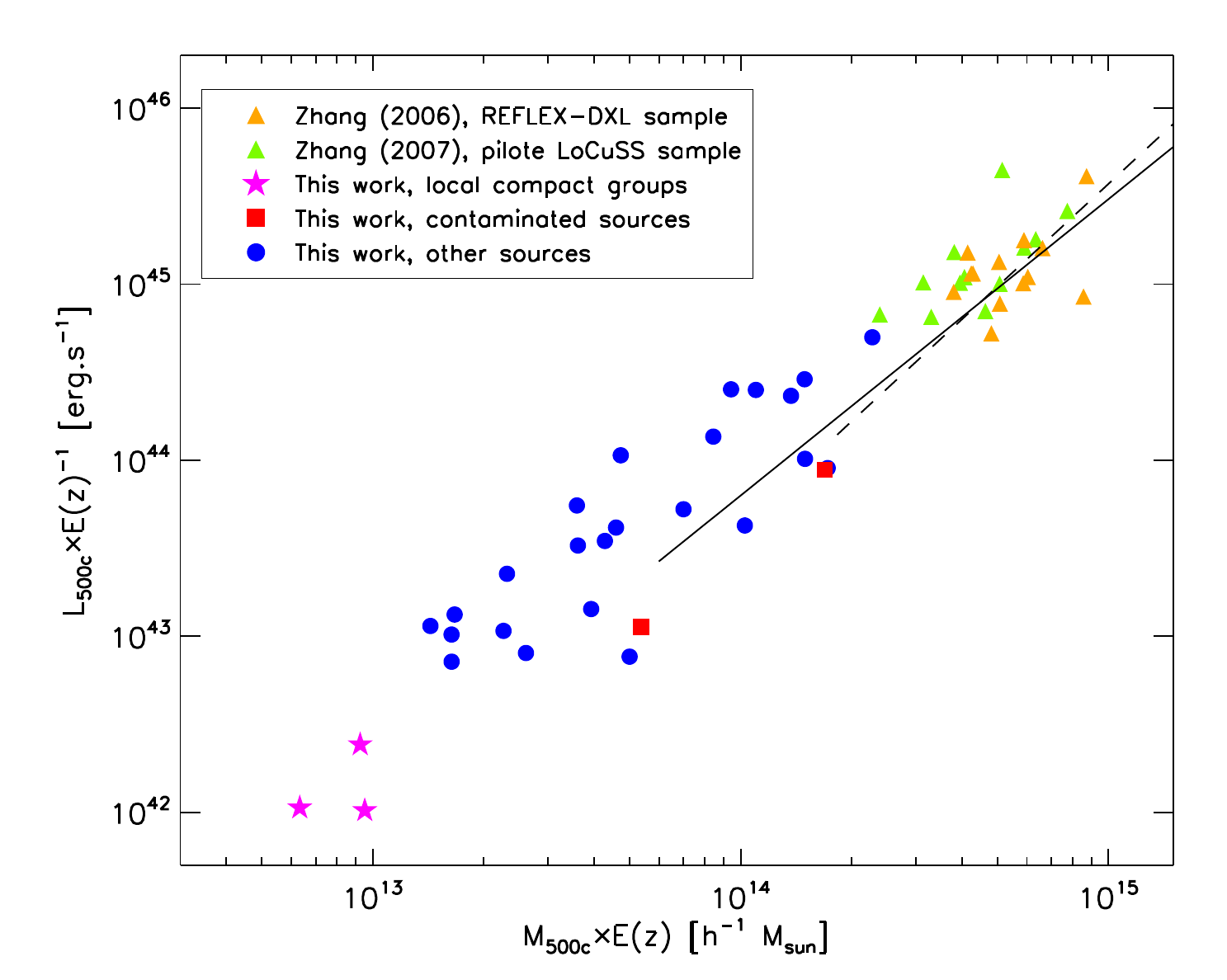}
    \vfill
    \caption{The Mass-Luminosity relation for the sample clusters. 
    Because of the large redshift range spanned by the data, the mass 
    and luminosity parameters are scaled assuming self-similar evolution 
    (factor $E(z)$, see Sec. \ref{ltsec}). 
    For comparison, the massive cluster samples from \citet{Zhang2006} 
    and \citet{Zhang2007} are also shown. 
    The dashed and solid lines also show the mean $M_{500}$-$L_X$ relation 
    infered from the $M_{500}$-$T$ and $L_X$-$T$ measured in \citet{arnaud2005} 
    and \citet{arnaudevrard} respectively for $T >$ 3.5 keV and $T >$ 2 keV.}
    \label{M-L}
    \end{center}
  \end{figure}

The correlation between cluster luminosities and masses derived 
from the spatial analysis is shown in Fig.~\ref{M-L}.
Once self-similar evolution is assumed, our data points
appear to be continuous with the massive cluster samples of 
\citet{Zhang2006} and \citet{Zhang2007} [respectively at $z\sim0.3$ 
and $z\sim0.2$]. Moreover, the scatter in the relation is relatively 
low  (i.e. almost comparable to that of the high mass sample).
This overall consistency is remarkable in the sense that, 
for half of the clusters, less than 500/300 photons
were available for the spatial/spectral fitting. This provides
an important indicator of the reliability of our measurements, 
and could be further interpreted as adding weight to the self-similar 
evolution hypothesis. 

The comparison between our data points and the indicative $M$-$L_{X}$ relation
derived from \citet{arnaud2005} and \citet{arnaudevrard} also suggests
some flattening of the correlation at the low mass end.
Our luminosity sampling, however, suffers significant Malmquist bias, 
and masses derived from a single isothermal $\beta$-model fitting
are known to be underestimated \citep{VikhMT}.
Both effects tend to contribute to the apparent flattening in the 
$M$-$L_X$ relation.

 \subsection{The $L_X$-$T$ relation}
 \label{ltsec}

\begin{figure}
    \begin{center}
    \includegraphics[width=8.6cm]{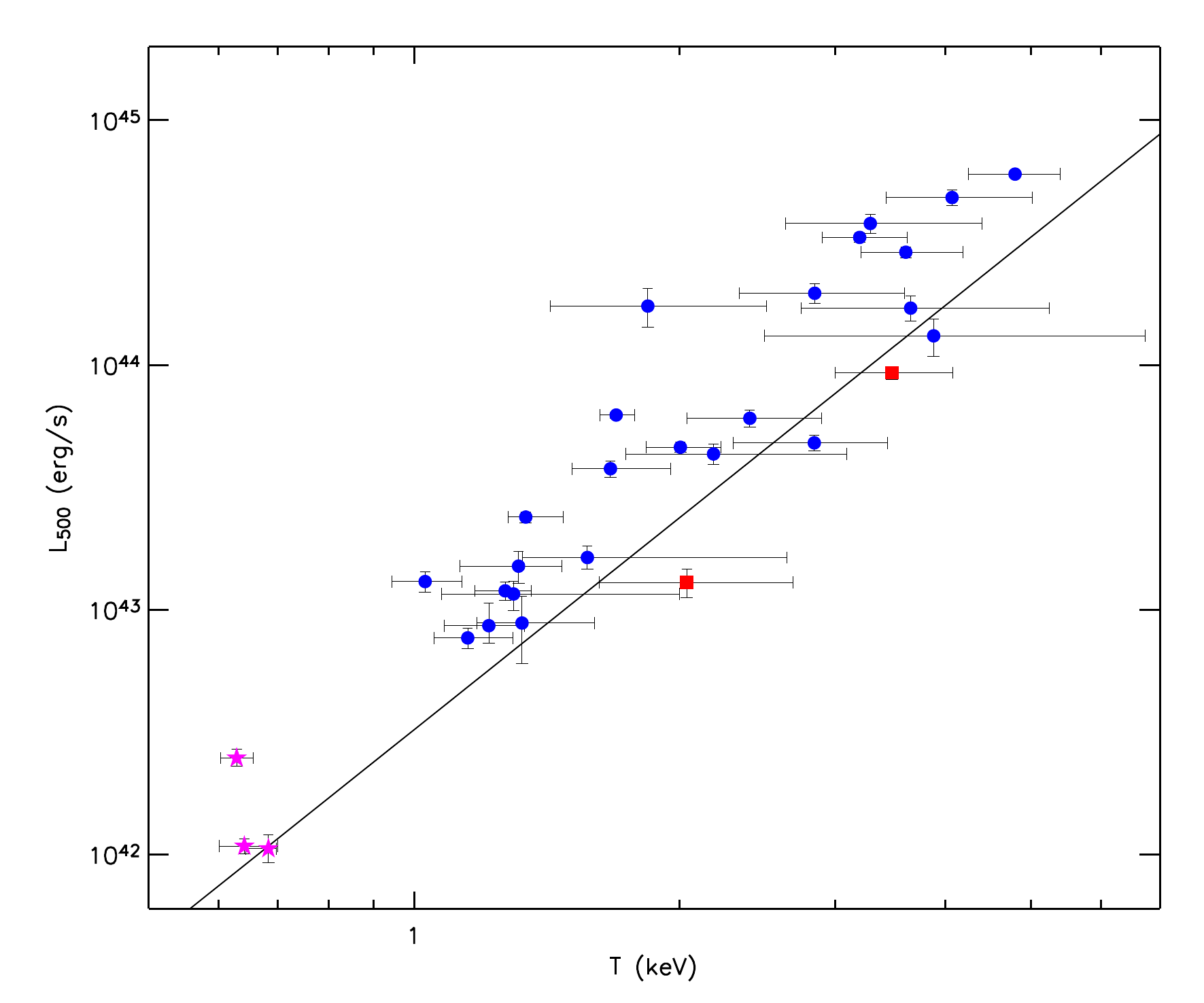}
    \caption{\label{rawlt}The $L_{500}$-$T$ relation.
    Same symbols as in Fig. \ref{M-L}.
     The sources that are
    used for the analysis of the $L_X-T$ evolution (circles) have
    redshifts in the range [0.14-1.05] and fall on average above the
    local $L_X$-T relation of \protect\cite{arnaudevrard} (shown by the
    solid line).}
    \end{center}
  \end{figure}

  The correlation between the observed luminosity and the temperature of our
  clusters is shown in Fig. \ref{rawlt}. The comparison with
  the local relation measured by \cite{arnaudevrard} suggests a positive
  redshift evolution of the luminosity at a given temperature.
  To test this, we performed an analysis of the enhancement
  factor, $F(z,T)=L_X/L_X(T,z=0)$, with respect to this local reference.
  In most of the intra-cluster gas models, $F$ is a simple function of
  redshift: the self-similar assumption yields $F(z)=E(z)$ where 
  $E(z) = H(z)/H_0$, is the evolution of the Hubble constant.
  More elaborate models that include non-gravitational physics (i.e. 
  cooling and heating) generally propose an evolution of the form  
  $F(z)=(1+z)^{\alpha}$ (e.g. \citealt{voit}).
  All these models assume that the entire cluster
  population follows a unique evolutionary track. In the present
  study,  we adopted this hypothesis, since the Malmquist bias
  prevents us from investigating the extent to which evolution 
  could be a function of cluster temperature.

    \begin{table}
    \begin{center}
    \begin{minipage}{7.5cm}
    \caption{\label{onePz}Trends in the literature for the
    modelling of F(z) as a power law of $(1+z)$. For comparison,
    constraints from the C1 sample were added, ignoring the
  selection effects.
    }
    \begin{tabular*}{7.5cm}{lc}
    \hline
    \hspace{.5cm}Ref. & power of (1+z) \\
    \hline
    \protect\cite{vikh02}           & $1.5\pm0.3$\\
    \protect\cite{Novicki}          & $2.1\pm1.1$\\
    \protect\cite{Ettori}           & $0.6\pm0.3$\\
    \protect\cite{lumb}         & $1.5\pm0.3$\\
    \protect\cite{Kotov}            & $1.8\pm0.3$\\
    \protect\cite{MaughanScaling}       & $0.8\pm0.4$\\
     C1 clusters with $0.1<z<0.4$ (14 sources)  & $2.3\pm0.8$\\
     C1 clusters with $0.4<z<1.1$ (10 sources)  & $1.3\pm0.5$\\
     All C1 clusters above $z=0.1$ (24 sources) & $1.5\pm0.4$\\
    \hline
    \end{tabular*}
    \end{minipage}
    \end{center}
  \end{table}

  To date, various attempts to constrain $F(z)$ as a power law in
  $(1+z)$ have yielded discordant results (see Table \ref{onePz}).
  Several explanations for this discrepancy have been invoked:
  poor statistics, deviations between the several available local
  reference relations, disparity in mass and redshift of the current high-z
  cluster samples (if $F$ is not a simple function of $z$), as
  well as biased samples.
  Although limited by photon statistics, the XMM-LSS
  cluster sample, with its well controlled selection function,
  provides an important opportunity to test the impact of
  the selection process in such studies.

  In a preliminary analysis, \cite{Williserr} measured a mean value
  $F=1.46$ from an  initial  XMM-LSS sample covering the $[0.30-0.43]$
  redshift range.
  Another estimate using the D1 subsample (\citealt{D1paper}, section
  3.4) also found a luminosity enhancement for the $z\sim0.3$ domain,
  and showed that it could be statistically significant only above
  $T\sim1.5$~keV.
  However, none of these studies fully addressed the impact of selection
  biases on the derived results.

  In the present article, involving a larger sample, we perform a
  self-consistent likelihood analysis of our data parameter space, for
  a given $F(z)$. We excluded from the
  analysis XLSSC-050 (contaminated by flares), XLSSC-018 (contaminated
  by an AGN), and the 3 local groups ($z\lesssim0.1$ and $T\lesssim1$~keV)
  for which our measurements could be affected by a large radial
  extrapolation and by X-ray emission from member galaxies.
  The available sample for this analysis is thus restricted to 24 objects.

 \subsubsection{The likelihood model}

\label{lhmodel}

 \begin{figure}
  \begin{pspicture}(0,0)(8,15.6)
    \vbox{\hspace{-0.3cm}\includegraphics[height=15.6cm]{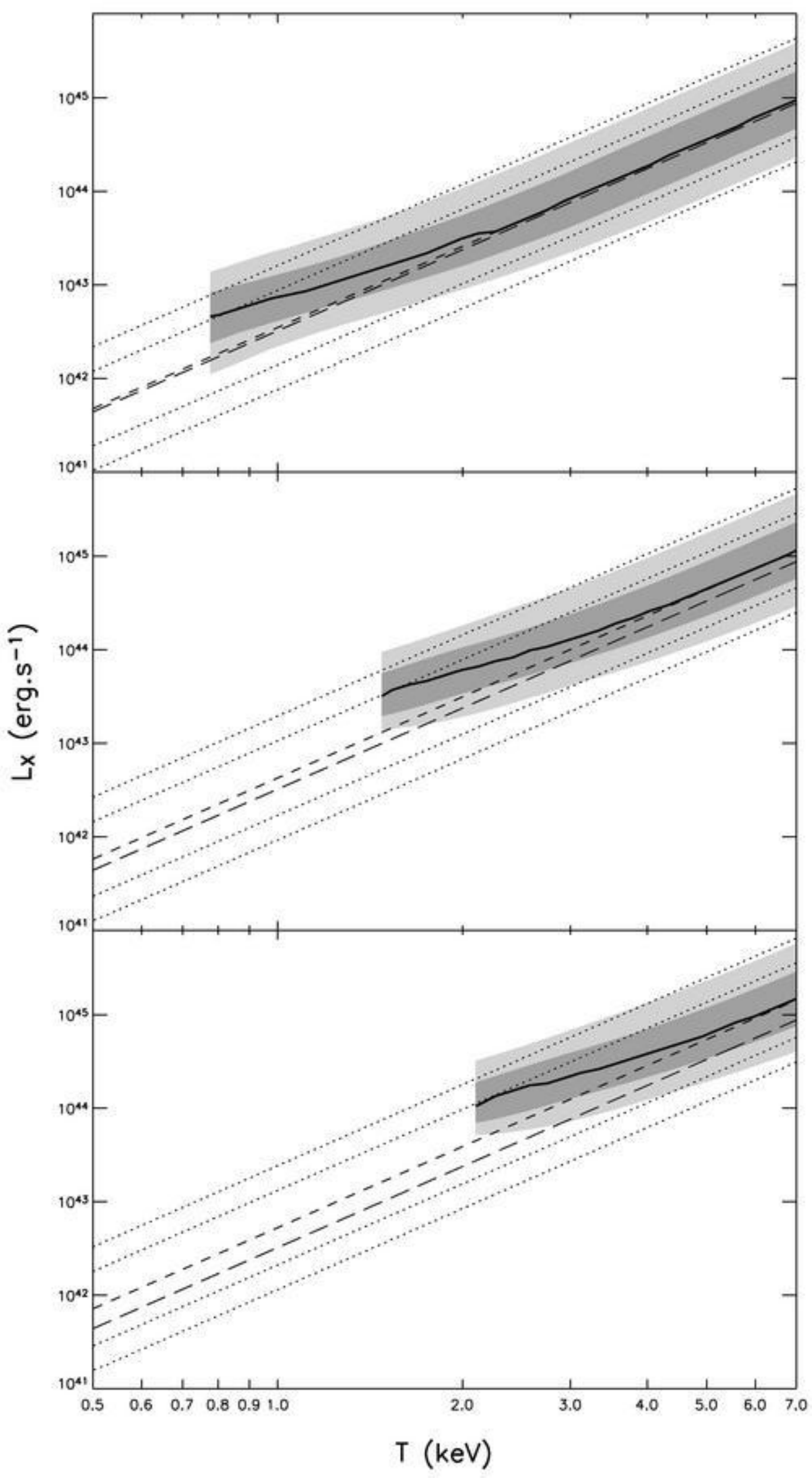}}
  \rput(-6.67,5.35){\large z\,=\,1.0}
  \rput(-6.67,10.18){\large z\,=\,0.5}
  \rput(-6.67,15.02){\large z\,=\,0.1}
  \end{pspicture}
 \caption{Comparison of the modelled $L_X$-T distribution
       with and without selection effects for several redshifts,
       assuming self-similar evolution.
       The light and dark shaded regions show, for each temperature, the
       luminosity interval than contains respectively 95.5 and 68.3\% of
       the expected detections. The solid line gives the maximum of the
       luminosity PDF for each temperature; it is cut at the
       temperature for which the detection probability (integrated
       over $L_X$) falls to 5\%.
       For comparison, the local \protect\cite{arnaudevrard} relation,
       evolved according to the self-similar model, is indicated by the short dash
       line, and the 1 and 2$\sigma$ bounds arising from our scatter model
       are shown by dotted lines. The long dash line is the local relation. \label{ltbias}}
\end{figure}

  Folding our selection function into the analysis of the $L_X-T$
  relation cannot be achieved by the usual $\chi^2$ fitting.
  Instead, we have to estimate the likelihood of our data,  given a $F(z)$ model.
  We thus  start by assuming  that cluster luminosities  are
  distributed by a log-normal probability density function (PDF)
  around the mean value $L_{mean}(T,z)=L(T,z=0)\times{F(z)}$, which we
  will call the true distribution:
  \begin{equation}
      \frac{dP_{true}(L_X|T,z)}{d\ln{L_X}}=\frac{1}{\sqrt{2\pi}\sigma_{\ln{L_X}|T}^{}}\exp{\left(-\frac{\ln{\left(L_X/L_{mean}\right)}^2}{2\sigma_{\ln{L_X}|T}^2}\right)}.
  \end{equation}
  An estimate of the scatter in the local $L_X$-T relation is provided by
  \cite{stanek}. The authors fitted a single dispersion value to the
  \cite{RB02} sample (thus assuming that it is the same from the group regime
  to large clusters) and obtained $\sigma_{\ln{T}|L_X}=0.25$. We  further assume that this scatter does not evolve with
  redshift. Combining the lognormal distribution with the power law shape of
  the mean $L_X-T$ relation, we get $\sigma_{\ln{L_X}|T}\sim2.7\times0.25\sim0.7$
  where 2.7 is the approximate $L_X-T$ slope in the \cite{RB02} sample.

  The normalised PDF for detection of a cluster with such $L_X$ and $T$
  is then the product of the `true' PDF with our selection function:
  \begin{equation}
     \frac{dP_{det}(L_X|T,z)}{d\ln{L_X}} = P_0(T,z) \times \frac{dP_{true}(L_X|T,z)}{d\ln{L_X}}\times f(L_X,T,z),
  \end{equation}
   where:
  \begin{equation}
    P_0(T,z) = \left(\int\left[\frac{dP_{true}(L_X|T,z)}{d\ln{L_X}}\times f(L_X,T,z)\right]\ d\left(\ln{L_X}\right)\right)^{-1}.
  \end{equation}
   We compute $f(L_X,T,z)$ (from Fig \ref{skycov}),  assuming a canonical value of 180~kpc for
   the cluster core radii.
   The resulting modification of the PDF is illustrated in Fig.~\ref{ltbias} at different redshifts assuming
   self-similar evolution: including selection effects in the
   likelihood model renders underluminous cool clusters undetectable.
   Note that for clusters with temperatures close to the detection threshold
   at any given redshift, the bias in mean $L_X$ at a given $T$ due to
   selection effects can be a factor 2 or more.

    \begin{figure*}
    \begin{center}
    \begin{pspicture}(0,0)(12,7.2)
    \hspace{-0.4cm}
    \hbox{
      \includegraphics[width=9cm]{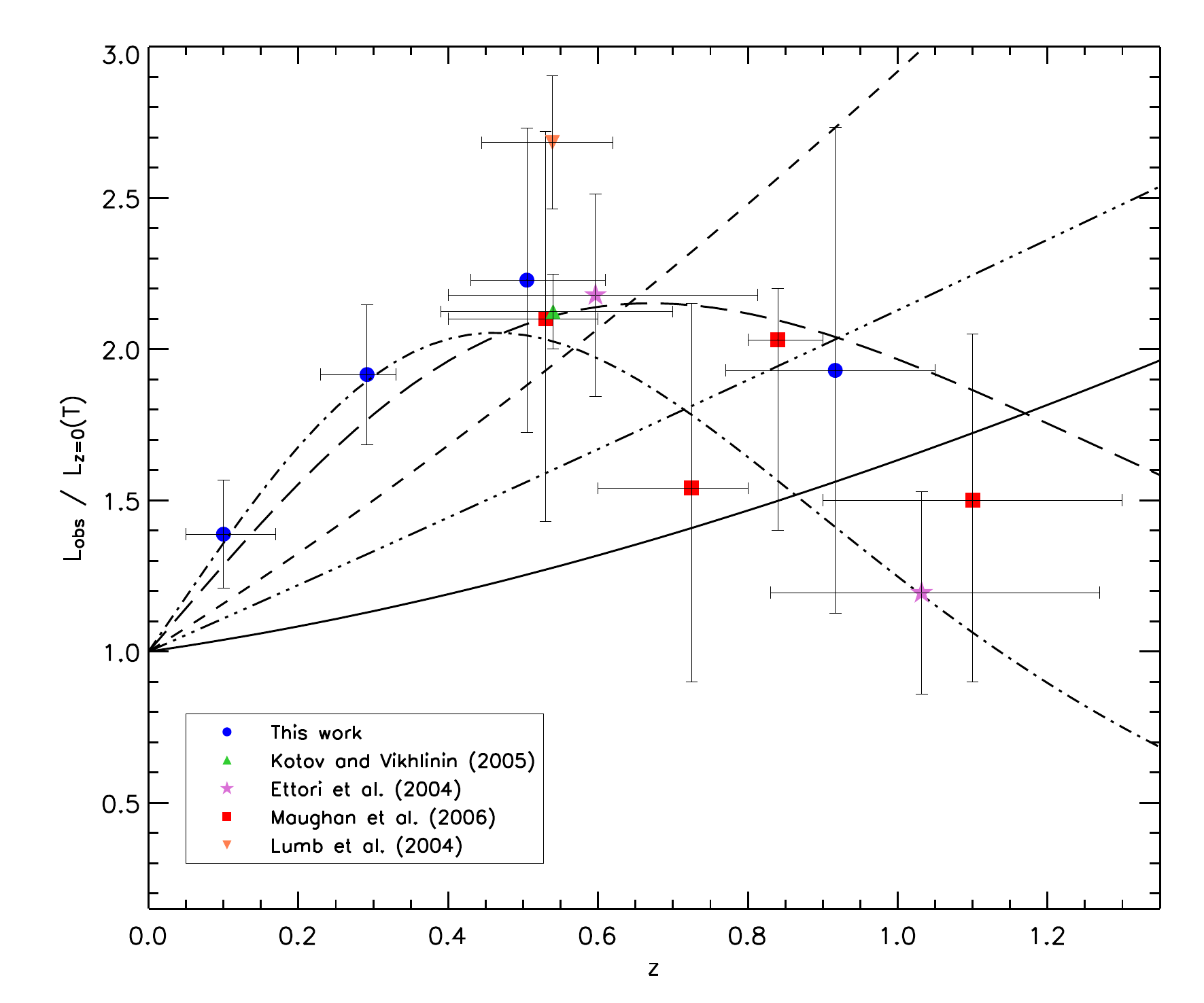}
      \includegraphics[width=9cm]{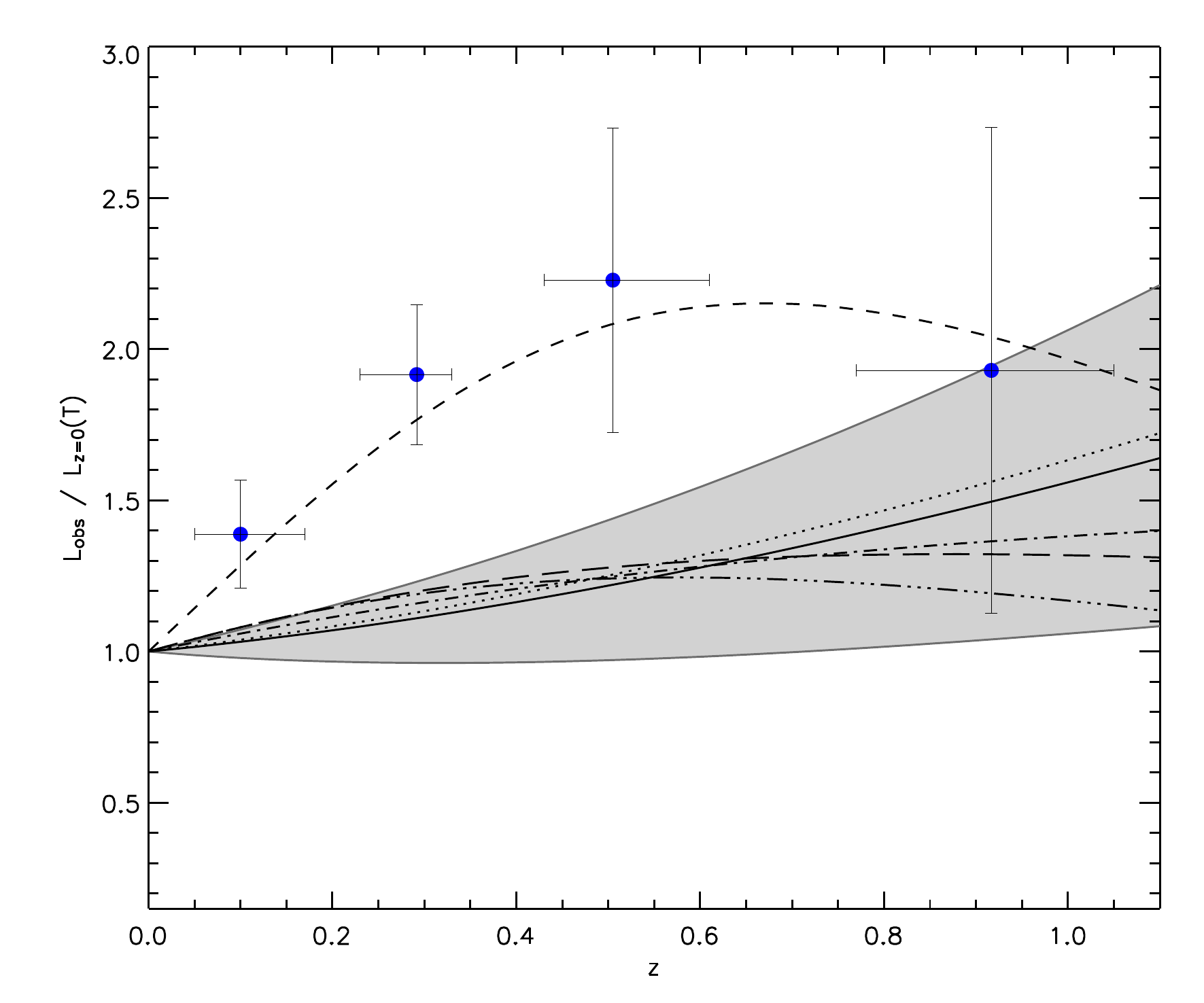}}
      \rput(-16.1,6.4){\large (a)}
      \rput(-6.96,6.4){\large (b)}
    \end{pspicture}
    \caption{\label{Levol}Evolution of the $L_X$-$T$ relation.
    {\bf(a)}: \underline{Raw analysis}. The data points from several studies are plotted (stacked
    into redshift bins for clarity). Whenever possible,
    we converted luminosity estimates from the other samples to
    $L_{500}$ using the information provided by the authors. The following
    differences remain: the luminosities of the
    \protect\cite{lumb} clusters are estimated within the virial radius;
    the data points of \protect\cite{MaughanScaling} also include the
    clusters from \protect\cite{vikh02} and have luminosities estimated
    within $R_{200}$. Overlaid are several enhancement factor fits
    from our baseline analysis: the (1+z) power law and ad-hoc 2
    parameter fits to our data alone (resp. short and long dashed lines), and
    the (1+z) power law and ad-hoc 2 parameter fits (resp. dot-dashed and
     3dot-dashed lines) fitted to the C1 clusters combined with
    the samples of \protect\cite{Kotov} and \protect\cite{Ettori}. 
    For comparison, the self-similar prediction is indicated by the solid line.
    {\bf(b)}: \underline{Taking into account selection effects}. The filled circles and short
    dashed line recalls the location of our raw data points, and best fit
    model from the preliminary analysis. The corrected enhancement
    factor fit for the 2 parameter model is shown as the long dashed line.
    The final 1 parameter fit and its $1\sigma$ confidence interval
    is displayed by the solid line and the shaded region. Expectations
    from several intra-cluster gas models are plotted for comparison:
    the self-similar predictions (dotted line) and two models
    by \protect\cite{voit}  including non gravitational physics (dot-dashed and
    3dot-dashed lines)
    }
    \end{center}
  \end{figure*}

  Practically, errors on the luminosity can be considered as negligible, since the fractional
  errors on temperature are usually much larger and the $L_X-T$ relation goes like
  $T^3$. The likelihood of an observed cluster C being drawn from a given $F(z)$ is
  thus obtained by marginalizing over the possible cluster temperatures, i.e.:
  \begin{equation}
      L(F|C) = \int_0^{+\infty} \left[\frac{dP_{det}(L_X|T,z)}{d\ln{L_X}}\right] \times P_C(T) dT,
  \end{equation}
  where $P_C(T)$, the temperature PDF of the cluster, is derived from the C-statistic
  distribution provided by the X{\sc{spec}}'s {\sc steppar} command.
  Finally, the likelihood of the enhancement factor  is computed as the product of the
  detection probabilities, for the given F(z), of all clusters pertaining to our
  sample:
  \begin{equation}
      L(F) = \prod_{i=0}^{N} L(F|C_i)
  \end{equation}

  To estimate errors on fitted parameters, we identify $-2\ln{L(F)}$
  with a $\chi^2$ distribution, as is asymptotically valid in the limit of
  large samples according to Wilks' theorem, and quote 68\% confidence
  intervals.
  The same identification is used to assess how much a given model deviates
  from the best fit $F(z)$ by correcting the measured $\Delta \chi^2$
  to the equivalent value that yields the same probability for 1 free
  parameter.

 \subsubsection{Raw fit}

  In a first step, we neglect the selection function in our formalism
  (i.e. effectively imposing $f(L_X,T,z)=1$ in the likelihood computation).
  This will both provide us with a reference point to assess the impact of
  the selection process, and allow for comparison with previous work,
  where the issue of sample selection has been ignored.

  As shown in Table~\ref{onePz}, fitting the usual $F(z)= (1+z)^\alpha$
  yields $\alpha=1.5\pm0.4$, which is consistent
  with  \cite{vikh02}, \cite{lumb} and \cite{Kotov}.
  Nevertheless, this simple model does not seem to correctly reproduce the
  observations  over the full redshift range: the values of $\alpha$
  obtained  over    $[z = 0.1-0.4]$ and
  $[z = 0.4-1.1]$ are incompatible at the one sigma level
  (see Table~\ref{onePz}). Interestingly, \cite{Ettori} noticed 
  a similar discrepancy within their own sample, but 
  did not consider its implications for their subsequent analysis.

  In Fig.~\ref{Levol}a, we display simultaneously our data
  points along  with those from previous studies. Error bars are large, but
  the points   suggest a non-monotonic evolution, with a maximum
  around $z\sim0.5$, thus excluding an enhancement factor of the form
  $F(z)=(1+z)^{\alpha}$.
  This may explain the diverging results of Table \ref{onePz}.
  Consequently, we fitted  an ad hoc two-parameter model of the form
  $F(z)=(1+z)^{\alpha}\times E(z)^{\beta}$.
  This model is intrinsically degenerate and would need very accurate
  data to individually constrain $\alpha$ and $\beta$, but has the ability
  to reproduce the apparent non-monotonic evolution.
  Since  cases $\alpha=0$,~$\beta=1$, and $\beta=0$ correspond to the
  self-similar and power law models respectively, we may evaluate through
  this parametrization how much the observations deviate from any of
  them.

  A fit over the C1 clusters yields $\alpha=4.7$ and $\beta=-5.4$ and excludes
  the self-similar evolution at the $3.5\sigma$ level. Our best $(1+z)^\alpha$
  model is however less than $0.4\sigma$ away from the best fit and the
  evidence of a non-monotonic evolution is thus rather weak from our data
  alone (the probability of such a deviation being real is roughly 45\%).
  This is not the case any longer when including also in the fit the
  data from \cite{Kotov} and \cite{Ettori}\footnote{We assumed a gaussian
  PDF  of the cluster temperature for these sources in our likelihood
  model}, for which we could convert the luminosity to $L_{500}$.
  A simple power law model is then ruled out at the $8.4\sigma$ level
  based on 59 sources.

\subsubsection{Taking into account selection effects}

\label{lhfitsel}
  With this comparison baseline, we can now turn to investigate the impact
  of our selection process on the observed $L_X$ enhancement (restoring the
  correct value of $f(L_X,T,z)$ in the likelihood).
  This task is tractable here because the XMM-LSS C1 selection function
  has been thoroughly assessed.

  Maximizing the likelihood over our ad hoc two-parameter model now
  yields $\alpha=1.3$ and $\beta=-1.3$. As can be seen in Fig.~\ref{Levol}b,
  the inferred evolution rate is considerably lower than with the raw fit.
  As a result, the self-similar expectation is now less than
  $7\times10^{-2}\sigma$ away from the best fit model, and the statistical
  evidence for a non-monotonic evolution is virtually null.
  To evaluate the remaining deviation from self-similarity, we switch back to
  a one parameter model by fixing $\beta=1$. The best fit value for $\alpha$
  is then $-0.07^{+0.41}_{-0.55}$, and the corresponding $F(z)$ is shown in
  Fig.~\ref{Levol}b.

  This latter analysis points to a negative evolution of the $L_X$-T 
  relation (i.e. $L(z|T)$ increases with $z$) over  the $0<z<1$  range. 
  These results quantitatively favour the self-similar hypothesis, although
  our current data set is still marginally consistent with no evolution at 
  all as is clear from Fig.~\ref{Levol}b.
  
\begin{figure}
       \begin{center}
       \includegraphics[width=8.8cm]{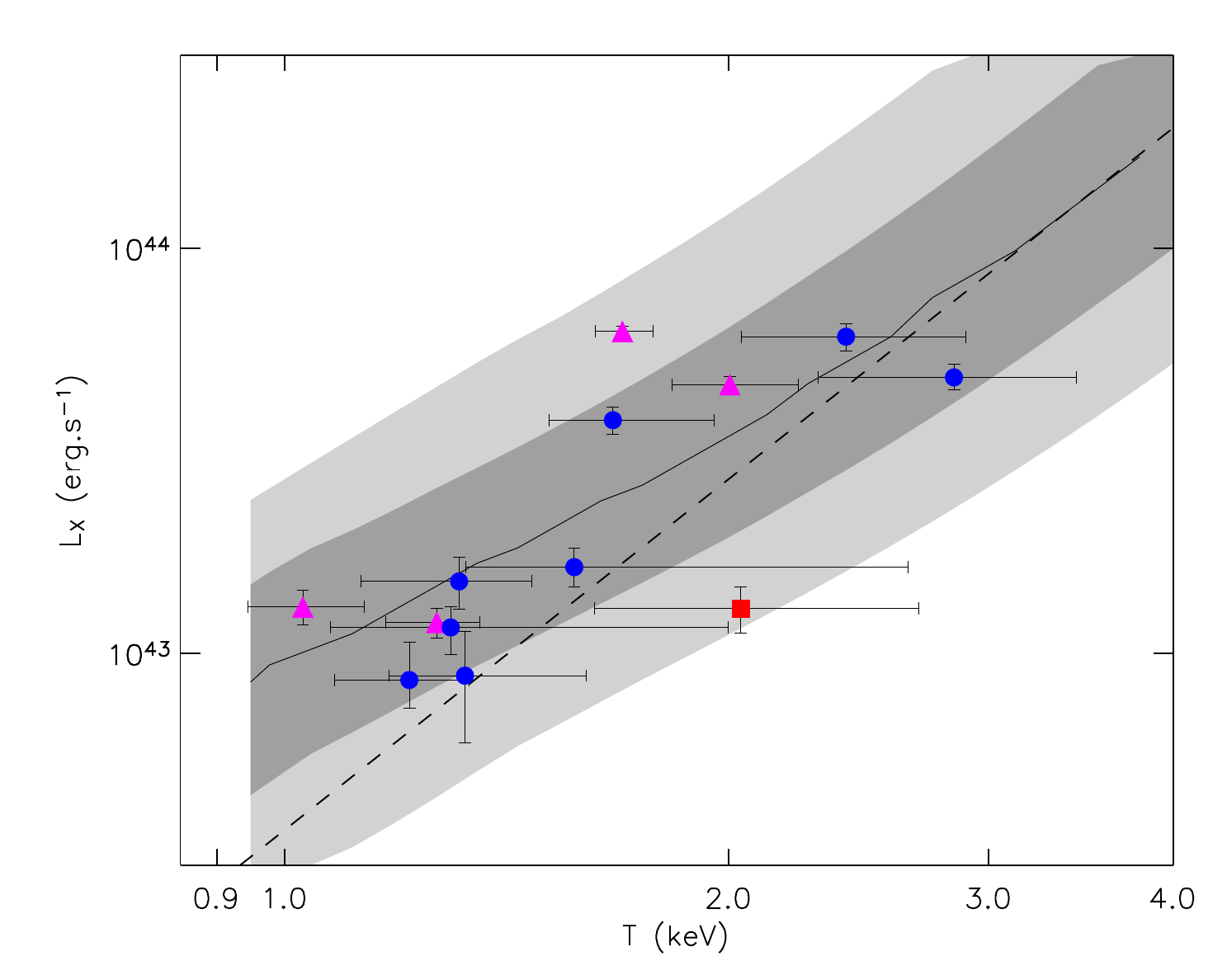}
       \caption{\label{zpoint3}Similar to figure \protect\ref{ltbias} for $z=0.3$,
       with the $0.2 < z < 0.4$, with C1 clusters overlaid. Again, the dashed
       line is the evolved model of  \protect\cite{arnaudevrard}.
       The  triangles indicate the D1 clusters presented by \protect\cite{D1paper}.
       The only cluster (XLSSC-018, square) that falls well below the
       \protect\cite{arnaudevrard} $L_X$-T relation, and at the limit of our
       $2\sigma$ contour, is likely to be contaminated by an AGN (see
       Appendix \ref{C1app}).}
       \end{center}
  \end{figure}

  This is illustrated in more detail in Fig. \ref{zpoint3} for the
  $z\sim0.3$  intermediate redshift,  where the 1-2~keV groups make up
  the peak of the XMM-LSS sensitivity (cf. section \ref{cosmo}).

  These results are still preliminary because of the poor
  statistics (in both the number of sources and the temperature estimates).
  Moreover, they are likely to be very sensitive to the choice of the local 
  $L_X$-$T$ reference (including scatter) due to the redshift-temperature 
  correlation within the sample.
  Nevertheless, our findings seriously question any attempt to assess
  the evolution of cluster scaling laws without modelling the full
  source selection process.

  In the rest of the article, we assume that the cluster scaling laws evolve
  following the self-similar model, as suggested by the above analysis.

\section{COSMOLOGICAL MODELLING}

\label{cosmo}
In this section, we compare the observed properties of the C1
sample with cosmological expectations based on the latest WMAP
constraints \citep{wmap}.

\subsection{The model}

\label{modeldesc}

We model the cluster population and its evolution  following
\cite{pipeline}, but with slightly modified assumptions as to the
cluster scaling laws.

The linear power spectrum $P(k)$ of dark matter fluctuations is
computed using an initial power law  of index $k^{n_s}$ and the
transfer function from \citet{BBKS}, taking into account the shape
parameter:
\begin{equation}
  \Gamma=\Omega_m h\times exp\left[-\Omega_b\left(1 + \sqrt{2h}/\Omega_m\right)\right]
\end{equation}
introduced by \citet{sugi}. The overall power spectrum is normalised 
at z=0 to $\sigma_8$, and the redshift evolution is obtained from the
linear growth factor \citep{carroll}. Based on this power
spectrum, we use the \citet{Jenkins} formalism to derive the dark
matter halo mass function $dn/dm(z)$. This procedure is an
accurate fitting of the mass function obtained from numerical
simulations, provided one defines the mass of the halo to be that
included inside $r_{200b}$, the radius  enclosing an overdensity
of 200 with respect to the mean background density.

In order to reproduce the C1 selection function, we need to
translate the mass parameter into observable parameters, which is
practically achieved by means of scaling laws. Usually, halo
masses in such relations are defined with respect to the critical
density of the Universe (and not to the background density).
Hence, the need to assume a model for the halo profiles to connect
the two mass definitions. For this purpose, we used NFW profiles
\citep{NFW} with a scaling radius $r_s$ provided by the model of
\citet{bullock} which relates $r_s$ to the virial mass of the halo
through the concentration parameter $c=r_{vir}/r_s$. The
 $M_{500}-M_{vir}$ conversion was obtained from
the formulae provided in the Appendix C of \citet{HuKravtsov},
with the estimate of the overdensity within the virial radius from
\cite{KitSut}.

The gas temperature within the dark matter halos is then
computed using the $M_{500}$-$T$ relation of \citet{arnaud2005},
i.e. a power low of slope $\alpha=1.49$, valid for clusters with
$T>4$~keV. To account for the steepening of the relation at lower
temperatures suggested by their data, we used a higher slope of
$\alpha=1.9$ below 3~keV. The lower halo mass limit is then fixed
by imposing $T~>~0.5$~keV. Bolometric luminosities are derived
using the $L_X$-$T$ relation of \citet{arnaudevrard}. Self-similar
evolution is assumed for both these scaling laws.

To account for the scatter  observed in cluster properties, we
encapsulate the dispersion of the M-T and L-T relations in the M-L
relation, for the sake of simplicity. We assign  the  X-ray
luminosity assuming a log-normal distribution. Following
\cite{stanek}, who measured $\sigma_{ln{M}|L}=0.37$, we use
$\sigma_{ln{L}|M}\sim0.37\times1.59\sim0.6$ where 1.59 is the
slope of their $M-L_X$ relation.
The total XMM-NEWTON EPIC count rate is estimated using the same
spectral model as in section \ref{spec}, with a fixed hydrogen
column density of $2.6\times 10^{20}$~cm$^{-2}$ (representative of our
field), folded through the EPIC response matrices for the THIN
filter in accordance with our observing mode. The  selection
function (Fig. \ref{skycov}) is finally applied assuming, as in
the previous section, a constant physical core radius of
$180$~kpc.

In what follows, given the still modest size of the current
cluster sample (some 30 objects over 5 deg$^2$) we restrict most
of the comparisons to qualitative visual ones.

\subsection{The redshift distribution}

Our  model (using $\sigma_8$ = 0.74, from WMAP3) predicts $6.2$ Class 1
clusters per square degree which, assuming Poisson noise alone
(i.e.$\pm 1.1$ for the current 5.2 \dd), is compatible
with our observed density of $5.4$ deg$^{-2}$ as
given by the objects listed in Table \ref{cluslist}.

\begin{figure}
    \begin{center}
    \includegraphics[width=8.6cm]{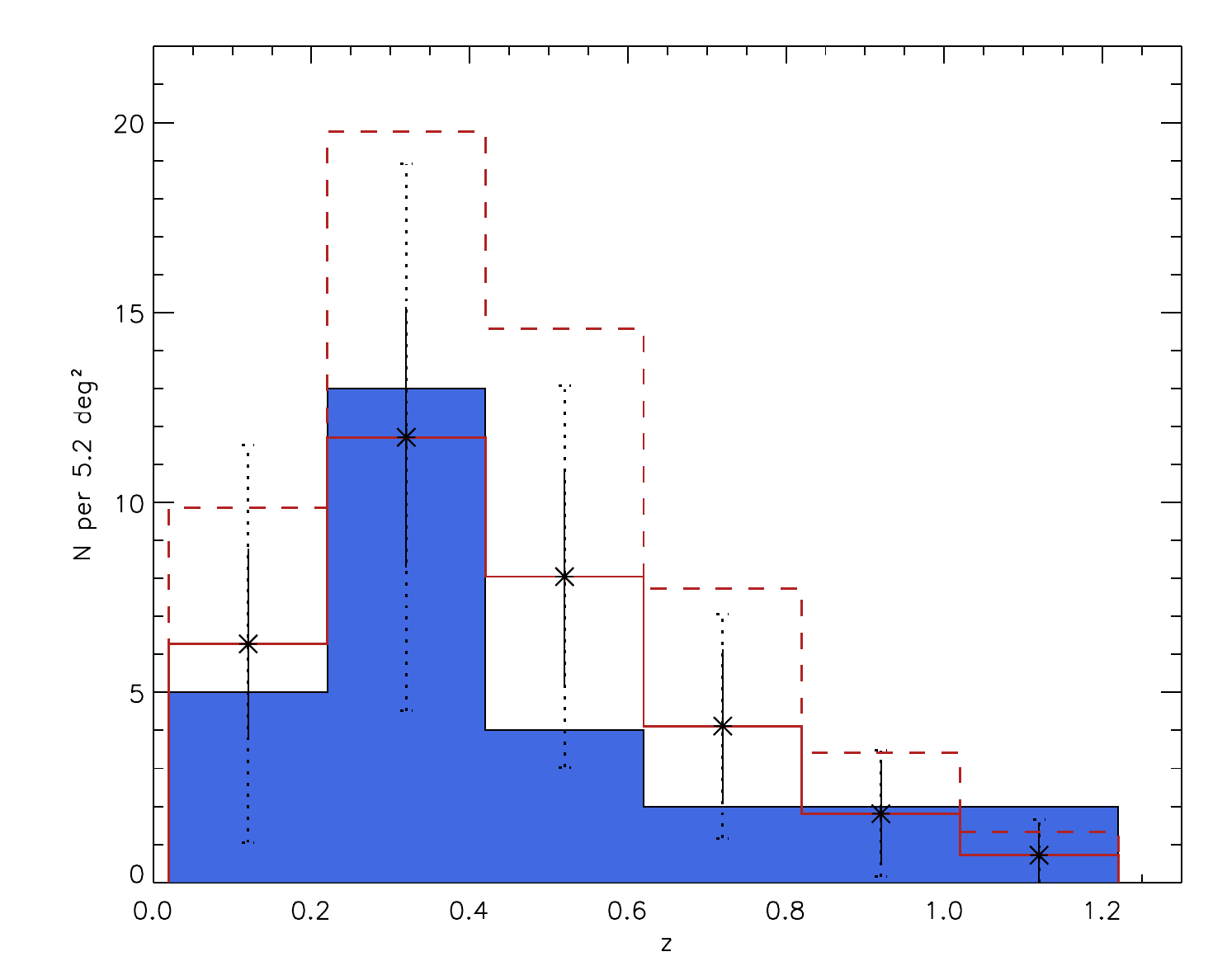}
    \caption{Redshift distribution of the C1 sample. The filled
    histogram shows the redshift distribution of our sample,
    while the solid line shows the expectations of our cosmological
    model (WMAP3: $\sigma_8 = 0.74$ and self-similar evolution for the 
    $L_X-T$ relation). Expected fluctuations around the mean model density
    due to shot noise and (shot noise + cosmic variance) are respectively
    displayed with the solid and dotted error bars. The dashed
    line shows the expectations for a model with WMAP $1^{st}$ year
    cosmological parameters ($\sigma_8 = 0.85$)   and a non-evolving
    $L_X-T$ relation; combining  this latter $\sigma_8$ value with 
    self-similar evolution would produce a normalisation even higher,
    by about 35\%.
     \label{dndz}}
    \end{center}
  \end{figure}

The observed redshift distribution of the C1 sources, shown in
Fig.~\ref{dndz}, shows good overall agreement with the model 
expectations, and suggests that we are crossing a void-like
region within $0.4<z<0.6$. 
As is evident from the errors in Fig.~\ref{dndz}, this apparent 
underdensity can be fully accounted for by statistical plus 
cosmic variance.
We estimated the cosmic variance using the formalism from 
\protect\cite{HuKravtsov}, under the assumption that the surveyed 
volume is enclosed within a top-hat sphere;
in the sensitivity regime of the XMM-LSS survey, the total uncertainty
on the $N(z)$ bins scales roughly as $2\sqrt{N}$.

Interestingly, the C1 cluster density predicted assuming 
$\sigma_8=0.85$ (as inferred by \citealt{wmap1} from the $1^{st}$
year WMAP data) is 14.6 deg$^{-2}$ and appears to be much higher
than required to match our observations. In this case, even using
a non-evolving $L_X-T$, which is roughly our  $1\sigma$ lower bound
from the previous section, our model predicts 10.7 C1 clusters per
deg$^2$. Our data are 3.7$\sigma$  (considering
Poisson fluctuations) below this latter model. Even including cosmic 
variance, the difference is approximately 2$\sigma$.
This suggests that increasing the area by a factor of 2 would already 
permit a stringent test of the value of $\sigma_8$ indicated by WMAP. 

In this context, we examine in detail in Sec.\ref{modelnz} some
parameters which play a role in the interpretation of the cluster number 
counts.

 \subsection{The Flux and luminosity distribution}
 
As our selection function does not depend on flux alone,
as assumed in previous generations of surveys, we have no direct 
estimate of the log(N)-log(S) relation. In this section, we provide 
the necessary information to enable the inter-comparison 
between our statistical distributions and those of past as well
as future surveys.

\begin{figure}
    \begin{center}
    \includegraphics[width=8.6cm]{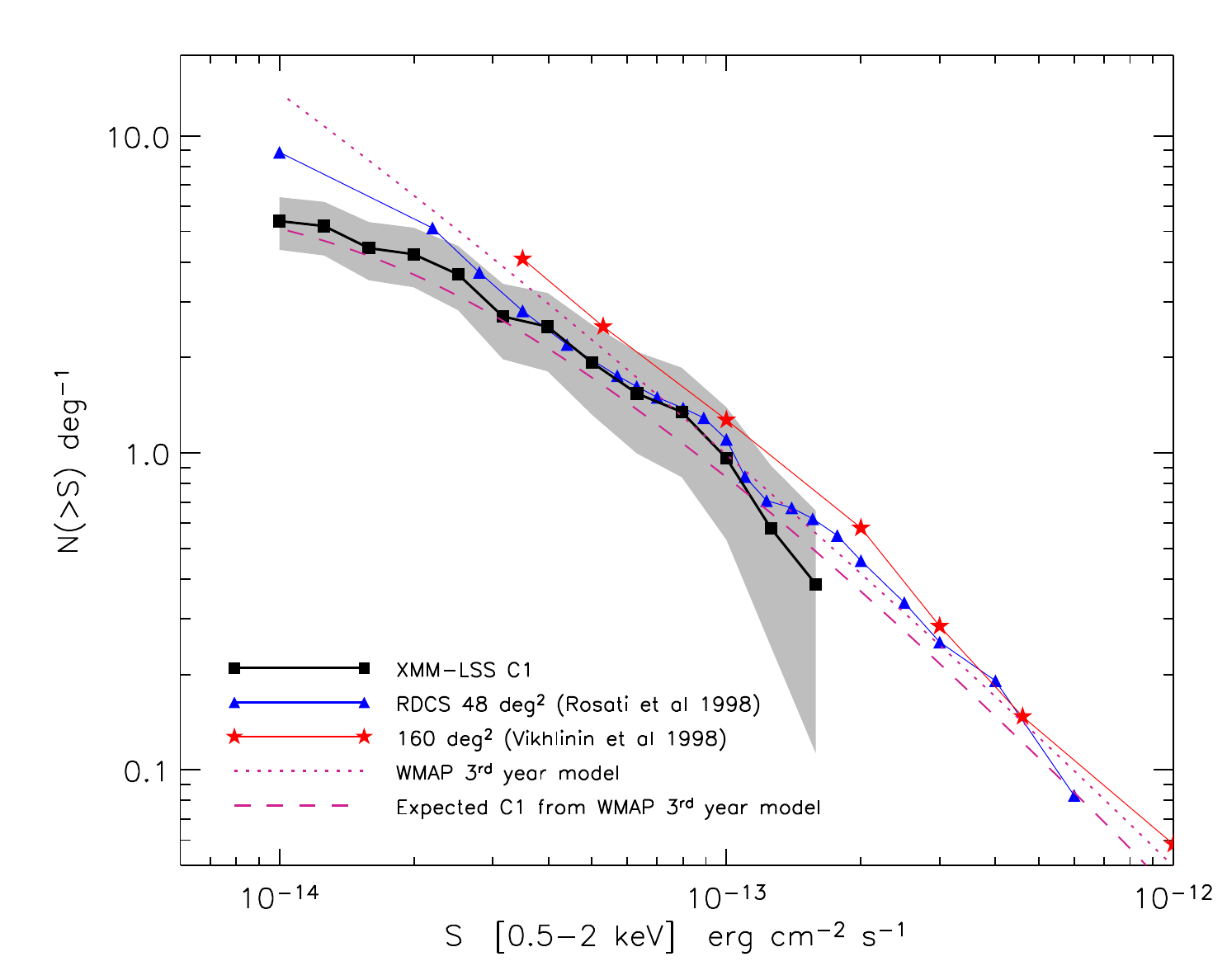}
    \caption{The C1 log(N)-log(S). The squares show the observed C1 number counts
    and the shaded region indicates the $1\sigma$ error region.
    Overlaid are the predictions from our model: the dashed line for C1
    clusters, and the dotted line for the whole cluster population
    (i.e. all clusters above a given flux).
    Since only a fraction of cluster fluxes are detected in 
    any survey, there is an important caveat in this comparison: 
    our model delivers total cluster fluxes 
    (i.e integrated to infinity) while our measurements are performed within $0.5~$Mpc 
    (which includes   $\sim 2/3$ of the total flux for a cluster with
    $R_c=180$~kpc and $\beta=2/3$).
    Previously, \protect\cite{160sqdeg} claimed to measure total
    fluxes in an unbiased manner, while \protect\cite{RDCS} estimated that
    they had recovered at least 80-90\% of the total flux.
    \label{lnls}}
    \end{center}
  \end{figure}

Our cosmological model can be both compared to the local log(N)-log(S),
and used to predict an expected distribution for our C1 sample . 
This is shown in Fig. \ref{lnls}. 
The raw log(N)-log(S) from our model is first compared with the 
estimates from \cite{RDCS} and \cite{160sqdeg} based on flux 
limited samples, and said to be corrected for selection effects 
as well as for incompleteness.
We find overall consistency, although our model suggests a 
somewhat higher total cluster density than the RDCS 
measurements above a flux of $\sim10^{-14}$ \flux. 
Furthermore, the predicted flux distribution, once folded with 
the C1 selection function, is consistent within the errors with 
the C1 log(N)-log(S) relation derived from the present sample.  
This adds credibility to our full selection process and 
log(N)-log(S) modelling.

\begin{figure}
    \begin{center}
    \includegraphics[width=8.6cm]{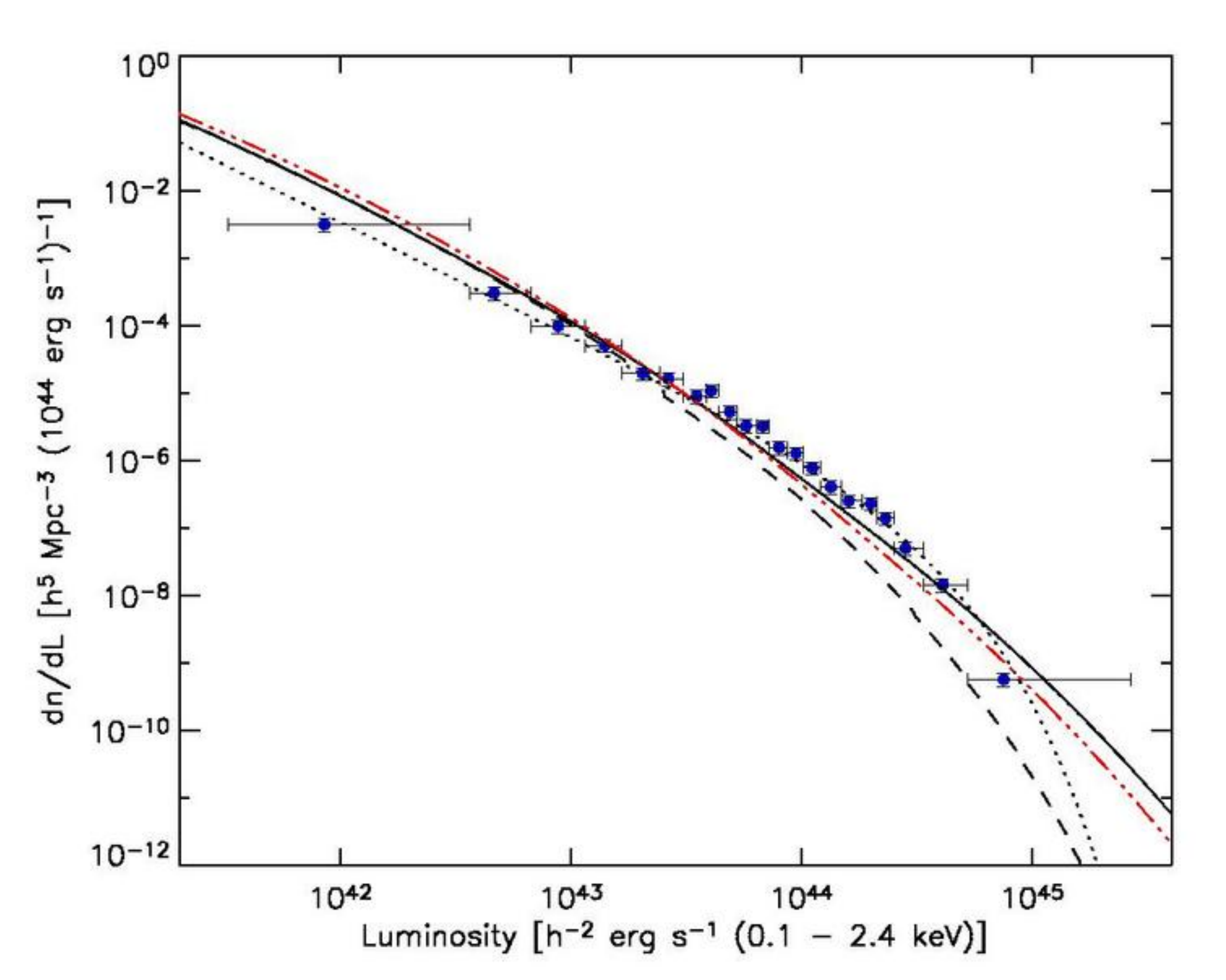}
    \caption{The local luminosity function predicted by our model, compared
     with the REFLEX data points \citep{reflex}.
    The dotted line is the best fit to the REFLEX data from the
    original article. The dashed line is the expectation from
    our model  assuming no scatter in the $M-L_X$ relation.
    The solid line shows  our model (including scatter)
    yielding a much better agreement. The 3dot-dashed line is the prediction of our model
    for a redshift of 0.8.\label{nL}}.
    \end{center}
  \end{figure}

Similarly, our sample is currently too small to derive the cluster 
luminosity function at any redshift, but this again can be predicted 
by our model. 
We first compute as a cross-check the local luminosity function and 
compare it with the estimates from the REFLEX survey (\citealt{reflex}), 
which constitutes the largest complete cluster sample currently available 
at low-redshift (Fig. \ref{nL}).
The model shows, on average, good agreement with the measured REFLEX 
luminosity function (which accounts for the survey sky coverage). 
It lies slightly below the REFLEX measurements for luminosities in the range
$5\times 10^{43}$-$5\times 10^{44}$\lum, behavior that is also present in the 
ROSAT Bright Cluster Sample \citep{ebeling00} as can be appreciated 
in the comparative plots of \citet{mullis04}.
We further emphasize that, as is conspicuous in Fig.~\ref{nL}, the 
inclusion in the model of scatter in the $M-L_X$ relation is 
necessary in order to obtain predictions that are compatible with the 
REFLEX constraints at the high $L_X$ end.
Finally, we note that the prediction of our model at $z=0.8$ (also 
displayed in Fig.~\ref{nL}) is very similar to the local one out 
to $L_{X} = 10^{44}$\lum, and about a factor of 2 lower at 
$L_{X} = 10^{45}$\lum\ (although still within the REFLEX error bars). 
Since the EMSS \citep{gioia90}, the question of the evolution of 
the cluster luminosity function has been controversial, especially 
for luminosities above $5\times 10^{44}$\lum\ [see a review by
\citet{mullis04} including several ROSAT samples]. 
The most recent determination of the cluster luminosity function 
for $L_{X}<10^{44}$\lum\ from the COSMOS-XMM data shows no evolution
\citep{finoguenov07}. Our calculations suggest that, interestingly, 
for a concordance cosmology, the observed luminosity function does 
not significantly evolve out to $z\sim0.8 for L_{X} < 10^{44}$\lum, 
while the cluster mass-function and scaling laws do. 
This can be regarded as a `conspiracy' between cosmology 
and cluster physics.

  \subsection{Mass-observable relations}

\begin{figure}
    \begin{center}
    \includegraphics[width=8.6cm]{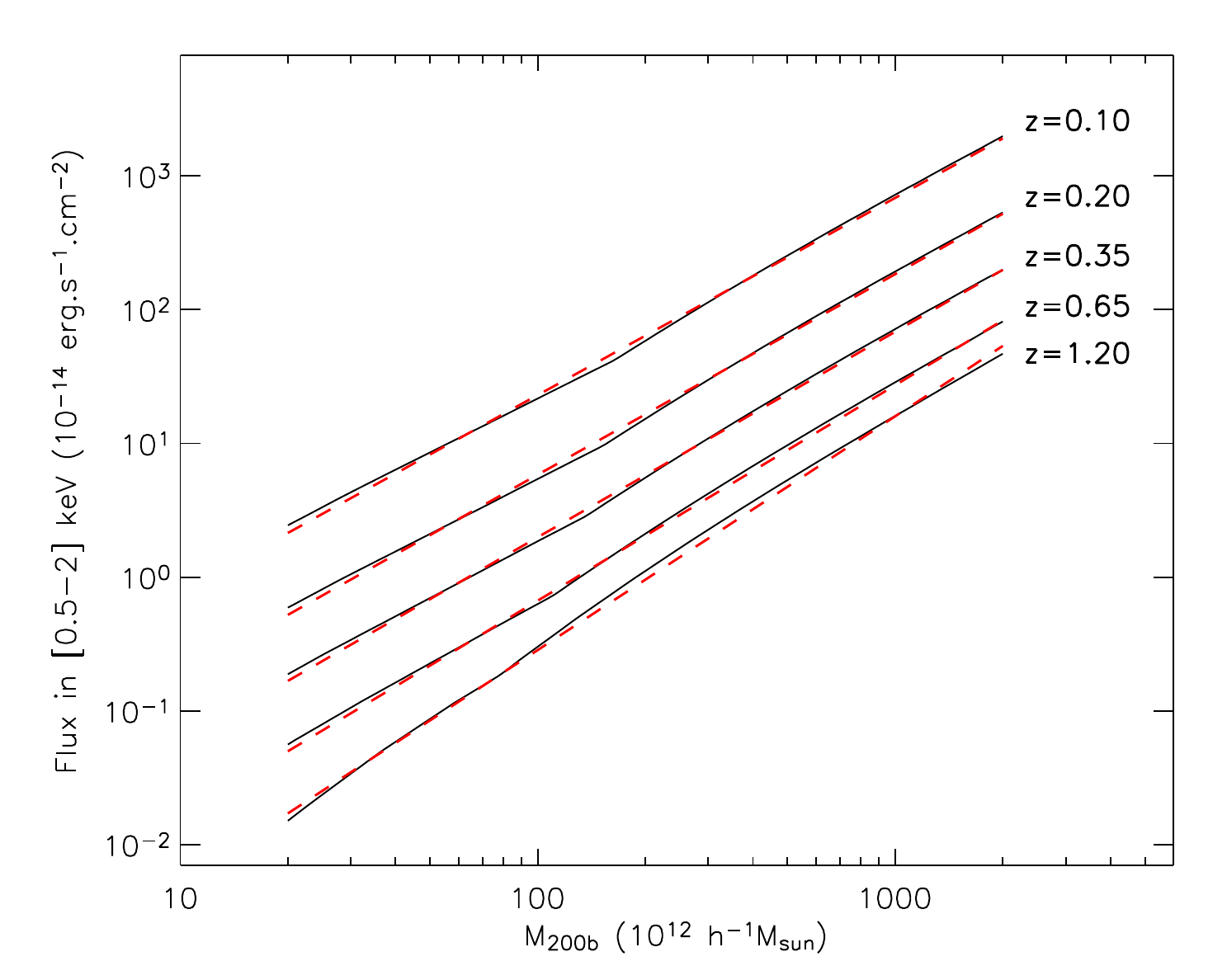}
    \caption{Calibration of the [0.5-2]~keV band total flux vs $M_{200b}$
    relation. The exact correlations derived from our model are shown for
    several redshifts with plain lines. The overplotted dashed lines
    show the recovered values using the fit presented in
    Table~\protect\ref{Mfit}.\label{M-F}}
    \end{center}
  \end{figure}

  Last but not least, our model provides us with a tool for deriving heuristic
  mass-observable relations.
  For this purpose, we adopt the very general parametrisation  introduced
  by \cite{Hu}:
  \begin{equation}
    OBS\ =\ OBS_0\left(\frac{M_d}{M_0}\right)^{p(z)}\times e^{A(z)}\label{Mf},
  \end{equation}
  where $M_d$ is the mass defined within some overdensity $d$, $OBS$ is the
  observable of interest (flux, count-rate) and $M_0$ and $OBS_0$ are reference values for both
quantities. In practice, we used $M_0=10^{12}~h^{-1}M_\odot$ and
$M_d = M_{200b}$ i.e. the mass enclosed within the radius
delineating an overdensity of 200 with respect to the mean
background density. Model data points for a large range of masses,
redshifts and fluxes (count-rates) were fitted with the above
formula as illustrated on Fig. \ref{M-F}. We found that the
functions $p(z)$ and $A(z)$  are sufficiently well described by
functionals of the form:
  \begin{equation}
     \left\{
     \begin{array}{l}
     p(z) = p_0 + p_1\,z\\
     A(z) = a_0\,z^{a_1}.\\
     \end{array}
     \right.
     \label{APz}
  \end{equation}

  \begin{table}
  \begin{center}
  \begin{minipage}{8.3cm}
  \begin{center}
    \caption{Parameters of the mass-observable calibration from our
cosmological model, as defined by equations \ref{Mf} and
\ref{APz}.   \label{Mfit}}
    \begin{tabular}{c|cc}
    \hline
    Parameter   &
    mass-flux relation\footnote{For a total source flux in the [0.5-2]~keV band in \flux} &
    mass$\,-\,$count-rate relation\footnote{For a MOS1+MOS2+PN count-rate in the [0.5-2]~keV band in \countr} \\
    \hline
    $OBS_0$       & $0.0329$  & $1.23\times10^{10}$\\
    $a_0$       & $-37.68$          & $-36.50$\\
    $a_1$       & $0.0645$          & $0.0680$\\
    $p_0$       & $1.45$            & $1.41$\\
    $p_1$       & $0.248$           & $0.269$\\
    \hline

    \end{tabular}
  \end{center}
  \vspace{-.2cm}
  \end{minipage}
  \end{center}
  \end{table}

  The best fit values for the Mass vs Count-rate and Mass vs Flux relations
are provided in Table \ref{Mfit}.  We do not give errors on the
parameters as the overall  accuracy of both fits with respect to
the model data is better than 15\% over the $0.05<z<1.2$ and
$2\times10^{13}<M<2\times10^{15}~h^{-1}M_\odot$ ranges. 
In practice, the intrinsic dispersion of the M-L relation also needs 
taking into account. 
Assuming $\sigma_{ln{L}|M} \sim0.6$ (Sec. \ref{modeldesc}) this 
translates into a dispersion of -45\% +82\% in the Flux$-$ or 
Count-rate$\,-\,$Mass relation. 
Thus ``perfect'' flux measurements may yield, via this formula, mass 
accuracies of the order of -60\% +100\%.
Such an empirical relation has obvious  useful practical usages. 
It should be noted however, that extrapolating the formulae in the 
current form above $z>1.2$ is not straightforward as a number of 
prominent low temperature lines ($O_{VII},~ Fe_{VII}, ~ Fe_{VIII}$) 
are redshifted below the 0.5~keV boundary, creating discontinuities 
that cannot be accounted for with the above simple functionals.

\section{Discussion}

We have shown that an accurate understanding of the sample
selection effects is essential for a proper study of the evolution
of the L-T relation of clusters of galaxies. Once these are allowed for,
our data are compatible with self-similar evolution but
cannot exclude no evolution at all. Subsequently, we have defined
a cosmological cluster evolutionary model to which we applied the
survey selection function.  It matches well our observations in
terms of cluster number density, redshift and flux distribution, 
hence providing a theoretical basis to further explore the 
ability of large XMM cluster surveys to constrain cosmology.

In this respect, one of  the critical issues is the role and the
number of cluster parameters involved in the cluster scaling
law evolution.  As a first step, we have shown that for our best
fitting model, it is possible to derive good  mass-observable
relations for the entire population of interest.  We realistically
explore below the impact of various parameters and hypotheses on
the cluster scaling laws, on  the selection function and,
subsequently, on cluster number counts. In this way we hope to
pave the way for future investigations.

  \subsection{Cluster shape parameters}

  Our selection function, as applied to our model,  assumes
  a fixed core radius of $R_c=180$~kpc.
  Since $R_c$ is related to the overall physical size of the
  system, one naturally expects this value to depend on the 
  cluster mass (or on any related parameter). 
  This should ideally be included in the cosmological analysis, 
  but we are currently lacking a well established $R_c-T$ relation. 
  The slope reported by \cite{Ota2004} is, for instance, much 
  stronger than the one observed in the \citet{sandersonp03} 
  sample. 
  \cite{Ota2002} also revealed that, due to cool cores, the core radius 
  distribution infered by performing a single $\beta$-model analysis 
  over the whole cluster population is actually double peaked and thus 
  ill-defined.
  
  Furthermore, another known observational feature of X-ray groups and
  clusters is the tendency of the $\beta$ parameter to drop with
  decreasing system temperature (see e.g. \citealt{JoFor}, or more 
  recently \citealt{OsmPon}) but this correlation remains poorly 
  defined, specially in the group regime.
  As our selection function is mainly surface brightness limited, we
  expect  the detection probability to depend on $\beta$ as well as 
  $R_c$ and so, two additional scaling relations are needed.

  One could in principle derive them from the observed sample,
  but, given the generally small number of collected photons,
  it is not possible to resolve precisely the $\beta-R_c$ degeneracy
  for most of  our sources. Furthermore, such a study would be complicated by
  the fact that our selection process is dependent on these parameters, thus
  requiring a sophisticated self-consistent modelling similar to what was done in
  Section~\ref{ltsec} for the $L_x-T$ relation.

  Given the small number of systems in the present sample, we
  postpone such a study to a future paper  investigating the
  evolution of mean cluster/group profiles by stacking images
  (Alshino et al in prep.)

  \subsection{The evolution of the $L_X-T$ relation}

The   constraints we were able to put on the  evolution of the L-T
relation ($\alpha=-0.07^{+0.41}_{-0.55}$ for the
$F(z)=E(z)\times(1+z)^\alpha$ model) are not only limited  by the
relatively modest size of the sample and the uncertainties on the
temperature measurements, but  also depend on the assumed
scatter of the relation (this parameter was fixed in the fit).

  \subsubsection{Impact of the assumed scatter value}
\label{scatterLT}

In order to quantify  the impact of the scatter in the
determination of the evolution of the L-T relation, we ran a
series of fits assigning different values to
$\sigma_{\ln{L}|T}$. We find that the $\alpha$ index describing
the deviation from self-similarity
 can be modelled by the following empirical formula:
  \begin{equation}
    \alpha = 1.78 - \left(\frac{\sigma_{\ln{L}|T}}{0.45}\right)^{1.41}\label{evolVSsig}.
  \end{equation}

The uncertainty on $\sigma_{\ln{T}|L}$ of $\sim 25\%$  quoted by
\cite{stanek} along with the $L_X-T$ slope uncertainty (say
between 2.5 and 3) yield an uncertainty of $\sim  20\%$ for
$\sigma_{\ln{L}|T}$ (same reasoning as in Sec. \ref{lhmodel}). 
This translates to a conservative confidence range of 
$[0.5-0.9]$ for $\sigma_{\ln{L}|T}$, which in turn gives fitted 
values of $\alpha$ in the range $[0.6,-0.87]$.
The corresponding deviation is thus larger than the statistical errors
from our fit, and precise constraints on $\sigma_{\ln{L}|T}$ are
definitely required in order to place firm and rigorous constraints on
$F(z)$. Given that the scatter in the $L_X-T$ relation could well
be a function of redshift and mass, one will have to wait for
large samples of high-z clusters with well monitored selection
effects in order to  undertake a fully self-consistent
determination of the evolution of the slope and scatter of the
$L_X-T$ relation.

  \subsubsection{Impact of the sample size and of the uncertainties on the temperature}

We have further investigated the extent to which the determination of
the slope of the evolution of the $L_X-T$ relation is conditioned
by the size of the cluster sample and by the magnitude of the
error bars on the temperature (in comparison, errors on the
luminosity are assumed to be negligible). We have thus simulated a
large number of random cluster catalogues corresponding to 5, 20,
and 64 \dd\ XMM-LSS type surveys. The realisations are drawn from
our cosmological model, reproducing our observed $n(z,T)$
distribution along with selection effects; self-similarity is
assumed for the $L_X-T$ evolution. Each cluster temperature is
assigned a mean error estimated from the spectral fitting
simulations presented in the appendix of \citet{Willis}, for   10
ks XMM exposures.  In addition, we have considered the possibility
of dedicated deep XMM follow-up observations providing a
temperature accuracy of 10\% for cluster subsamples of various
sizes. For each sample, the slope was fitted following the method
described in Sec \ref{lhmodel} and \ref{lhfitsel}. Results are
summarized  in Table \ref{simula}.

\begin{table}
  \begin{center}
  \begin{minipage}{8.3cm}
  \begin{center}
    \caption{Expected accuracy in the determination of the evolution of
    the $L_X-T (z) \propto E(z)(1+z)^{\alpha}$ relation for
    various sizes of XMM-LSS type surveys, estimated from simulations.
    The temperature measurements come either directly
    from the 10 ks XMM exposures, with corresponding accuracy. Or, part of them
    are improved by subsequent deep XMM observations, assuming an
    uncertainty of 10\%; the last
    column gives the number of such clusters to undergo deep XMM pointings. $\sigma_{\alpha}$
    is the mean 1$\sigma$ error on $\alpha$  for a survey realisation. \label{simula}}
    \begin{tabular}{c|ccc}
    \hline
     Area & Temperature accuracy & $\sigma_{\alpha}$&  $N_{re-obs}$ \\
     \hline
     5 \dd\ & from the survey & 0.59& -\\
     20 \dd\ &  "  " & 0.28 & -\\
     64 \dd\ & " " & 0.15 & -\\
     \hline
     5 \dd\ & 10\% for the $z>0.8$ clusters & 0.58 & 2 \\
    20 \dd\ &  "  " &0.25& 14 \\
    64 \dd\ & " " & 0.14& 41 \\
    \hline
     5 \dd\ & 10\% for the $z>0.5$ clusters & 0.50 & 9 \\
    20 \dd\ &  "  " &0.22& 45 \\
    64 \dd\ & " " & 0.13& 139 \\
    \hline

    \end{tabular}
  \end{center}
  \vspace{-.2cm}
  \end{minipage}
  \end{center}
  \end{table}

 The simulations show that a significant improvement can be
reached by increasing the sample size as the accuracy on $\alpha$
scales roughly as the square root of the surveyed area. This is a
very noticeable result, given that 2/3 of  our clusters have no
more than 500 counts available for the spectral fit. The spectral
accuracy is of the order of 20\% below 2~keV, and 50\% around 5~keV
\citep{Willis}, the latter concerning mostly distant clusters. The
simulations further show that increasing the accuracy of the
temperature measurements  does not yield a very significant
improvement on the slope of the $L_X-T$ relation - compared to the
amount of time that would be necessary to obtain accurate
temperatures for the $0.5<z<1$ hot clusters. This is mostly due to
the fact that the dispersion in the $L_X-T$ relation itself is
large. Quantitatively, applying the $\sqrt{area}$ rule of thumb on
Fig. \ref{Levol}b, shows that a 10 \dd\ area (or any sample with a
similar size and controlled selection effects) has the potential
to exclude the non evolution hypothesis. But even with 600
clusters (i.e. a 64 \dd\ area, and 140 very well measured clusters
above $z>0.5$) it seems difficult to, for example, discriminate 
between the two modified-entropy models of cluster evolution 
proposed by \citet{voit}.

\subsubsection{Working on samples close to the detection limit}

Fig. \ref{zpoint3} shows that for the $0.2<z<0.4$ range, most of
our clusters are detected in a $L_X-T$ region where the selection
effects are significant, as is to be expected, given that lower
mass halos are much more numerous than massive ones. As a result,
overluminous clusters will be overrepresented in our sample.
For the present data set, this is also true at any
redshift, and this situation is easily understandable as a result of
the combined effects of (1) the high efficiency of {\sc Xamin}
close to the detection limit (it was designed to enable the
construction of the largest possible controlled cluster sample),
and  (2) the steepness of the cluster mass function. Generally
speaking, given the contradictory former results obtained from
data based on heterogeneous ROSAT-selected samples, our findings
suggest that graphs such as Fig. \ref{zpoint3} should be first
carefully constructed, when studying the cluster $L_X-T$ relation
at any redshift.

  \subsection{Modelling the cluster number density}
\label{modelnz}

While most of the  ingredients linking cosmology to the expected
cluster distribution are rather well established (mass functions, mean
scaling relations for massive clusters) many details of the
mass-observable issues are still pending. Among these, we would
cite : the evolution of cluster scaling laws, the scaling laws in
the group regime, and the role of the scatter in all relations. Our
analysis has shown that our current data are compatible with
self-similar evolution for the L-T relation\footnote{ In this
respect we stress that combining a fully self-similar $M-T$ 
relation with a $L_X-T$ relation whose slope is not the one 
predicted by the self-similar model (only gravitational physics) 
implies that the $L_X-T$ and $M-L_X$ relations cannot simultaneously
evolve in a self-similar way. The point seems to have been 
overlooked in the cluster detailed studies so  far}. 
In this section we thus focus on two other specific aspects, 
namely the slope of the M-T relation for the groups and the 
scatter in the M-L relation. We illustrate below the sensitivity 
of the cluster number counts to these quantities by means of our 
model.

  \subsubsection{The M-T relation in the group regime}

\begin{figure}
    \begin{center}
    \includegraphics[width=8.6cm]{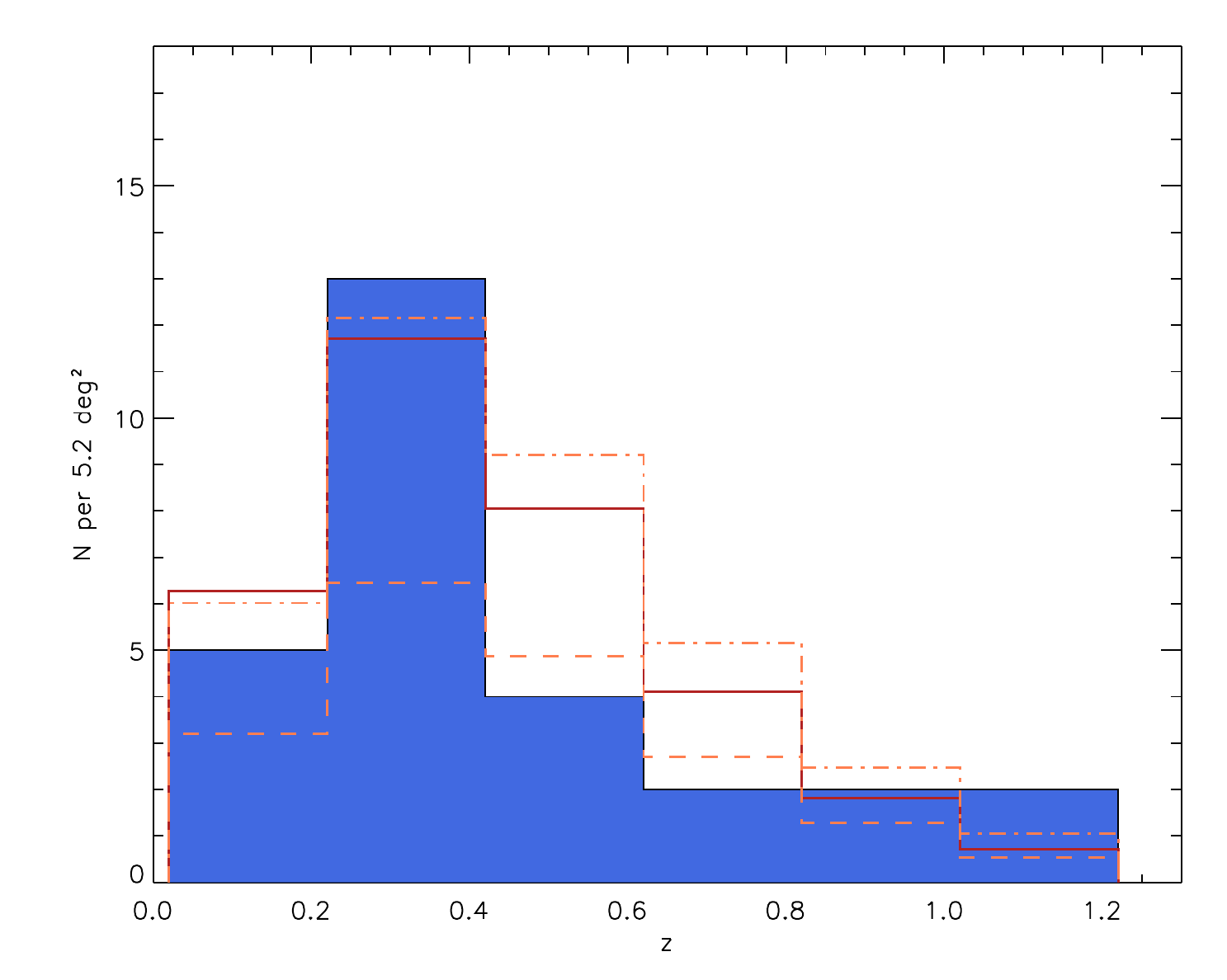}
    \caption{Impact of the steepening of the $M-T$ relation in the
    group regime on the modelling of the cluster number density.
    The full line corresponds to our fiducial model (broken power law, Sec. \ref{modeldesc}),
    giving 6.2 clusters per deg$^2$.
    If  we assume the \protect\cite{VikhMT}
    result (single slope of 1.5, dash line), the cluster density drops to 3.6 deg$^{-2}$.
    The relation of \citet{arnaud2005}, motivated by a possible steepening
    of the mass function below 3.5~keV, (single slope of 1.7, dash-dot line)
    seems to be closer to our observations and yields a cluster density of
    6.9 deg$^{-2}$.\label{steep}}
    \end{center}
  \end{figure}

Our model assumes that the $M-T$ relation for both massive clusters 
and groups can be modelled by a broken power law.
Recent measurements [\cite{VikhMT} and \cite{arnaud2005}] agree 
very well in the high mass regime, but yield contradictory results 
for groups. 
\cite{VikhMT} claimed that the previously reported steepening 
of the relation (by e.g. \citealt{FRB}) resulted from a bias 
toward low mass, due to an incorrect modelling of the density 
profile at large radius. 
However, their assertion relied on only two low-temperature systems. 
On the other hand, \cite{arnaud2005} found a steepening at low 
temperature based on a larger sample of groups, but their analysis 
assumes a NFW profile at large radius. 
As this point is currently still controversial, we adopted an 
intermediate solution: a  steepening to 1.9 of the relation
below 3~keV (note that a change in the   $L_X-T$ relation for
groups would have similar impact).  
Fig.~\ref{steep} illustrates how the various hypotheses
on the shape of the $M-T$ relation in the low-mass regime impact
on the cluster number counts. Our assumed broken power-law model provides the
best fit on the basis of visual inspection, but the difference
to the \cite{arnaud2005} model is not statistically
significant with the current sample size.

  \subsubsection{Impact of the scatter}

\begin{figure}
    \begin{center}
    \includegraphics[width=8.6cm]{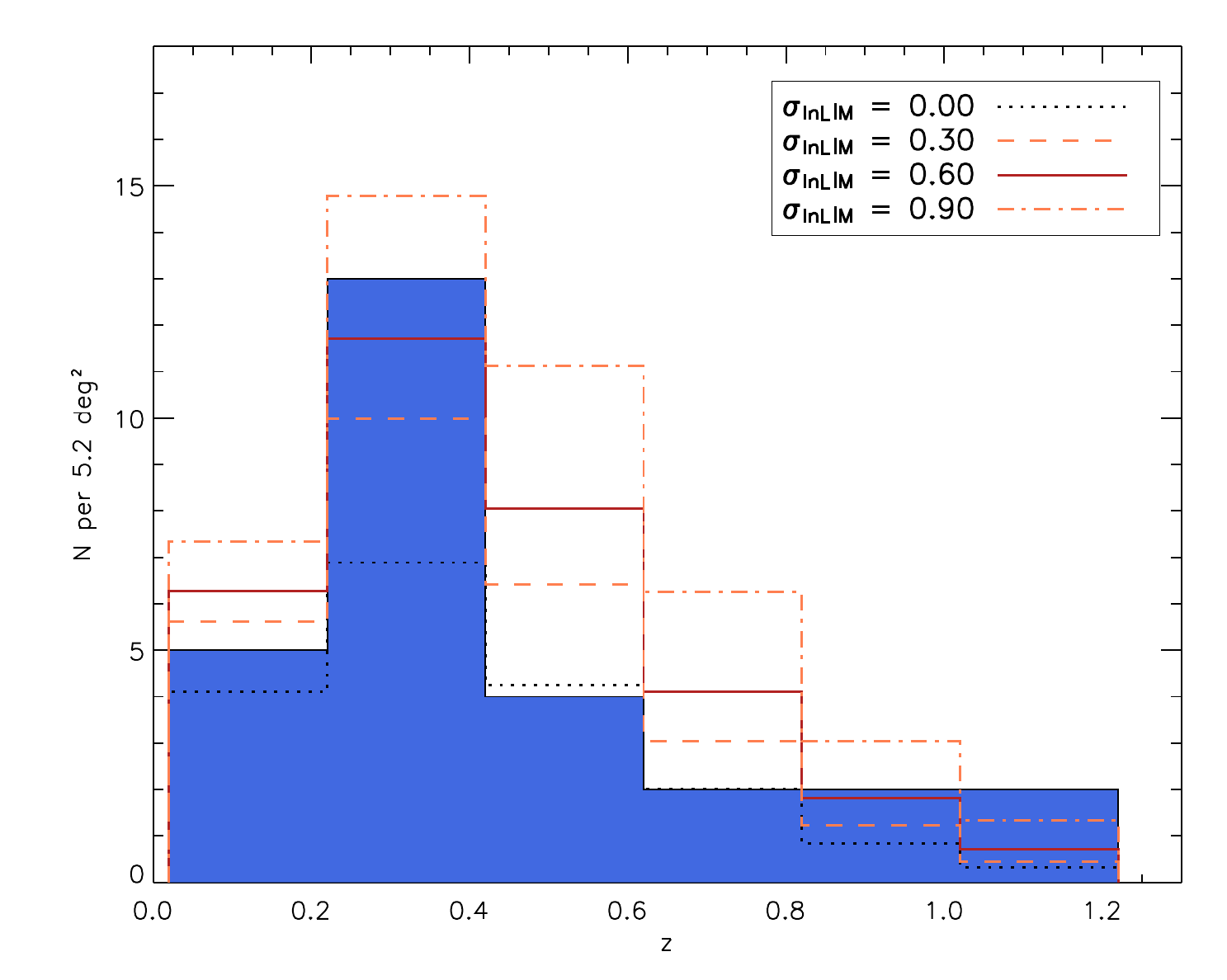}
    \caption{Impact of scatter in the $M-L_X$ relation on the
    predicted C1 redshift distribution. For given scatter values, the predicted
    total number of clusters per deg$^2$
    are respectively : 3.6 (no scatter),
    5.1 (0.3), 8.4 (0.9); our model gives
    6.2 (0.6) .
    \label{sca}}
    \end{center}
  \end{figure}

We discussed above how the scatter in the $L_X-T$ relation influences
the determination of the evolution of the relation from the
observations (Sec. \ref{scatterLT}) . In our model, all sources of
dispersion are  encapsulated in the $M-L_X$ relation, since
the determination and understanding of dispersion in
$L_X-T$ and $M-T$, and thus  $M-L_X$,  and of their possible evolution,
is still in its infancy. In this respect, we note
that measurements of scatter other than those of \cite{stanek} are available:
lower values for $\sigma_{\ln{L}|M}$ have been found by \citealt{Zhang2006} and
\citealt{Zhang2007} (respectively 0.41 and 0.33), but they are
dominated by higher temperature systems which are known to have more
regular properties. In terms of the accuracy with which the scatter
is determined, the typical errors estimated by \cite{stanek}
lead to an uncertainty in $\sigma_{\ln{L}|M}$  of the order of 20\%.

In Fig.~\ref{sca}, we show the impact on the C1 redshift distribution
of  neglecting the scatter in $M-L_X$, as well as the influence of 
plausible values of scatter. 
If the luminosities are symmetrically scattered around the expected 
value for a given mass, the net effect on our cluster counts 
(approximately the number of clusters detected
above a given luminosity  at a given redshift) is an increase of
the  detected cluster density since, due to the steepness of
the mass function, substantially more low-mass clusters become
detectable than high-mass cluster undetectable.

\subsection{Cosmology  and cluster survey self-calibration}

\begin{figure}
    \begin{center}
    \includegraphics[width=8.6cm]{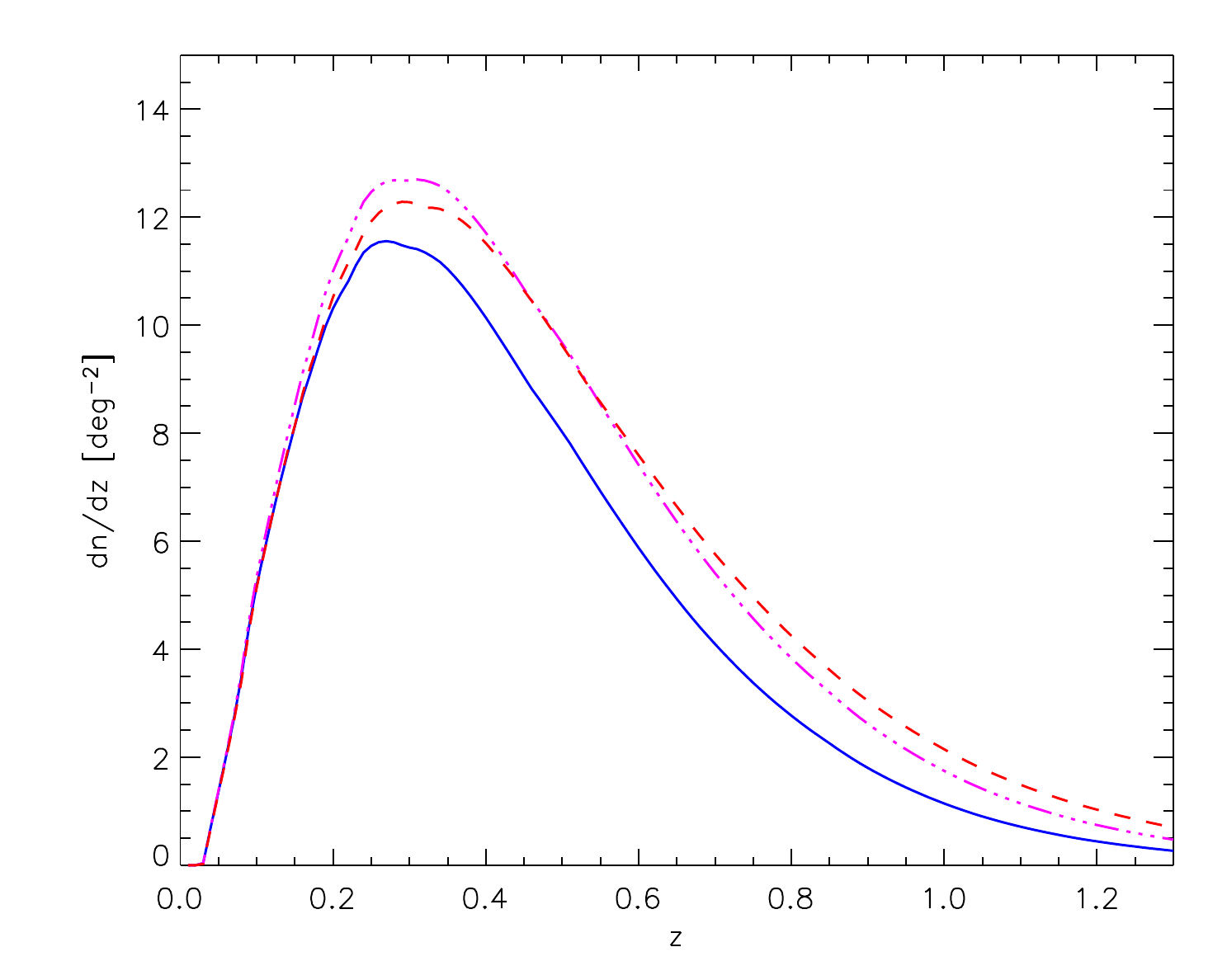}
    \caption{Dark energy vs scaling laws for our modelled cluster
    reshift distribution. The solid line shows the prediction for
    the \protect\cite{wmap} cosmology and self-similar evolution of
    the scaling laws. Assuming instead $F(z)=E(z)\times(1+z)^{0.34}$
    for the $L_X$-T evolution, which is our $1\sigma$ higher bound
    from section \ref{ltsec}, yields the dotted-dashed line. As a
    comparison, the dashed line shows the distribution obtained by
    keeping self-similar evolution but changing the dark energy
    equation of state from -1 to -0.6.\label{wevol}}
    \end{center}
  \end{figure}

We have demonstrated that taking into account survey selection
effects is not only critical for the proper modelling of the
cluster number counts (hence cosmology) but also for the
determination of the   cluster evolutionary scaling laws, which in
turn, also impact  on $dn/dz$ (Fig. \ref{dndz}). Further, we
explored the impact on $dn/dz$ of the slope of the $M-T$ relation
(Fig. \ref{steep}) and of the dispersion in the scaling laws (Fig.
\ref{sca}). This illuminated,  by means of  real data, the
magnitude of some of the degeneracies between cosmology and
cluster evolution. However, the number of cosmological free parameters
that we have explored in the present study is very limited
(basically, only $\sigma_8$, Fig. \ref{dndz}) and a more general
analysis should, among others, include constraints on the Dark
Energy parameter $w$. For instance, in Fig. \ref{wevol} we
exhibit the predictions of our model for different values of $w$
and $\alpha$ (evolution parameter of $L_X-T$). A clear degeneracy
is apparent, adding to those already reviewed.

In order to cope with this critical issue, it has been suggested
that X-ray (or SZ) cluster number counts, in principle, have the
ability to {\em self-calibrate} even an evolving mass-observable
relation, because the number density of clusters as a function of
redshift has a fixed functional form given by cosmological
simulations (\citealt{levine02};  \citealt{Hu}; \citealt{majumdar03}).
This suggests the possibility of simultaneously solving for the
cosmology and fitting an ad hoc parametrized mass-observable relation, 
provided that a sample of several thousands of clusters is available.

While our - still somewhat phenomenological - best fitting 
model, allowed us to derive  flux$-$ and count-rate$\,-\,$mass 
relations, several caveats still have to be pointed out. 
We have noticed that the flux-mass relation is very sensitive to 
the details of the X-ray emitting plasma (e.g. emission lines) and
also, obviously, to the way fluxes are estimated  in shallow XMM
exposures; in earlier studies, even the temperature-dependence 
of the bremsstrahlung continuum was ignored (e.g. \citealt{majumdar03}). 
Even neglecting measurement uncertainties, our mass$\,-\,$observable 
mapping exhibit $1\sigma$ uncertainties in the range [-60\%,+100\%] 
coming solely from the dispersion of the $M-L_X$ relation.

In this respect, while vastly increasing the number of clusters
may be useful, additional independent observations on
cluster evolutionary physics will constitute a crucial input. This
will improve the modelling of the mass-observable relations, and thus
the constraints on the mass function. In particular, deeper XMM
pointings will provide accurate temperatures and, consequently, a
better estimate of the dispersion in the L-T relation. Weak-lensing,
using the CFHTLS images, will provide independent mass measurements (Berge
et al in prep.). 

\begin{figure}
    \begin{center}
    \vspace{-0.4cm}
    \includegraphics[width=8.6cm]{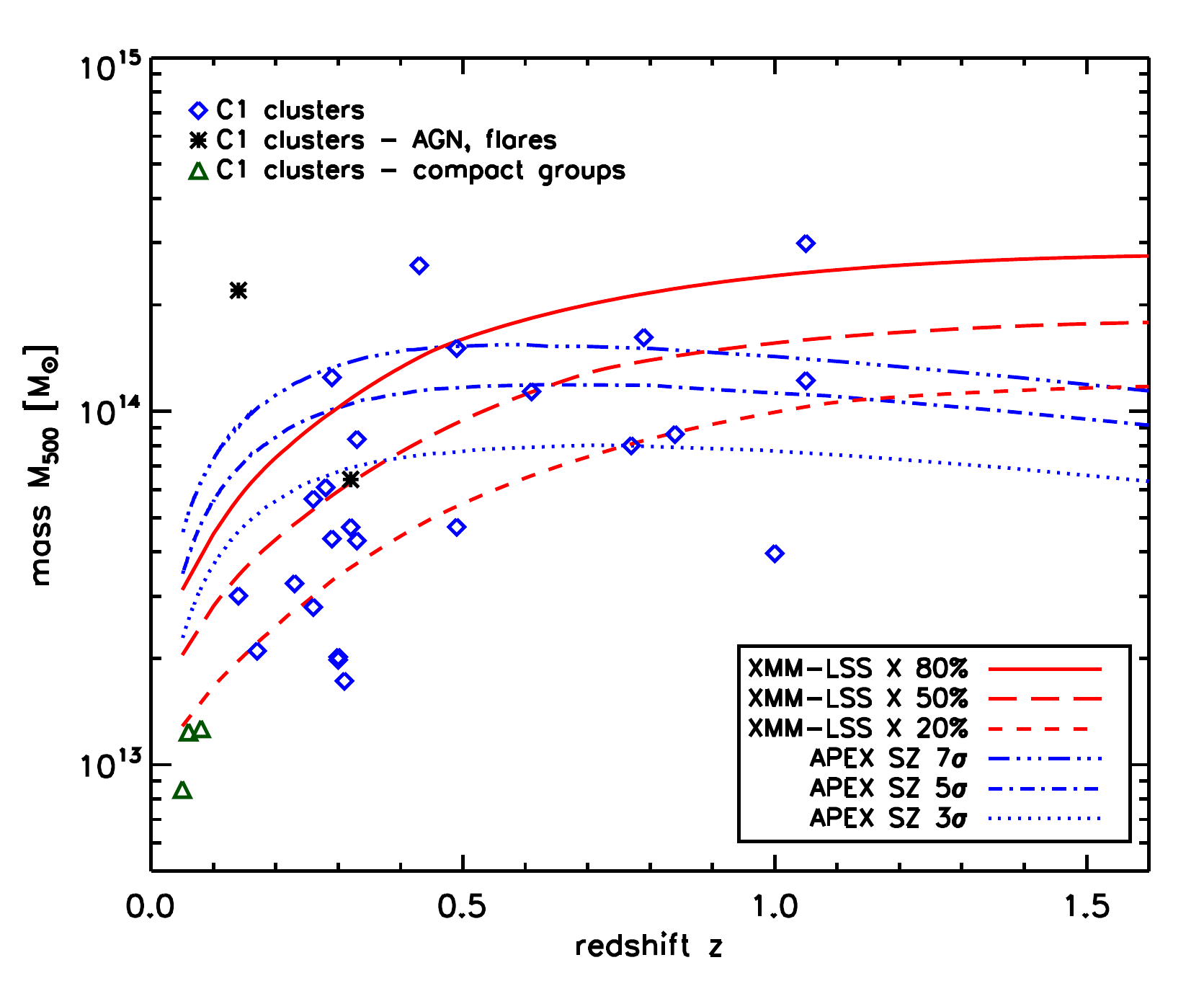}
    \vspace{-0.4cm}
    \caption{Comparison of the XMM-LSS and APEX-SZ sensitivity
    in terms of limiting mass.
    The red lines show various probability detection thresholds
    for the C1 clusters. The blue lines are the predictions for the $10
    \mu K$ APEX survey (see text). The X-ray and SZ curves make
    use of the best fitting model to our current data set (Sec.
    \ref{modeldesc}).
    \label{multilamb}}
    \end{center}
  \end{figure}

In the future, the XMM-LSS field will be covered by
Sunyaev-Zel'dovich (SZ) observations [OCRA (One Centimeter 
Receiver Array), AMiBA (Array for MIcrowave Background Anisotropy) 
and APEX (Atacama Pathfinder EXperiment)]. 
Since the SZ decrement is an integral of the cluster pressure 
($n_{e}T_{e}$) along the line of sight, and is independent of $z$, 
it provides an especially interesting complement to the X-ray 
emissivity (scaling as $n_{e}^{2}T_{e}^{1/2}$ 
for moderately hot clusters).
Comparing our limiting mass as a function of redshift 
with those of these SZ 
surveys is all the more informative as we
introduced a new class-oriented selection to define the XMM-LSS 
cluster sample.
This is done, as a test case, in Fig.~\ref{multilamb} for the
XMM-LSS and the APEX-SZ surveys [\citet{ApexSZ}, \citet{ApexSZ2}].
The SZ limit has been obtained under the assumption that APEX will 
observe at two frequencies (150 and 220 GHz) with a 1\arcmin\ beam and 
10~$\mu$K/beam white noise in each channel.
We have included the effect of contamination by primary Cosmic 
Microwave Background (CMB) anisotropies. Given the spatial spectrum 
of the CMB, the instrumental noise and cluster profiles, 
we compute the expected SZ flux limit as a function of cluster size 
using matched filters. We then convert from SZ fluxes and 
sizes to masses and redshifts, using the same cosmology and 
scaling laws as in our X-ray modelling.
Details of the computation are given in \cite{melin05}.

The XMM-LSS mass limit of $10^{14}~M_\odot$ around $z=1$ for the C1 
selection, appears comparable to the prediction for the deepest SZ 
surveys to date. In addition, the XMM-LSS uncovers more low-mass 
systems below $z<0.5$.
According to our current model, the APEX-SZ cluster density
detected at 3$\sigma$ is 3.8~deg$^{-2}$ out to $z=1$, to be compared
with 5.4~deg$^{-2}$ for the observed C1 X-ray selection. In a
forthcoming paper (Melin et al in prep.) we explore the cosmological
constraints expected from a few hundred clusters, whose
masses are determined by a joint X-ray/SZ/weak-lensing analysis.

\section{SUMMARY AND CONCLUSION}

\begin{itemize}

\item This work presents a well-controlled  XMM cluster sample
over 5~deg$^2$ with  a density of 5.4~deg$^{-2}$ at medium X-ray
sensitivity (down to $\sim10^{-14}$ \flux\ for the extended
sources in question). A complementary sample of about the same 
size, but with less well defined selection criteria is in 
preparation (Adami et al).
We provide positions, redshifts, fluxes, luminosities, temperatures 
and masses along with X-ray and optical images.

\item The selection is based on well-defined criteria pertaining
to spatial extent properties, and are similar to a surface brightness
(rather than flux) limit.

\item The resulting cluster redshift distribution extends  out to
$z=1.2$ and peaks at $z\sim 0.3$. Half of the clusters have
temperatures in the range 1-2~keV, occupying the $0.2<z<0.4$ interval.
This intermediate population, the building blocks of the present
day clusters, is systematically unveiled  by the XMM-LSS survey.
It is also the first time that a wide-area blind X-ray survey has
provided reliable cluster temperature measurements.

\item  We demonstrate that taking into account the sample
selection effects is vital for a proper determination of the
evolution of the $L_X-T$ relation. This is due in large part to
the steepness of the cluster mass function, which results in sources
clustering close to the detection limit, and of these the overluminous 
systems are preferentially detected. This may explain
the often contradictory results obtained with heterogeneous
samples selected form ROSAT surveys.

\item Modelling the selection effects carefully, our data
appear to point to approximately self-similar evolution. A sample twice as
large should allow exclusion of the no-evolution hypothesis, at better
than the 1$\sigma$ level. Using extensive simulations, we find
that increasing the sample size is more efficient than increasing
the accuracy of the temperature measurements for constraining the
evolution; this is due to the large intrinsic dispersion of the
$L_X-T$ relation.

\item Our results suggest a higher normalisation of the cluster
log(N)-log(S) at faint fluxes than previously obtained by deep
ROSAT surveys.

\item  Cluster masses were estimated from surface brightness
profile fitting,  assuming hydrostatic equilibrium. 
Our sample follows surprisingly well (given the very modest XMM 
exposures of $\sim$ 10 ks and our rudimentary modelling of the 
mass profile) the local $L_X-M$ relation, when evolved self-similarly.

\item Self-consistent cosmological modelling of the cluster
population, convolved with the accurately determined
survey selection function, confirms that the properties of our
current data set are compatible with the concordance cosmology
along with cluster self-similar evolution.

\item This model then allowed us to investigate several
degeneracies arising from cluster physics, regarding the
cosmological interpretation of the number counts. In particular, we
stress the influence of the, still poorly determined, scatter in
the cluster scaling laws.

\item The present study led us to investigate many of the issues
raised when attempting to perform precision cosmology using real cluster
data, including the need for
precise selection effects and a realistic error budget. 
Thanks to the 
temperature information obtained from the detected photons for all C1 
clusters, this work constitutes one of the first attempts to break, 
in a self-consistent way, the degeneracy between cosmology and cluster 
evolution, in the analysis of the cluster number counts. The
next step will be to apply cosmological modelling to
the full XMM-LSS area ($10~deg^2$ - to be completed in 2007) adding
input from the combined weak lensing and SZ survey as well as from
the 3-dimensional cluster distribution.

\item All data presented in this paper \--\ cluster images taken
at X-ray and optical wavebands in addition to detailed results for
the spectral and spatial analyses \--\ are available in
electronic form at the XMM-LSS cluster online database: {\tt
http://l3sdb.in2p3.fr:8080/l3sdb/}.

\end{itemize}

\section*{acknowledgements}

We are grateful to the referee, C. Collins, for his constructive 
remarks on the manuscript.
The results presented here are based on observations obtained with
XMM-Newton, an ESA science mission with instruments and
contributions directly funded by ESA Member States and NASA. The
cluster optical images were obtained with MegaPrime/MegaCam, a
joint project of CFHT and CEA/DAPNIA, at the Canada-France-Hawaii
Telescope (CFHT) which is operated by the National Research
Council (NRC) of Canada, the Institut National des Sciences de
l'Univers of the Centre National de la Recherche Scientifique
(CNRS) of France, and the University of Hawaii.  This work is
based in part on data products produced at TERAPIX and the
Canadian Astronomy Data Centre as part of the Canada-France-Hawaii
Telescope Legacy Survey, a collaborative project of NRC and CNRS.
AG acknowledges support from Centre National d'Etudes Spatiales.
The Italian members of the team acknowledge financial contribution
from contract ASI-INAF I/023/05/0. AD, OG, EG, PGS and JS
acknowledge support from the ESA PRODEX Programme `XMM-LSS', and
from the Belgian Federal Science Policy Office for their support.
HQ acknowledges partial support from the FONDAP Centro de
Astrofisica. We thank the IN2P3/DAPNIA computer centre in Lyon.

\bsp

  \appendix

  \section{Notes on spatial fitting and photometric accuracy}

  \begin{figure}
    \begin{center}
    \includegraphics[width=8.6cm]{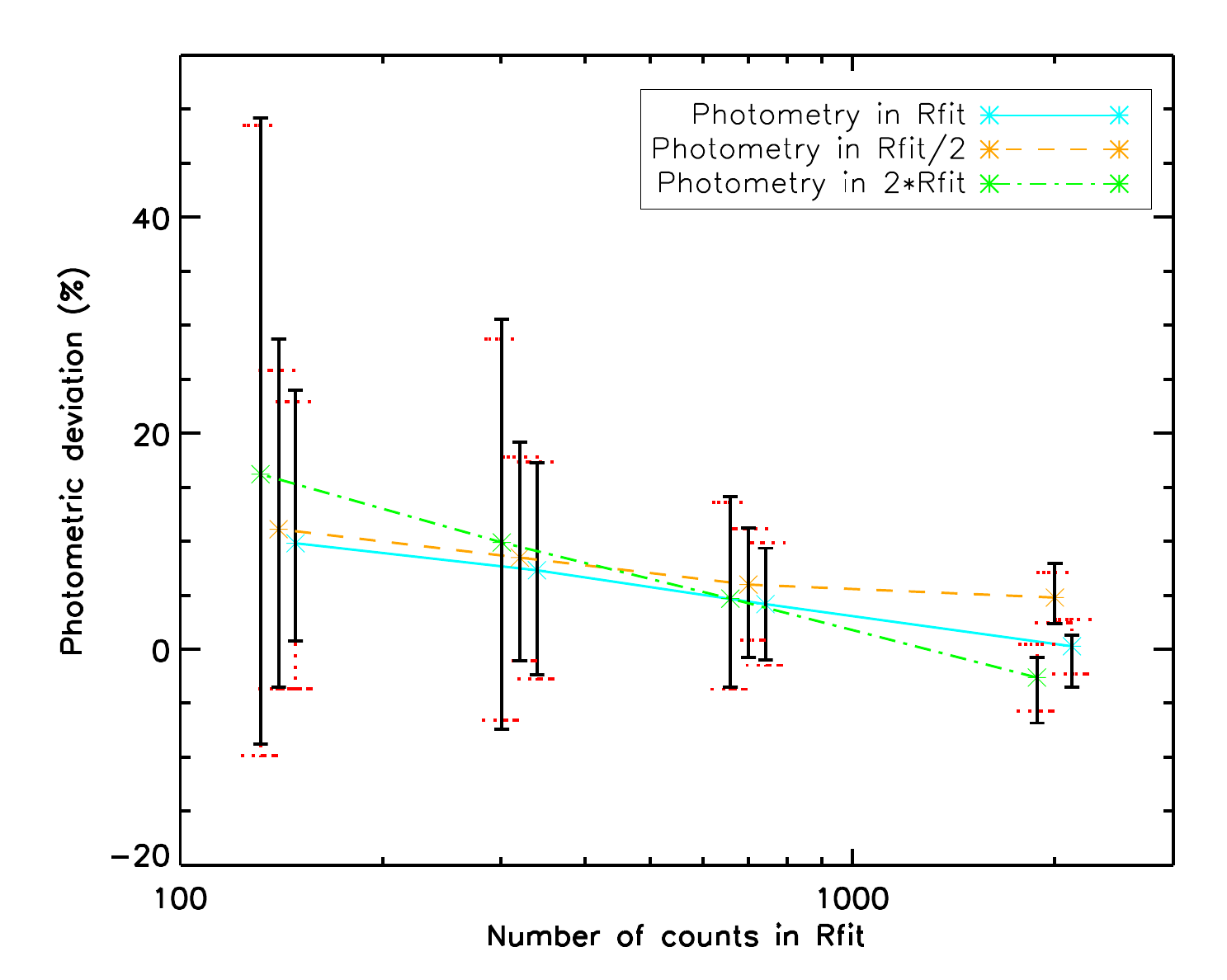}
    \caption{Photometric accuracy of our spatial fitting
    procedure using simulations. The data points show the
    mean deviation from the true value in our simulation
    as a function of available signal (number of counts
    within spectral fitting radius $R_{fit}$), and the
    extrapolation factor (flux measurement within $R_{fit}$,
    $R_{fit}/2$, and $2*R_{fit}$). The solid error bars with
    short hats show the dispersion among measured values,
    while the dotted error bars with large hats gives the
    mean value of the $1\sigma$ bounds estimated by the
    procedure.\label{simresult}}
    \end{center}
  \end{figure}
  \label{lumacc}

  In this appendix, we test the accuracy of our luminosity
  measurement procedure, running the spatial fitting algorithm
  on simulated images.

  For this purpose, we created $10^4s$ XMM pointing simulations with
  vignetted and particle background levels set to the mean observed
  values of \cite{read2003}, and point sources distributed according
  to the log(N)-log(S) of \cite{moret}. We then randomly included in
  these simulations extended sources with $\beta=2/3$ and several
  $R_c$ and count rate values (respectively from 10 to 50\arcsec
  and 0.005 to 0.1~\countr).
  A $\beta$-profile was fitted to all the simulated sources
  that were detected by the first pass of our pipeline (see \citealt{pipeline}
  for details) and for which the spatial fitting procedure detects at
  least three radial bins above $3\sigma$.

  The simulations showed that that below 400 detected counts in the spatial
  fitting radius, the degeneracy between $\beta$ and $R_C$ is large, but that
  the integrated flux (within the fitting radius) is well modelled by the
  best fitting ($\beta$,$R_C$) combination. Results of the simulations are
  presented in Fig. \ref{simresult}.

  The measurements are almost unbiased: less than a few percent in general,
  and up to $\sim10\%$ for very faint sources (100-200~counts). This small
  offset can be interpreted as a weak Eddington bias, as it is of the
  order of Poisson fluctuations, and probably results from our requirement
  of recovering at least 3 significant bins (which means retaining only
  the clusters that appear brighter among the faintest ones).
  This shouldn't be an issue for the C1 clusters as they are generally
  brighter (see Table \ref{cluslist}).
  The scatter around the mean value is very low for bright sources
  and mildly increases (up to $\sim15\%$) for the faintest ones.
  Estimating the luminosity within a different radius than the fitting
  one does not result in significant differences (although the scatter
  increases) as long as the profile is not extrapolated too far out, which
  is the case for our sample.
  Finally, the mean estimated errors are of the same order as the dispersion
  of the best fit value, assessing our confidence intervals.

  Other sources of error have already been considered by \cite{D1paper}
  appendix A:
  the impact of neglecting the errors on the temperature (which affects our
   estimate of $R_{500}$) was shown to be smaller than a few percent,
  while the contribution from undetected weak AGNs is probably lower than the
  percent level (which is comforted by the present simulations in which AGNs
  are included).

  We thus conclude that the possible systematics generated with our fitting
  procedure are rather low, especially compared to the statistical errors.

\section{Individual characteristics of the C1 galaxy clusters}
\label{C1app}

  \begin{figure*}
    \begin{center}
    \vspace{-0.2cm}
    \hbox{
    \includegraphics[width=8.6cm]{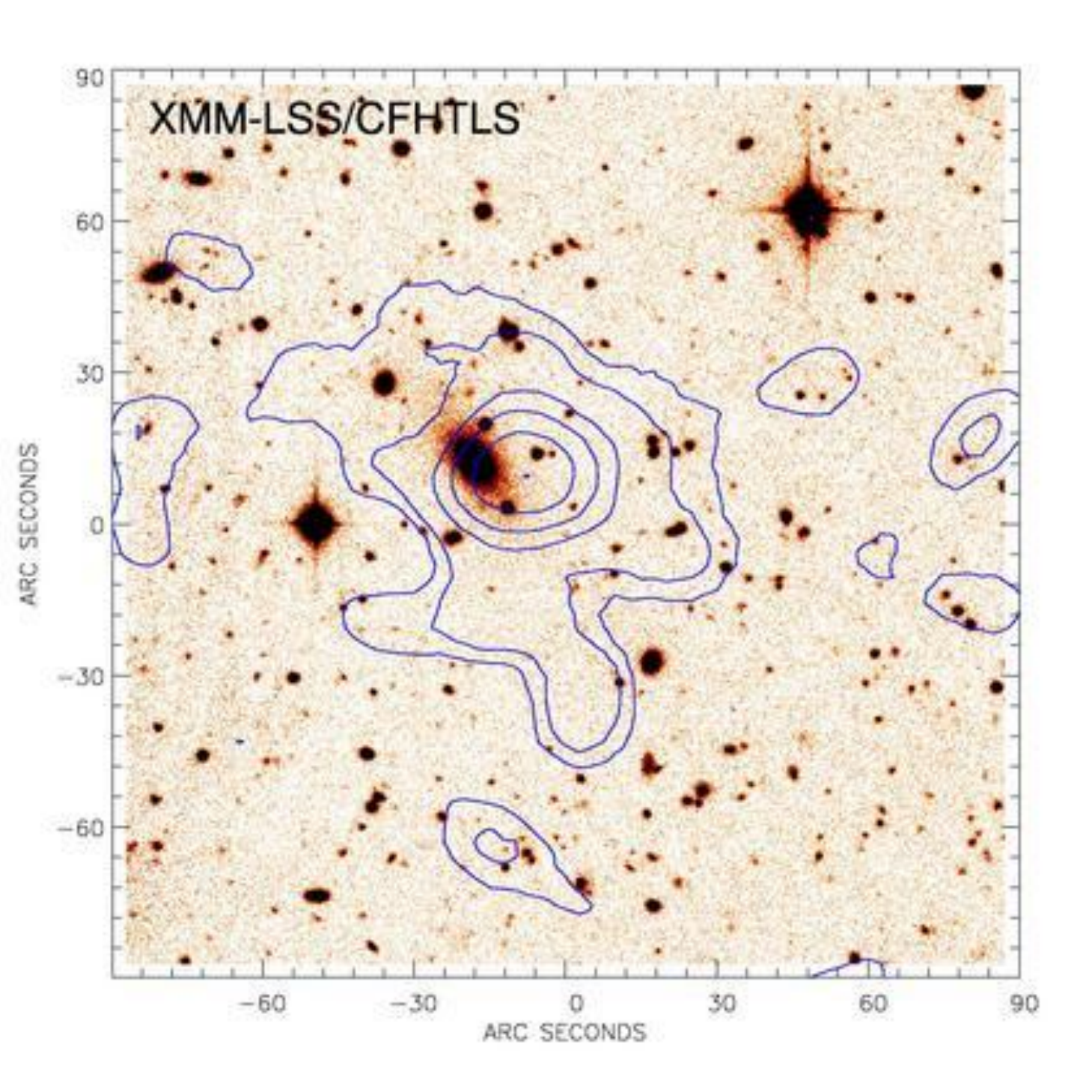}
    \includegraphics[width=8.6cm]{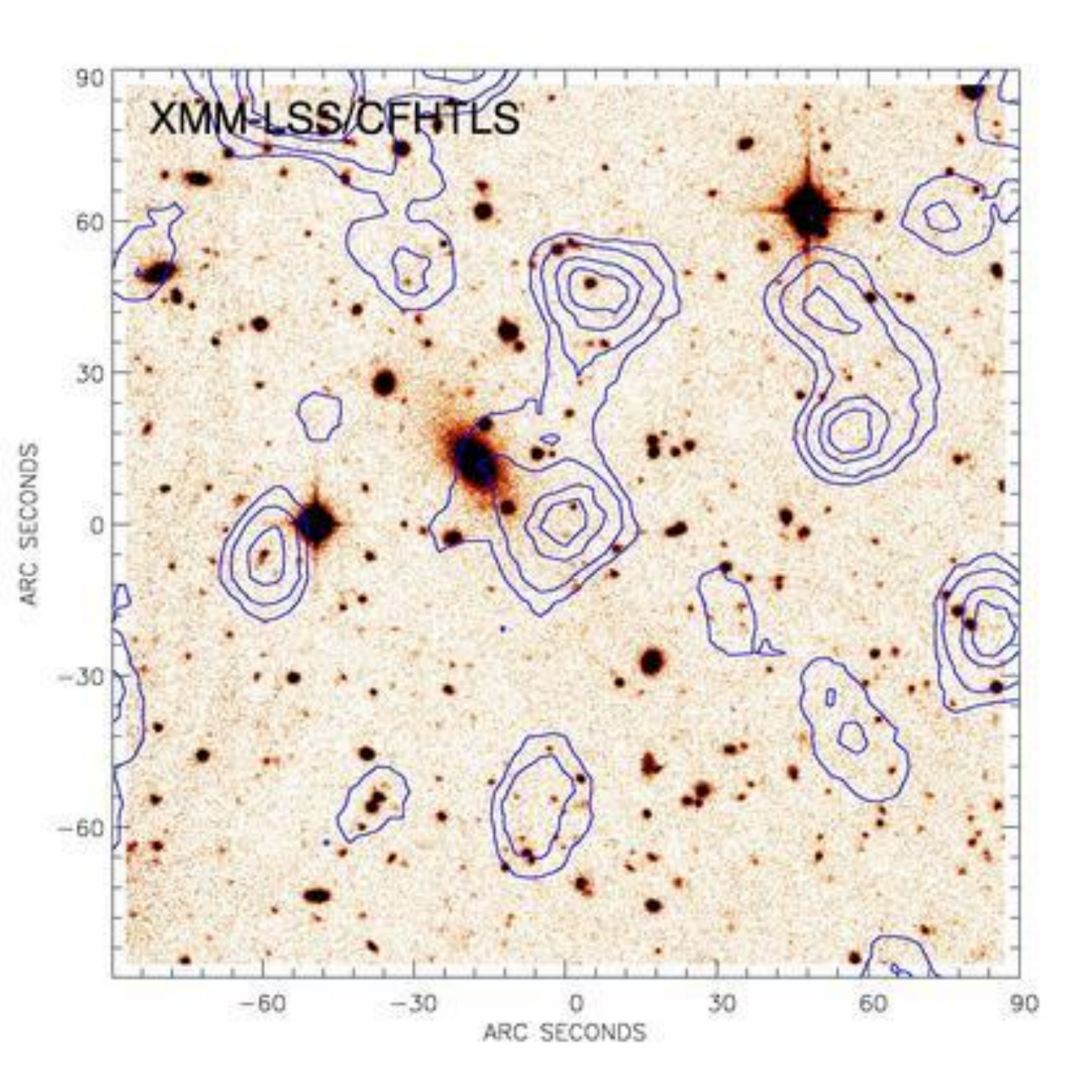}
    }
    \vspace{-0.2cm}
    \caption{Gaussian smoothed X-ray emission from source XLSSC-018 at z=0.32
    in bands [0.5-2]~keV (left) and [2-10]~keV (right). The emission in the soft band seems
    slightly offset from the cluster cD galaxy. In the hard band, the offset
    becomes much larger.\label{lssc18}}
    \end{center}
   \end{figure*}

Notes on individual sources:
\begin{itemize}
 \item {\bf XLSSC-028}:\\
  The velocity information obtained to date on this object does not allow us  to
  firmly conclude on its redshift.
  The two bright galaxies on which the X-ray emission is centred have a redshift
  of 0.08; their spectra are typical of elliptical galaxies, without emission
  lines. No other galaxies with this redshift have been measured in the field,
  but a number of z=0.3 objects are found within 500 kpc of the X-ray centroid.
  The measured X-ray temperature   is 1.5 keV for z=0.3 and, the luminosity is
  found to be $1.6e43$\lum, which puts the object close to the observed $L_X$-T relation.
  At z = 0.08, the source  is about a factor of 17 fainter, and corresponds to a
  temperature of 0.75 keV  which put it also exactly on the $L_X$-T relation.
 \item {\bf XLSSC-018}:\\
  The first measurement of temperature of this system ($\sim2.7_{-0.9}^{+2.5}$~keV, published
  in \cite{Willis}) was one of the hottest in our sample at a redshift around 0.3.
  Yet, the cluster doesn't seem massive, neither in the X-ray nor in the optical.
  The emission in the [2-10]~keV appears to be  significantly displaced from
  the low energy centroid (Fig.~\ref{lssc18}), and is
  possibly associated with an optical counterpart. This suggests that a fraction
  of the emission could be due to an AGN.
  Excluding the probable contaminated region from the spectral fit lowers by a factor
  of two the number of available photons, but surprisingly leads to much tighter error
  bars around a temperature of $2.0$~keV (as reported in Table\ref{cluslist}).
  We regard this as strong evidence that the cluster's X-ray emission suffers AGN
  contamination, and consequently discard this source from the $L_X-T$
  analysis.
 \item {\bf XLSSC-006}:\\
  This source was already studied in \citet{Willis}, and assigned a luminosity of 
  $(4.5\pm0.3)\times 10^{44}$\lum\ which is incompatible with our estimated 
  $(6.0\pm0.2)\times10^{44}$\lum. In that paper, the luminosity was 
  derived from the X\textsc{spec} spectral fitting measurement, to which 
  a correction factor was applied in order to extrapolate to $r_{500}$. 
  We interpret the discrepancy as resulting from the combination of two effects. 
  Firstly, X\textsc{spec} uses a background estimate comming from an annulus 
  surrounding the source, while in our spatial fitting procedures, we model it 
  over the whole pointing. As a consequence, the mean background level 
  used within X\textsc{spec} actually happened to be higher than our fitted 
  value due to residual contamination of the bright XLSSC-006 cluster in 
  this local annulus. 
  Additionaly, the observed source count-rate in the [0.5-2]~keV band (on 
  which our spatial fitting luminosity estimate is based), is found to be 
  higher within X\textsc{spec} than the one infered from the best spectral 
  fitting model.\\
\end{itemize}

Information on other C1 sample clusters is already published by 
\citet{ivanhz}, \citet{Willis}  and \citet{D1paper}.

\section{The C1 nearby galaxies}
\label{C1g}

These sources were discarded from the C1 galaxy cluster sample based 
on an obvious link between the main X-ray emission, and the presence
of a nearby galaxy on the same line-of-sight. A significant fraction 
of the total X-ray emission can however originate from another object
in the field. The source list is given in Table~\ref{nearby}, and 
details regarding the origin of the X-ray emission are given below: 

\begin{table}
\begin{center}
\caption{The nearby galaxy sample. XMM pointing identifiers refer to the
XMM-LSS internal naming convention as described in \protect\cite{FullCatalog};
the location of each pointing on the sky is shown in figure \ref{skydist}.
\label{nearby}}
\begin{tabular}{cccc}
\hline
Source name  &  Pointing  &   R.A.  &   Dec.  \\
             & 		  & (J2000) & (J2000) \\
\hline
 XLSS J022528.7-040041 & B03 & 36.3699 & -4.0115  \\
 XLSS J022251.4-031151 & B11 & 35.7146 & -3.1975  \\
 XLSS J022659.2-043529 & G06 & 36.7469 & -4.5916  \\
 XLSS J022430.4-043617 & G08 & 36.1268 & -4.6048  \\
 XLSS J022617.6-050443 & G16 & 36.5735 & -5.0788  \\
\hline
\end{tabular}
\end{center}
\end{table}

\begin{itemize}
\item {\bf XLSS J022528.7-040041}:\\
The nearby galaxy lying a the centre of the X-ray isophotes is alreday 
known as APMUKS(BJ) B022258.83-041412.5. The X-ray emission from this 
galaxy certainly originates from its interaction with two satellite companions, 
as conspicious in the images of Fig.~\ref{C1img}.
\item {\bf XLSS J022251.4-031151}:\\
This source is likely to result from the confusion of a point source
with the nearby elliptical galaxy APMUKS(BJ) B022018.98-032531.1.
It is associated with a radio source detected in \cite{tasse}.
\item {\bf XLSS J022659.2-043529}:\\
The nearby galaxy lying at the centre of the X-ray isophotes is already 
known as MCG~-01-07-011. We cannot exclude it to be the dominant galaxy 
of a poor nearby group, as its aspect is reminiscent of the compact group 
XSLC-011 at z~0.05.
\item {\bf XLSS J022430.4-043617}:\\
This source is likely to result from the confusion of a point source
with a nearby spiral galaxy. The galaxy is already known as 
6dF~J0224300-043614, and has a redshift of z=0.06916.
\item {\bf XLSS J022617.6-050443}:\\
The nearby galaxy lying a the centre of the X-ray isophotes 
(APMUKS(BJ) B022347.26-051811.5) is detected as a weak radio source in 
\cite{tasse}. As for, XLSS J022659.2-043529, we cannot exclude the possibilty 
that it is the dominant galaxy of a small group. Another option would be that 
it is part of the recently identified XBONG class (see e.g. \citealt{xbong}).
\end{itemize}


\addtocounter{section}{-1}
\stepcounter{figure}
\begin{figure*}
  \begin{pspicture}(0,0)(17.8,18.4)
  \vbox{
    \vbox{\includegraphics[height=9.2cm]{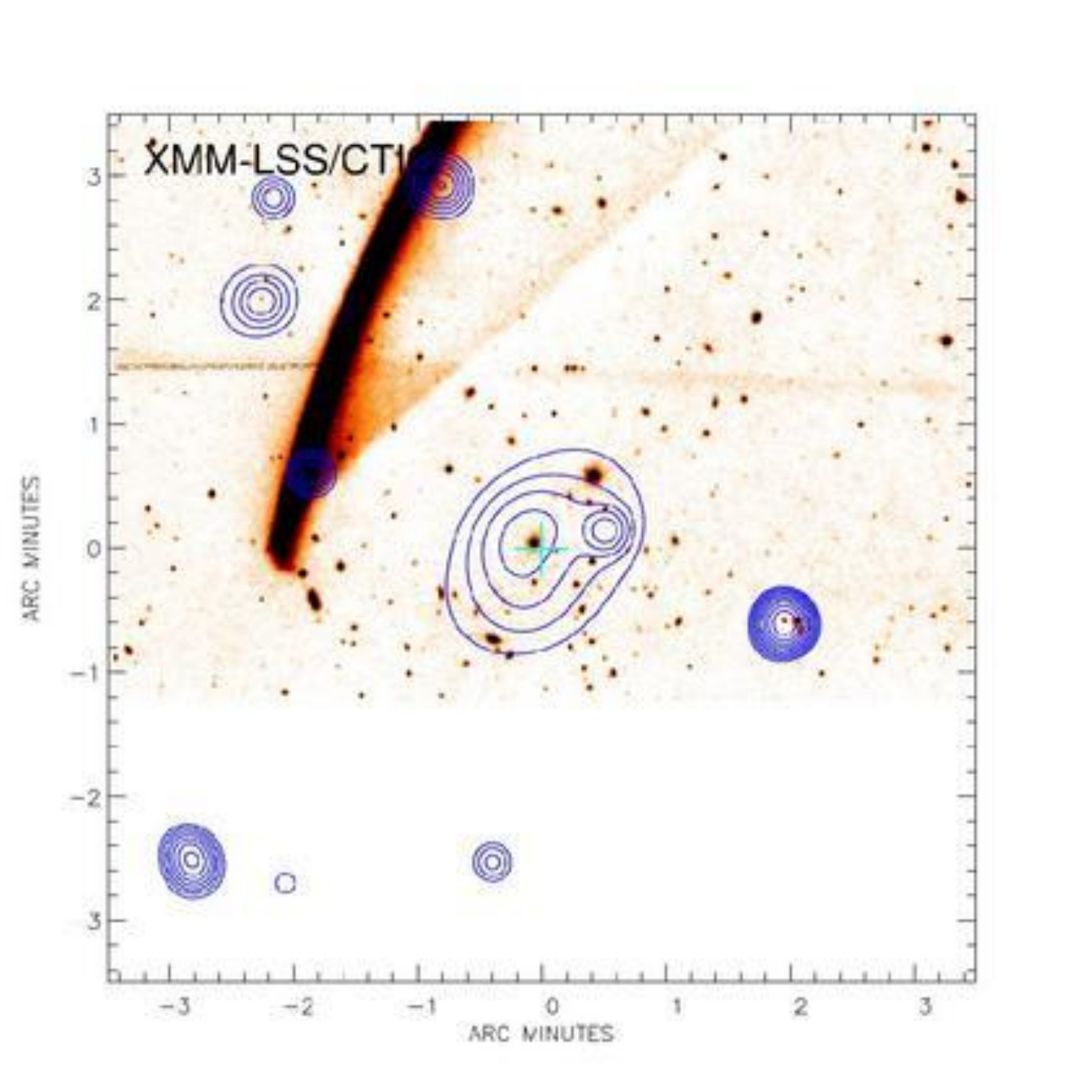}
    \includegraphics[viewport=0 0 420 450,clip,width=7.85cm]{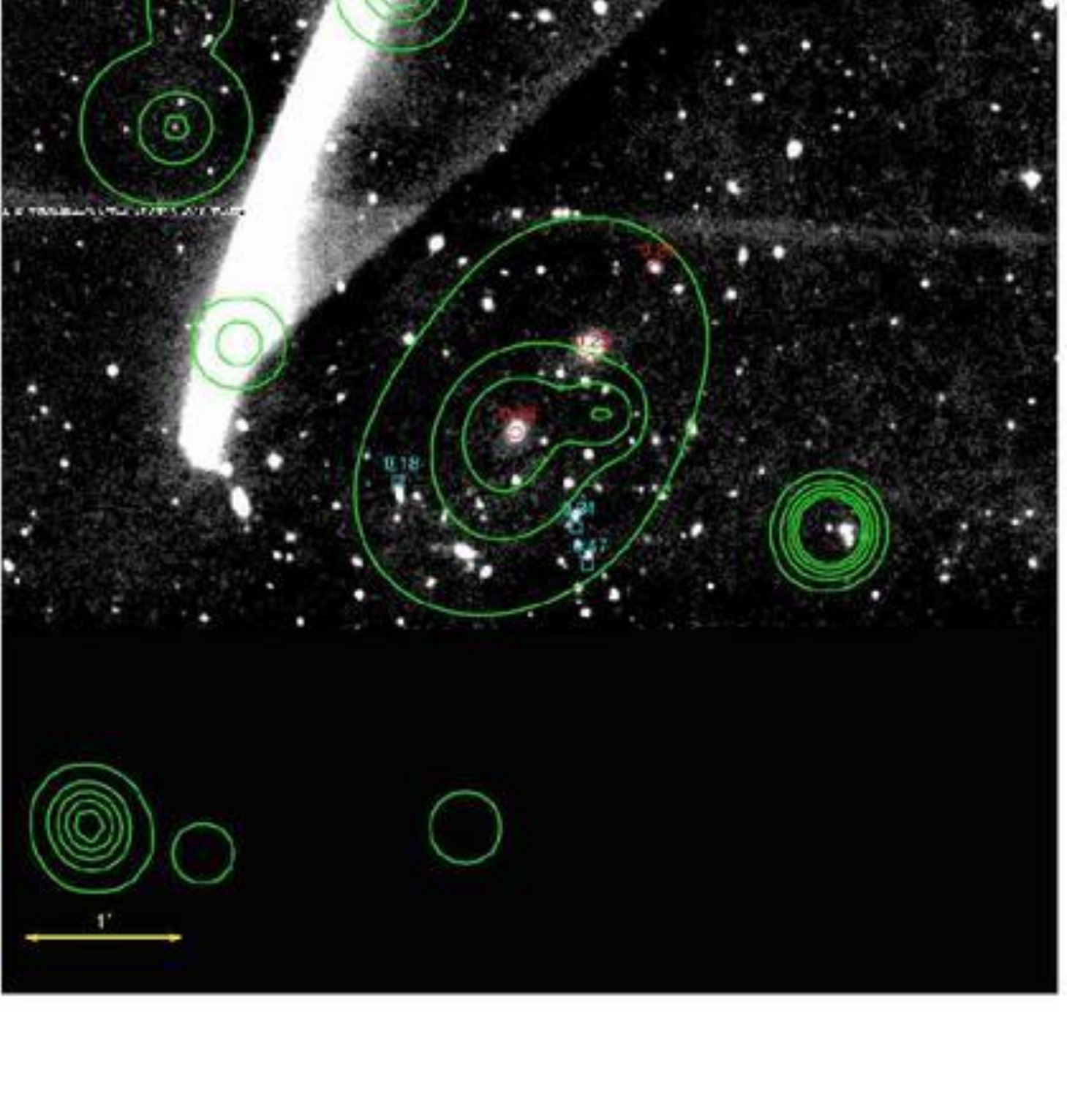}}
    \vbox{\includegraphics[width=9.2cm]{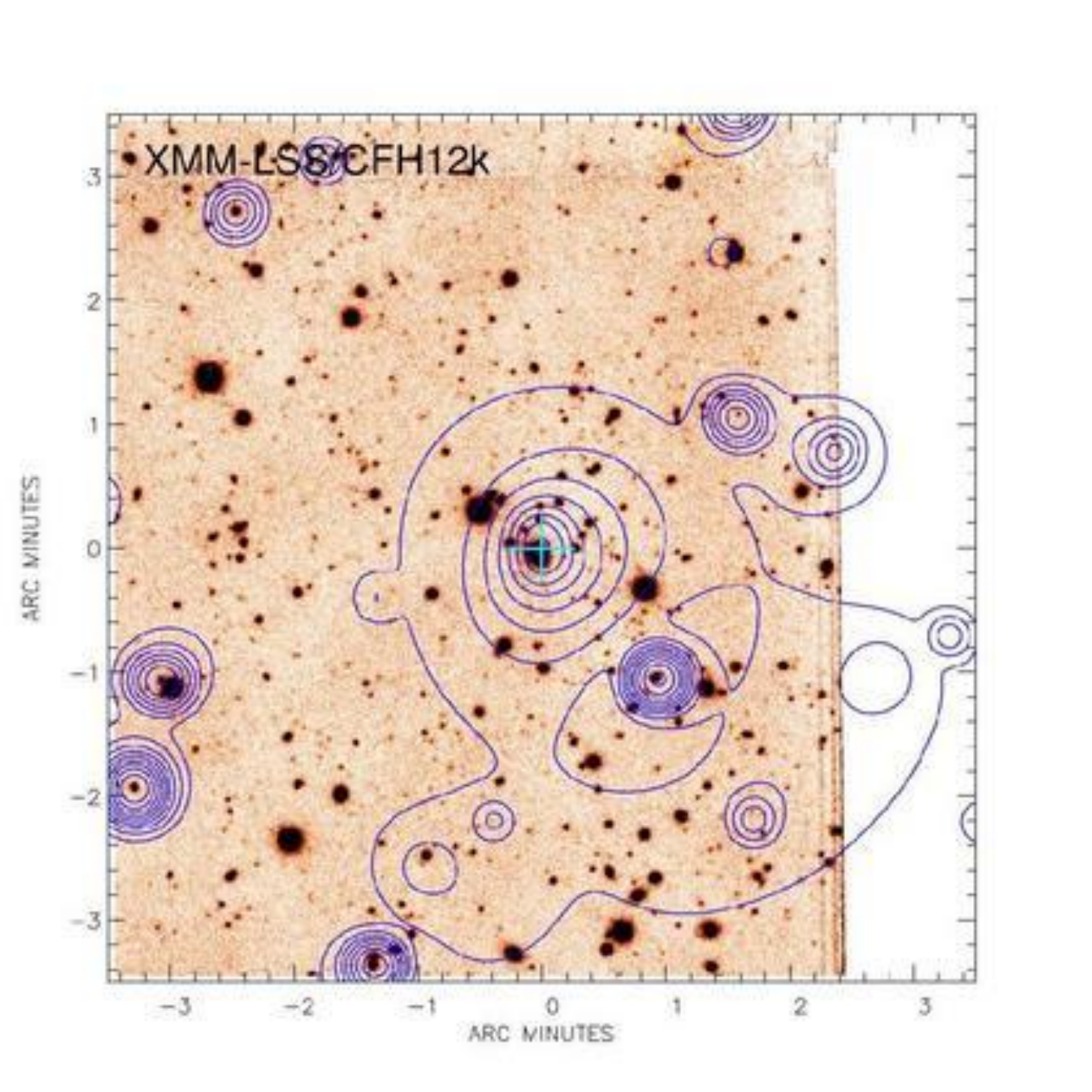}
    \includegraphics[viewport=0 0 420 450,clip,width=7.85cm]{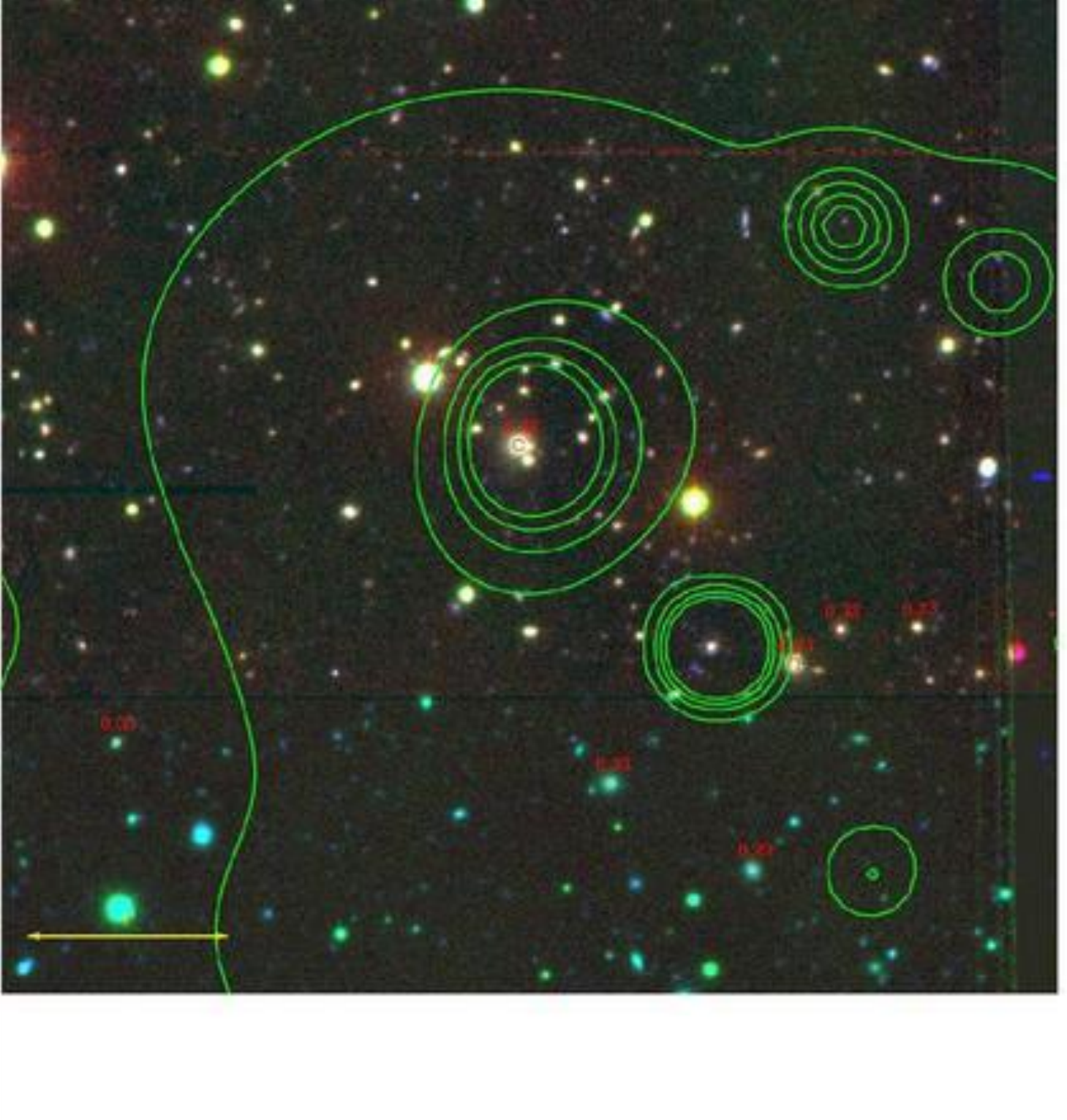}}}
  \rput(-8.6,13.7){\large (a)}
  \rput(-8.6,4.8){\large (b)}
  \end{pspicture}
 \caption{Images of the C1 clusters (sorted by right ascension). Left: X-ray/I band overlay of the central
 7\arcmin. Right: true colour image with X ray contours and measured redshift in
 the central 1.5~Mpc. (a): XLSSC-039. (b): XLSSC-023.}
\end{figure*}

\begin{figure*}
  \begin{pspicture}(0,0)(17.8,18.4)
  \vbox{
    \vbox{\includegraphics[height=9.2cm]{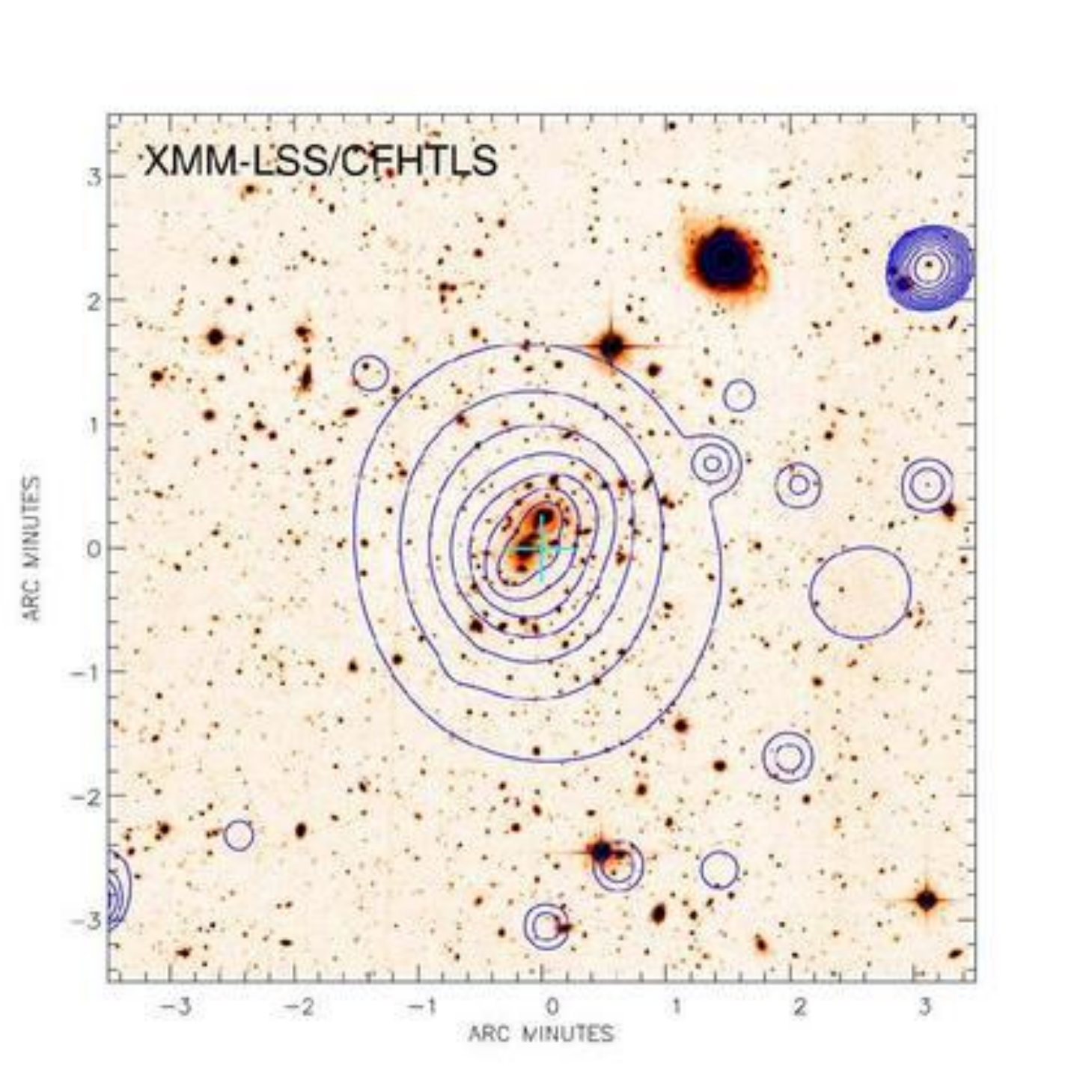}
    \includegraphics[viewport=0 0 420 450,clip,width=7.85cm]{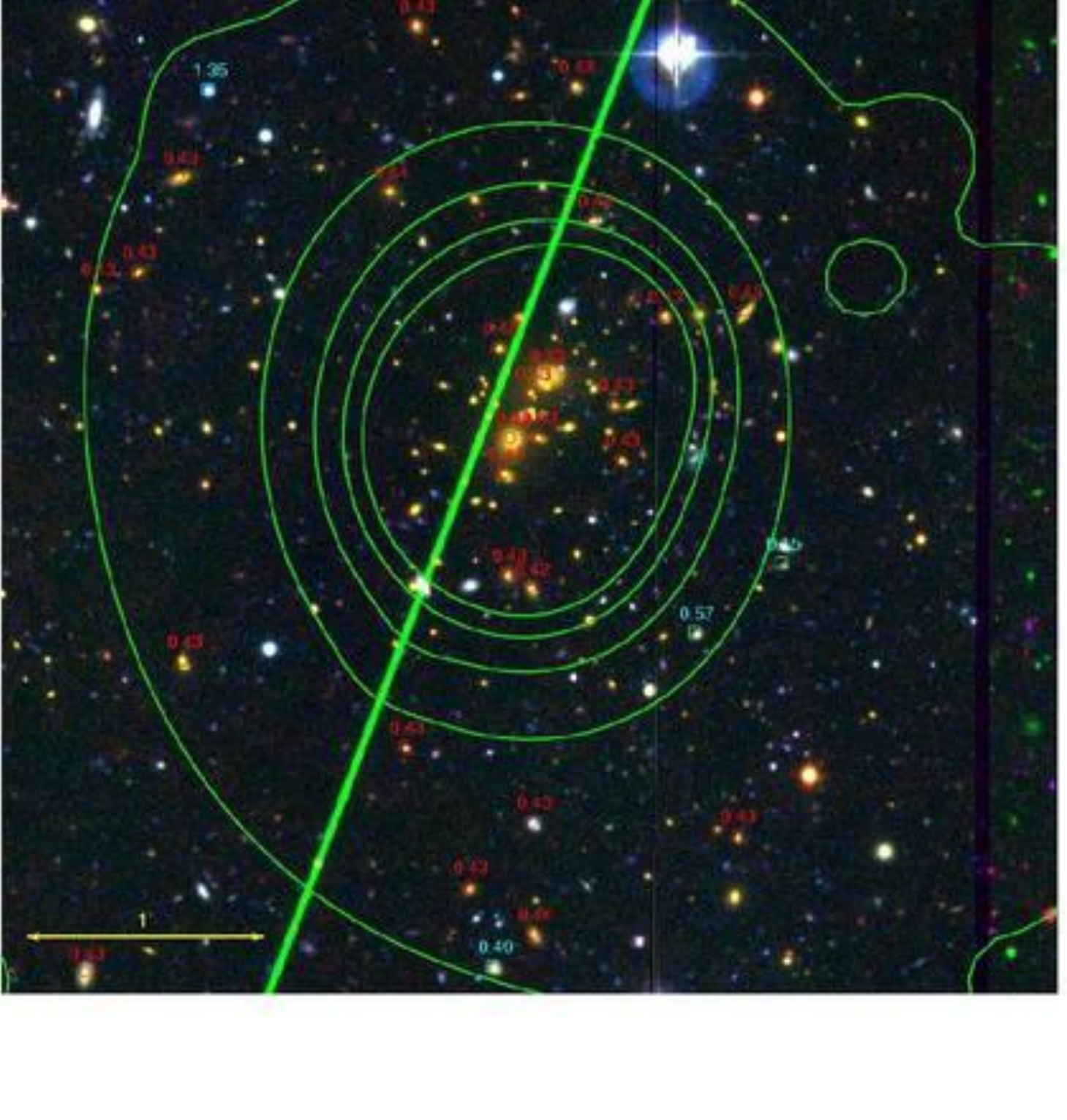}}
    \vbox{\includegraphics[width=9.2cm]{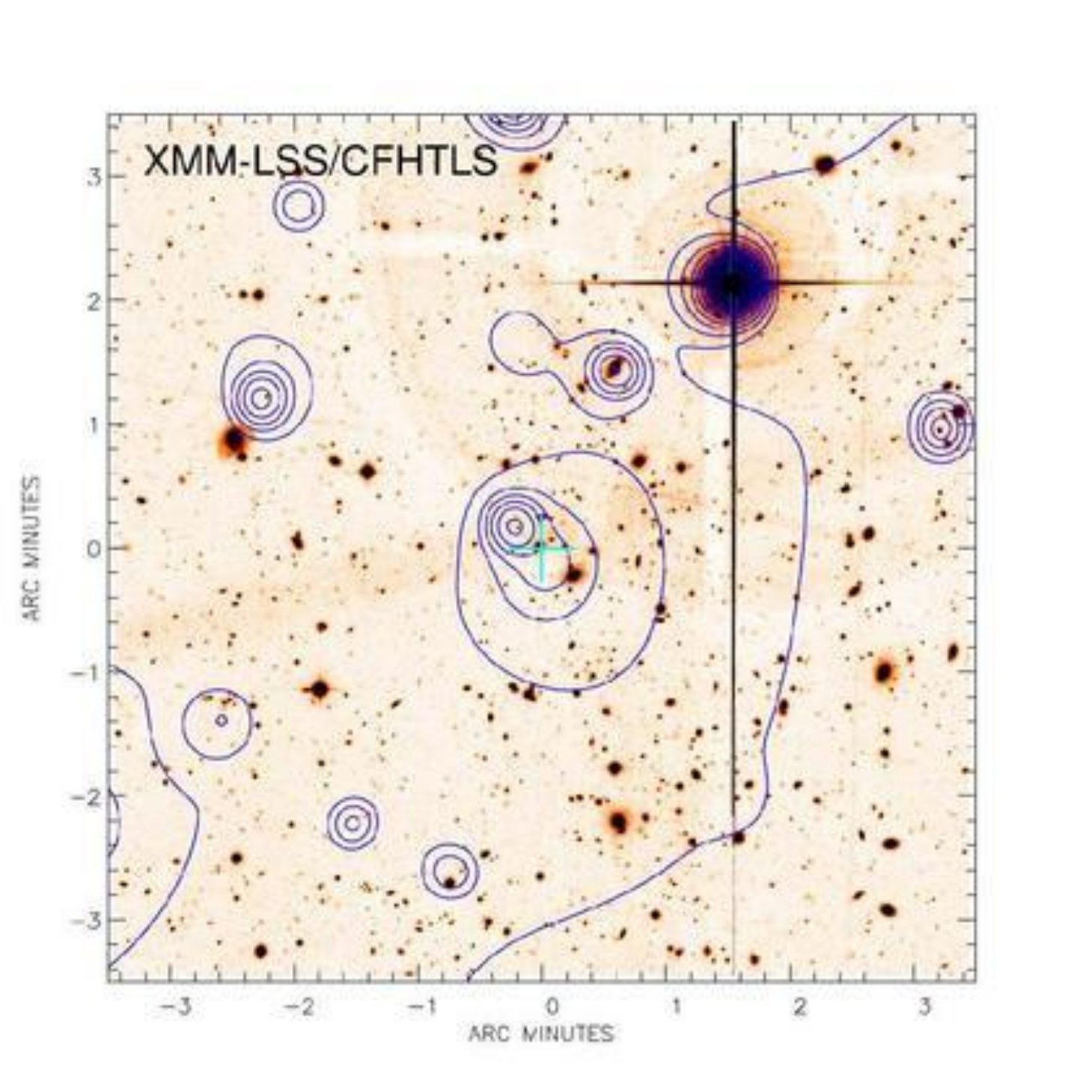}
    \includegraphics[viewport=0 0 420 450,clip,width=7.85cm]{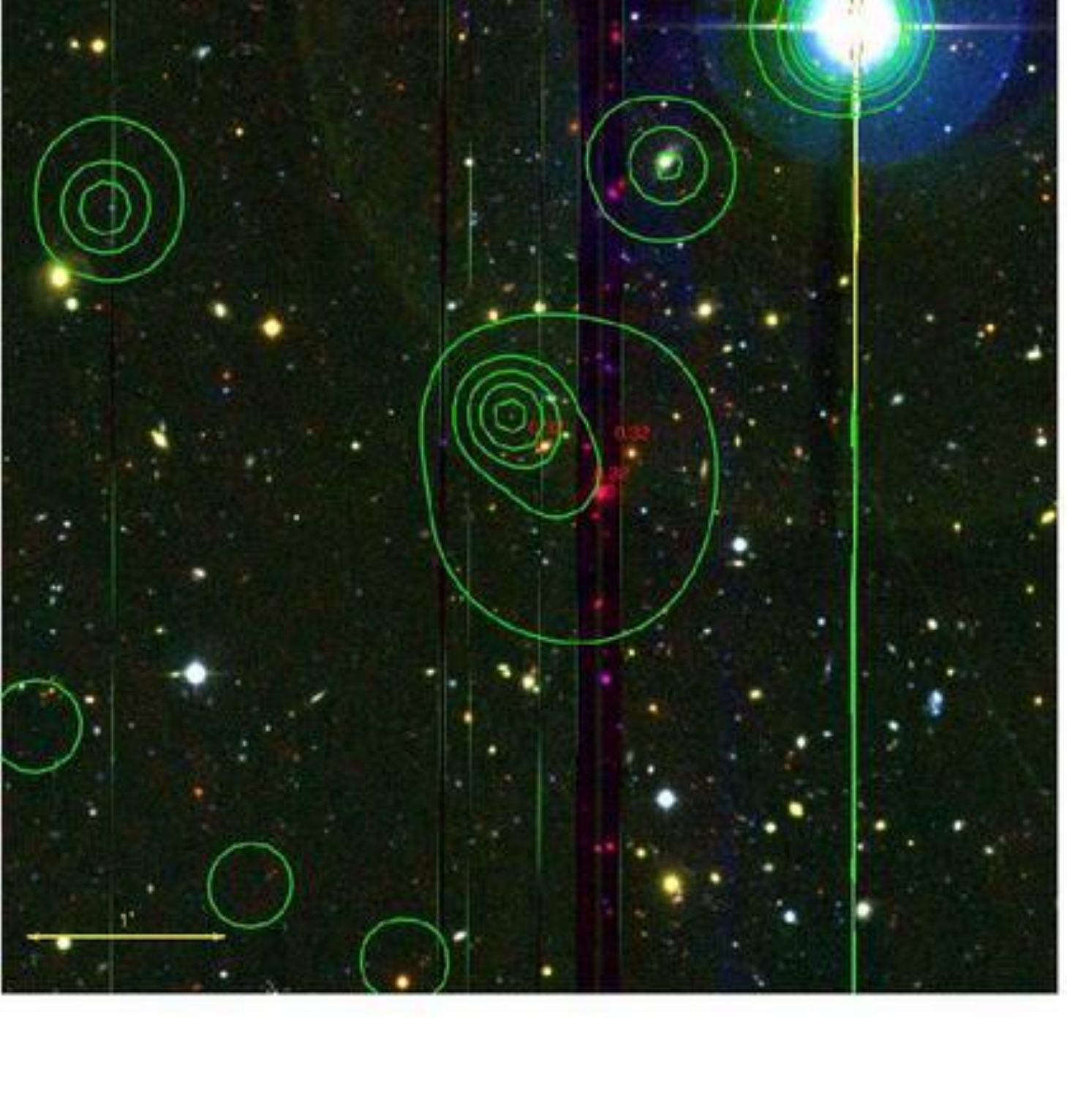}}}
  \rput(-8.6,13.7){\large (a)}
  \rput(-8.6,4.8){\large (b)}
  \end{pspicture}
 \contcaption{Images of the C1 clusters. (a) XLSSC-006. (b): XLSSC-040.}
\end{figure*}

\begin{figure*}
  \begin{pspicture}(0,0)(17.8,18.4)
  \vbox{
    \vbox{\includegraphics[height=9.2cm]{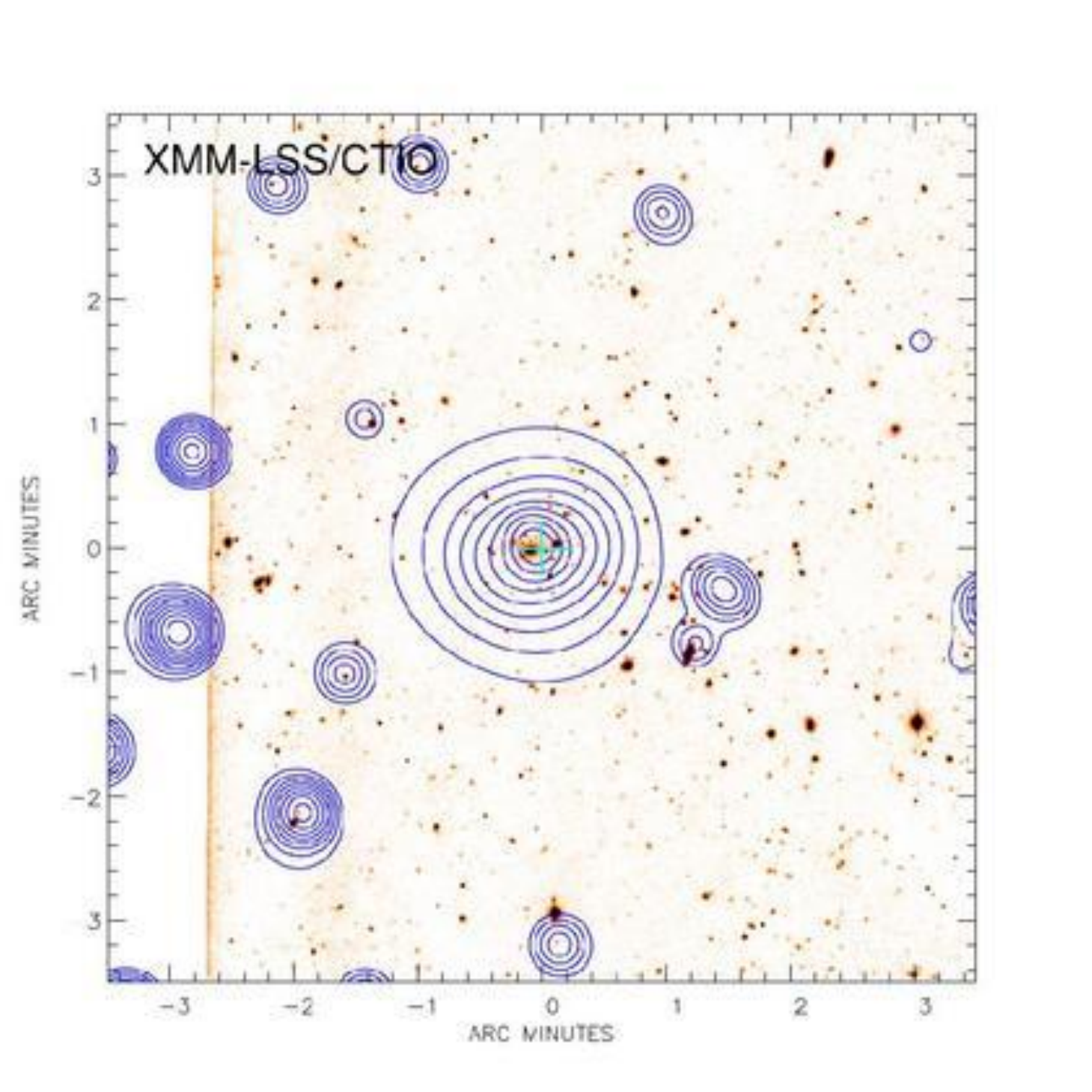}
    \includegraphics[viewport=0 0 420 450,clip,width=7.85cm]{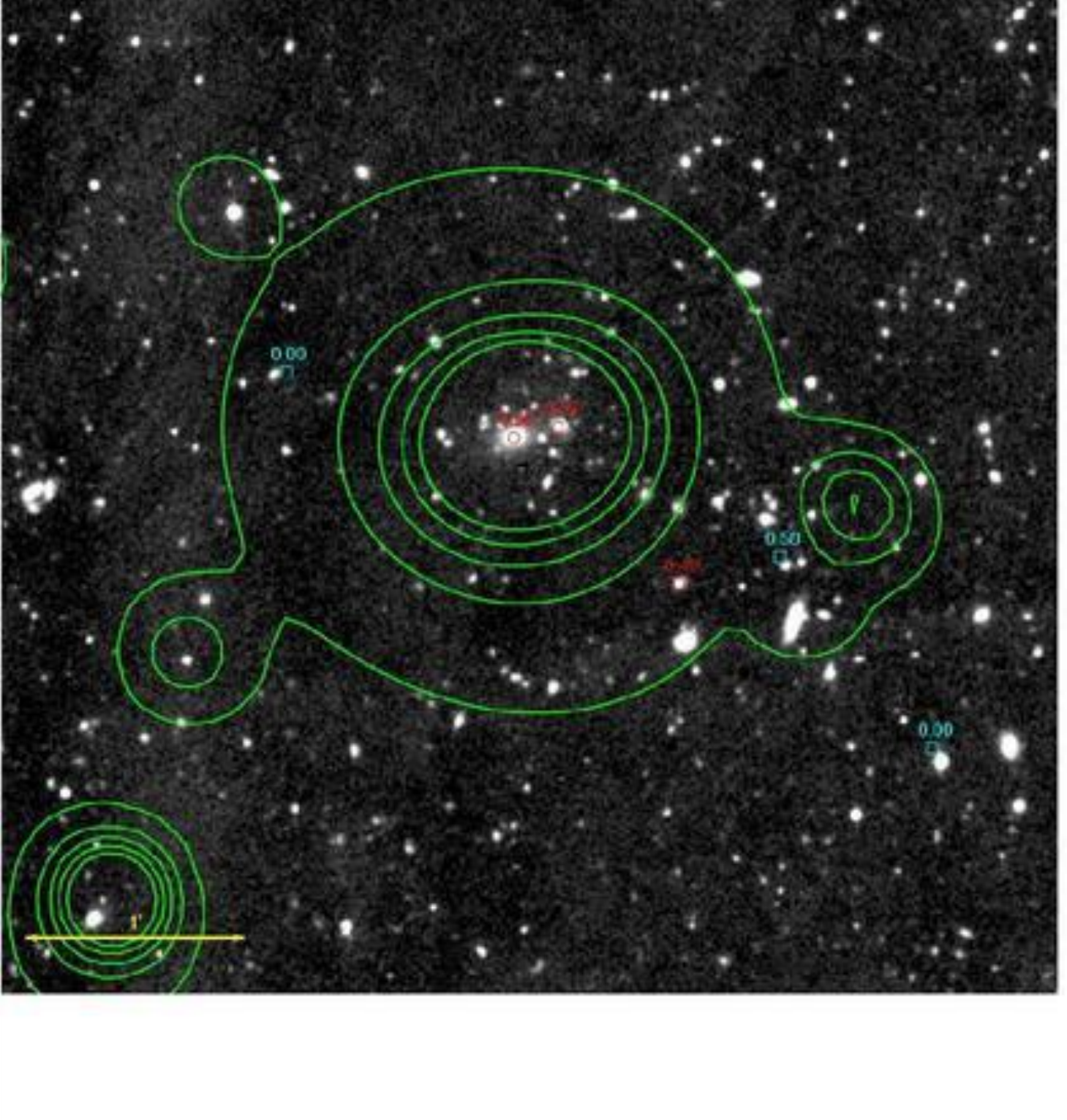}}
    \vbox{\includegraphics[width=9.2cm]{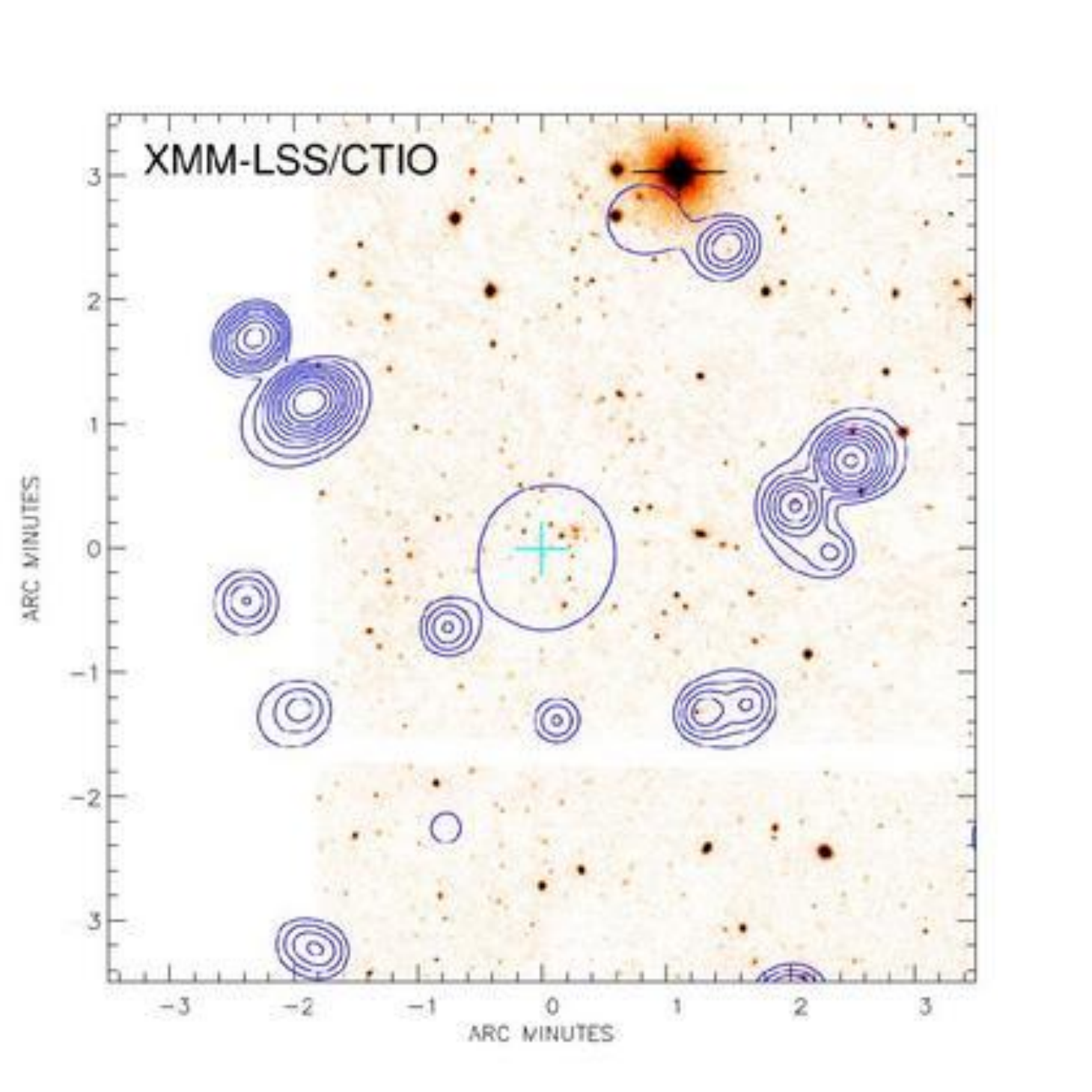}
    \includegraphics[viewport=0 0 420 450,clip,width=7.85cm]{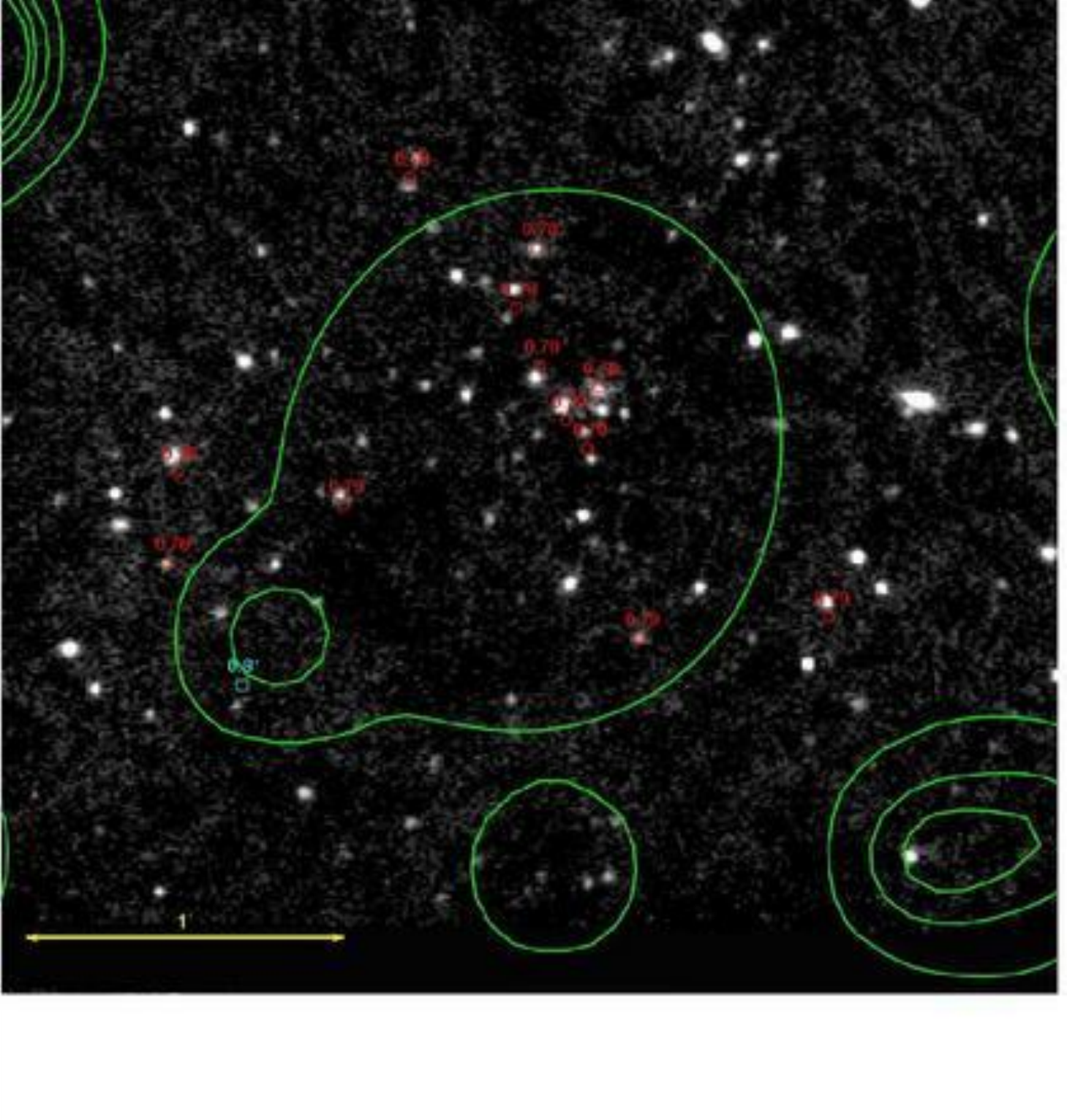}}}
  \rput(-8.6,13.7){\large (a)}
  \rput(-8.6,4.8){\large (b)}
  \end{pspicture}
 \contcaption{Images of the C1 clusters. (a) XLSSC-036. (b): XLSSC-047.}
\end{figure*}

\begin{figure*}
  \begin{pspicture}(0,0)(17.8,18.4)
  \vbox{
    \vbox{\includegraphics[height=9.2cm]{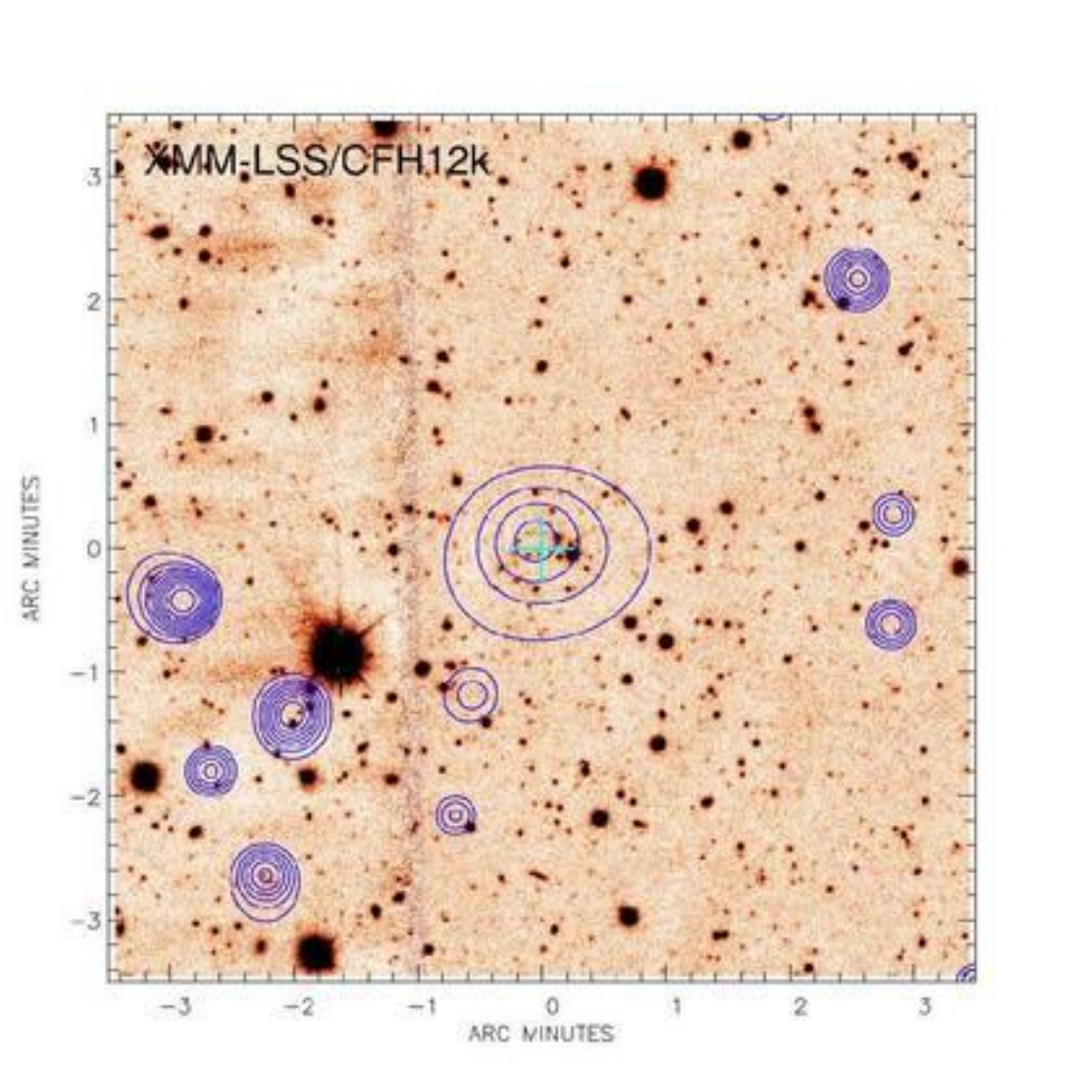}
    \includegraphics[viewport=0 0 420 450,clip,width=7.85cm]{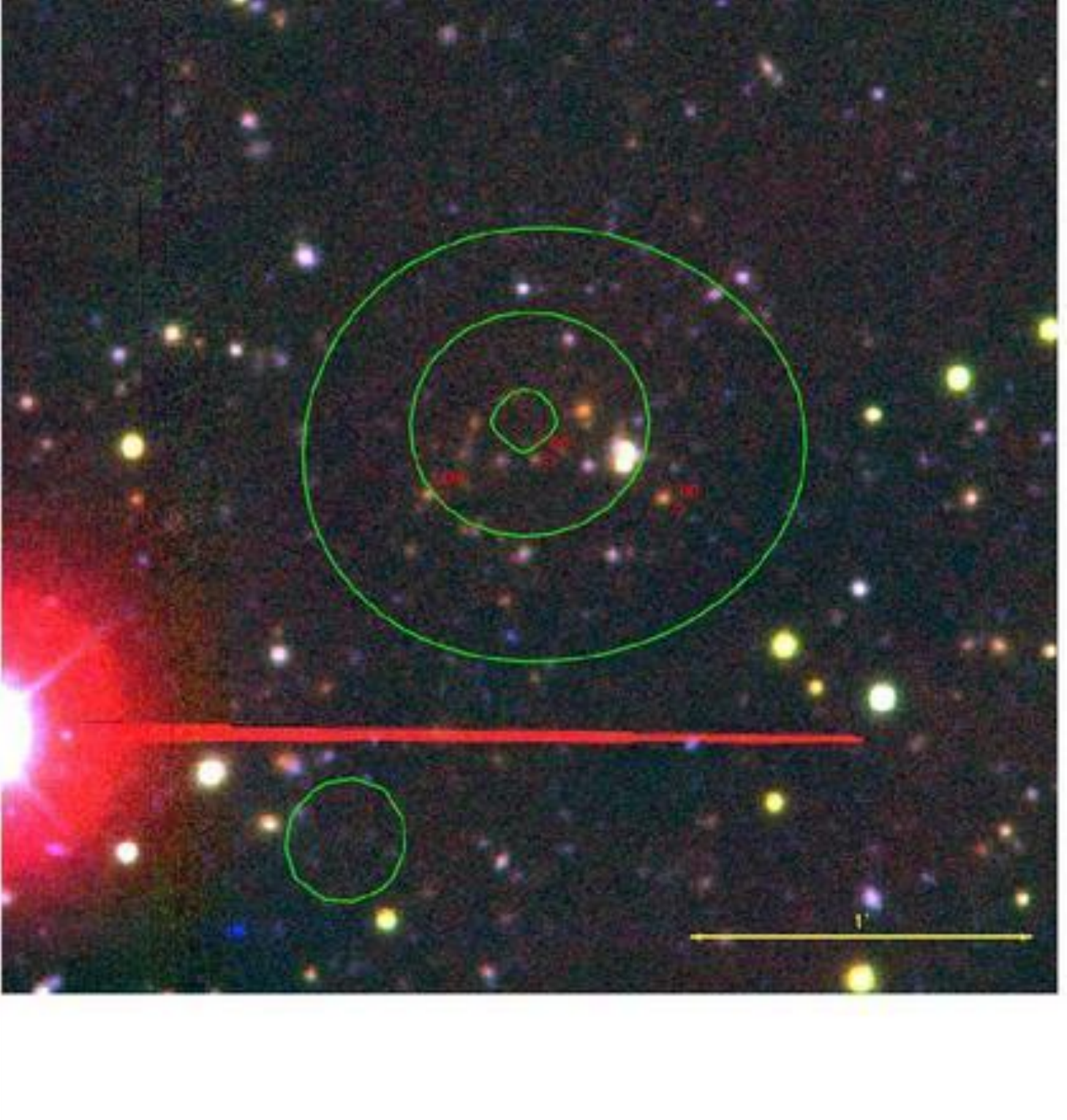}}
    \vbox{\includegraphics[width=9.2cm]{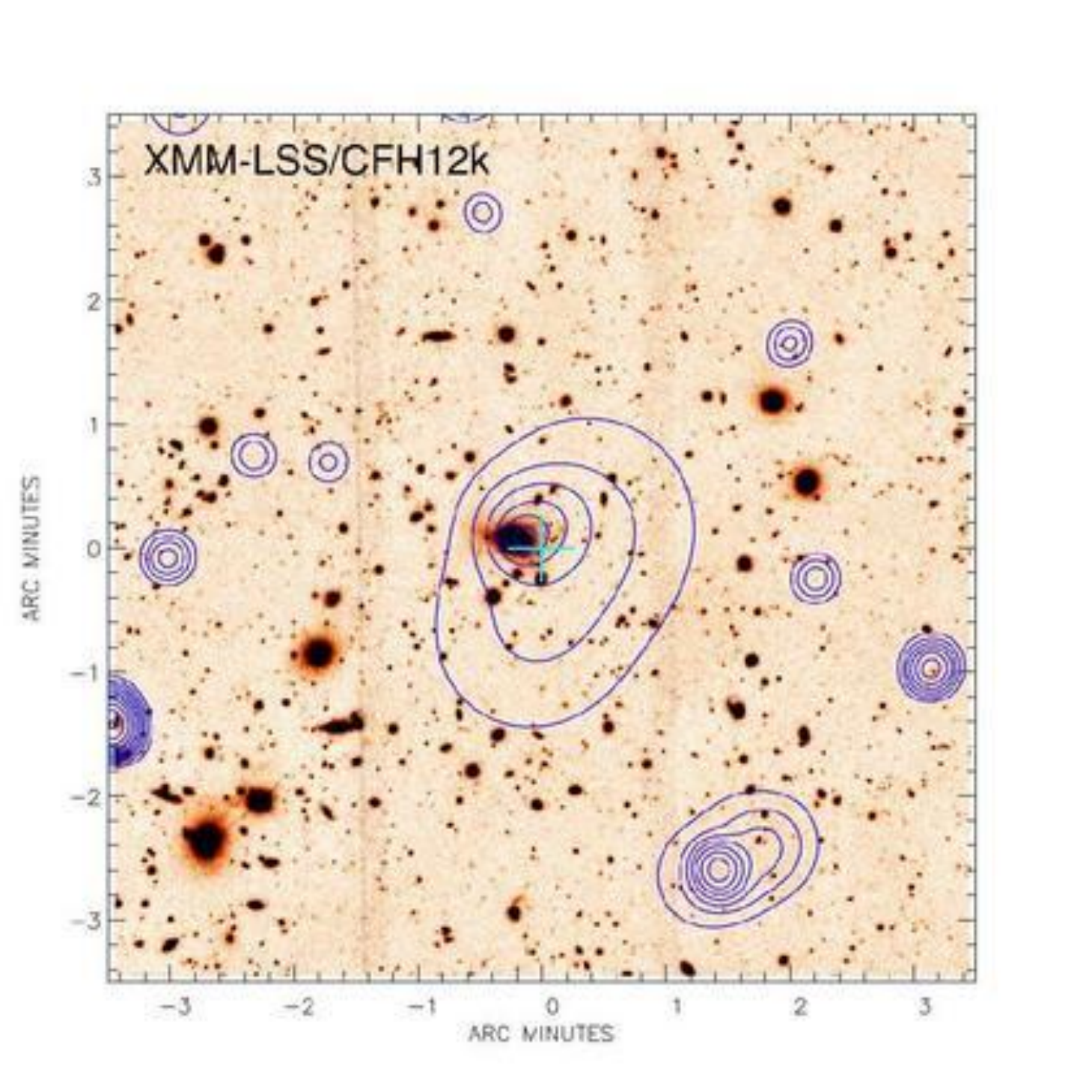}
    \includegraphics[viewport=0 0 420 450,clip,width=7.85cm]{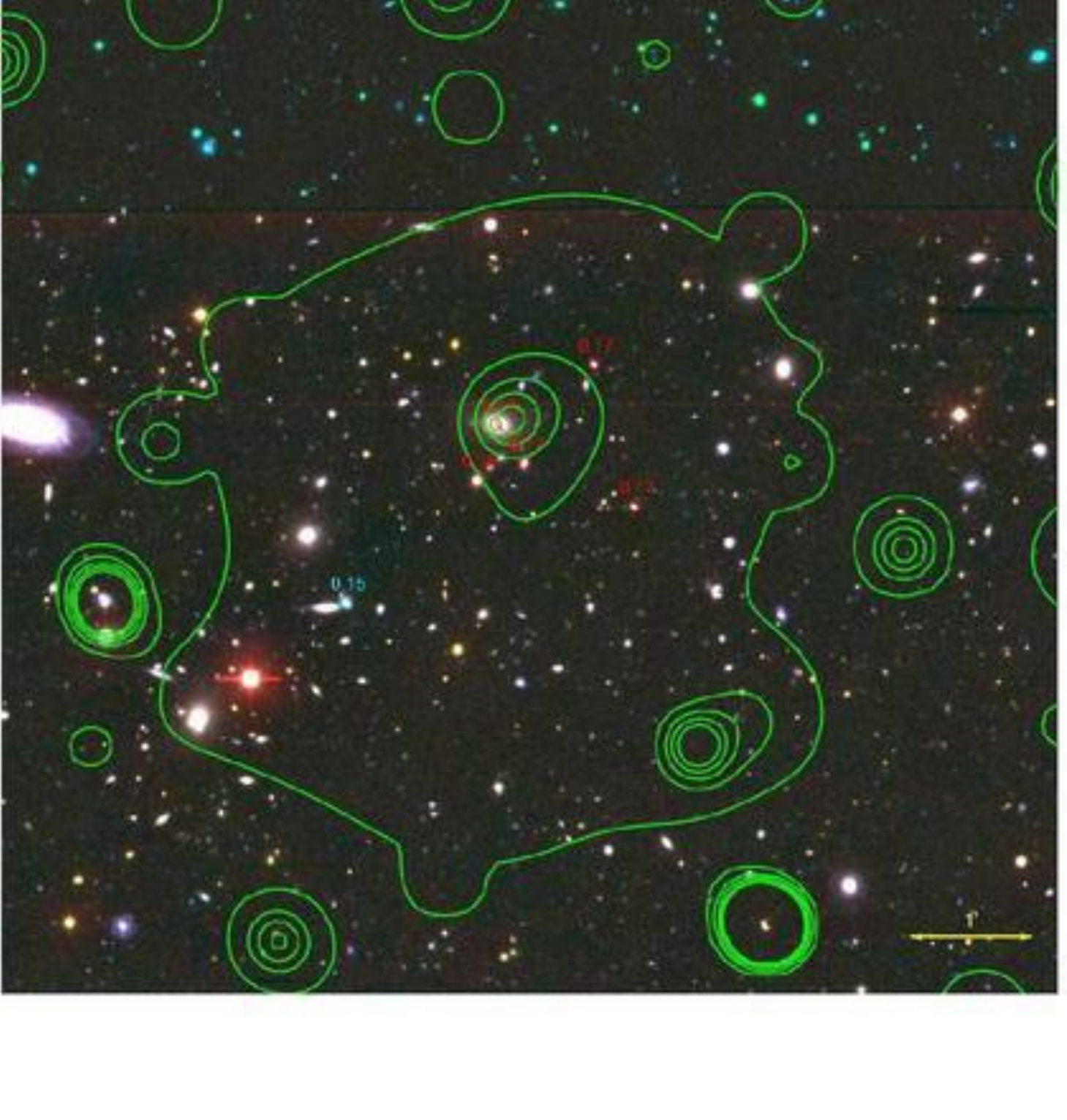}}}
  \rput(-8.6,13.7){\large (a)}
  \rput(-8.6,4.8){\large (b)}
  \end{pspicture}
 \contcaption{Images of the C1 clusters. (a) XLSSC-048. (b): XLSSC-035.}
\end{figure*}

\begin{figure*}
  \begin{pspicture}(0,0)(17.8,18.4)
  \vbox{
    \vbox{\includegraphics[height=9.2cm]{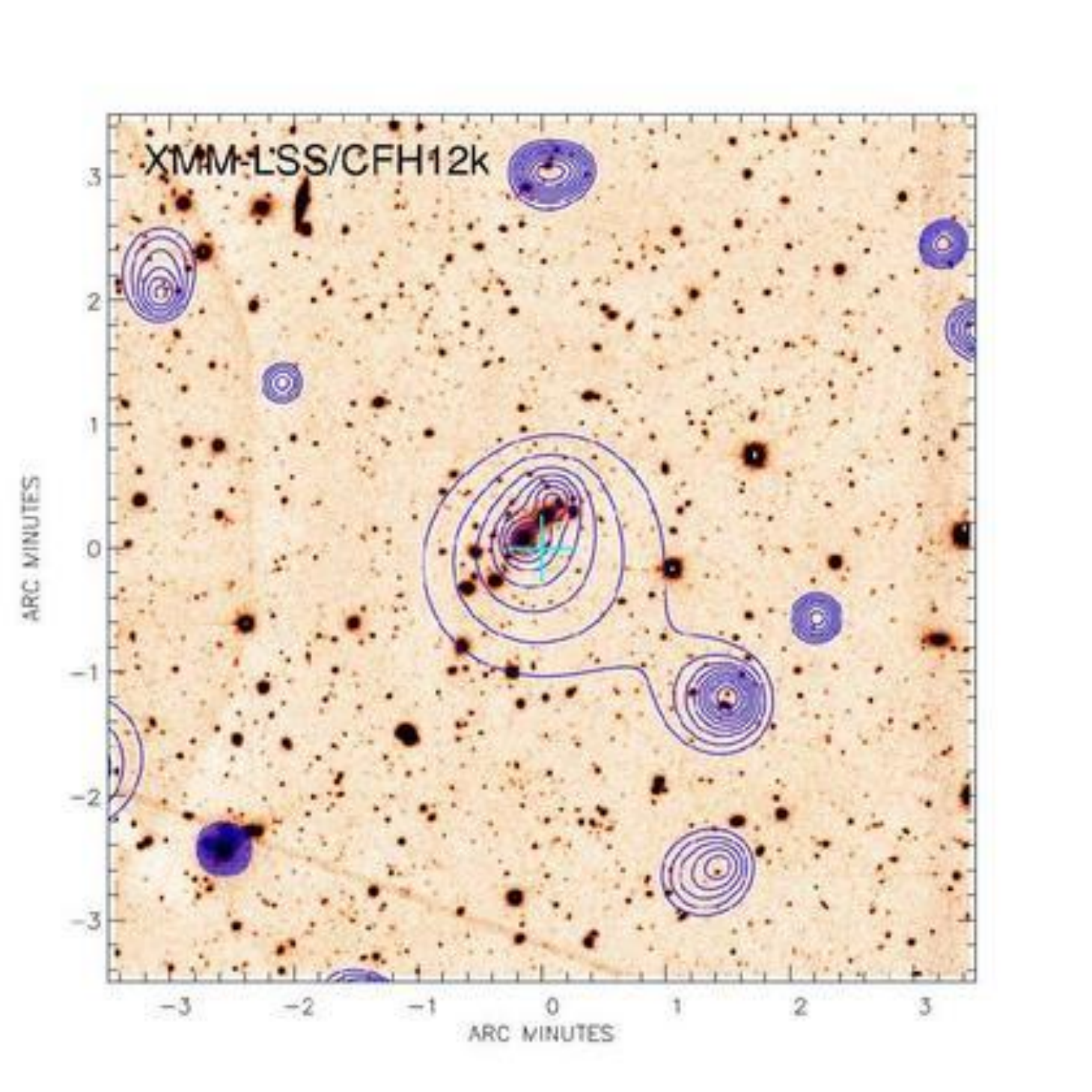}
    \includegraphics[viewport=0 0 420 450,clip,width=7.85cm]{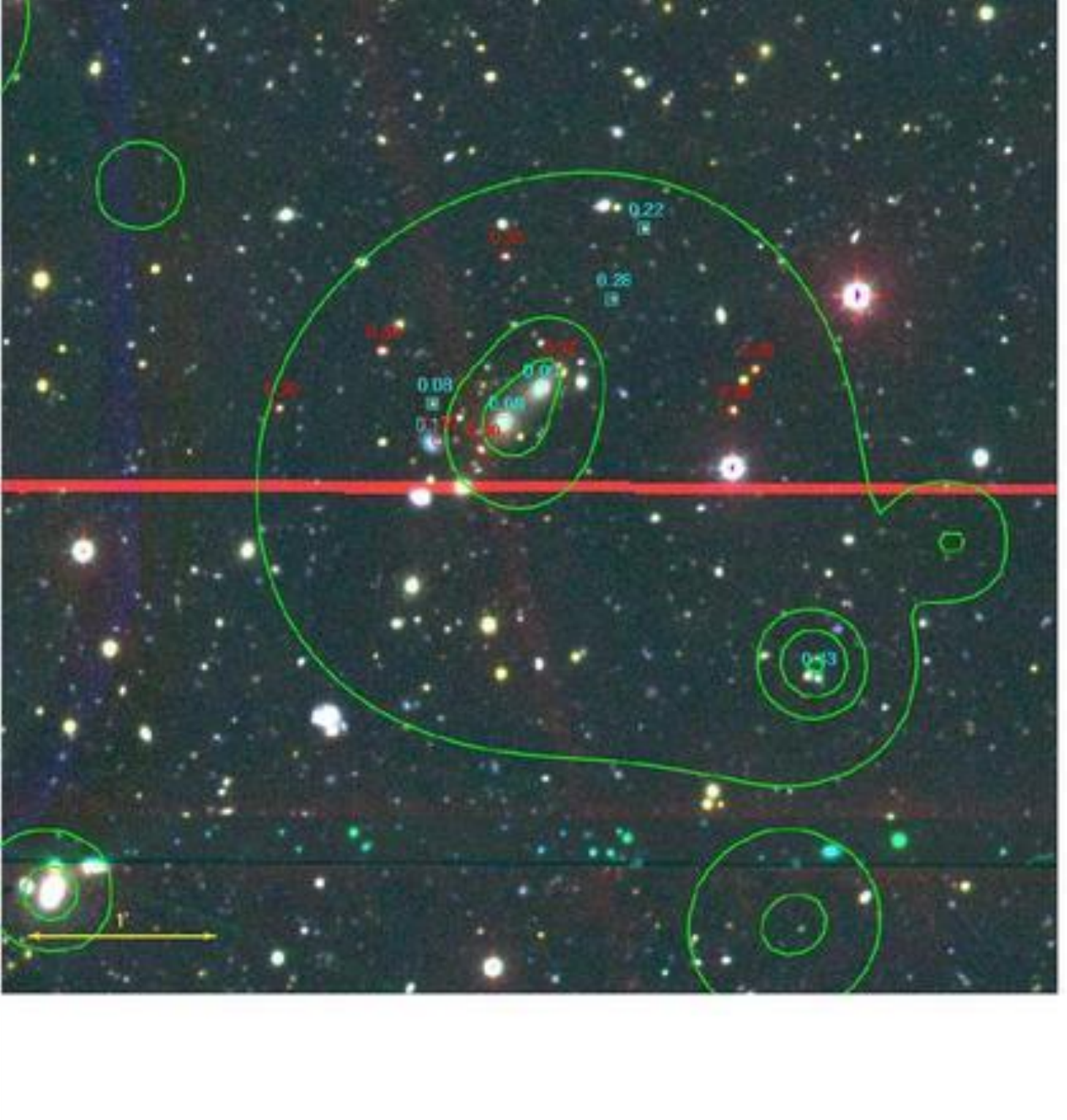}}
    \vbox{\includegraphics[width=9.2cm]{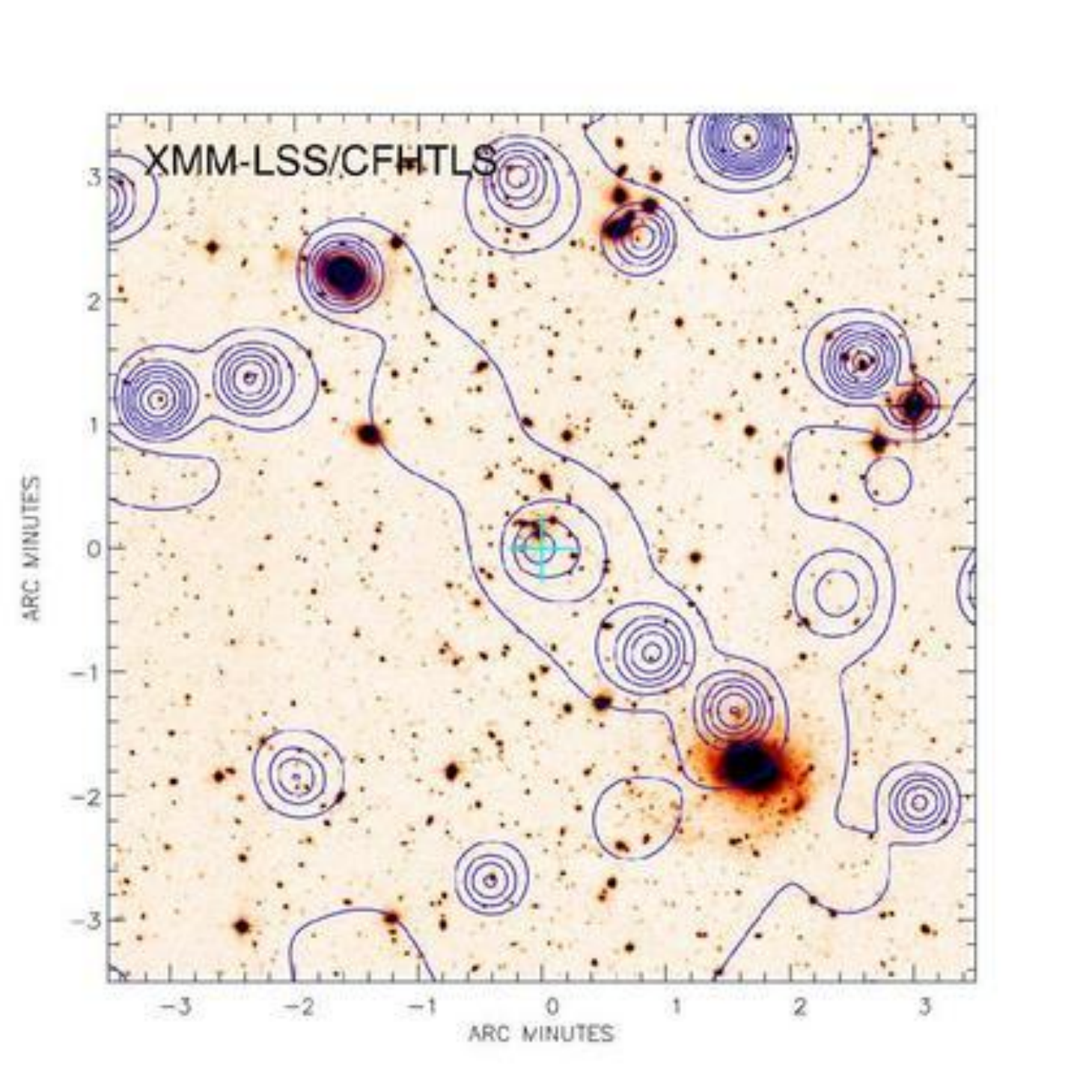}
    \includegraphics[viewport=0 0 420 450,clip,width=7.85cm]{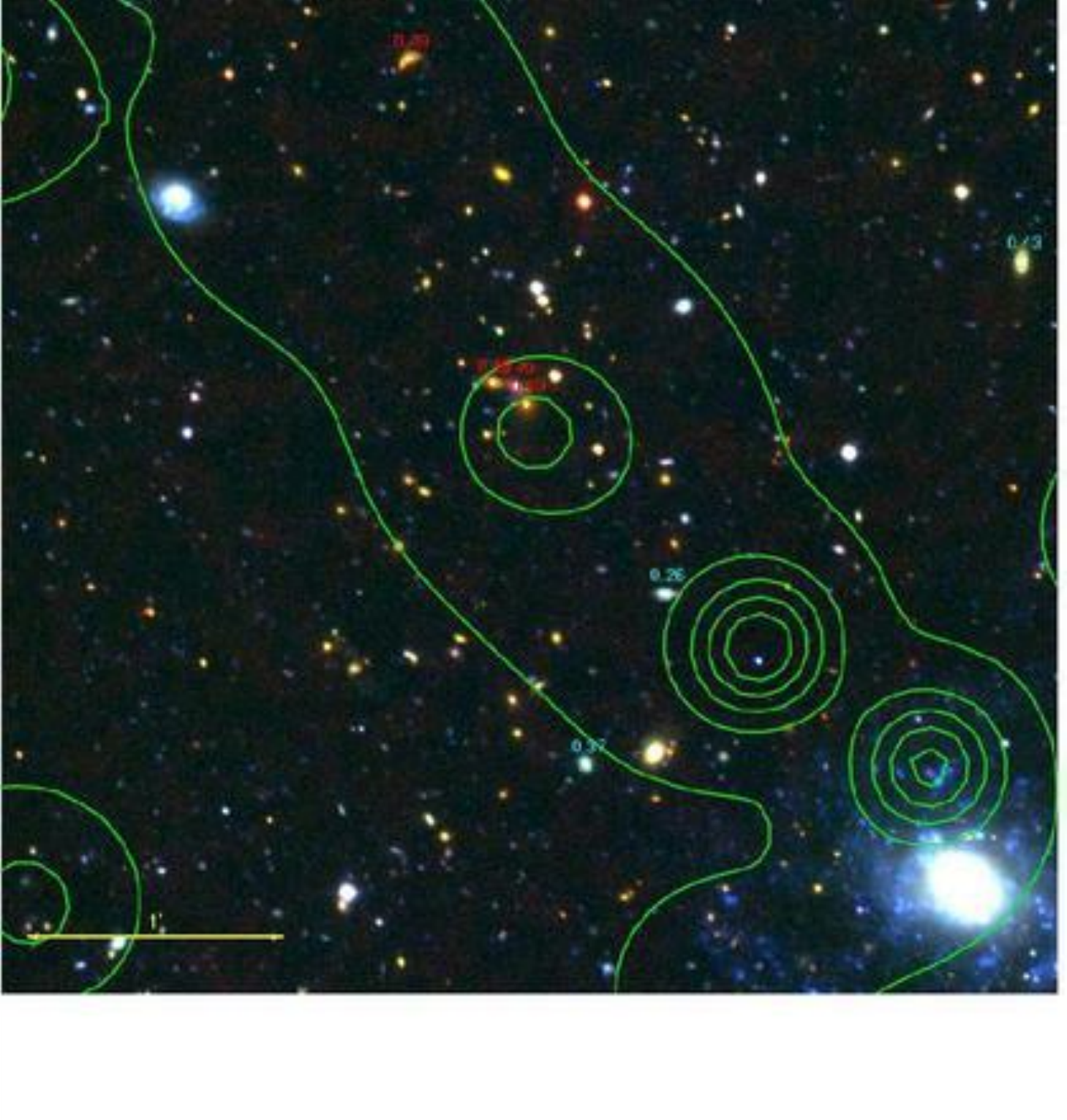}}}
  \rput(-8.6,13.7){\large (a)}
  \rput(-8.6,4.8){\large (b)}
  \end{pspicture}
 \contcaption{Images of the C1 clusters. (a) XLSSC-028. (b): XLSSC-049.}
\end{figure*}

\begin{figure*}
  \begin{pspicture}(0,0)(17.8,18.4)
  \vbox{
    \vbox{\includegraphics[height=9.2cm]{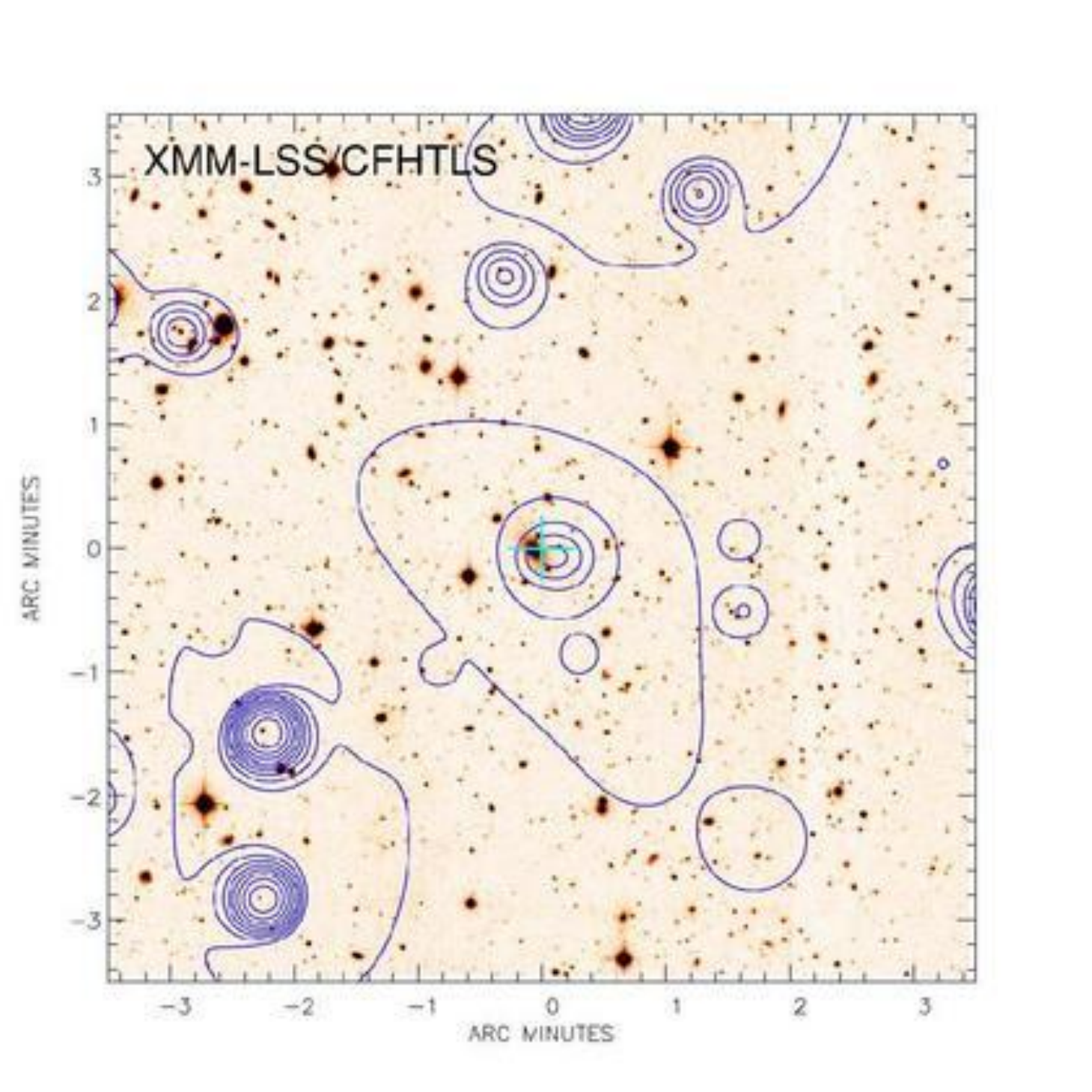}
    \includegraphics[viewport=0 0 420 450,clip,width=7.85cm]{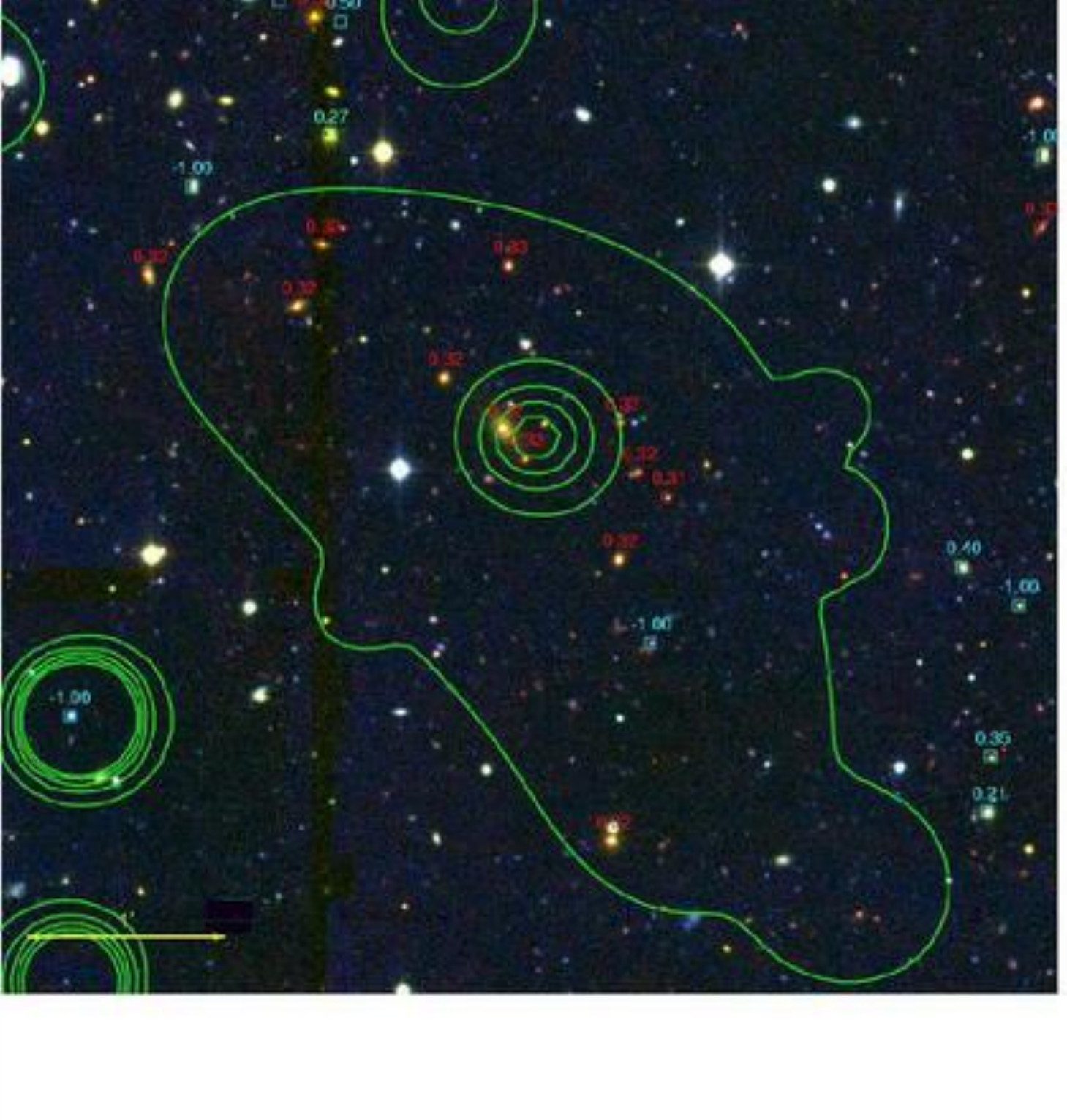}}
    \vbox{\includegraphics[width=9.2cm]{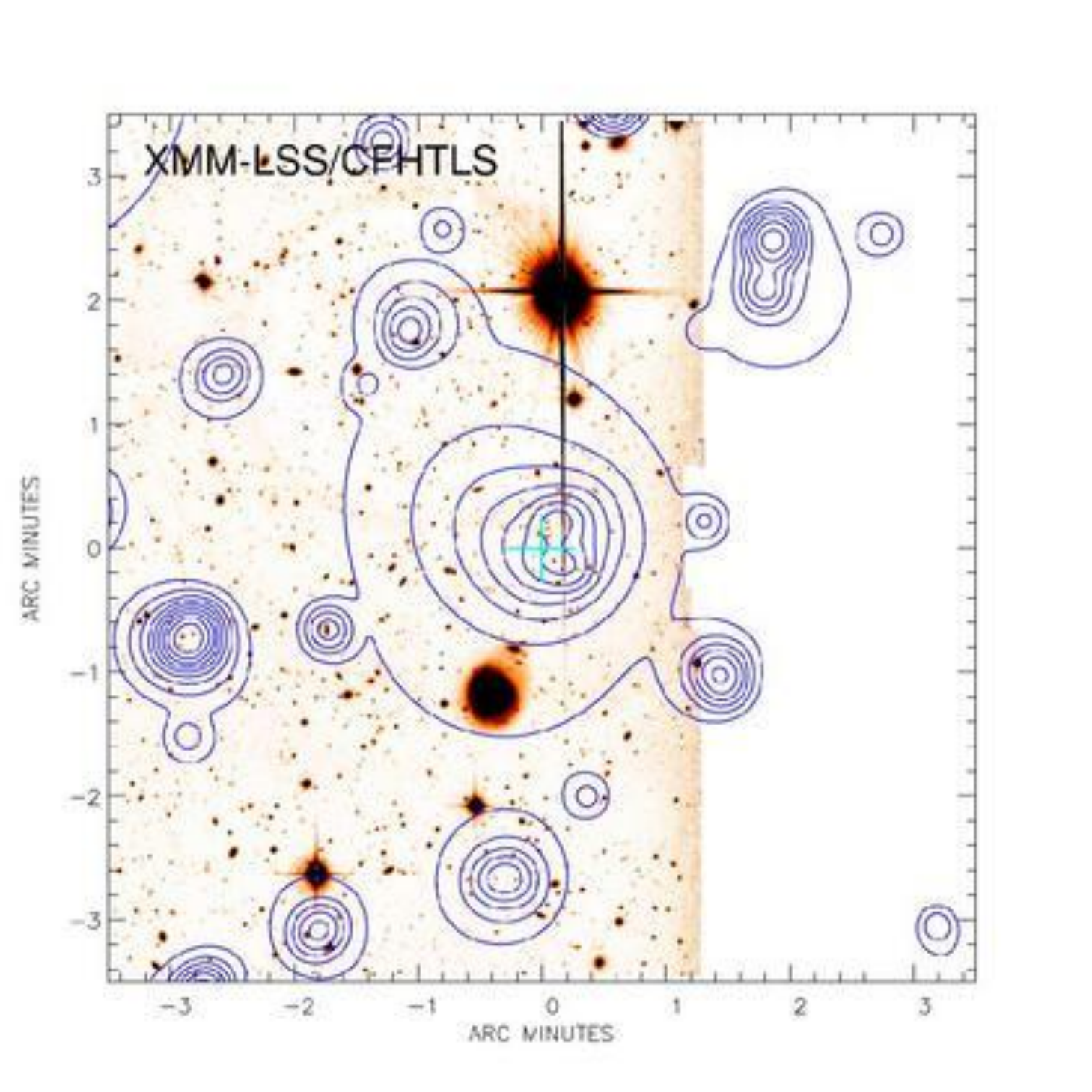}
    \includegraphics[viewport=0 0 420 450,clip,width=7.85cm]{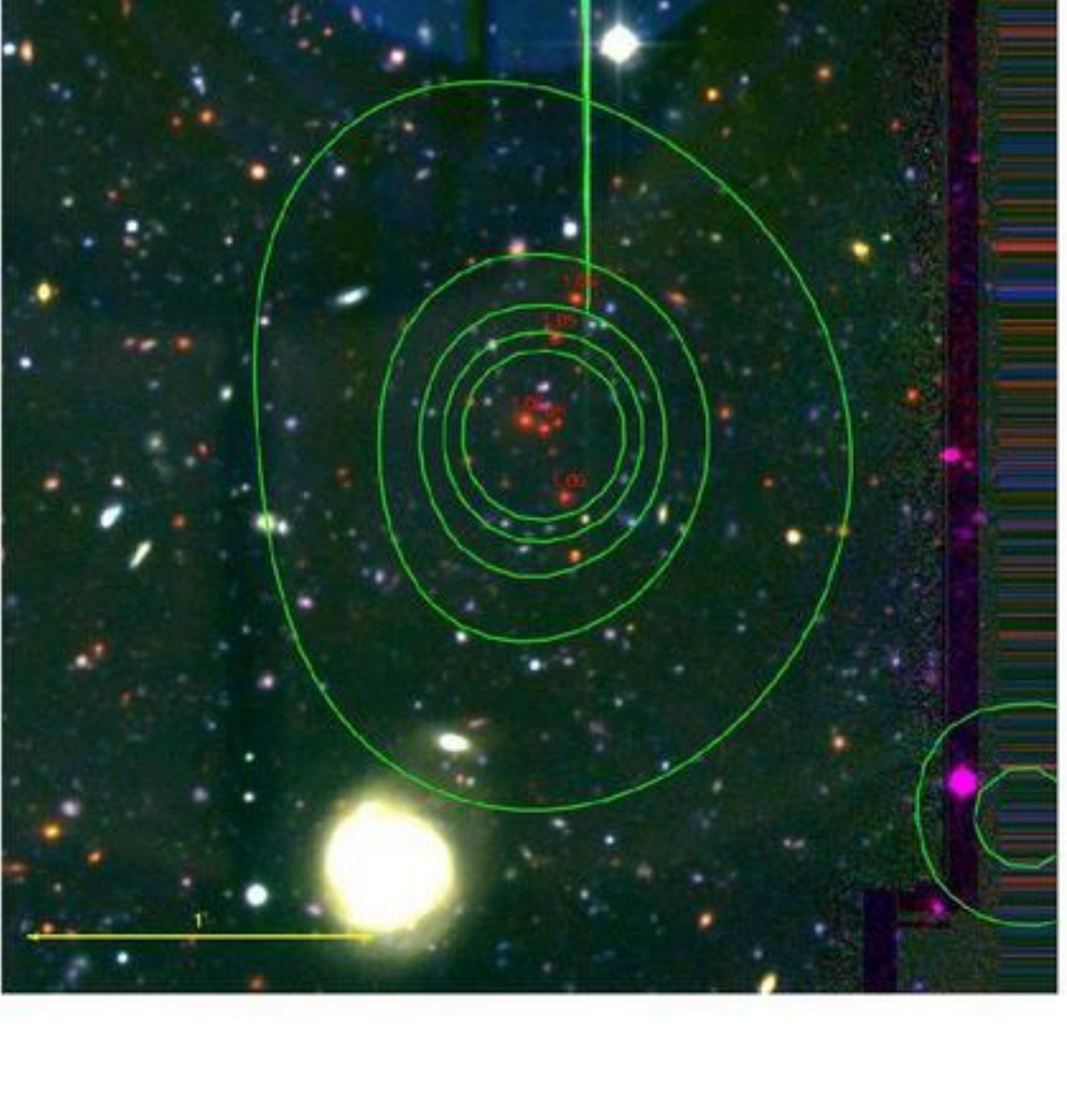}}}
  \rput(-8.6,13.7){\large (a)}
  \rput(-8.6,4.8){\large (b)}
  \end{pspicture}
 \contcaption{Images of the C1 clusters. (a) XLSSC-018. (b): XLSSC-029.}
\end{figure*}

\begin{figure*}
  \begin{pspicture}(0,0)(17.8,18.4)
  \vbox{
    \vbox{\includegraphics[height=9.2cm]{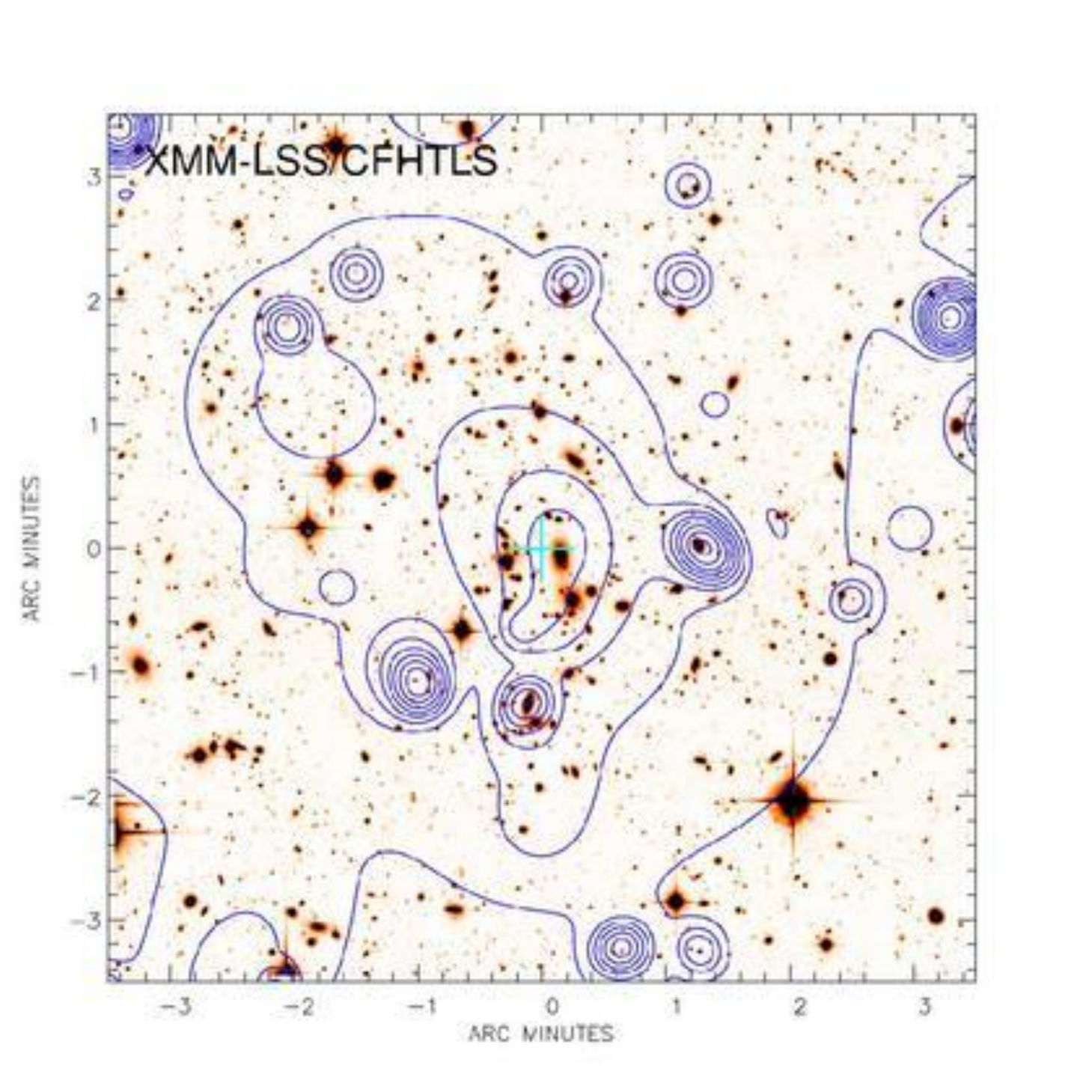}
    \includegraphics[viewport=0 0 420 450,clip,width=7.85cm]{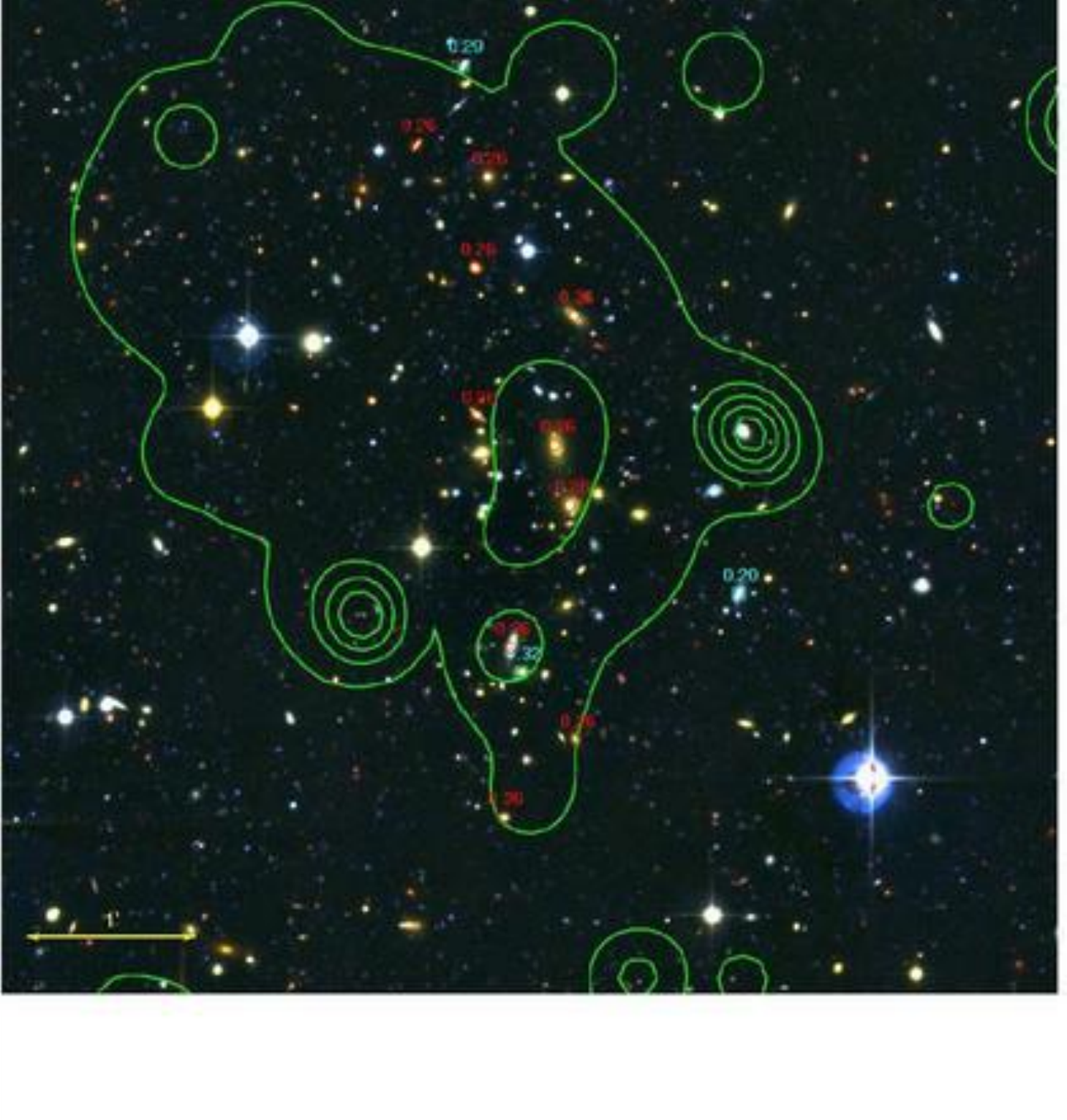}}
    \vbox{\includegraphics[width=9.2cm]{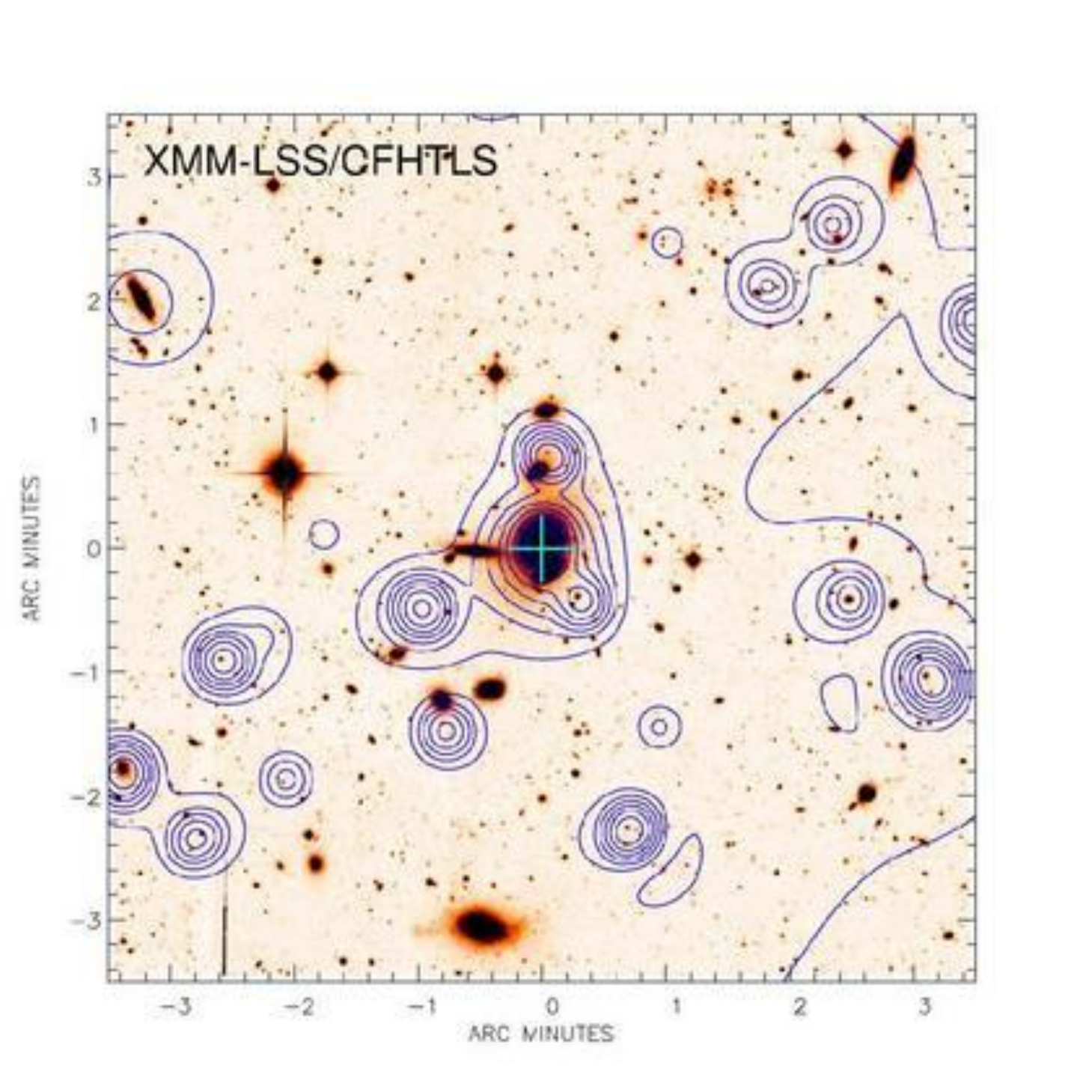}
    \includegraphics[viewport=0 0 420 450,clip,width=7.85cm]{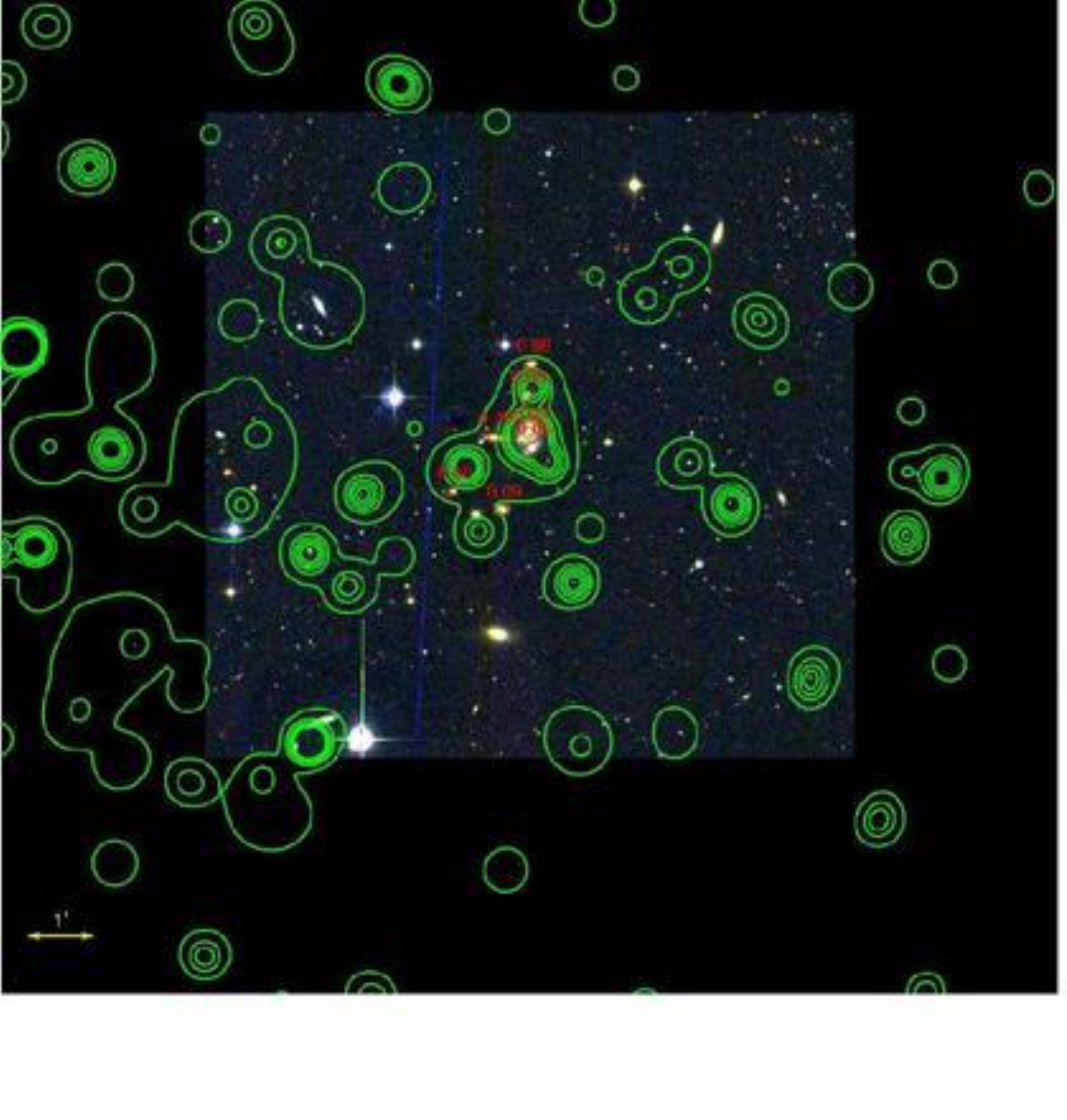}}}
  \rput(-8.6,13.7){\large (a)}
  \rput(-8.6,4.8){\large (b)}
  \end{pspicture}
 \contcaption{Images of the C1 clusters. (a) XLSSC-044. (b): XLSSC-021.}
\end{figure*}

\begin{figure*}
  \begin{pspicture}(0,0)(17.8,18.4)
  \vbox{
    \vbox{\includegraphics[height=9.2cm]{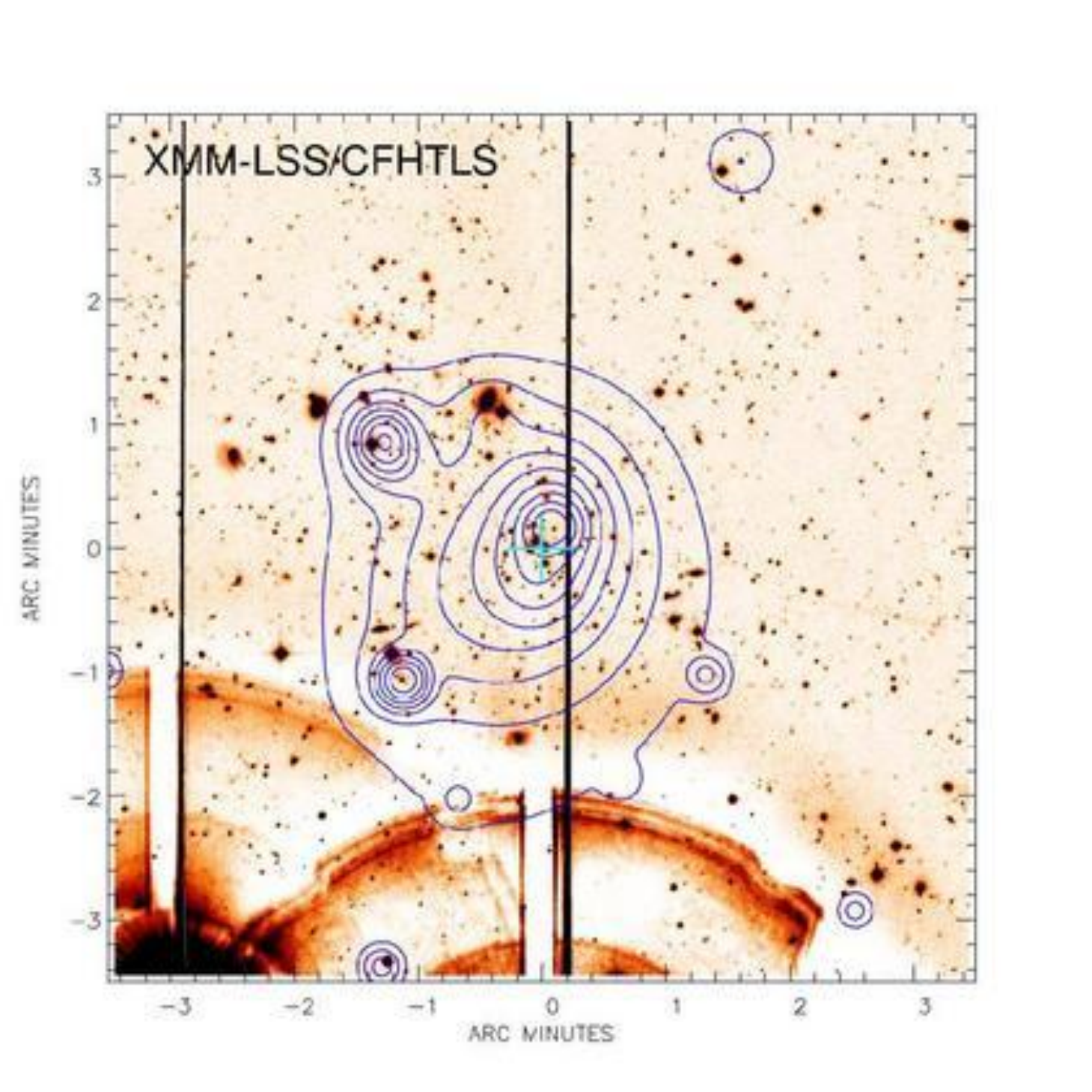}
    \includegraphics[viewport=0 0 420 450,clip,width=7.85cm]{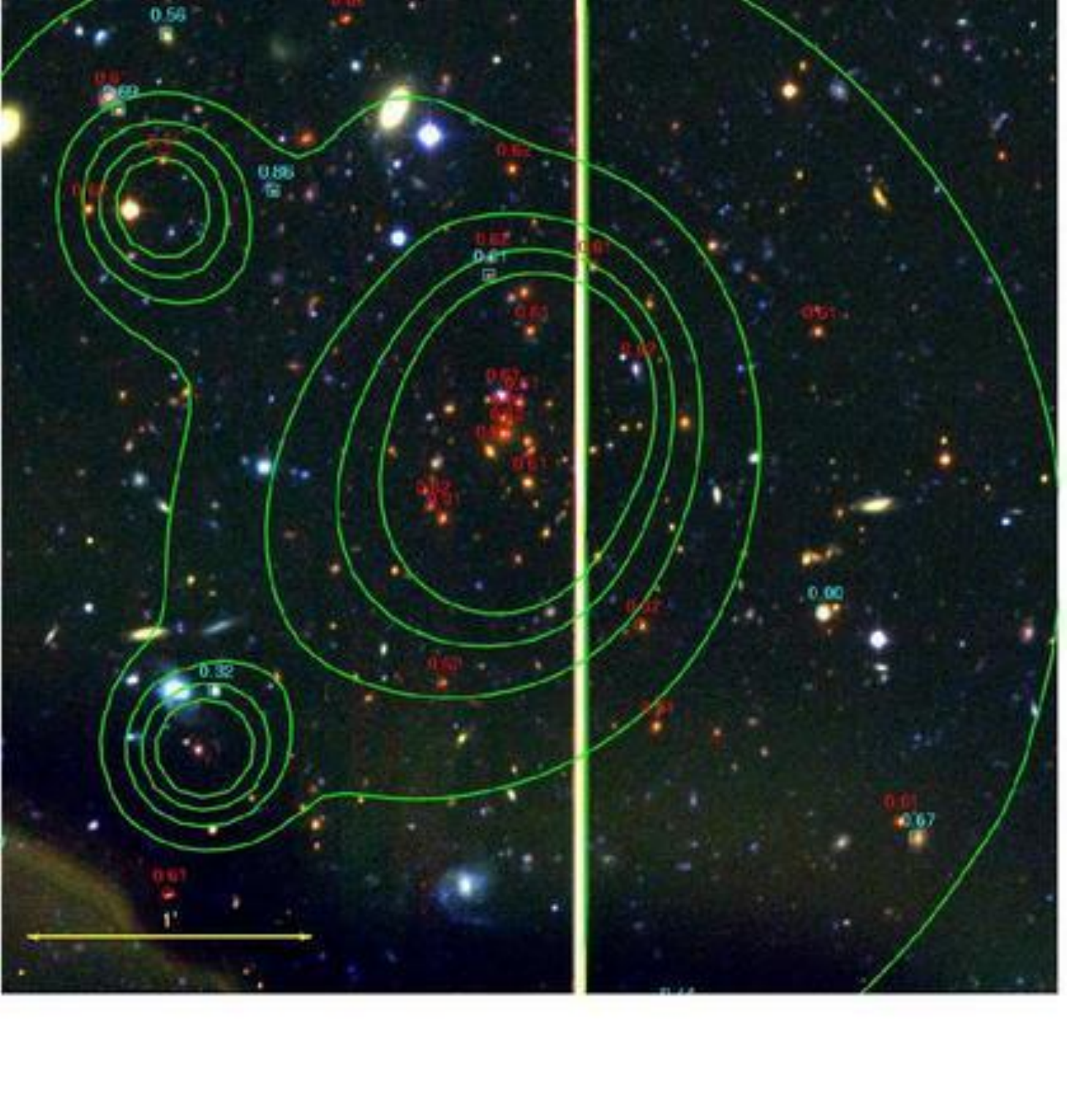}}
    \vbox{\includegraphics[width=9.2cm]{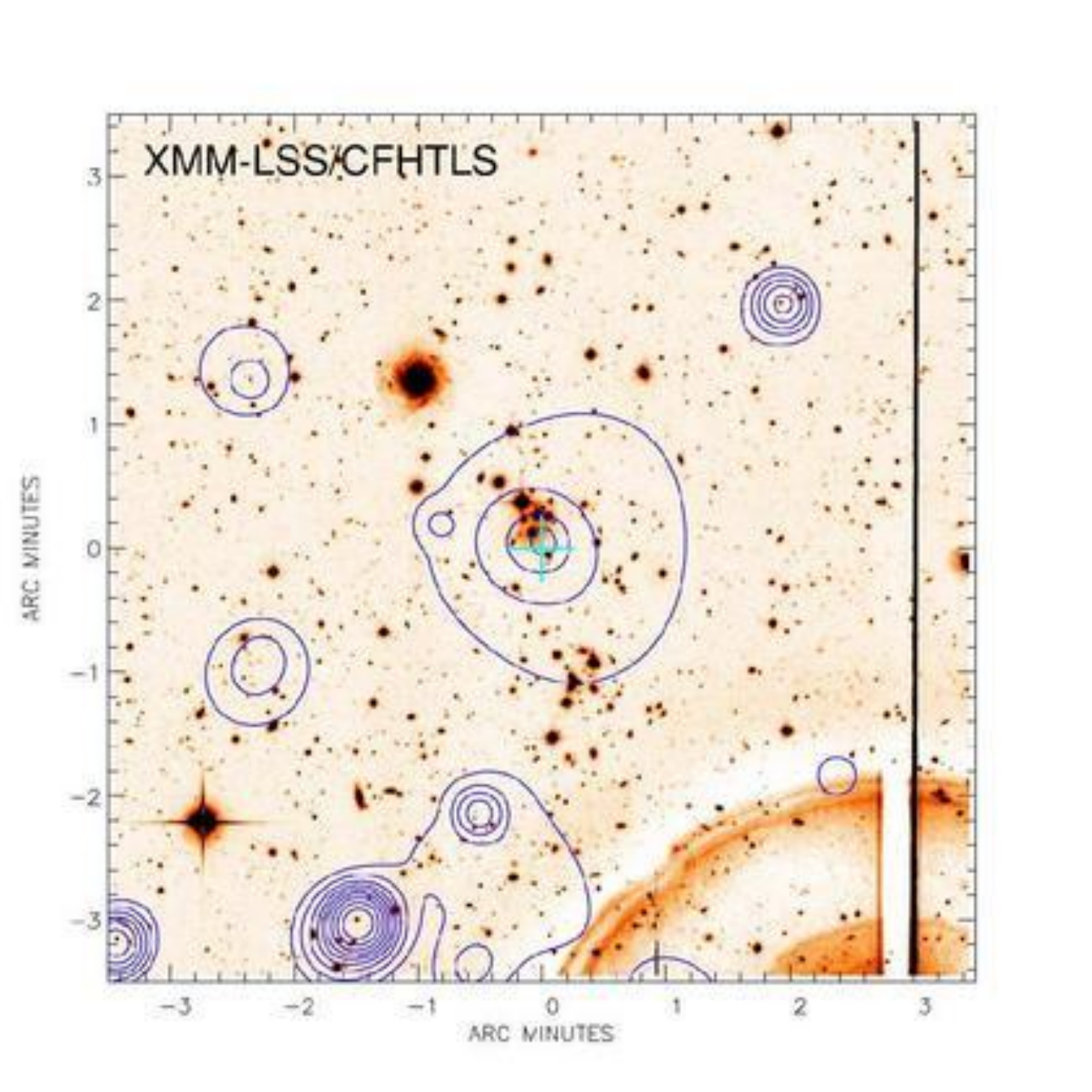}
    \includegraphics[viewport=0 0 420 450,clip,width=7.85cm]{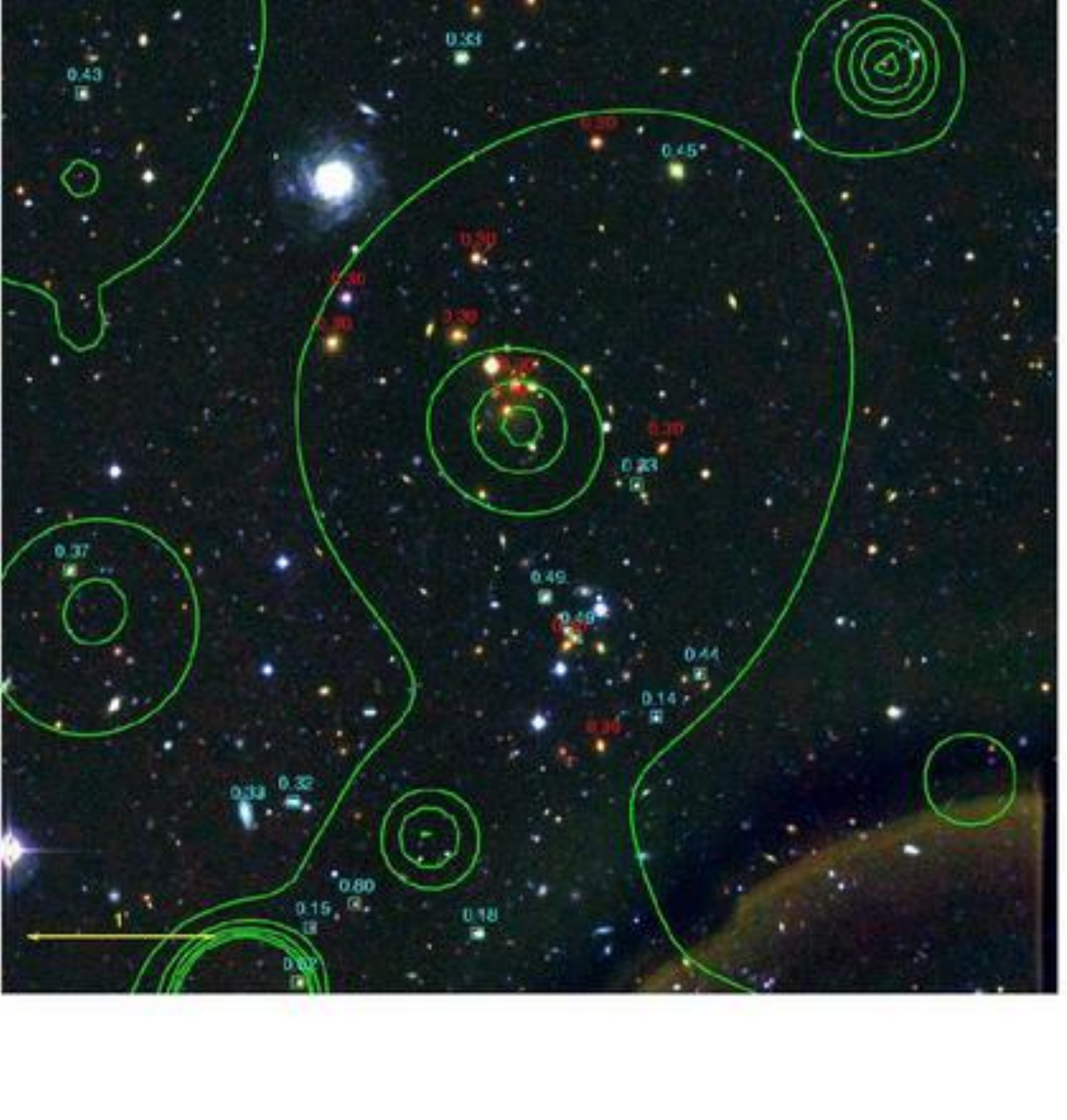}}}
  \rput(-8.6,13.7){\large (a)}
  \rput(-8.6,4.8){\large (b)}
  \end{pspicture}
 \contcaption{Images of the C1 clusters. (a) XLSSC-001. (b): XLSSC-008.}
\end{figure*}

\begin{figure*}
  \begin{pspicture}(0,0)(17.8,18.4)
  \vbox{
    \vbox{\includegraphics[height=9.2cm]{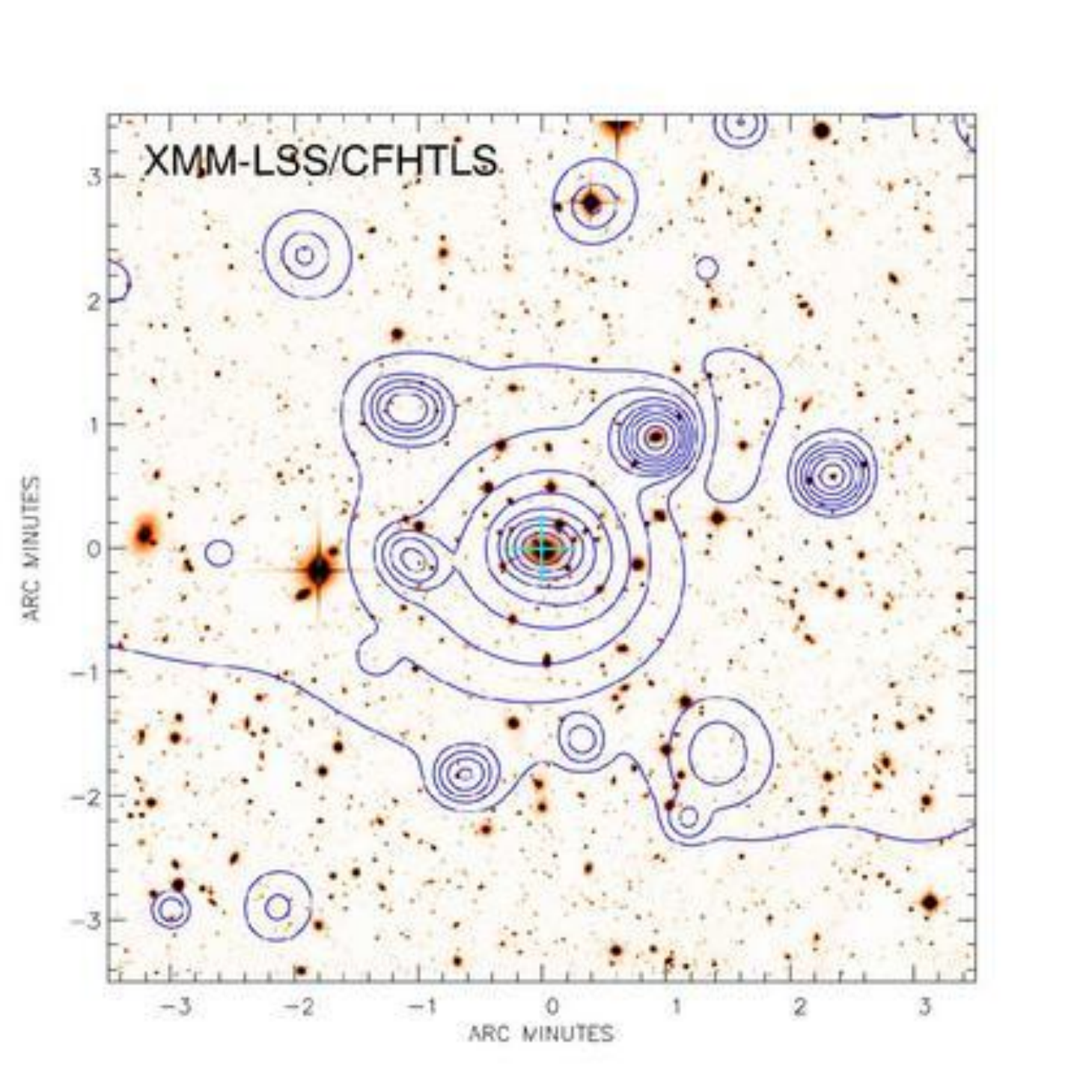}
    \includegraphics[viewport=0 0 420 450,clip,width=7.85cm]{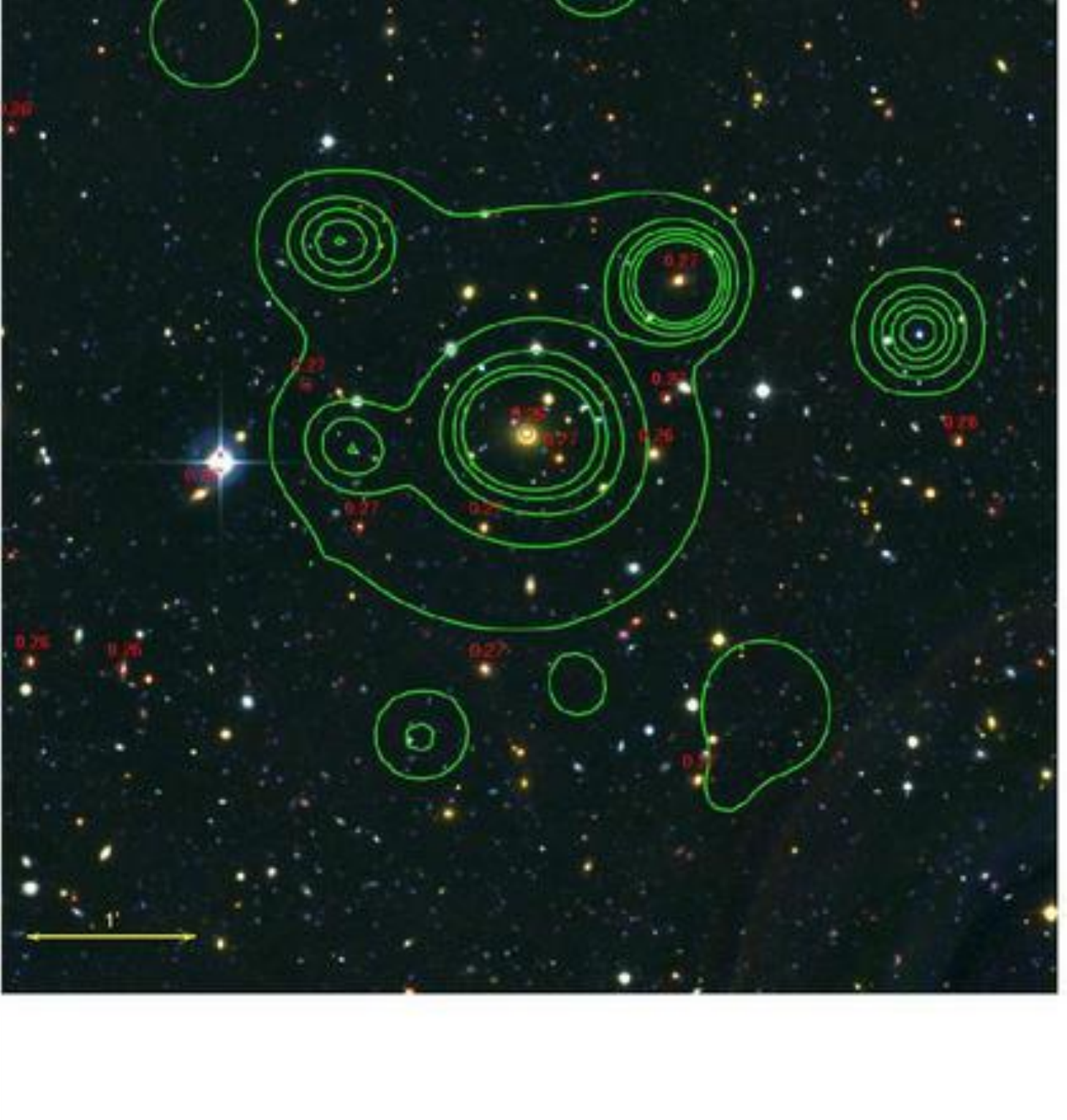}}
    \vbox{\includegraphics[width=9.2cm]{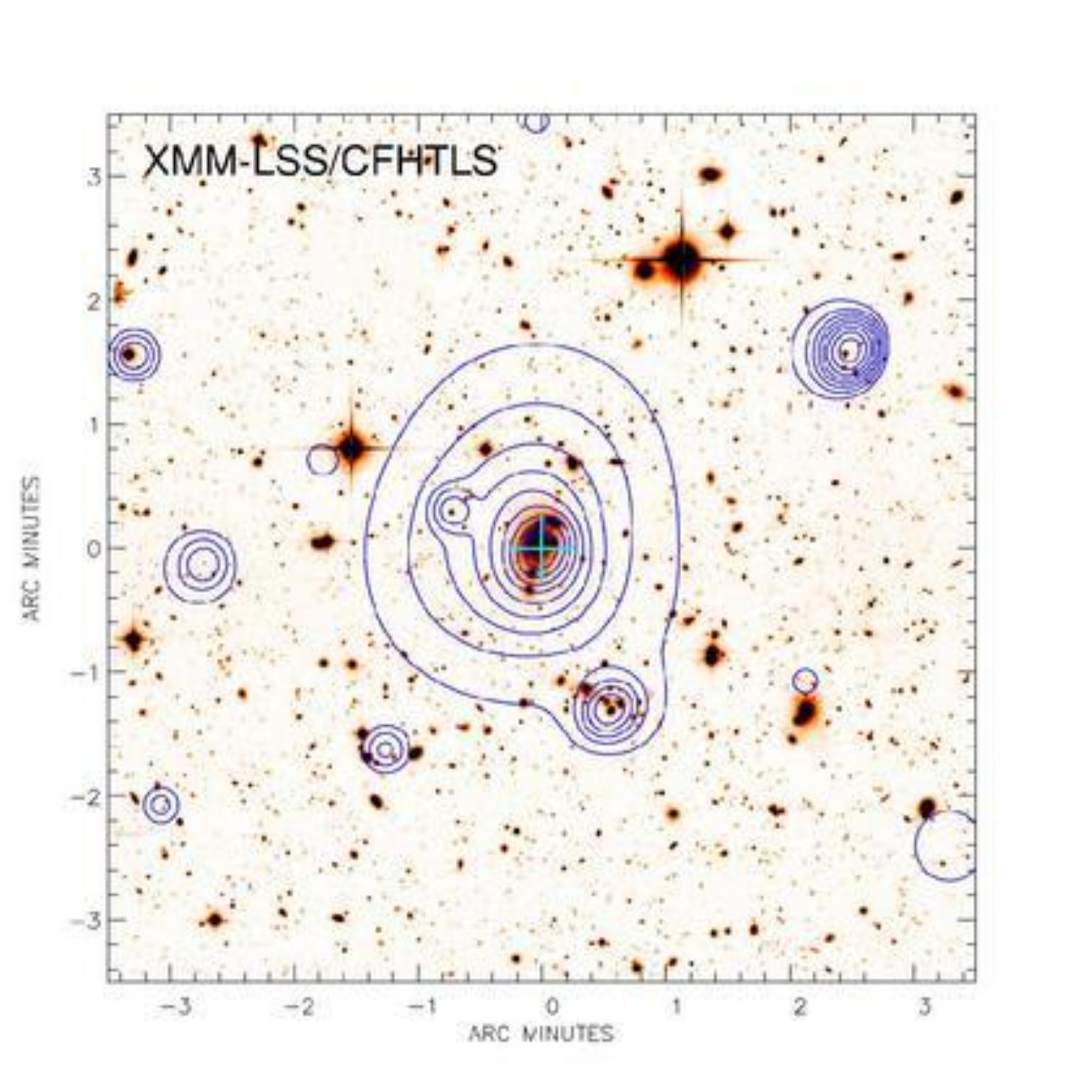}
    \includegraphics[viewport=0 0 420 450,clip,width=7.85cm]{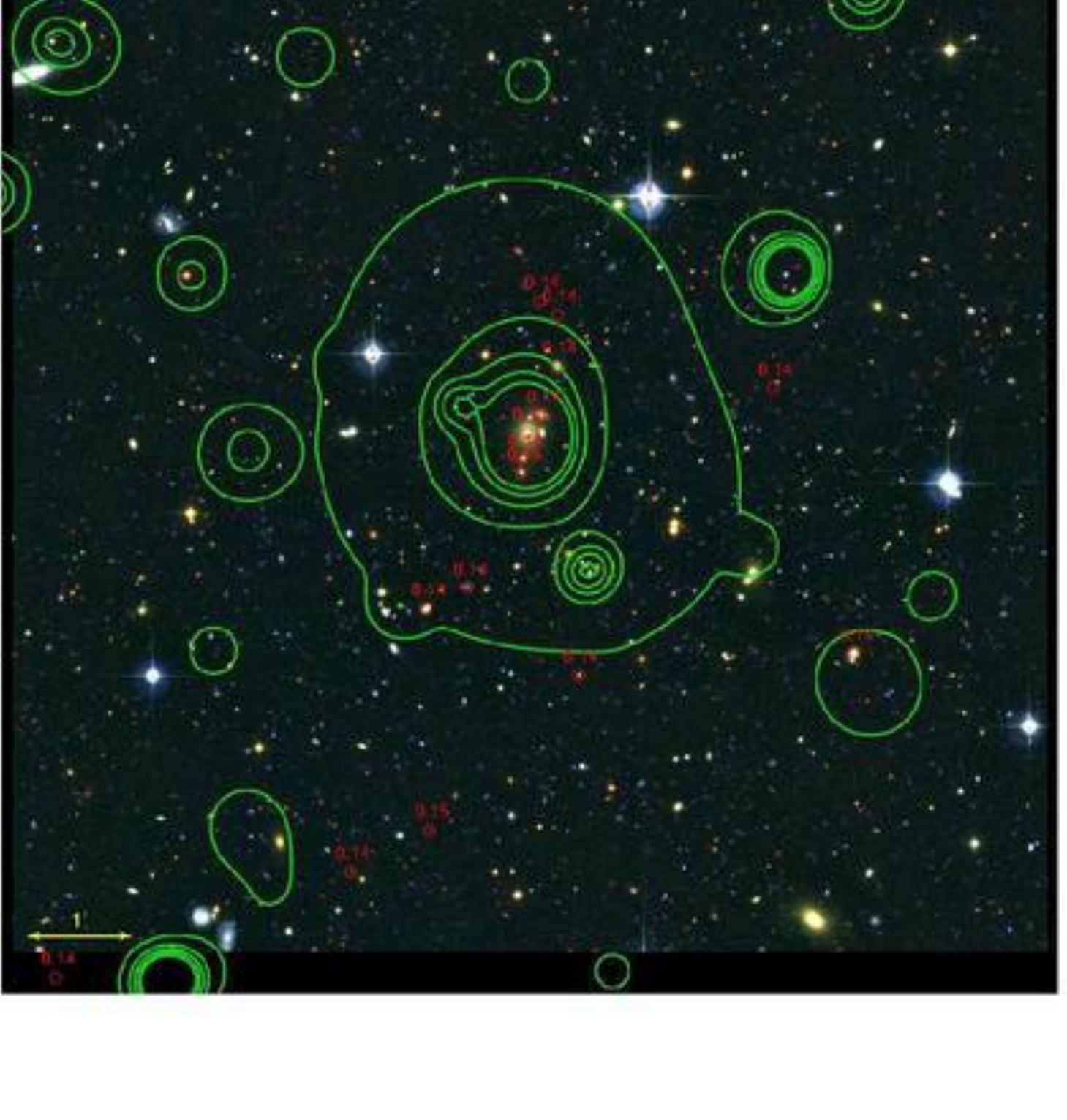}}}
  \rput(-8.6,13.7){\large (a)}
  \rput(-8.6,4.8){\large (b)}
  \end{pspicture}
 \contcaption{Images of the C1 clusters. (a) XLSSC-025. (b): XLSSC-041.}
\end{figure*}

\begin{figure*}
  \begin{pspicture}(0,0)(17.8,18.4)
  \vbox{
    \vbox{\includegraphics[height=9.2cm]{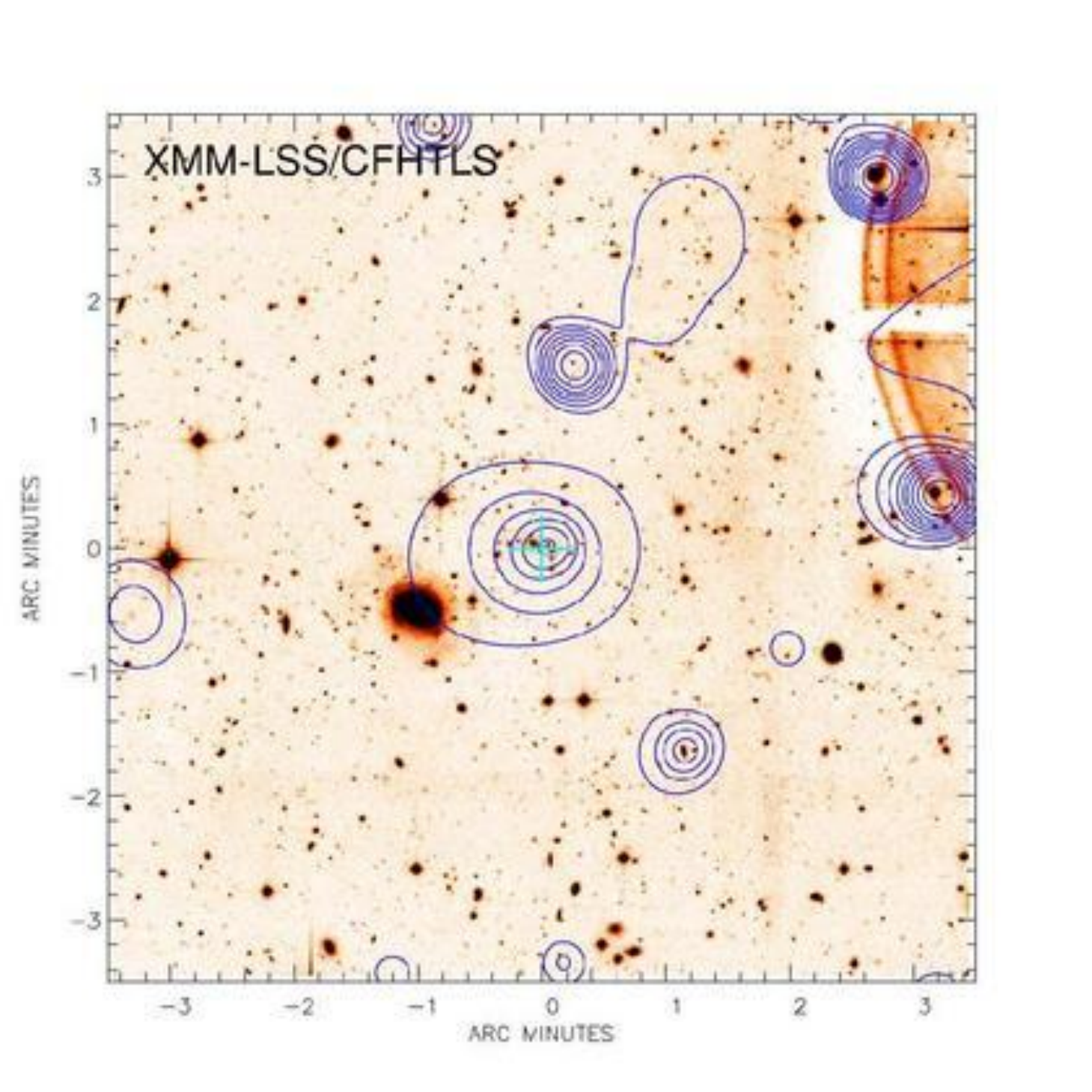}
    \includegraphics[viewport=0 0 420 450,clip,width=7.85cm]{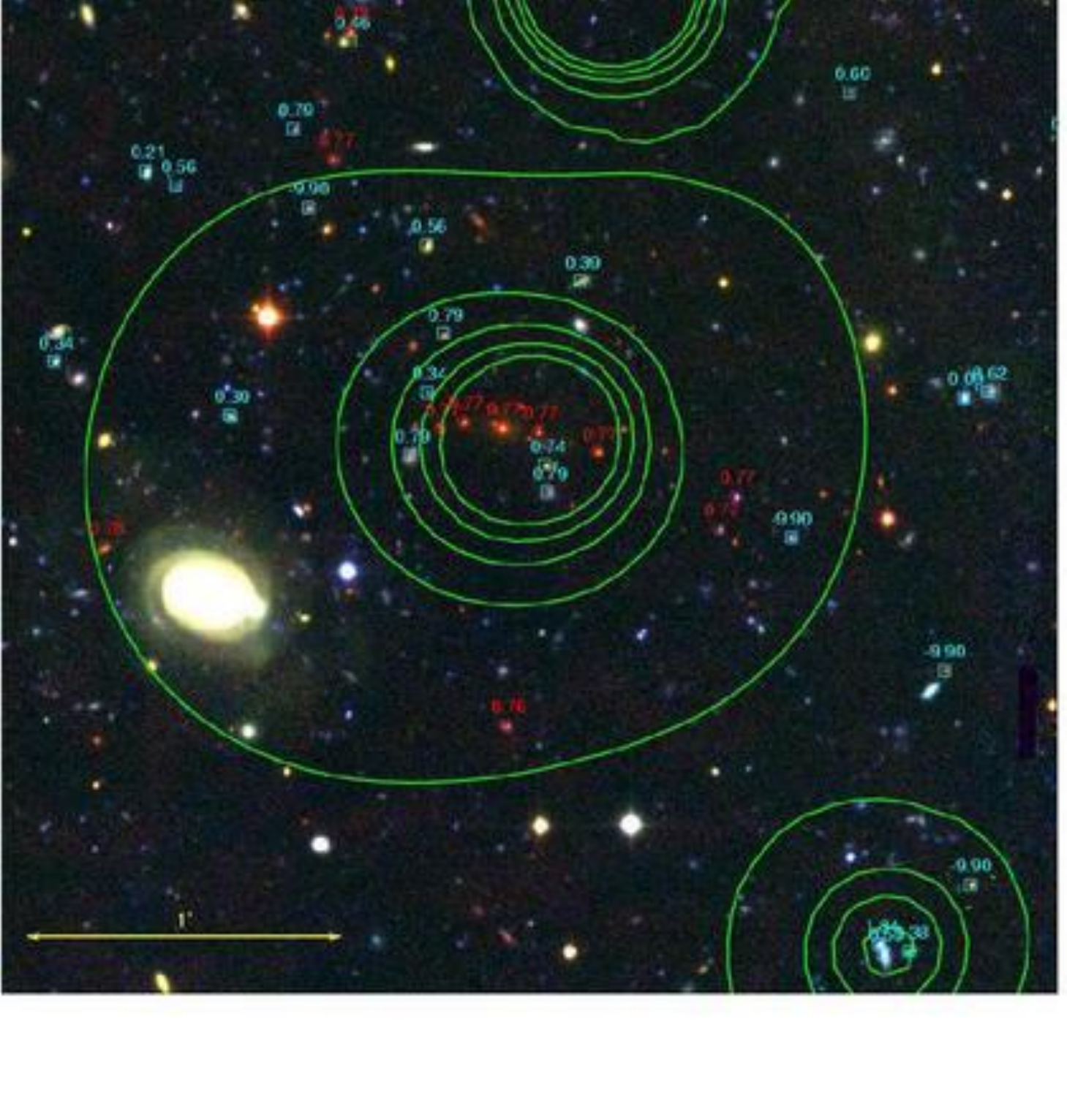}}
    \vbox{\includegraphics[width=9.2cm]{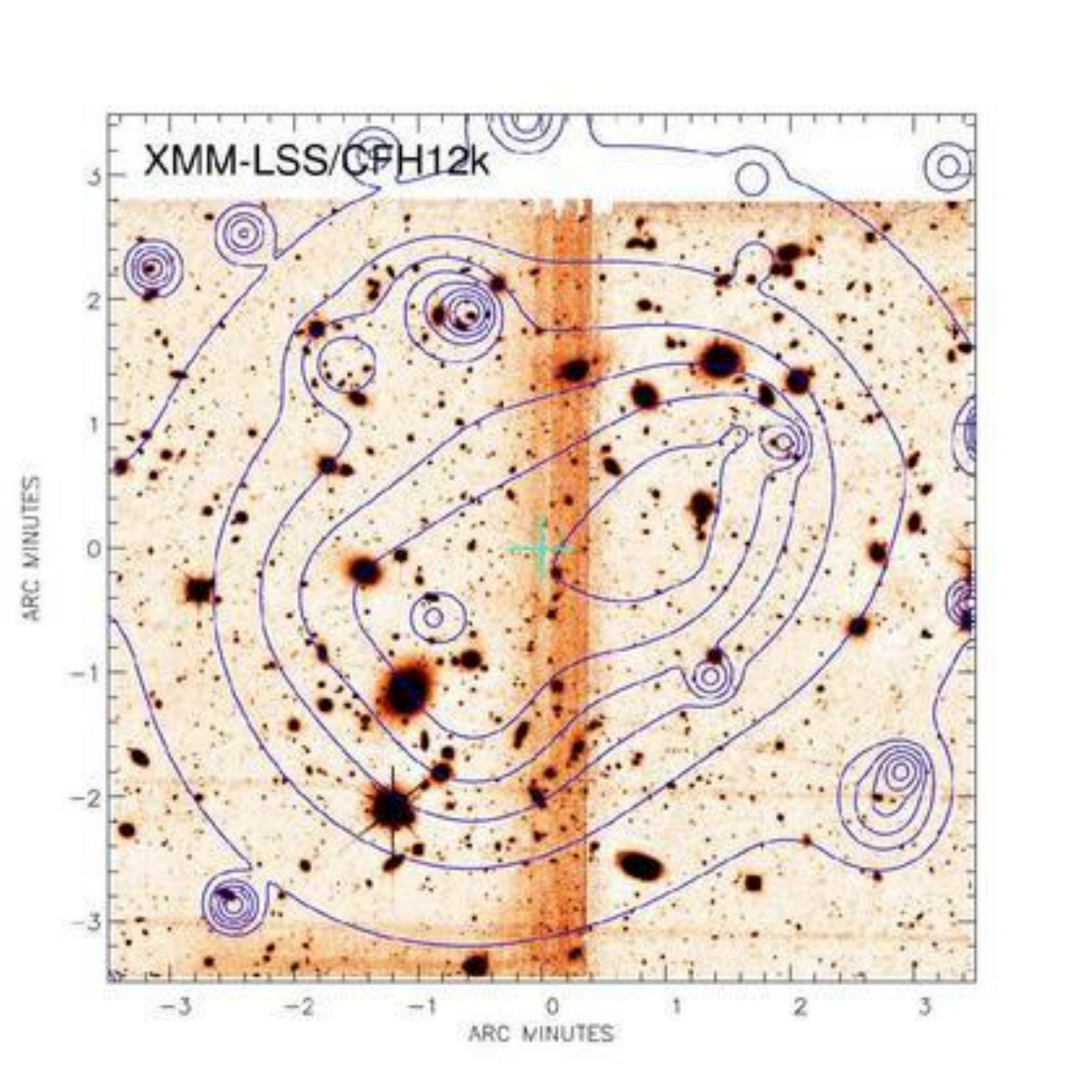}
    \includegraphics[viewport=0 0 420 450,clip,width=7.85cm]{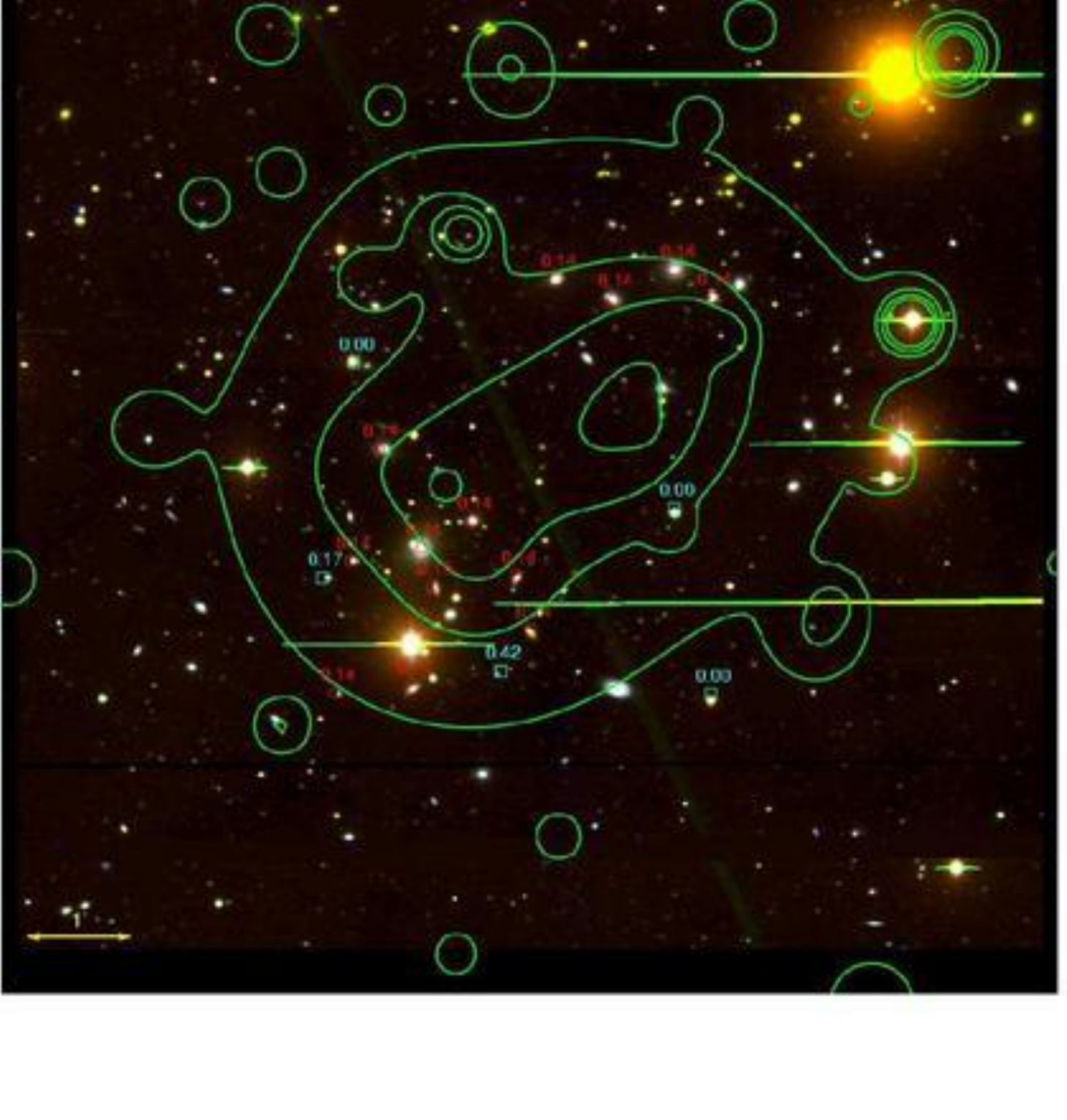}}}
  \rput(-8.6,13.7){\large (a)}
  \rput(-8.6,4.8){\large (b)}
  \end{pspicture}
 \contcaption{Images of the C1 clusters. (a) XLSSC-002. (b): XLSSC-050.}
\end{figure*}

\begin{figure*}
  \begin{pspicture}(0,0)(17.8,18.4)
  \vbox{
    \vbox{\includegraphics[height=9.2cm]{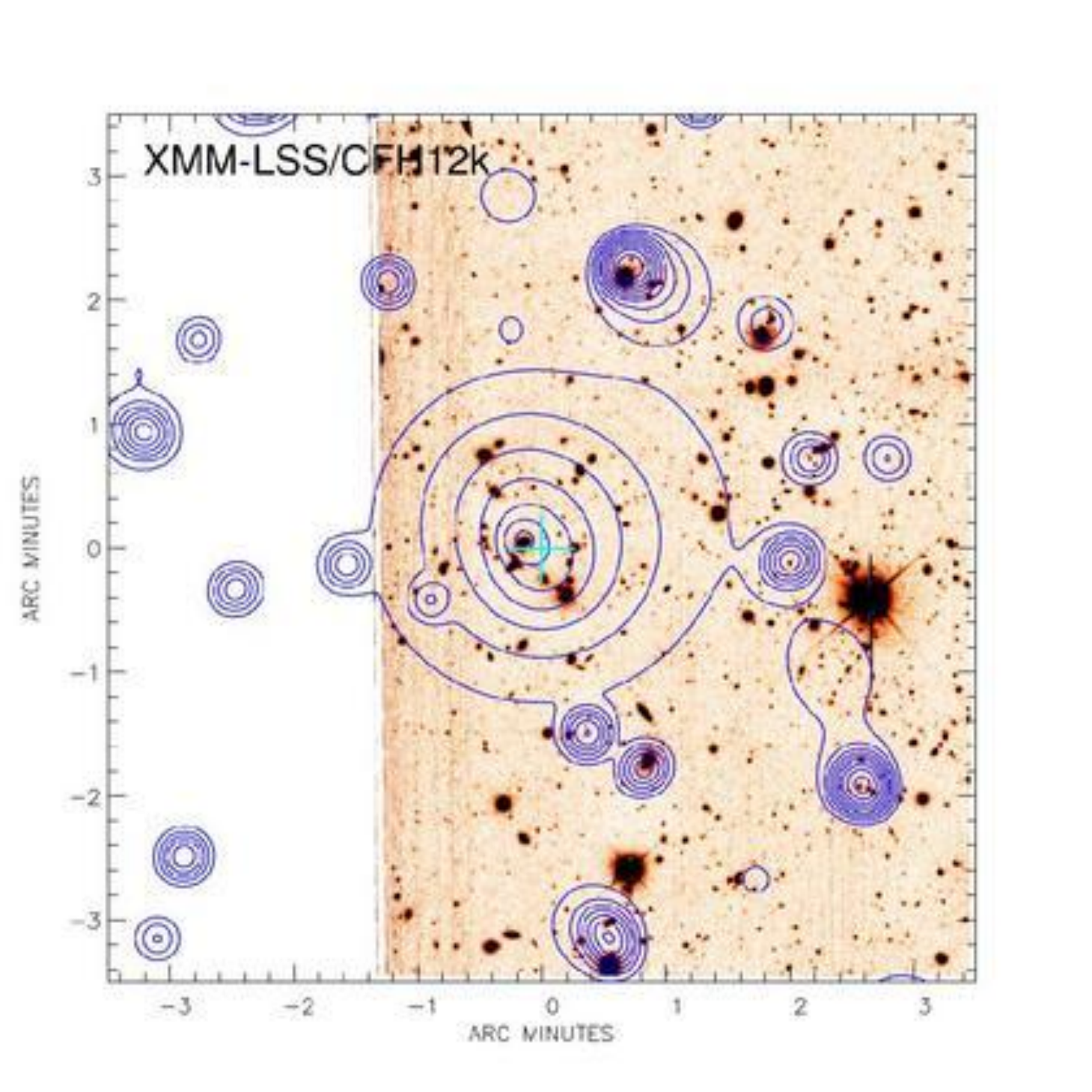}
    \includegraphics[viewport=0 0 420 450,clip,width=7.85cm]{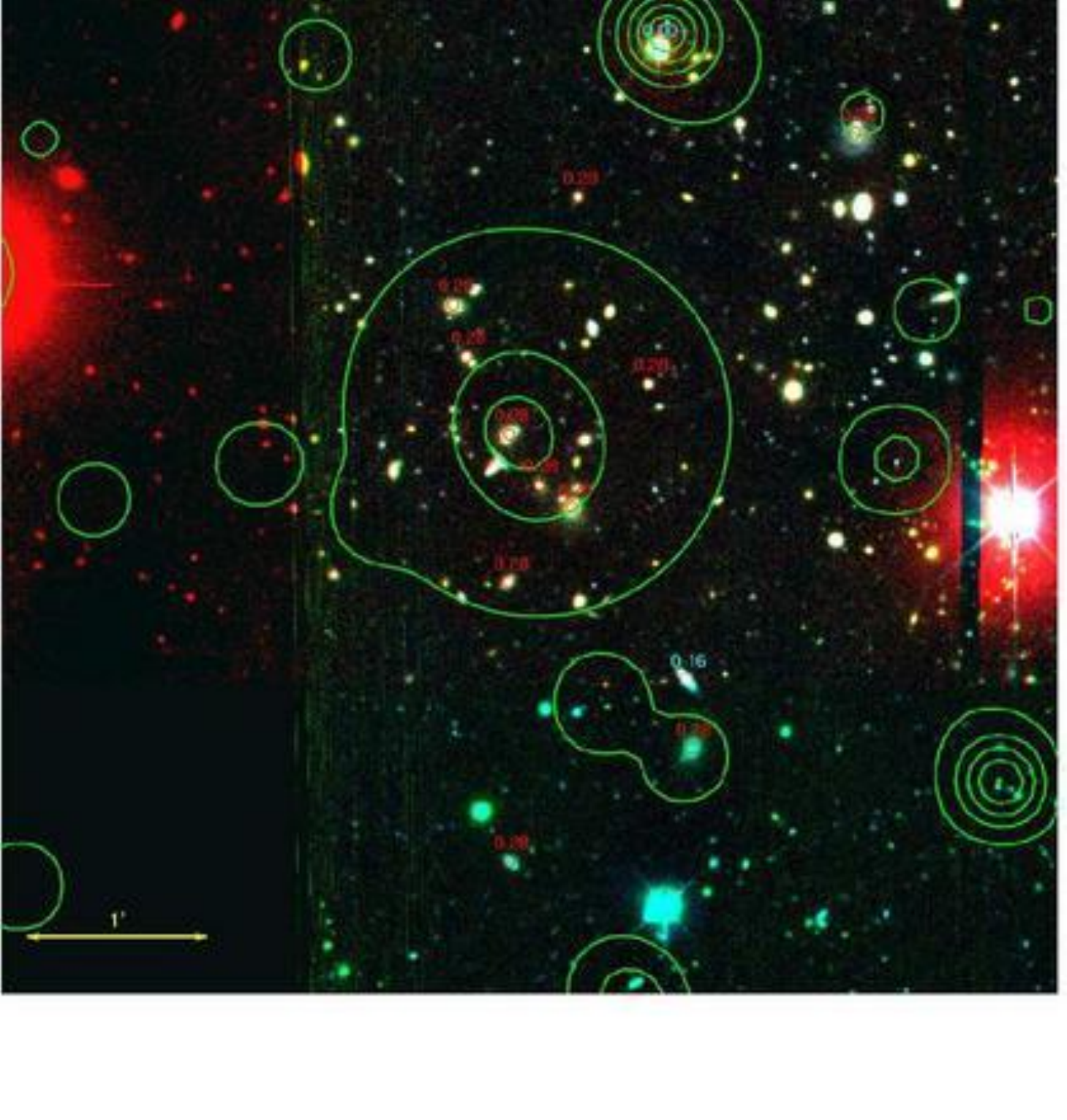}}
    \vbox{\includegraphics[width=9.2cm]{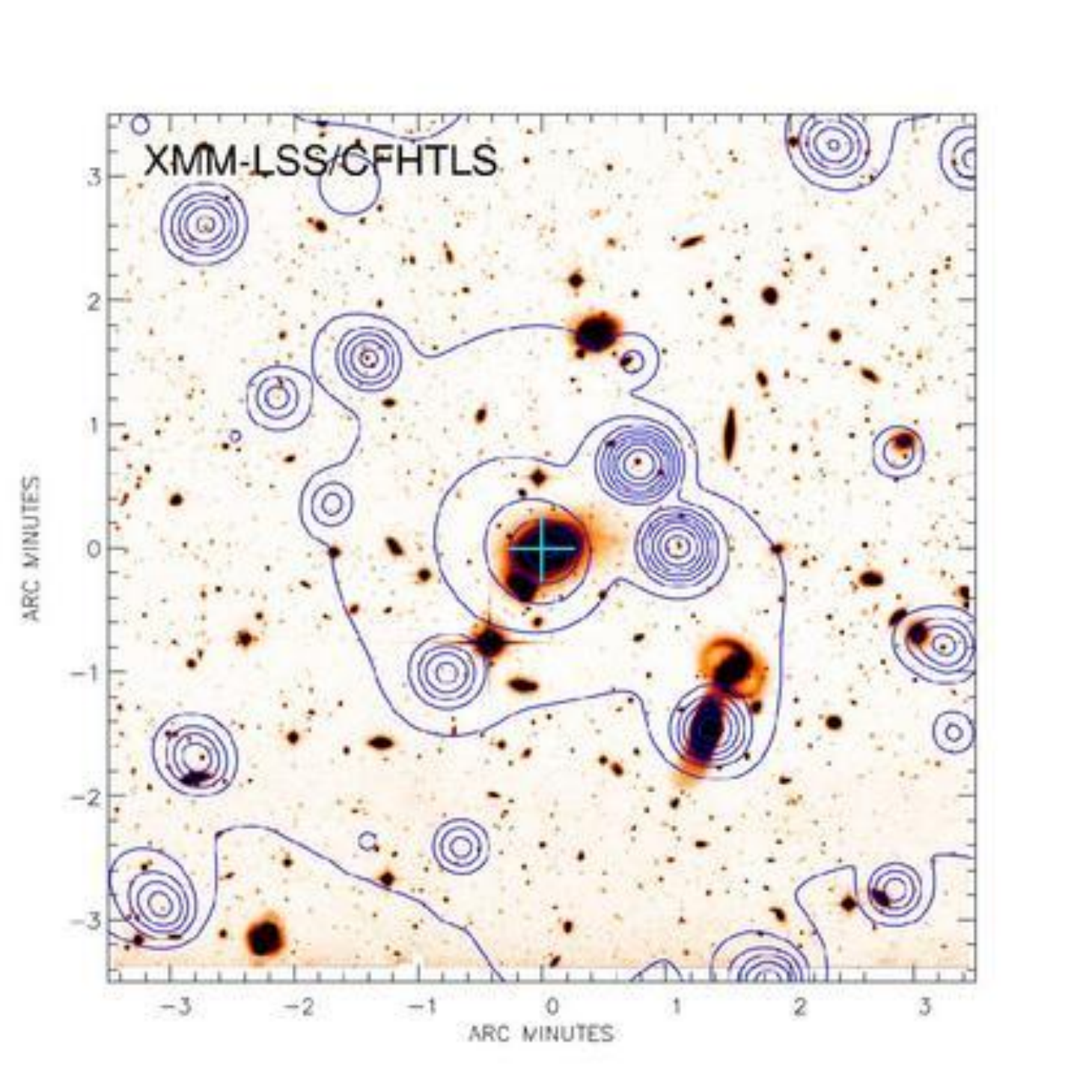}
    \includegraphics[viewport=0 0 420 450,clip,width=7.85cm]{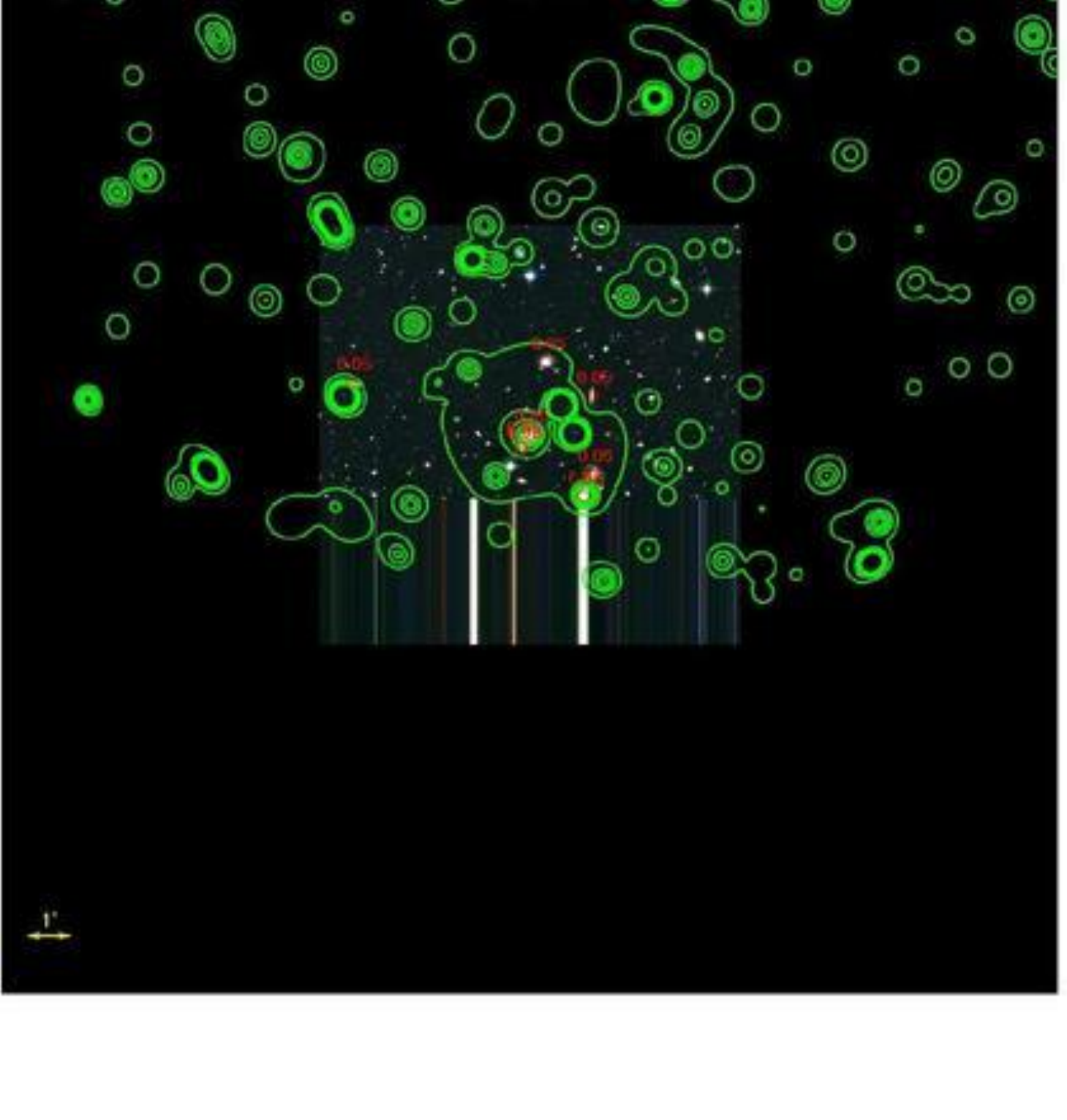}}}
  \rput(-8.6,13.7){\large (a)}
  \rput(-8.6,4.8){\large (b)}
  \end{pspicture}
 \contcaption{Images of the C1 clusters. (a) XLSSC-051. (b): XLSSC-011.}
\end{figure*}

\FloatBarrier

\begin{figure*}
  \begin{pspicture}(0,0)(17.8,18.4)
  \vbox{
    \vbox{\includegraphics[height=9.2cm]{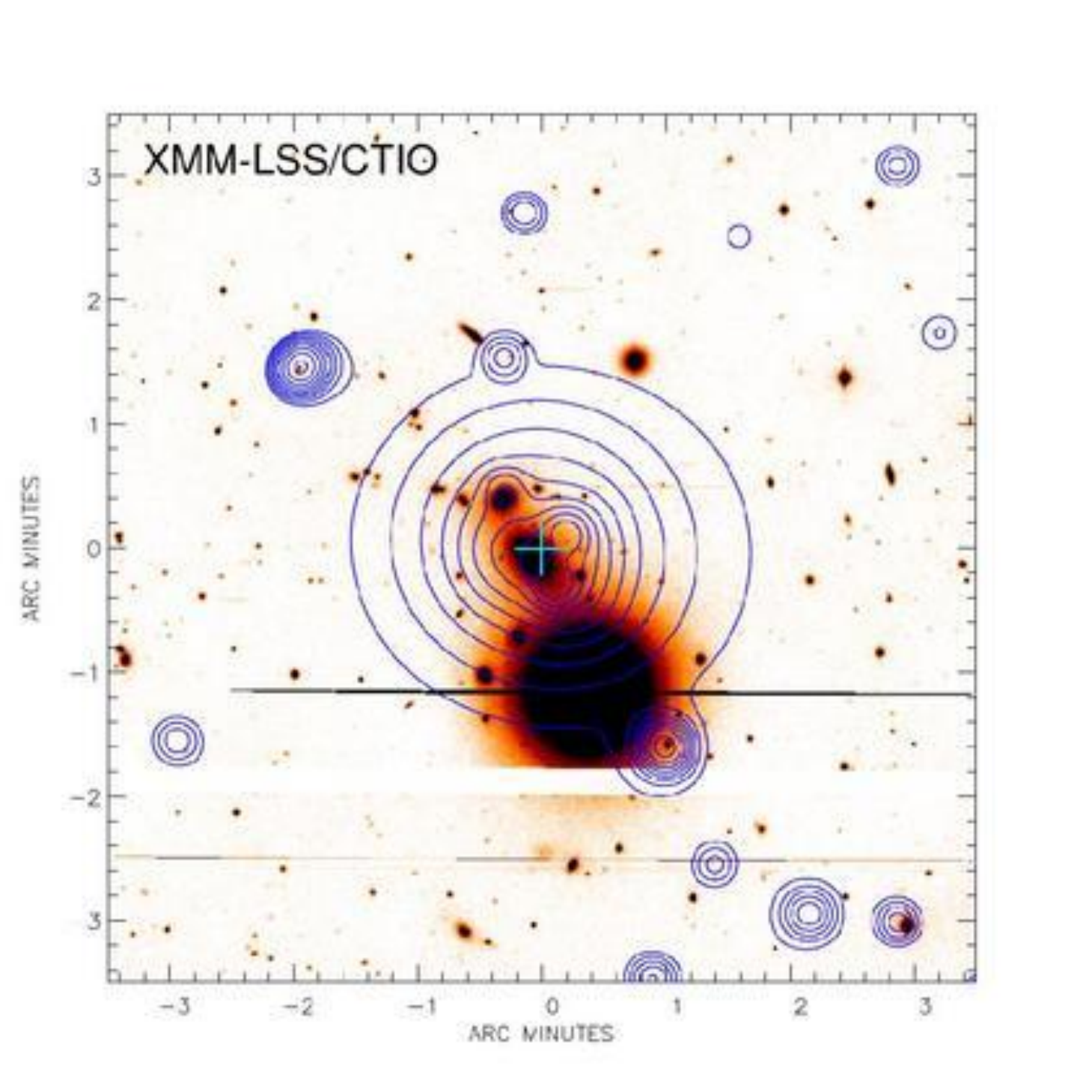}
    \includegraphics[viewport=0 0 420 450,clip,width=7.85cm]{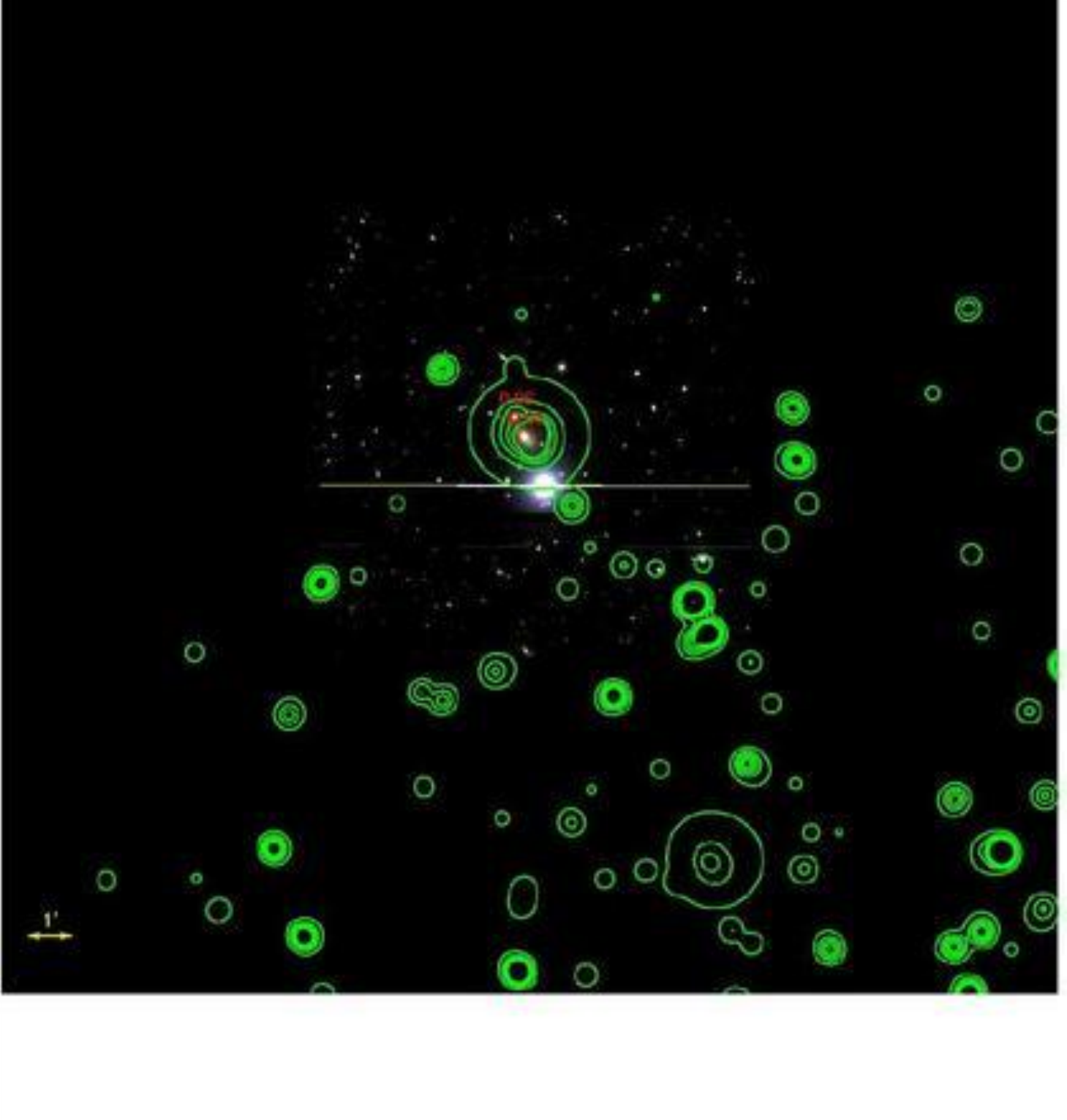}}
    \vbox{\includegraphics[width=9.2cm]{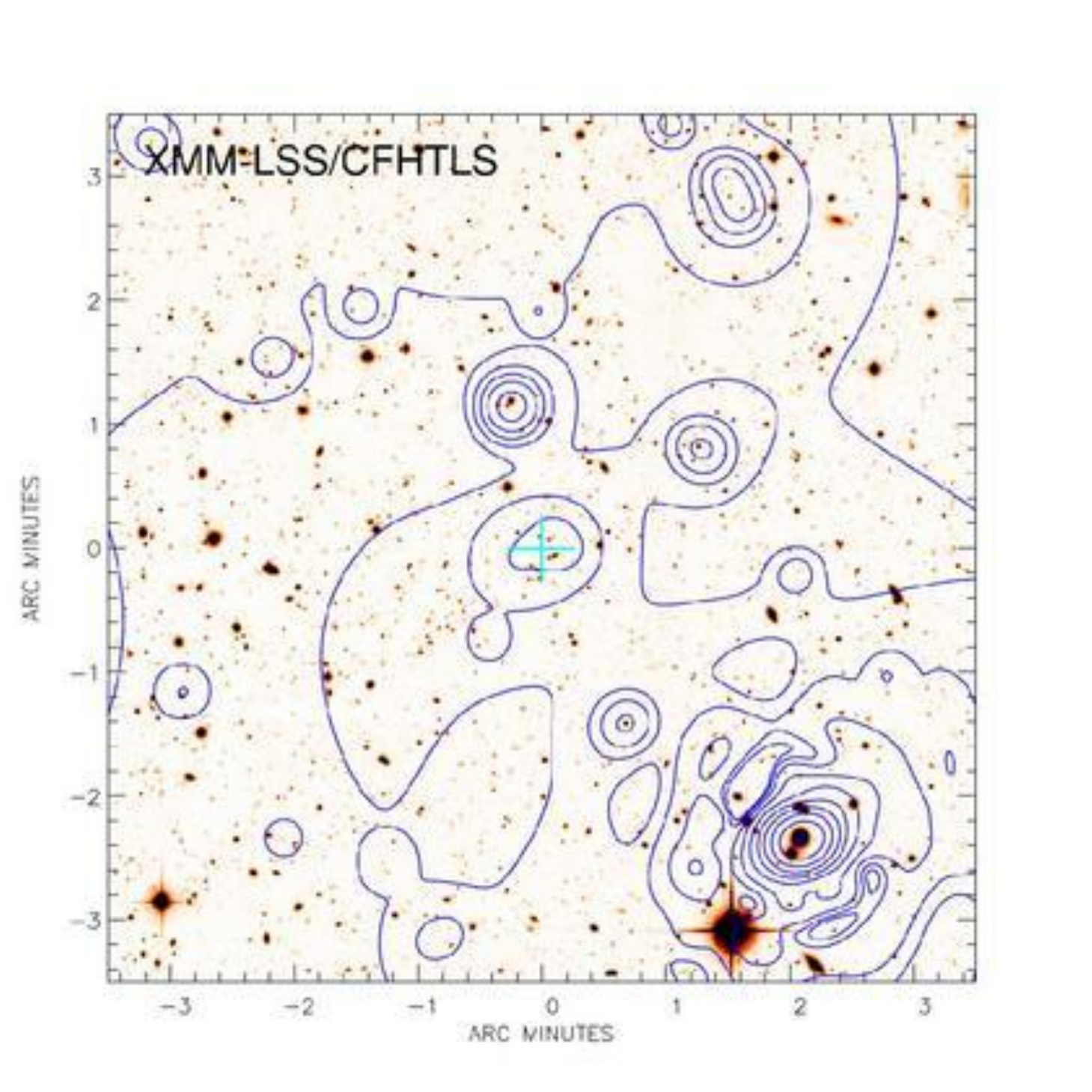}
    \includegraphics[viewport=0 0 420 450,clip,width=7.85cm]{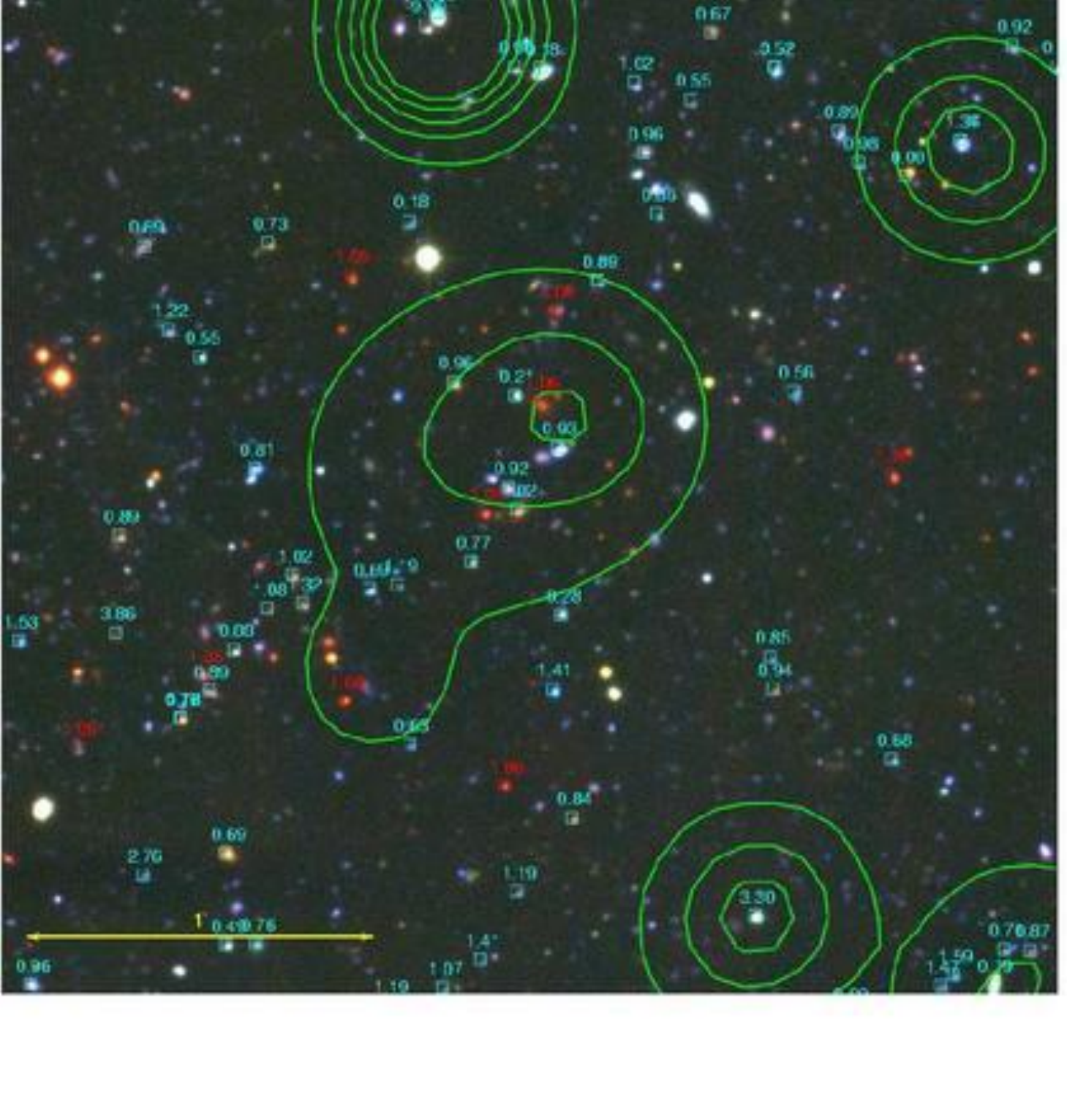}}}
  \rput(-8.6,13.7){\large (a)}
  \rput(-8.6,4.8){\large (b)}
  \end{pspicture}
 \contcaption{Images of the C1 clusters. (a) XLSSC-052. (b): XLSSC-005.}
\end{figure*}

\begin{figure*}
  \begin{pspicture}(0,0)(17.8,18.4)
  \vbox{
    \vbox{\includegraphics[height=9.2cm]{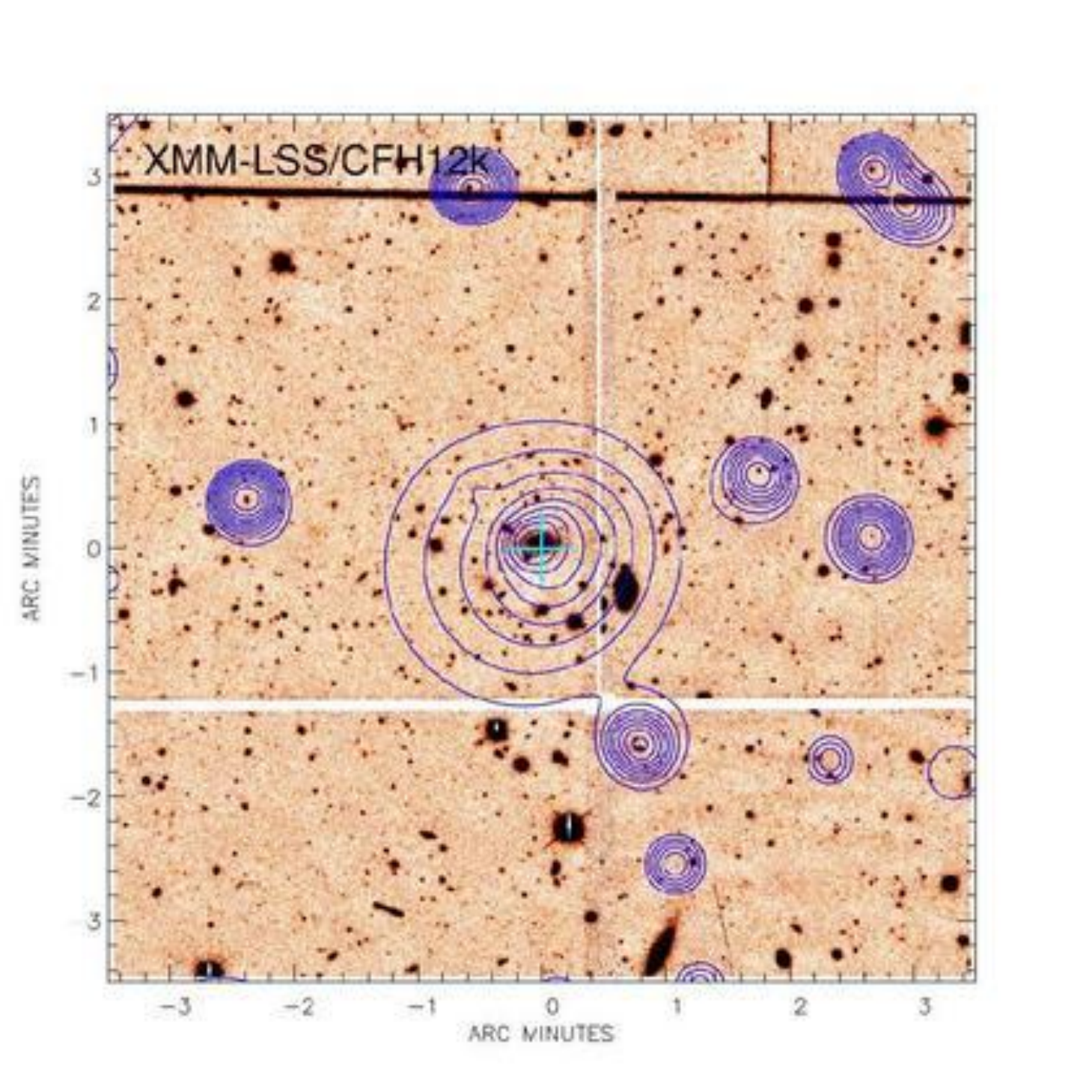}
    \includegraphics[viewport=0 0 420 450,clip,width=7.85cm]{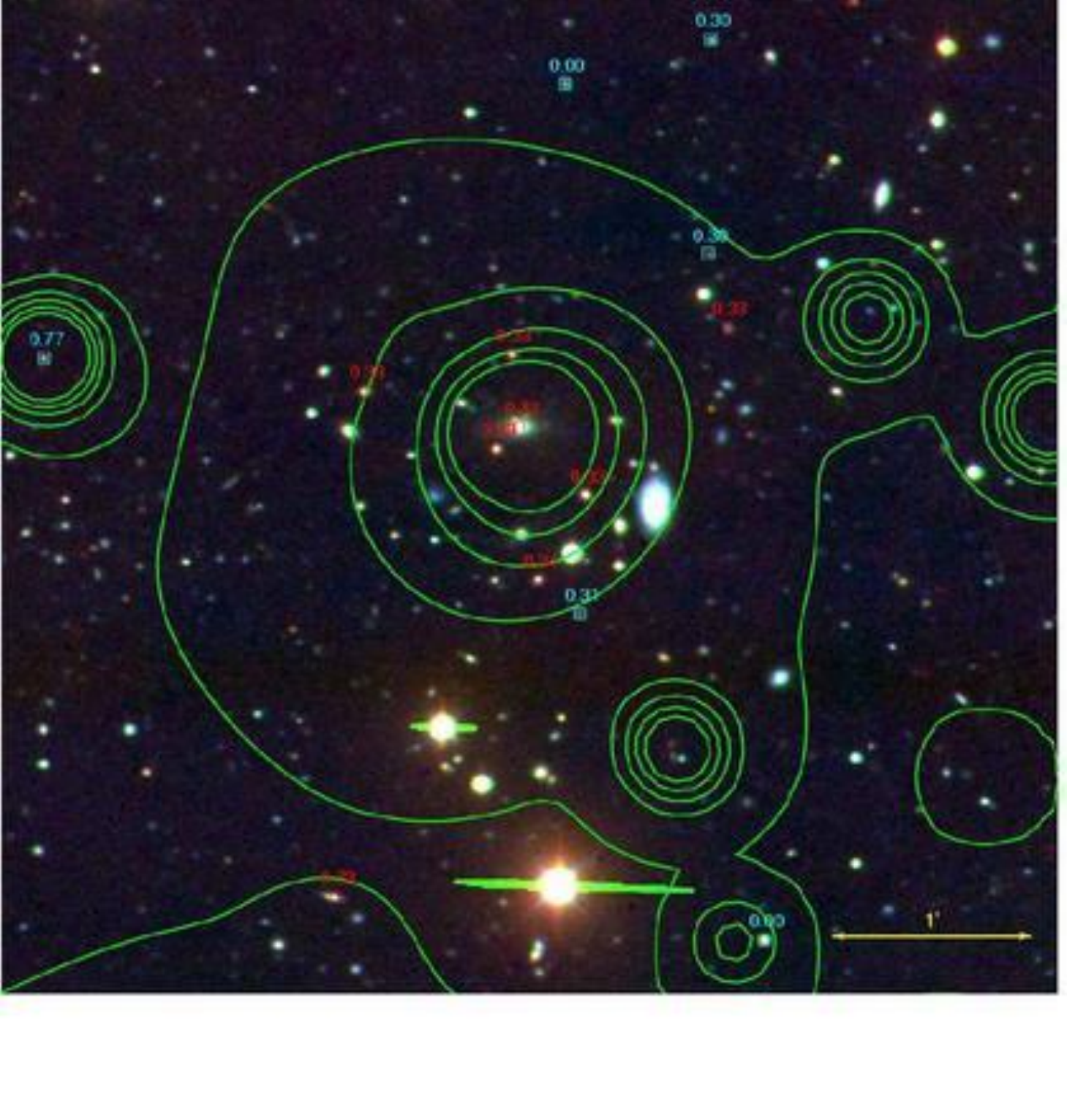}}
    \vbox{\includegraphics[width=9.2cm]{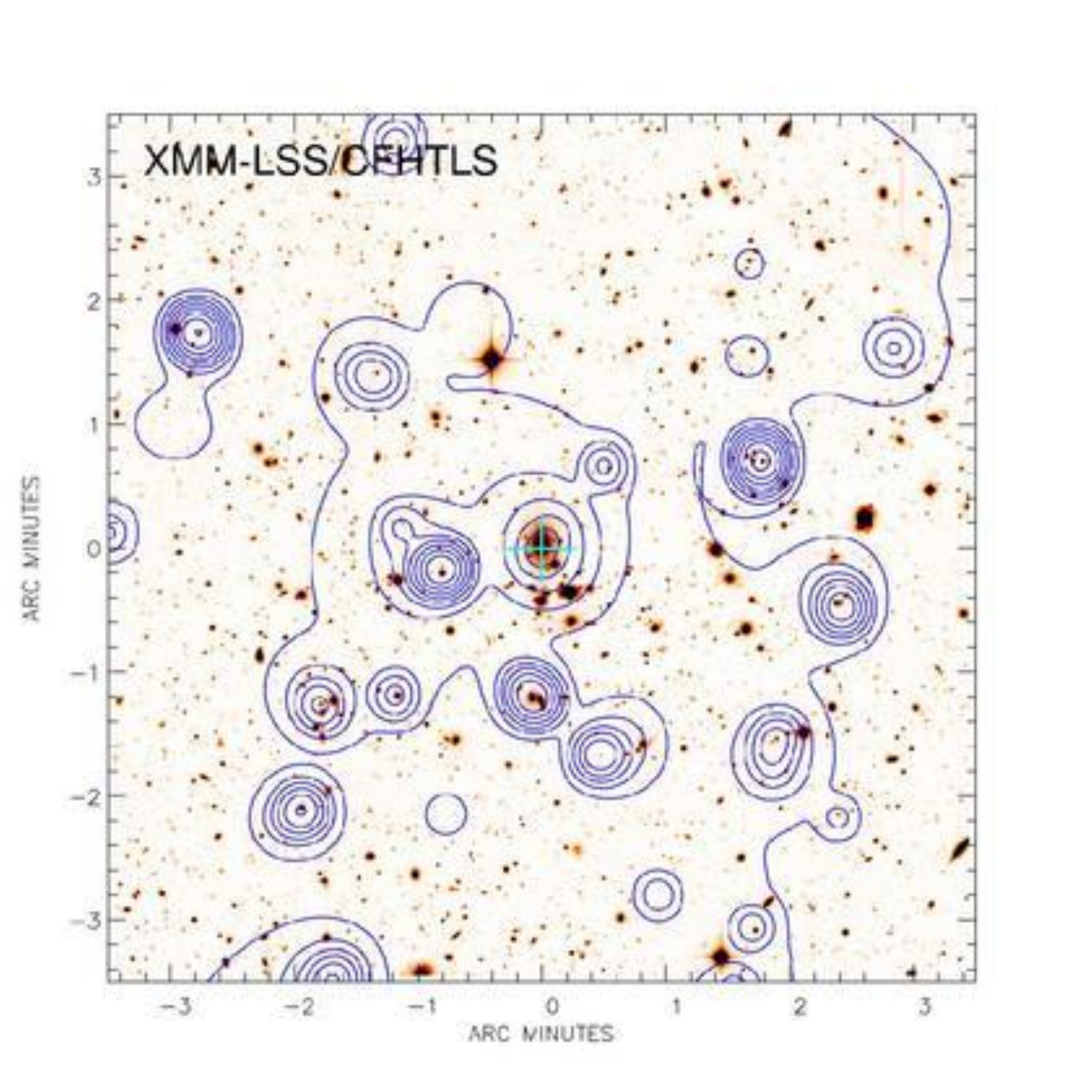}
    \includegraphics[viewport=0 0 420 450,clip,width=7.85cm]{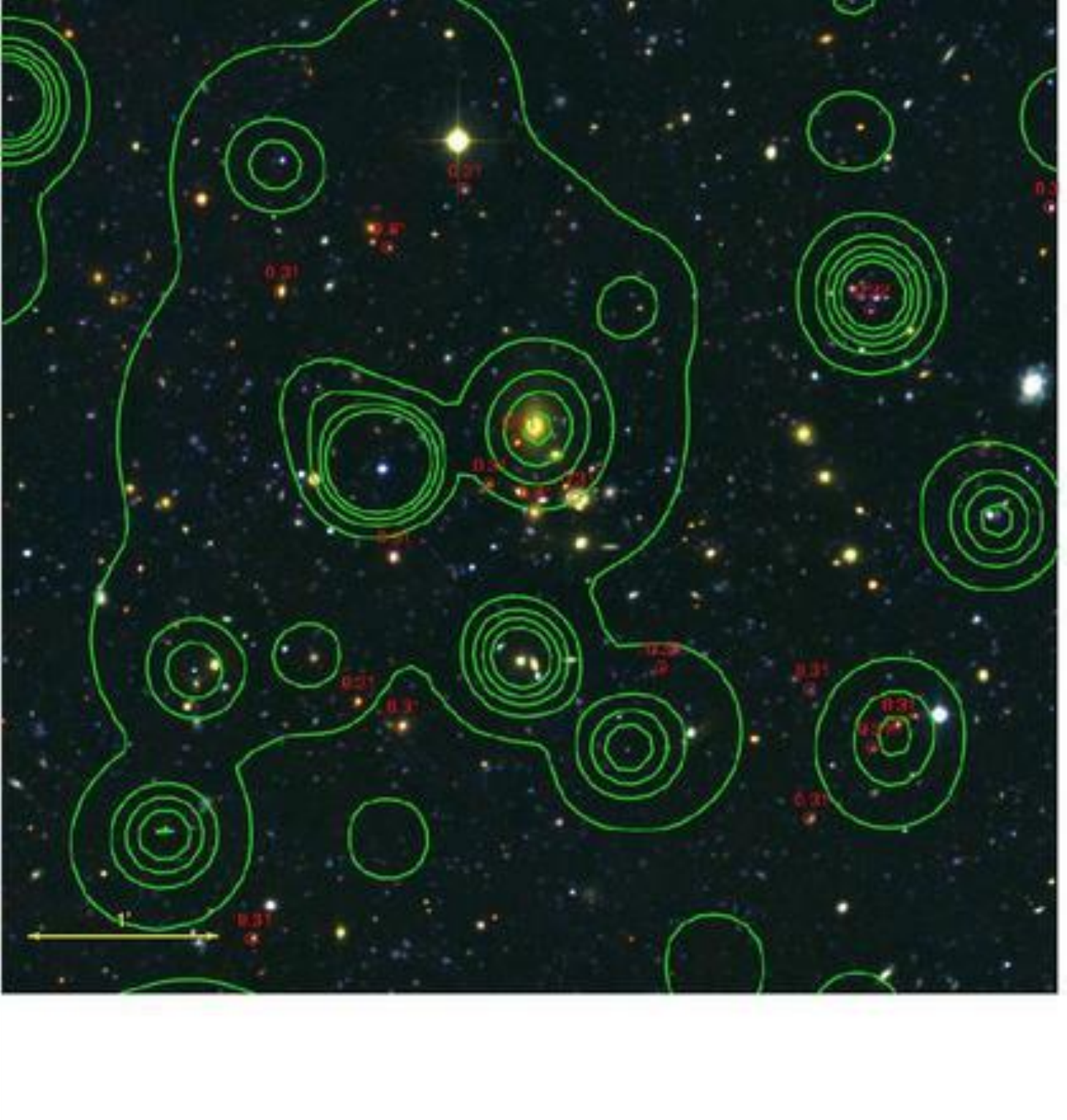}}}
  \rput(-8.6,13.7){\large (a)}
  \rput(-8.6,4.8){\large (b)}
  \end{pspicture}
 \contcaption{Images of the C1 clusters. (a) XLSSC-010. (b): XLSSC-013.}
\end{figure*}

\begin{figure*}
  \begin{pspicture}(0,0)(17.8,18.4)
  \vbox{
    \vbox{\includegraphics[width=9.2cm]{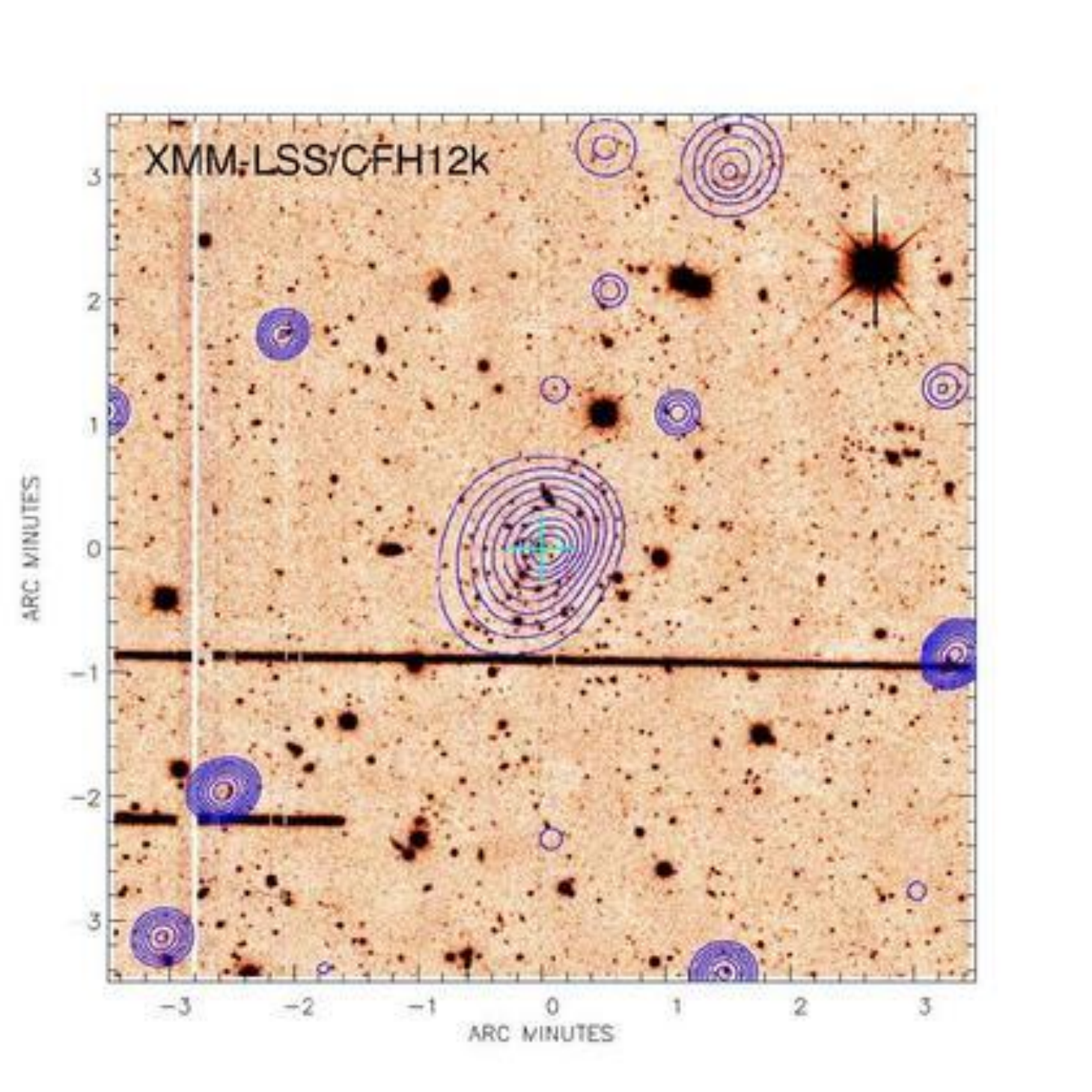}
    \includegraphics[viewport=0 0 420 450,clip,width=7.85cm]{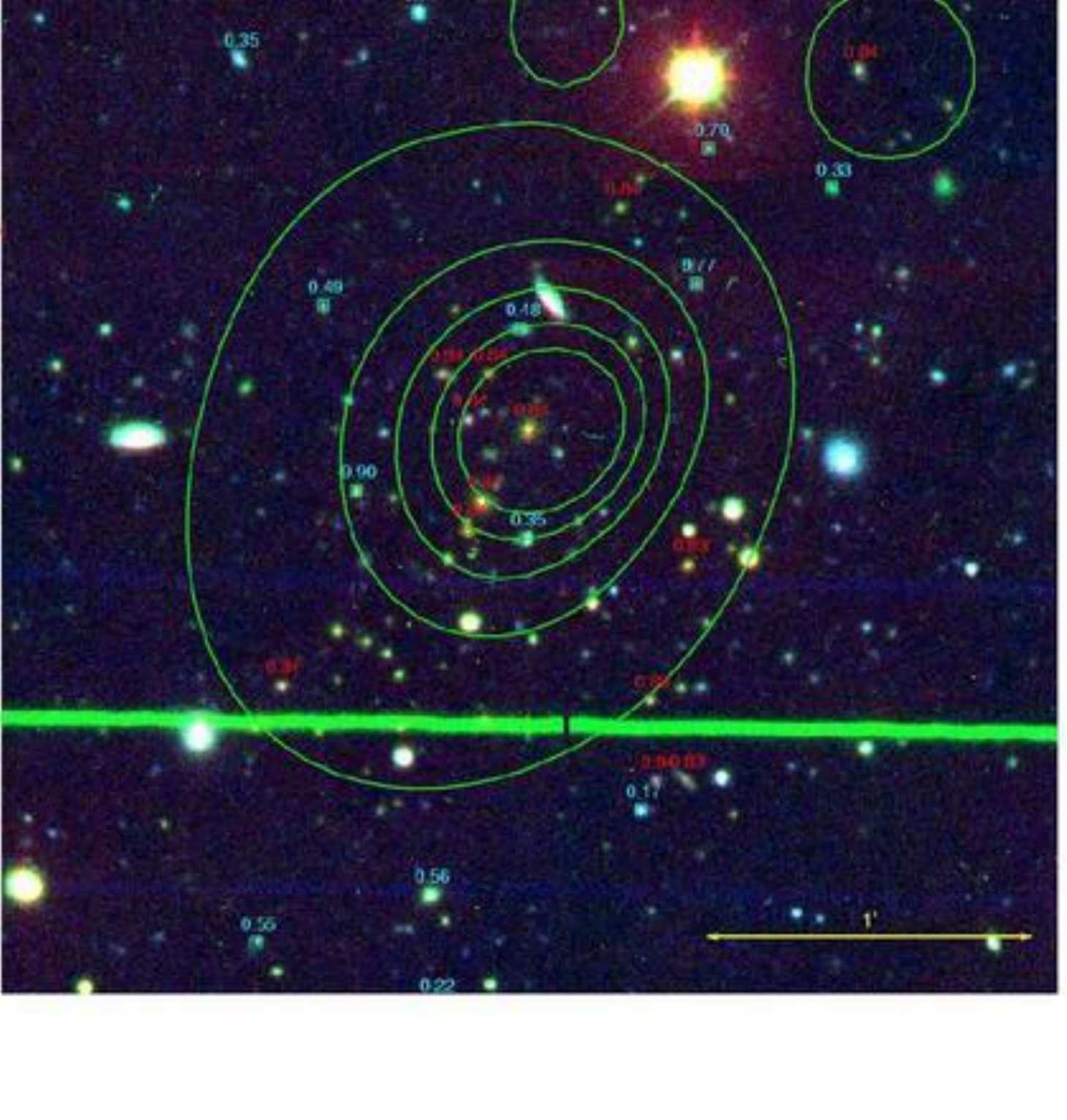}}
    \vbox{\includegraphics[height=9.2cm]{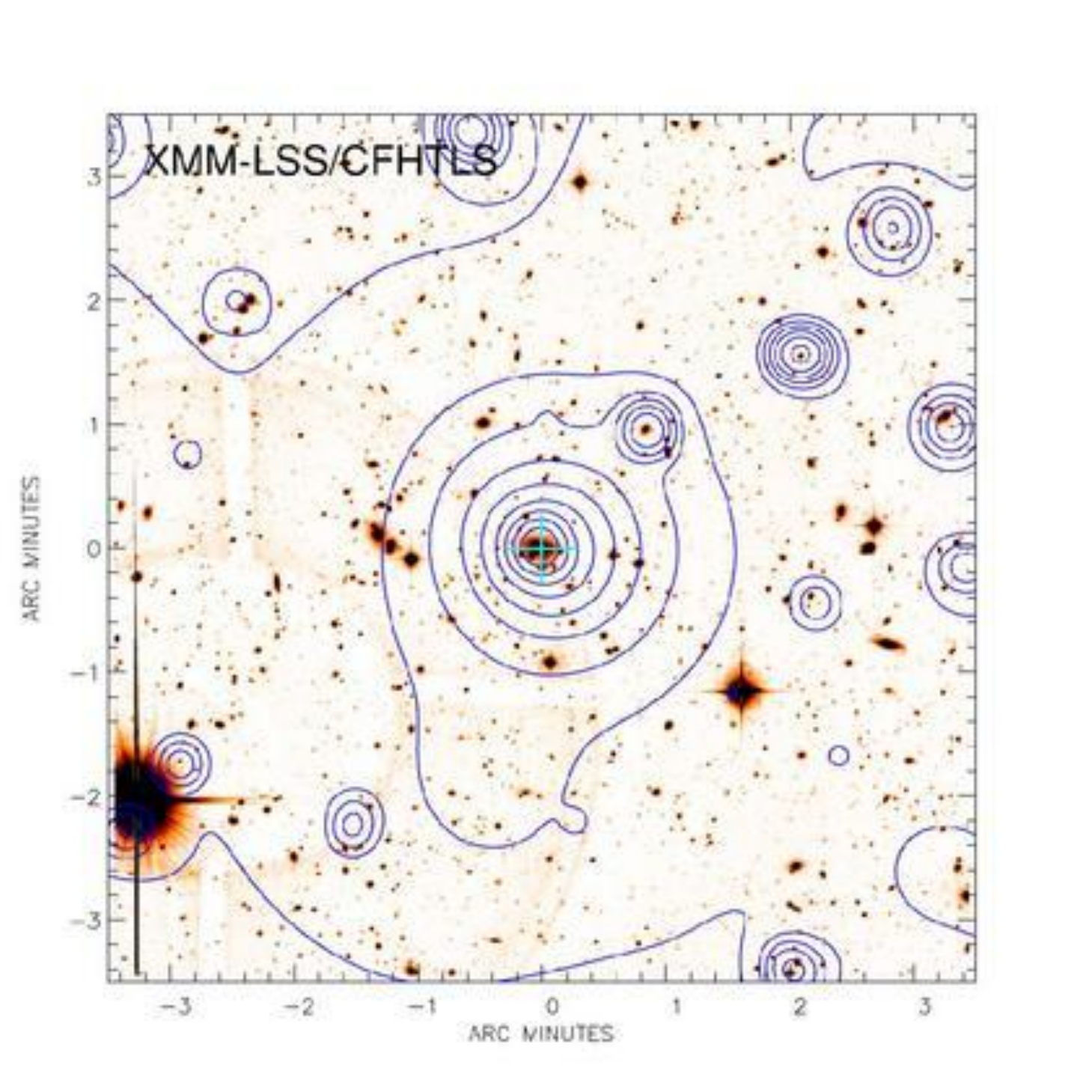}
    \includegraphics[viewport=0 0 420 450,clip,width=7.85cm]{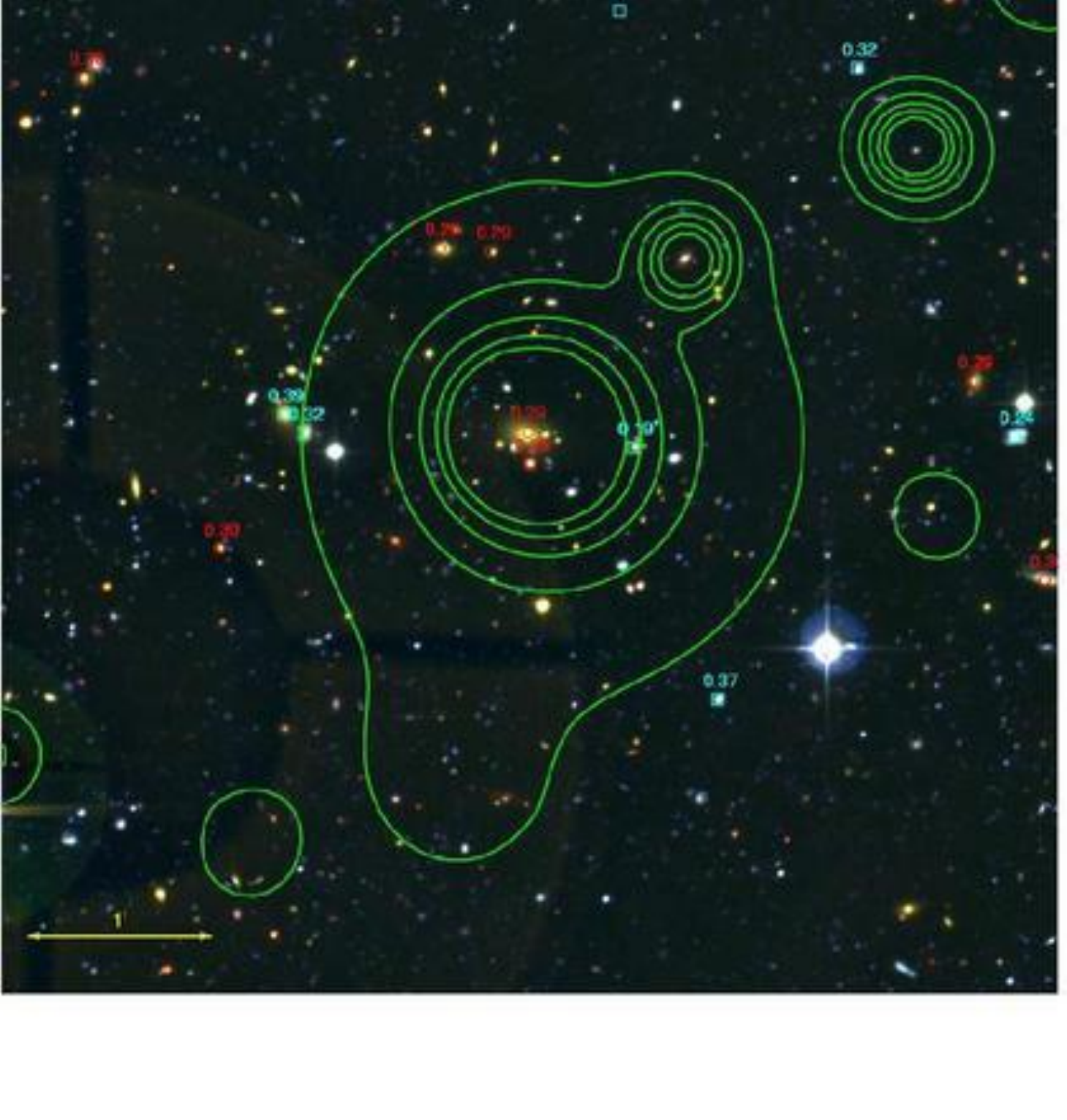}}}
  \rput(-8.6,13.7){\large (a)}
  \rput(-8.6,4.8){\large (b)}
  \end{pspicture}
 \contcaption{Images of the C1 clusters. (a) XLSSC-003. (b): XLSSC-022.}
\end{figure*}

\begin{figure*}
  \begin{pspicture}(0,0)(17.8,9.2)
    \vbox{\includegraphics[height=9.2cm]{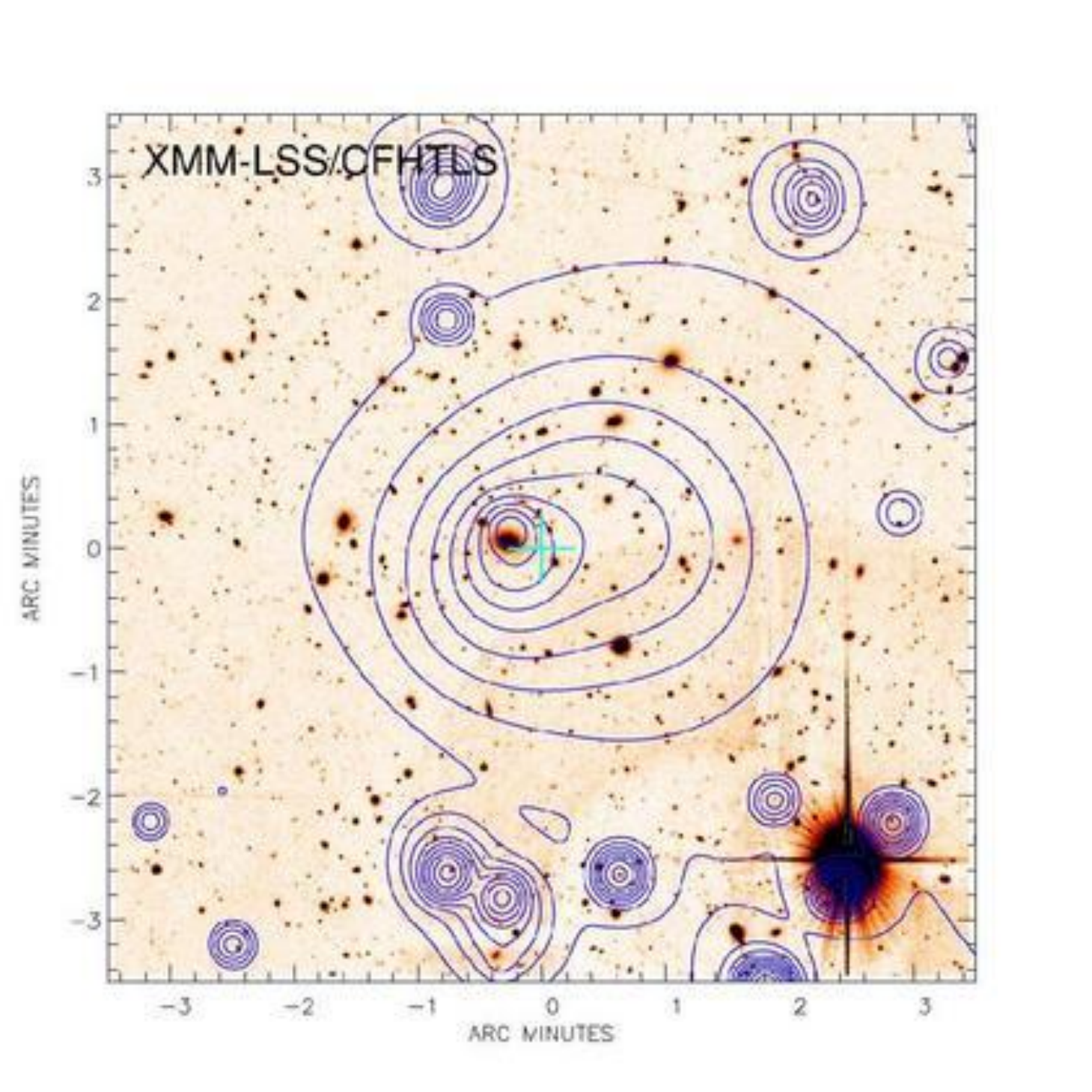}
    \includegraphics[viewport=0 0 420 450,clip,width=7.85cm]{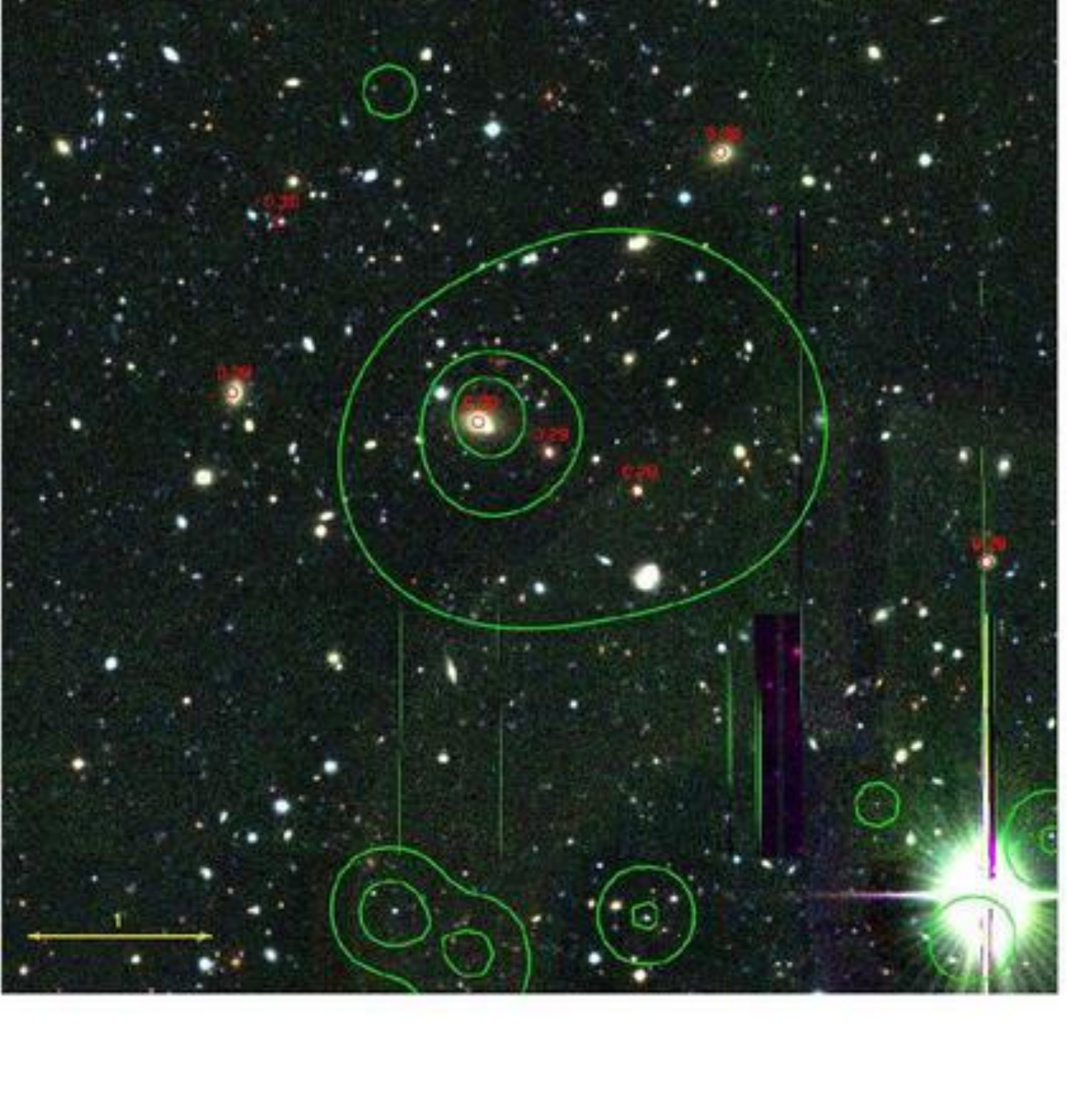}}
  \end{pspicture}
 \contcaption{Images of the C1 clusters: XLSSC-027.}
\end{figure*}

\stepcounter{section}

\begin{figure*}
  \begin{pspicture}(0,0)(17.8,9.2)
    \vbox{\hspace{-0.1cm}\includegraphics[height=9.2cm]{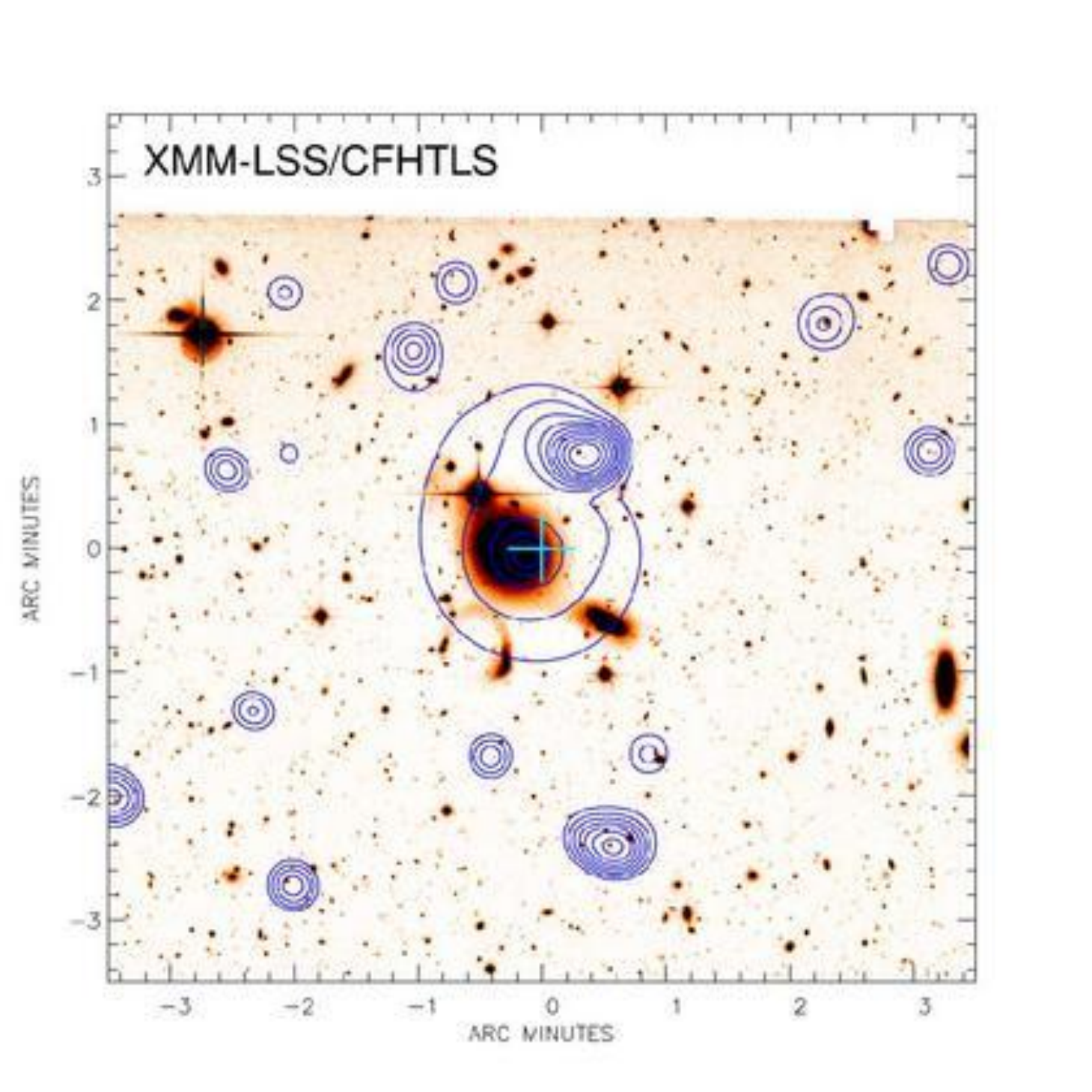}
    \hspace{-0.4cm}\includegraphics[height=8.2cm]{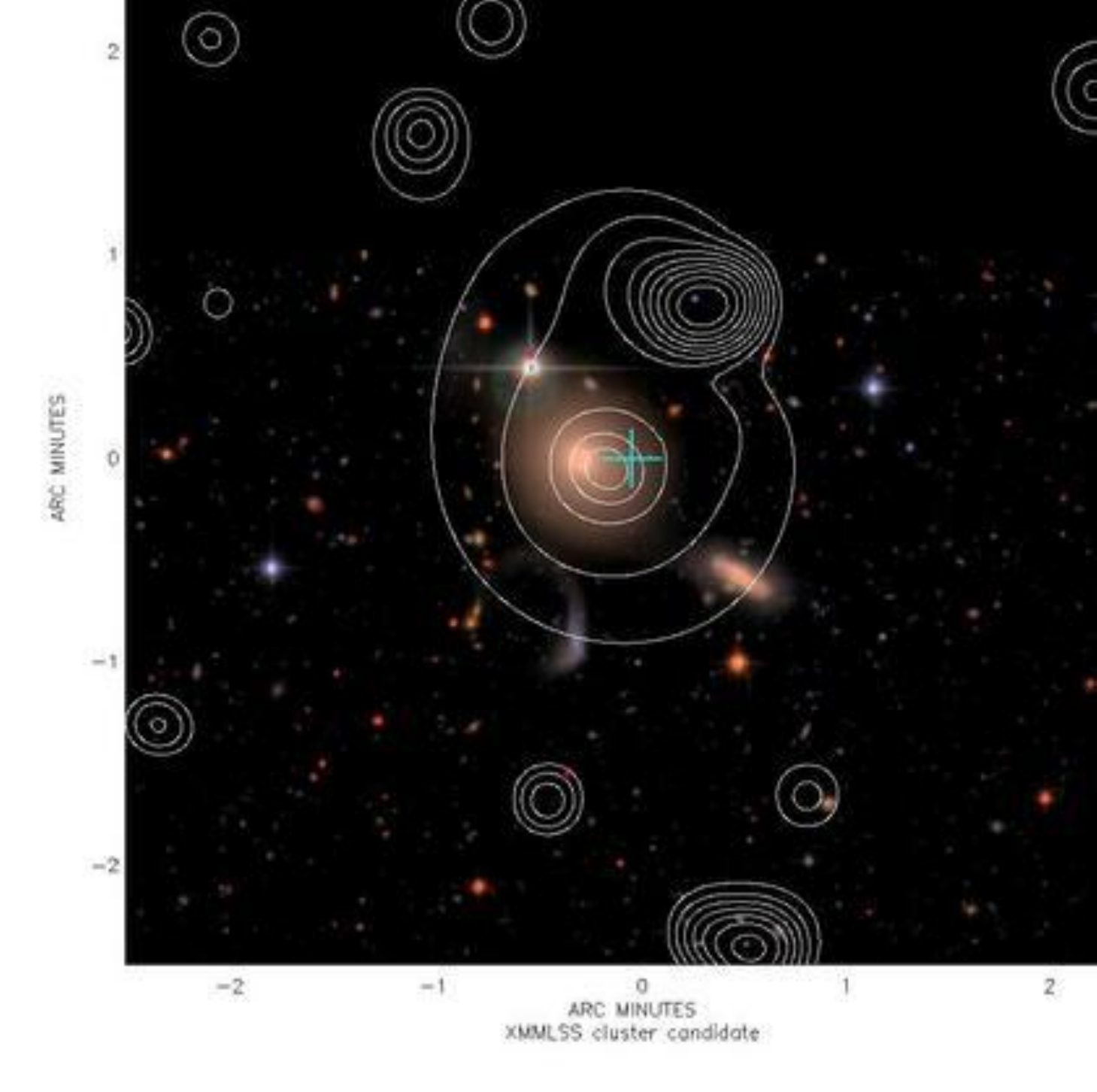}}
  \end{pspicture}
 \caption{Images of the C1 nearby galaxies: source XLSS J022528.7-040041. 
 Left: 7\arcmin wide I band image overlayed with X-ray contours. Right: 5\arcmin 
 wide true color image (g,r,i) overlayed with X-ray contours.\label{C1img}}
\end{figure*}

\begin{figure*}
  \begin{pspicture}(0,0)(17.8,18.4)
  \vbox{
    \vbox{\hspace{-.1cm}\includegraphics[height=9.2cm]{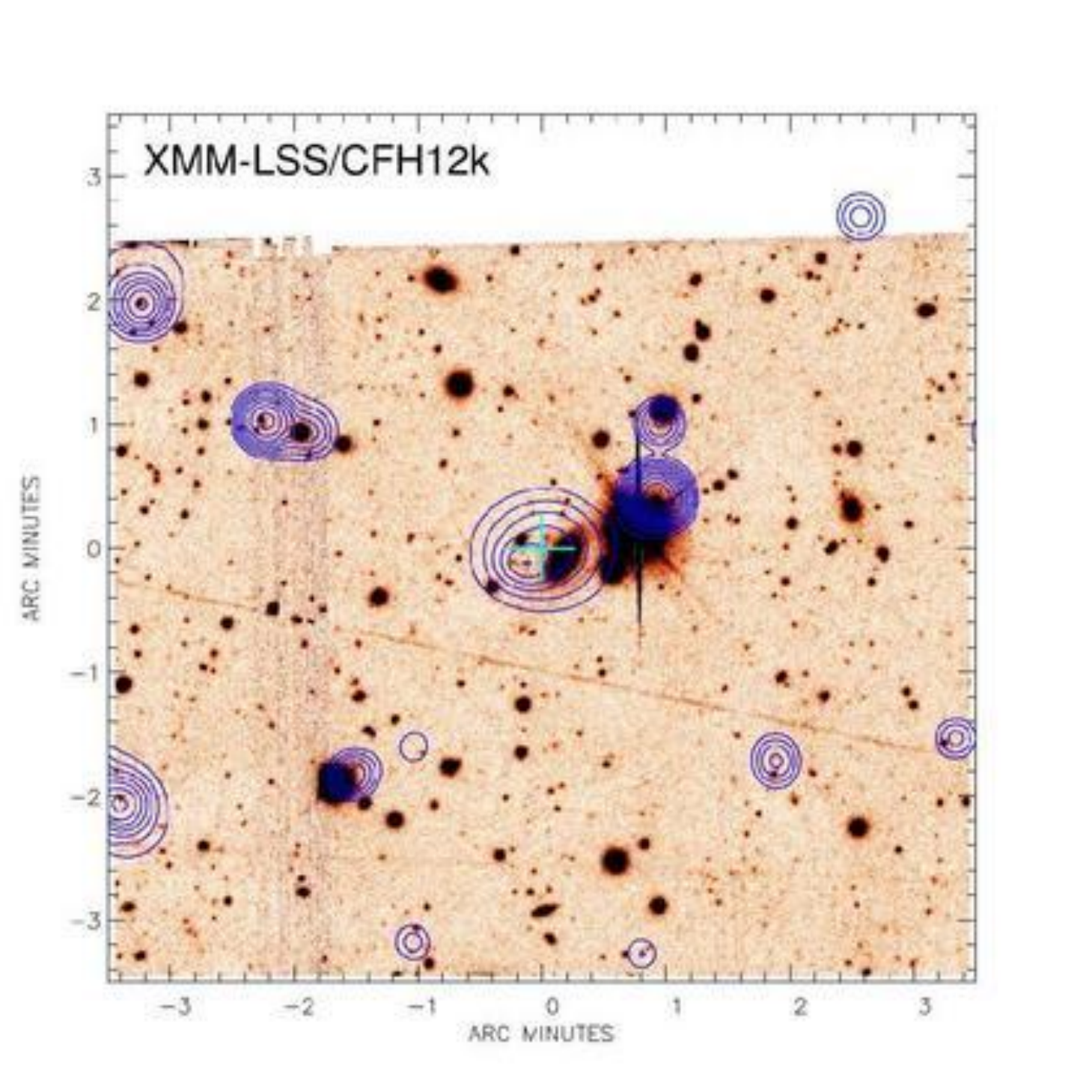}
    \hspace{-0.4cm}\includegraphics[height=8.2cm]{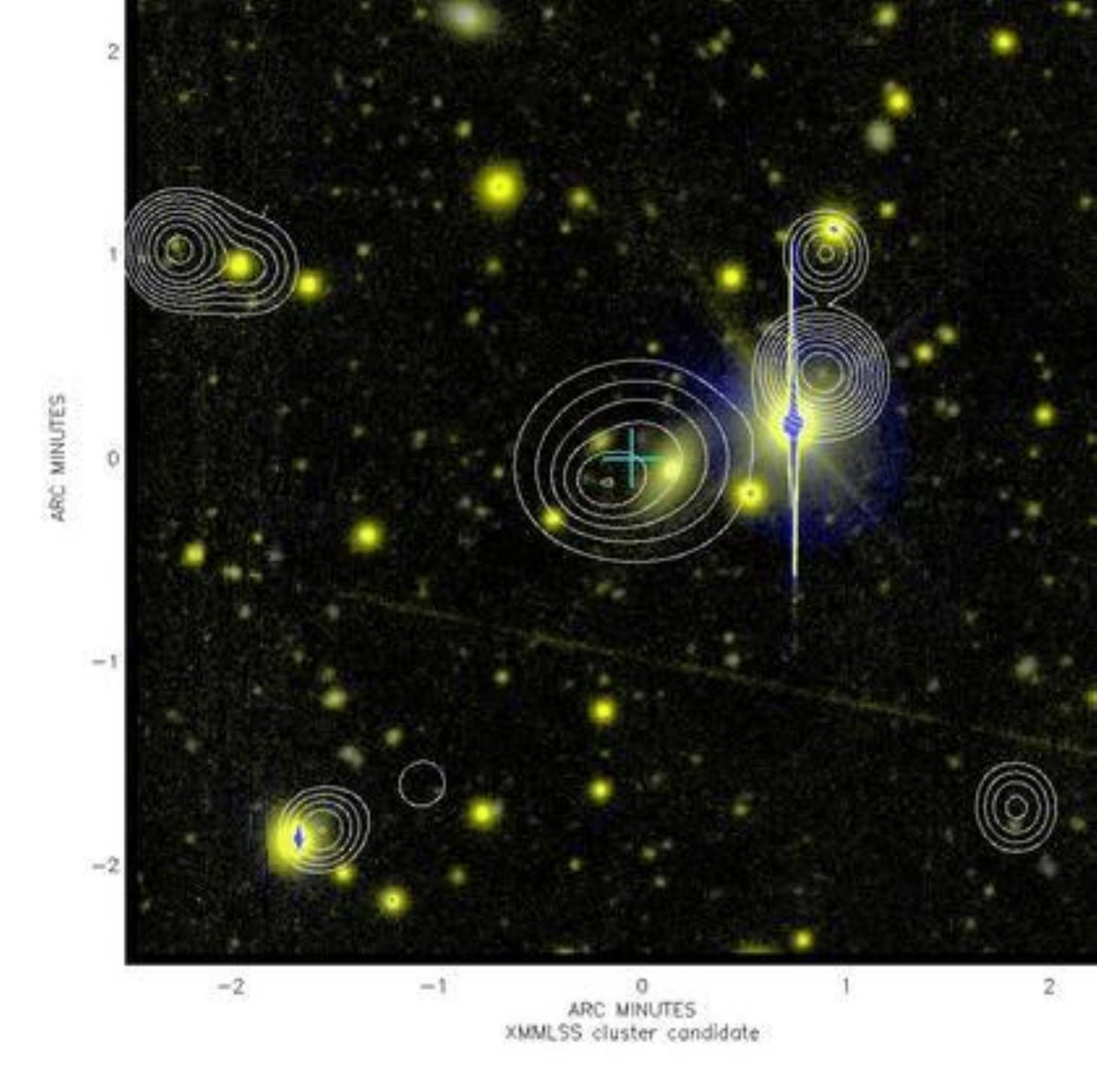}}
    \vbox{\hspace{-.1cm}\includegraphics[height=9.2cm]{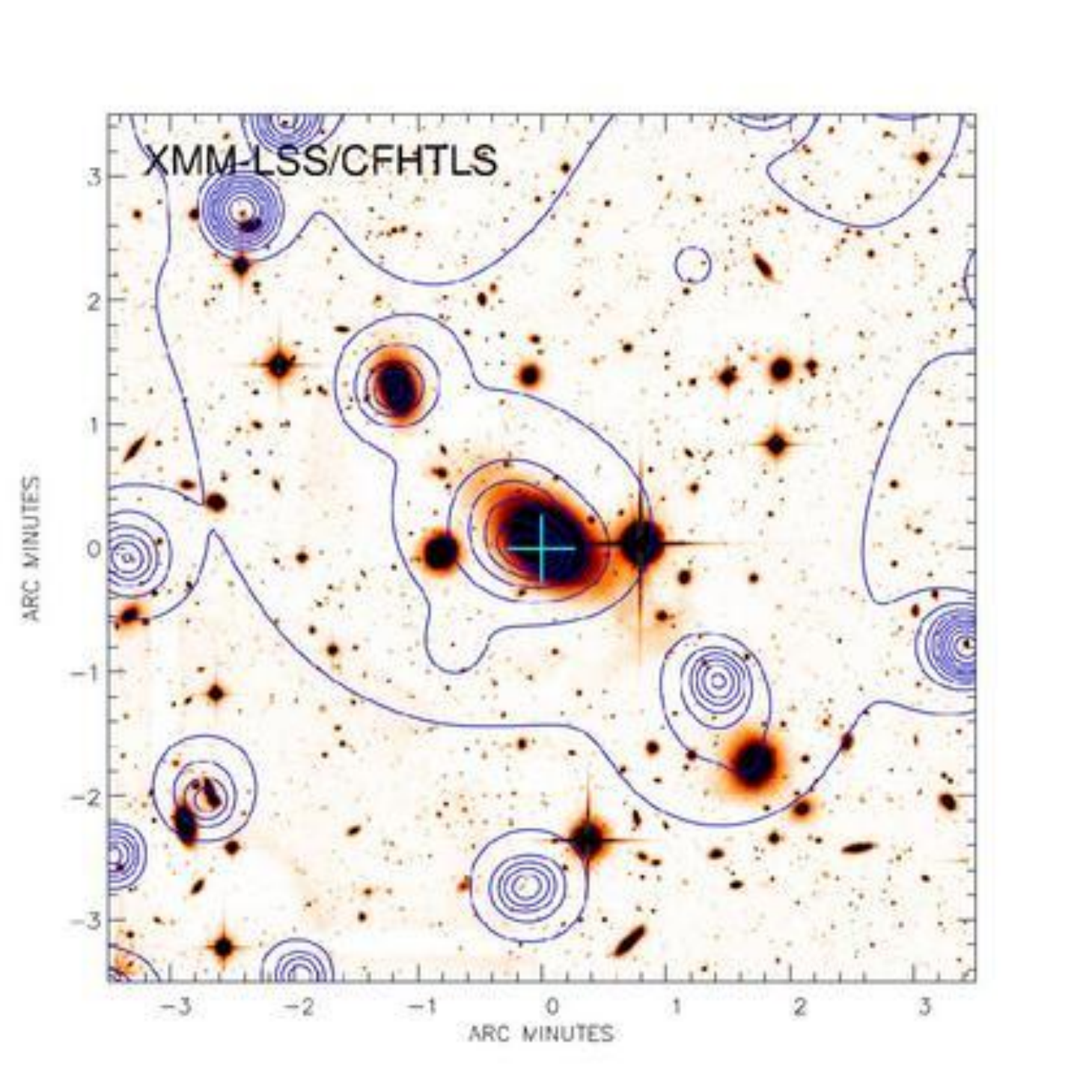}
    \hspace{-0.4cm}\includegraphics[height=8.2cm]{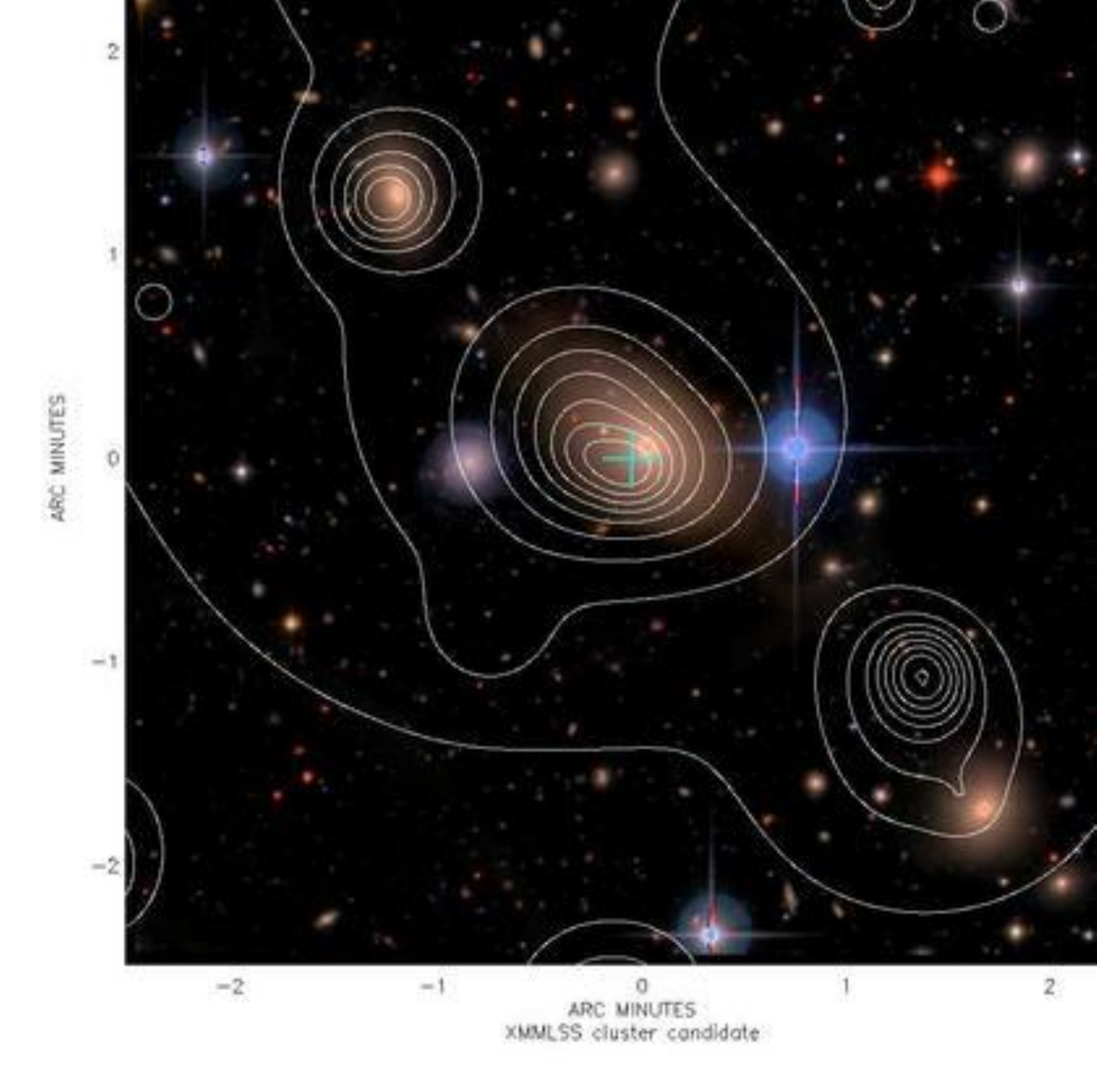}}}
  \rput(-8.65,13.7){\large (a)}
  \rput(-8.65,4.8){\large (b)}
  \end{pspicture}
 \contcaption{Images of the C1 nearby galaxies. (a) XLSS J022251.4-031151. (b): XLSS J022659.2-043529.}
\end{figure*}

\begin{figure*}
  \begin{pspicture}(0,0)(17.8,18.4)
  \vbox{
    \vbox{\hspace{-.1cm}\includegraphics[height=9.2cm]{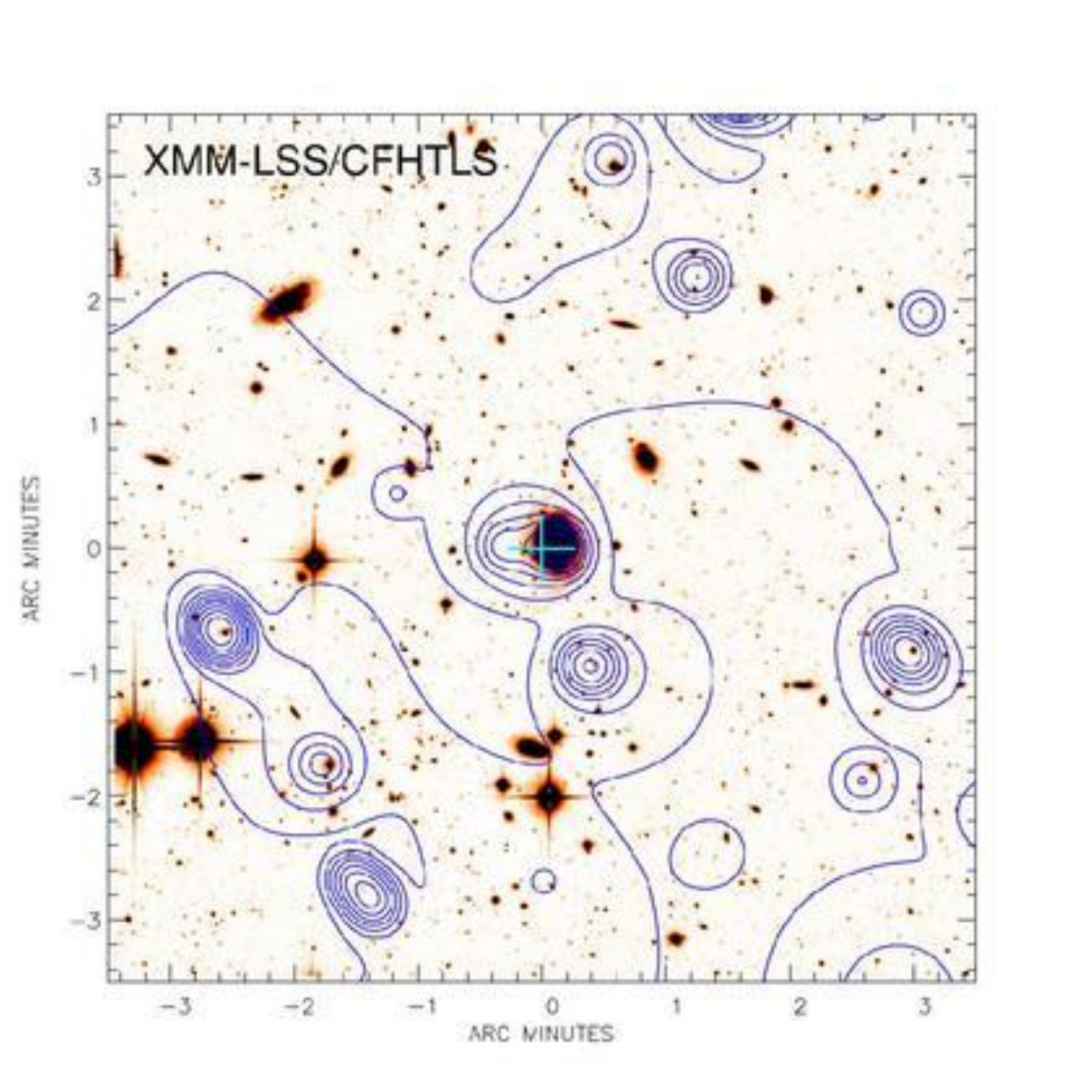}
    \hspace{-0.4cm}\includegraphics[height=8.2cm]{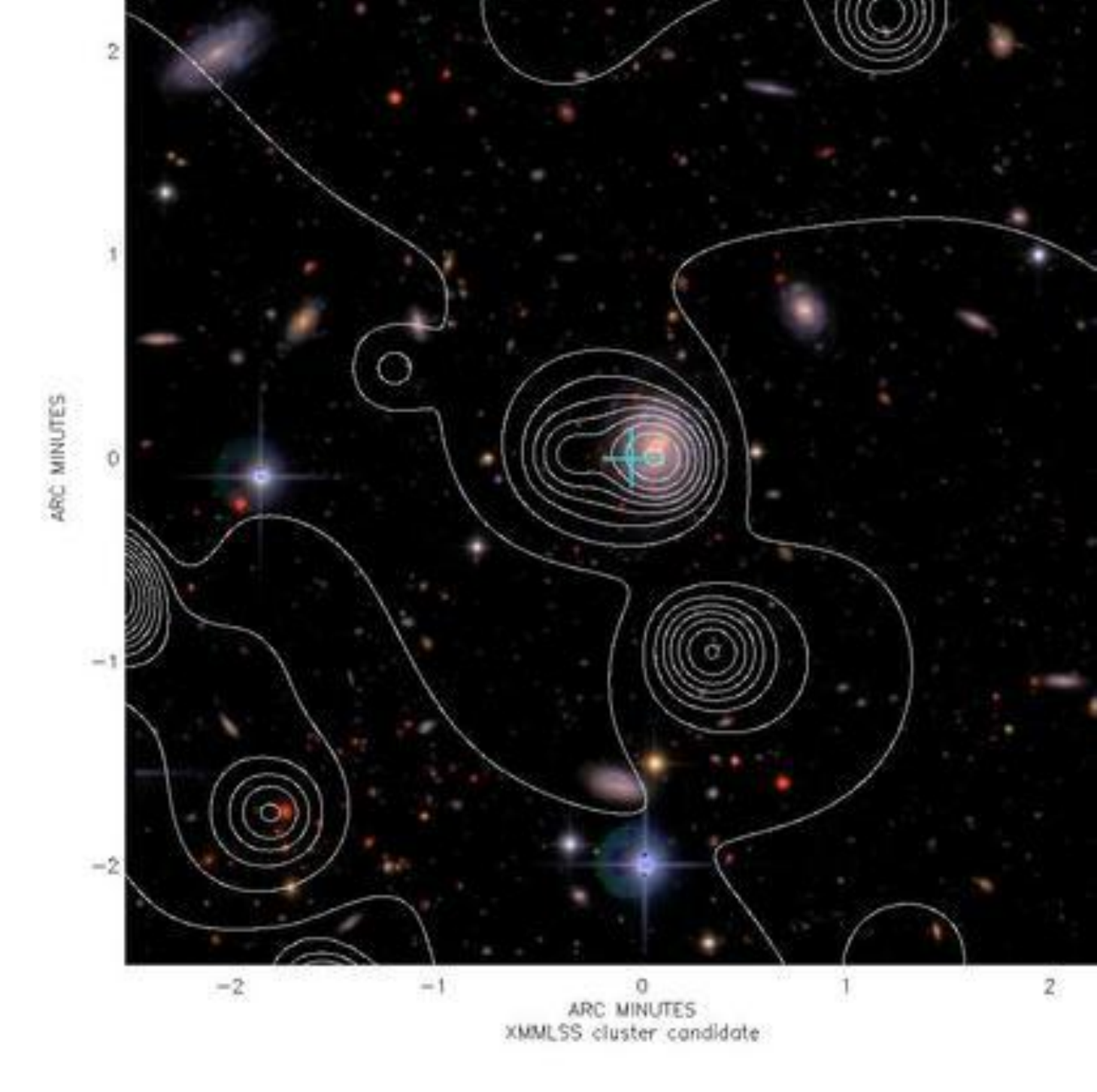}}
    \vbox{\hspace{-.1cm}\includegraphics[height=9.2cm]{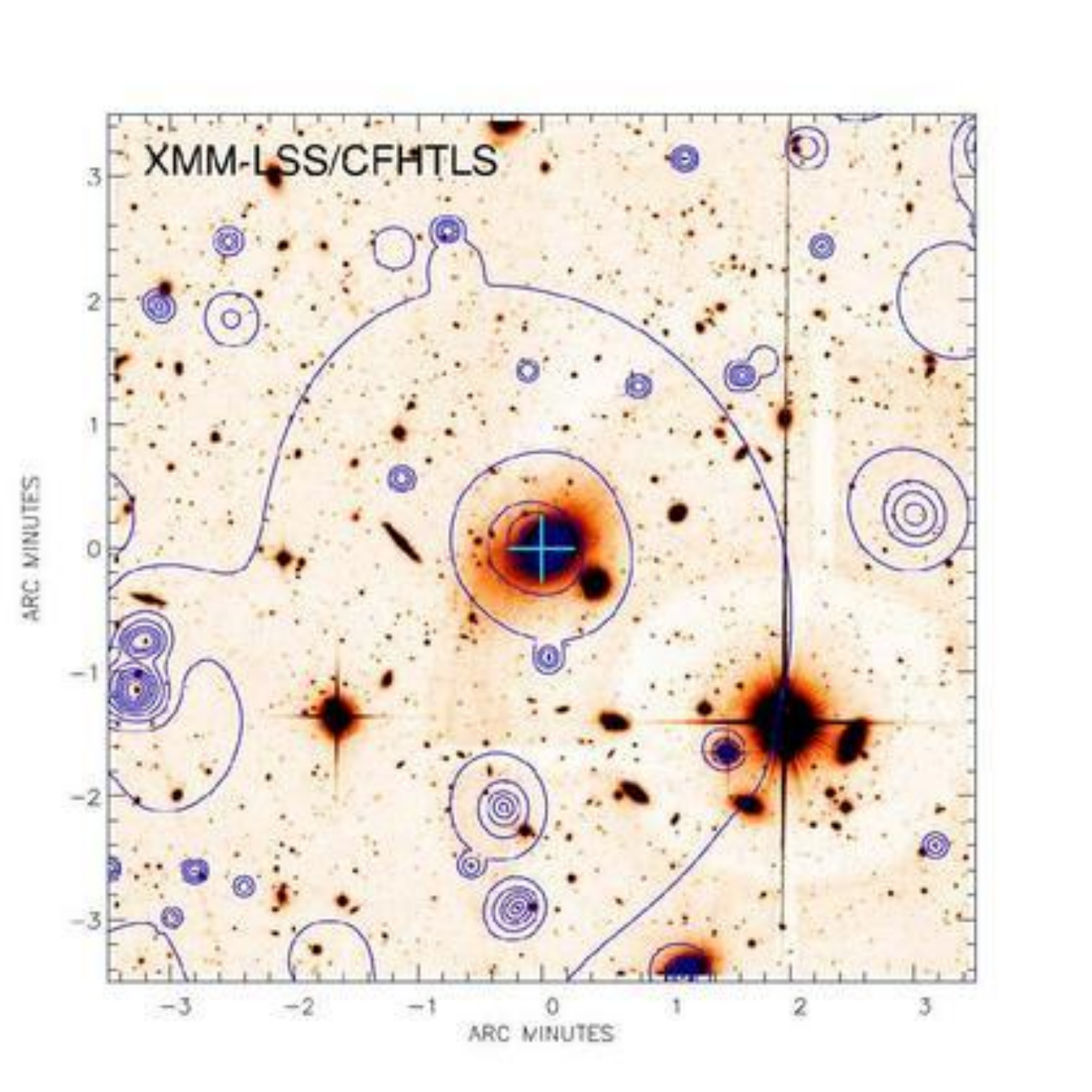}
    \hspace{-0.4cm}\includegraphics[height=8.2cm]{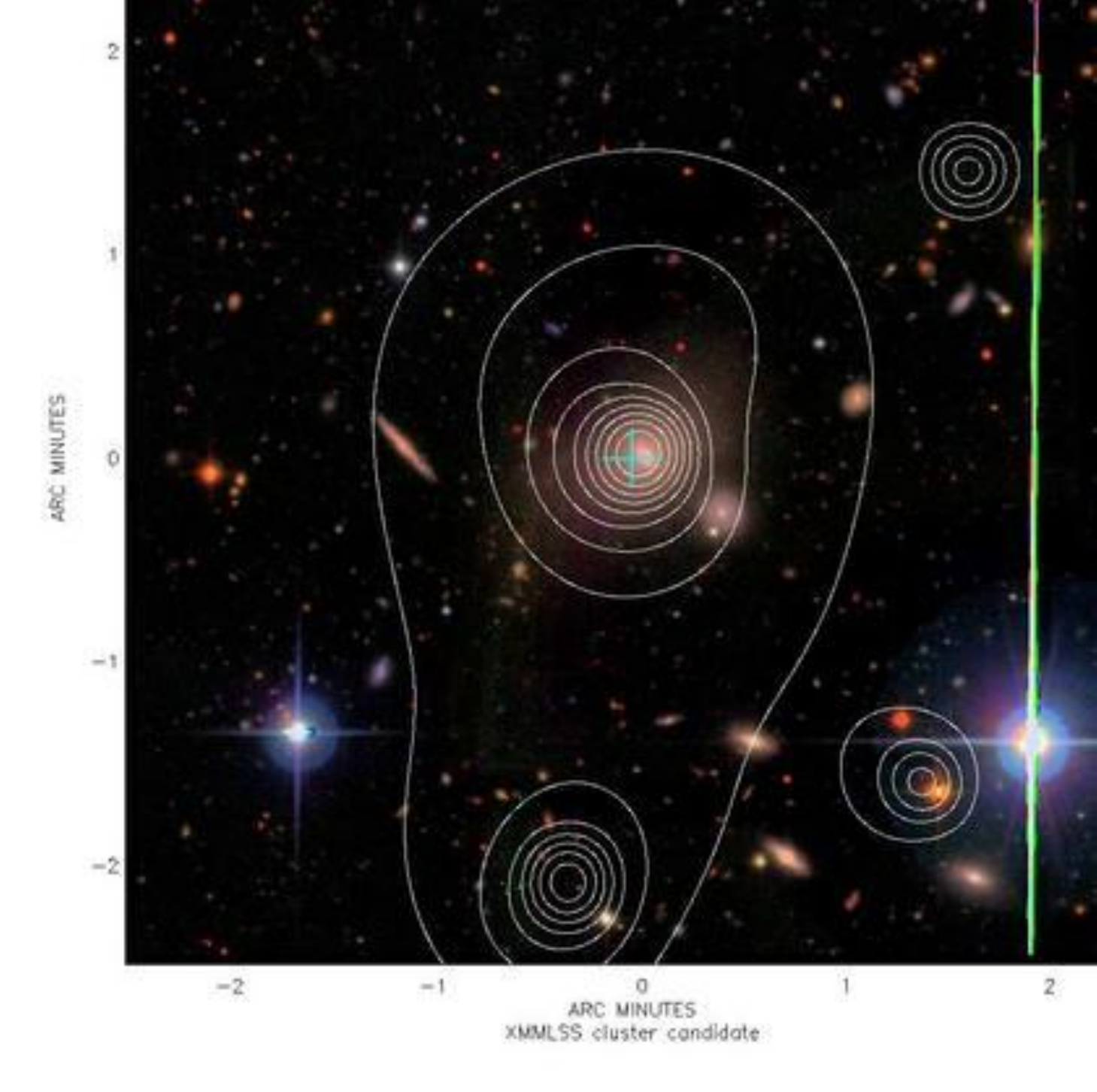}}}
  \rput(-8.65,13.7){\large (a)}
  \rput(-8.65,4.8){\large (b)}
  \end{pspicture}
 \contcaption{Images of the C1 nearby galaxies. (a) XLSS J022430.4-043617. (b): XLSS J022617.6-050443.}
 \label{lastpage}
\end{figure*}

\end{document}